\documentclass[preprint]{aastex61}

\usepackage{CJK} 
\usepackage{amsmath}
\usepackage{amssymb}
\usepackage{epsfig}
\usepackage{apjfonts}
\usepackage{natbib}
\usepackage{hyperref}
\usepackage{epstopdf}
\usepackage{verbatim}
\usepackage{enumitem}
\usepackage{color}
\setlist[enumerate]{itemsep=-1mm}
\usepackage{soul, xcolor}
\setstcolor{blue}
\usepackage{mathtools}
\usepackage{bigints}
\usepackage{tikz}

\makeatletter
\global\let\tikz@ensure@dollar@catcode=\relax
\makeatother

\shorttitle{Young Brown Dwarfs in Taurus And a Reddening-Free Classification}
\shortauthors{Zhang et al}
\accepted{by ApJ, 2018 Feb 20}

\begin{document}

\begin{CJK*}{UTF8}{gbsn}
% ------------------ Title ------------------
\title{The Pan-STARRS1 Proper-Motion Survey for Young Brown Dwarfs in Nearby Star-Forming Regions. I. Taurus Discoveries and A Reddening-Free Classification Method for Ultracool Dwarfs}

% ------------------ Authors ------------------
\author{Zhoujian Zhang (张周健)}
\affiliation{Institute for Astronomy, University of Hawaii at Manoa, Honolulu, HI 96822, USA}
\affiliation{Visiting Astronomer at the Infrared Telescope Facility, which is operated by the University of Hawaii under contract NNH14CK55B with the National Aeronautics and Space Administration.}

\author{Michael C. Liu}
\affiliation{Institute for Astronomy, University of Hawaii at Manoa, Honolulu, HI 96822, USA}
\affiliation{Visiting Astronomer at the Infrared Telescope Facility, which is operated by the University of Hawaii under contract NNH14CK55B with the National Aeronautics and Space Administration.}

\author{William M. J. Best}
\affiliation{Institute for Astronomy, University of Hawaii at Manoa, Honolulu, HI 96822, USA}
\affiliation{Visiting Astronomer at the Infrared Telescope Facility, which is operated by the University of Hawaii under contract NNH14CK55B with the National Aeronautics and Space Administration.}

\author{Eugene A. Magnier}
\affiliation{Institute for Astronomy, University of Hawaii at Manoa, Honolulu, HI 96822, USA}

\author{Kimberly M. Aller}
\affiliation{Institute for Astronomy, University of Hawaii at Manoa, Honolulu, HI 96822, USA}
\affiliation{Visiting Astronomer at the Infrared Telescope Facility, which is operated by the University of Hawaii under contract NNH14CK55B with the National Aeronautics and Space Administration.}

\author{K. C. Chambers}
\affiliation{Institute for Astronomy, University of Hawaii at Manoa, Honolulu, HI 96822, USA}

\author{P. W. Draper}
\affiliation{Department of Physics, Durham University, South Road, Durham DH1 3LE, UK}

\author{H. Flewelling}
\affiliation{Institute for Astronomy, University of Hawaii at Manoa, Honolulu, HI 96822, USA}

\author{K. W. Hodapp}
\affiliation{Institute for Astronomy, University of Hawaii at Manoa, Honolulu, HI 96822, USA}

\author{N. Kaiser}
\affiliation{Institute for Astronomy, University of Hawaii at Manoa, Honolulu, HI 96822, USA}

\author{R.-P. Kudritzki}
\affiliation{Institute for Astronomy, University of Hawaii at Manoa, Honolulu, HI 96822, USA}

\author{N. Metcalfe}
\affiliation{Department of Physics, Durham University, South Road, Durham DH1 3LE, UK}

\author{R. J. Wainscoat}
\affiliation{Institute for Astronomy, University of Hawaii at Manoa, Honolulu, HI 96822, USA}

\author{C. Waters}
\affiliation{Institute for Astronomy, University of Hawaii at Manoa, Honolulu, HI 96822, USA}

% ------------------ Abstract ------------------
\begin{abstract}
We are conducting a proper-motion survey for young brown dwarfs in the Taurus--Auriga molecular cloud based on the Pan-STARRS1 3$\pi$ Survey. Our search uses multi-band photometry and astrometry to select candidates, and is wider (370~deg$^{2}$) and deeper (down to $\approx$3~M$_{\rm Jup}$) than previous searches. We present here our search methods and spectroscopic follow-up of our high-priority candidates. Since extinction complicates spectral classification, we have developed a new approach using low-resolution ($R \approx 100$) near-infrared spectra to quantify reddening-free spectral types, extinctions, and gravity classifications for mid-M to late-L ultracool dwarfs ($\approx 100-3$~M$_{\rm Jup}$ in Taurus). We have discovered 25 low-gravity (\textsc{vl-g}) and the first 11 intermediate-gravity (\textsc{int-g}) substellar (M6--L1) members of Taurus, constituting the largest single increase of Taurus brown dwarfs to date. We have also discovered 1 new Pleiades member and 13 new members of the Perseus OB2 association, including a candidate very wide separation (58~kAU) binary. We homogeneously reclassify the spectral types and extinctions of all previously known Taurus brown dwarfs. Altogether our discoveries have thus far increased the substellar census in Taurus by $\approx 40\%$ and added three more L-type members ($\approx 5-10$~M$_{\rm Jup}$). Most notably, our discoveries reveal an older ($>$10~Myr) low-mass population in Taurus, in accord with recent studies of the higher-mass stellar members. The mass function appears to differ between the younger and older Taurus populations, possibly due to incompleteness of the older stellar members or different star formation processes.
\end{abstract}

% ----------------- start of body ------------------------------------------

% ----------------------- Section -----------------------
\section{Introduction}
\label{sec:introduction}

Thanks to extensive surveys of open clusters, star-forming regions, and the solar neighborhood over the past two decades, large samples of brown dwarfs have been discovered, which bridge low-mass stars \citep[$\gtrsim 70$~M$_{\rm Jup}$, set by the hydrogen-burning limit;][]{Dupuy+2017} and giant planets \citep[$\lesssim 13$~M$_{\rm Jup}$, set by the deuterium-burning limit; e.g.,][]{Spiegel+2011}. As the low-mass end of the initial mass function (IMF hereafter), substellar objects are essential to answering fundamental questions about star formation: What is the lowest-mass star that can form? Is there a universal IMF? Early-L dwarfs with effective temperatures of $\lesssim 2100$~K and ages of $\lesssim 10$~Myr have masses of $\lesssim 13$~M$_{\rm Jup}$ \citep[based on the DUSTY evolutionary models by][]{Chabrier+2000}, firmly in the planetary-mass regime. Therefore, young brown dwarfs are also of interest to understanding the formation mechanisms of gas-giant planets.

Nearby star-forming regions are ideal for discovering young brown dwarfs and free-floating planets, since their substellar and planetary-mass members are bright enough to be easily detected by current all-sky surveys. Located at the distance of $\approx 130-160$~pc \citep{Zeeuw+1999}, the Taurus-Auriga molecular cloud (Taurus hereafter) is one of the closest laboratories in the solar neighborhood. Taurus has regions of high extinction ($A_{\rm V}$ $\lesssim 30$~mag), as the natal molecular cloud has not fully dispersed since its most recent star formation activity around $\lesssim 10$~Myr ago \citep[e.g.,][]{Kraus+2009}. Although a large fraction of objects in Taurus are believed to have young ages ($\lesssim 10$~Myr) based on spectroscopic analysis and their positions on the HR diagram, recent studies of lithium abundances, disk fractions, and kinematic distributions of Taurus members have found the existence of a dispersed older population of stars \citep[$\approx 10-40$~Myr old; e.g.,][]{Sestito+2008, Daemgen+2015, Kraus+2017}. Constructing a complete sample of substellar and planetary-mass members and studying their age, spatial, and kinematic distributions are fundamental to comprehensively understanding the Taurus star formation history.

Extensive surveys in Taurus from X-ray to infrared wavelengths have discovered 76 substellar objects spanning spectral types of M6--L2 and masses of $\approx 70-6$~M$_{\rm Jup}$ \citep[e.g.,][]{Briceno+2002, Guieu+2006, Luhman+2006, Slesnick+2006, Luhman+2009, Luhman+2010, Luhman+2017, Quanz+2010, Rebull+2010, Rebull+2011, Esplin+2014, Best+2017}. The least massive objects discovered so far have masses of $\approx 6$~M$_{\rm Jup}$: one L dwarf by \cite{Luhman+2009} and two by \cite{Best+2017}. Some previous surveys searched Taurus for brown dwarfs over a large area but focused on disk-bearing objects, thus limited to relatively bright infrared emitters \citep[e.g.,][]{Luhman+2009, Luhman+2010, Rebull+2010, Rebull+2011, Esplin+2014}, while other searches have focused on deeper fields with fainter limiting magnitudes but narrower search regions \citep[e.g.,][]{Briceno+2002, Quanz+2010}. Therefore, the current substellar regime in Taurus is likely still incomplete, and a new brown dwarf survey is warranted, with both wider and deeper data.
  
We present initial results from a proper-motion search for young brown dwarfs and free-floating planets in Taurus based on the Pan-STARRS1 (PS1) $3\pi$ Survey \citep{Chambers+2016}. PS1 has mapped a large fraction of the sky (30,000~deg$^{2}$ with $\delta \geqslant -30^{\circ}$) and has obtained stacked images reaching down to $\approx 3$~M$_{\rm Jup}$ in Taurus,  allowing us for the first time to explore the entire Taurus region with such low mass sensitivity. In addition, a recent PS1 survey by \citeauthor{Best+2017} (\citeyear{Best+2017}; see also \citealt{Best+2015}) serendipitously discovered two planetary-mass ($\approx 6$~M$_{\rm Jup}$) members in Taurus, which were missed by previous surveys because those searches adopted too narrow search area, shallow depth, or conservative selection criteria \citep[Section~5 in][]{Best+2017}. This serendipity suggests that members with even lower masses could be found with a dedicated search based on PS1. In order to cover the full optical--near-infrared spectral energy distribution (SED) for substellar candidates, we also incorporate several other surveys, including {\it the Wide-Field Infrared Survey Explorer} \citep[{\it WISE};][]{Wright+2010}, the Two Micron All Sky Survey \citep[2MASS;][]{Skrutskie+2006}, and the UKIRT Infrared Deep Sky Survey Galactic Cluster Survey DR9 \citep[UGCS;][]{Lawrence+2007}.

The layout of this paper is as follows. We establish our selection criteria for substellar candidates based on photometry and astrometry in Section~\ref{sec:selection} and describe our preliminary spectroscopic follow-up in Section~\ref{sec:NIRspec}. In order to precisely measure intrinsic magnitudes, colors, and luminosities of our candidates, we develop a new classification method in Section~\ref{sec:rfSPT_extinction} to quantitatively determine reddening-free spectral types, extinctions, and gravity classifications of mid-M to late-L dwarfs through low-resolution near-infrared spectroscopy. In Section~\ref{sec:application_discussion}, we apply our classification method to identify new members of Taurus, Pleiades, and Per~OB2, from our candidates, and to reclassify previously known low-mass objects in Taurus and the field. We investigate the HR diagram and spatial and kinematic distributions of all Taurus members to explore the star formation history of Taurus. Several individual systems related to brown dwarf formation theories are also discussed. Finally, a brief summary comes in Section~\ref{sec:conclusion}.

% ----------------------- Section -----------------------
\section{Candidate Selection}
\label{sec:selection}

\subsection{Known Taurus Census}
\label{subsec:known_members}

Our candidate selection is based on previously known Taurus members from \citeauthor{Luhman+2017} (\citeyear{Luhman+2017}; their Table~1) and \cite{Best+2017}\footnote{As we were finalizing the revised version of this paper, \cite{Esplin+2017} reported 18 additional low-mass members of Taurus, which are not included in our analysis. We independently discover two of the objects, PSO~J064.6887+27.9799 (SpT$_{\omega} =$L0.7$\pm$1.1; SpT$_{\rm E17} =$M9.25$\pm$0.5) and PSO~J065.1792+28.1767 (SpT$_{\omega} =$L1.6$\pm$1.3; SpT$_{\rm E17} =$M9.25$\pm$0.5), where SpT$_{\omega}$ is our spectral classification (Section~\ref{subsec:rfspt}) and SpT$_{\rm E17}$ is from \cite{Esplin+2017}. In addition, one of our new discoveries, PSO~J062.4648+20.0118 (SpT$_{\omega} =$M9.1$\pm$0.9; SpT$_{\rm E17} =$M8.25$\pm$0.5), was classified as a non-member by \cite{Esplin+2017} due to its relative faintness compared to other Taurus members with similar spectral types. However, we classify this object to be an older Taurus object, as its HR diagram position is close to the 30~Myr isochrone (Section~\ref{subsec:color}). All these three objects have \textsc{vl-g} surface gravities based on our gravity classification (Section~\ref{subsec:youth_logistics}).}. Our list of Taurus members contains 415 objects in total, spanning B9--L2 in spectral type. \cite{Kraus+2017} recently collected all disk-free objects that were suggested as candidate Taurus members by previous studies and reassessed their membership. Most of the disk-free members in our list are also included in the \cite{Kraus+2017} sample, except for five objects as secondary components in close (separation$\leqslant10\arcsec$) binary systems, four objects as companions to Class II or earlier-type sources, and four discoveries by \citeauthor{Luhman+2017} (\citeyear{Luhman+2017}; two objects) and \citeauthor{Best+2017} (\citeyear{Best+2017}; two objects) that were found after the \cite{Kraus+2017} work.

There are 140 overlapping objects between the \cite{Kraus+2017} sample and our adopted member list, and 136 of these were assessed as confirmed or likely Taurus members (``Y'' or ``Y?'') by \citeauthor{Kraus+2017} (\citeyear{Kraus+2017}; their Table~8). The four exceptions are 2MASS~J04182909+2826191, 2MASS~J04184023+2824245, 2MASS~J04190197+2822332, and 2MASS~J04223075+1526310, which have unknown membership (``?'' for the former two) or are likely nonmembers (``N?'' for the latter two). We keep them in our list, as they are not confirmed nonmembers (``N''). In addition, there are 256 objects covered by \cite{Kraus+2017} but not by our list, and 14 of them have $\geqslant$M6 spectral types. Among these 14 substellar objects, 2 objects were assessed as likely or confirmed Taurus members, 3 objects were likely or confirmed nonmembers, and the remaining 9 objects have unknown membership based on \cite{Kraus+2017}.

\subsection{Input Databases and Search Area}
\label{subsec:area}

\begin{figure*}[t!]
\begin{center}
\includegraphics[height=5.5in]{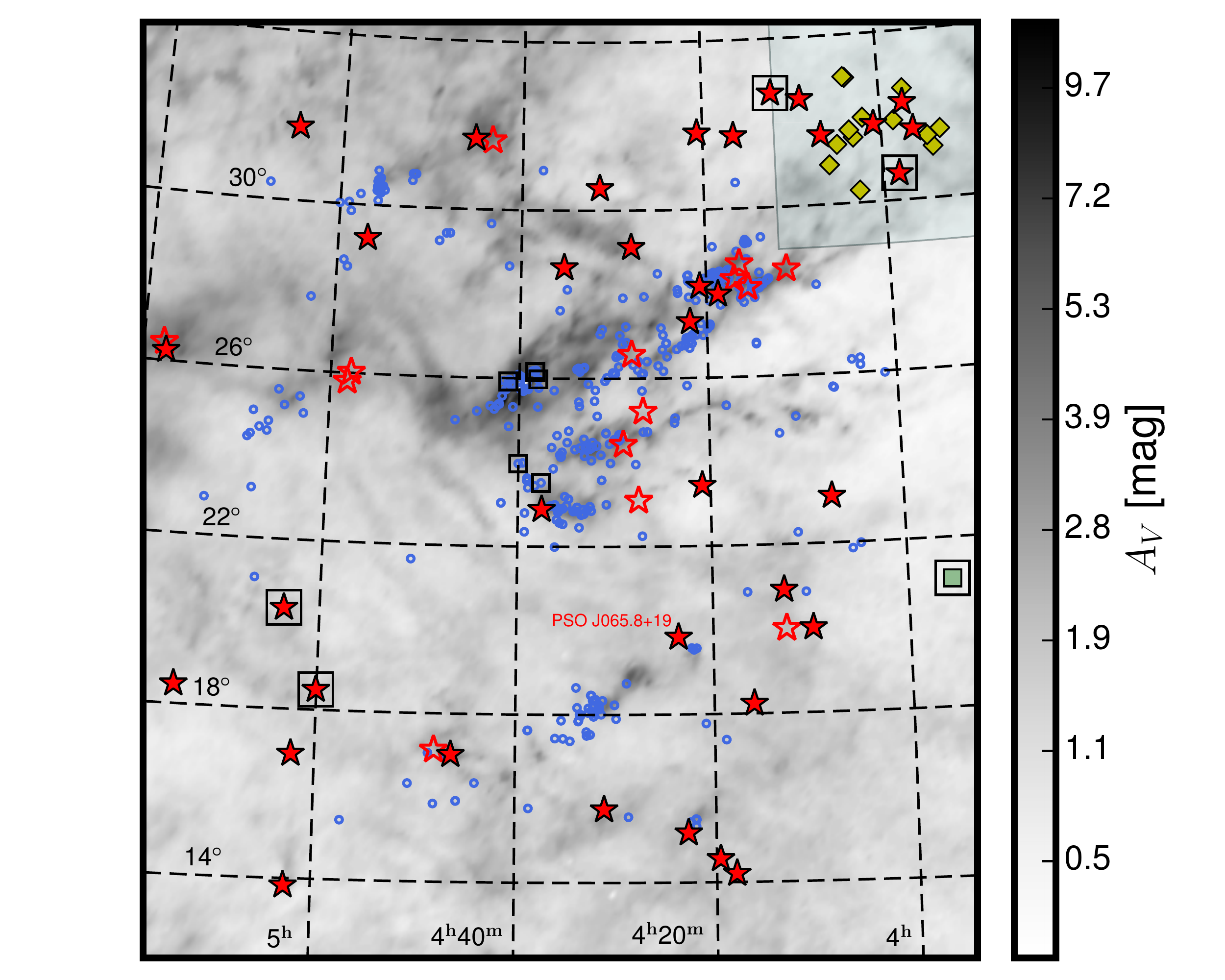}
\caption{Our search area in Taurus (370~deg$^{2}$), overlaid with our newly identified members in Taurus ({\it red} stars), Pleiades ({\it dark green} square), and Per~OB2 ({\it olive} diamonds), and previously known Taurus members ({\it blue} circles). The sky map uses a Cassini projection to show the $V$-band extinction ($A_{\rm V}$) from \cite{Schlafly+2014}. We use open stars to show our [M4,M6) discoveries, as they lack the gravity classifications needed for a firm membership assessment. The 10 objects (5 of our new discoveries, including the 1 probable new Pleiades member and 5 known $\geqslant$M6 members), with proper motion different from the mean Taurus motion by $>2\sigma$ (Section~\ref{subsubsec:hiPM}), are shown by open squares. We label our unusually bright L0 $\textsc{vl-g}$ dwarf discovery, PSO~J065.8+19 (Section \ref{subsec:brightL0}). The overlapping region between Taurus and Per~OB2 is noted by a \emph{light cyan} rectangle at the northwest corner. }
\label{fig:ExtMap}
\end{center}
\end{figure*}

We start our search by mining the PS1 Processing Version 3 database (2016 August, the final version of PS1 reprocessing prior to public release) using the Desktop Virtual Observatory \citep[DVO;][]{Magnier+2004, Magnier+2016} in an area centered at $\alpha=4^{\rm h}36^{\rm m}$, $\delta=23^{\circ}00\arcmin$ (J2000.0) with dimensions of $20^{\circ} \times 20^{\circ}$ ($\approx 370$~deg$^{2}$; Figure~\ref{fig:ExtMap}), thereby covering the entire extent of the Taurus star-forming region \citep[e.g.,][]{Kenyon+2008}. Another star-forming region, Pleiades \citep[$\approx$136~pc;][]{Melis+2014}, located in $\alpha=3^{\rm h}20^{\rm m}$--$3^{\rm h}40^{\rm m}$ and $\delta=22^{\circ}$--$28^{\circ}$~(J2000.0), is west of our search area. In addition, Perseus OB2 Association \citep[Per~OB2 hereafter;][]{Bally+2008}, with a distance of $\approx 318$~pc \citep{Zeeuw+1999}, is actually overlapped with the northwest part of our search region (Figure~\ref{fig:ExtMap}), as its geometry is $\alpha=3^{\rm h}20^{\rm m}-4^{\rm h}12^{\rm m}$, $\delta=29^{\circ}-38^{\circ}$ (J2000.0). Therefore, our search can discover new Pleiades and Per~OB2 members as well.

The PS1 database reports different types of photometry, and we use chip photometry and the forced warp photometry \citep[warp photometry hereafter;][]{Magnier+2016} in this work. Chip photometry is obtained by averaging fluxes from all individual chip exposures of an object, and warp photometry is calculated by fitting the point-spread function (PSF) in a stacked image constructed by warping all individual detections of an object into a united frame. The former provides more accurate photometry for an well-detected object, while the latter can achieve a greater depth with slightly lower accuracy. We choose between the chip photometry and warp photometry for each object following \cite{Best+2018}. We use the chip photometry for a PS1 band if it is brighter than the threshold value suggested by \citeauthor{Best+2018} (\citeyear{Best+2018}; their Table~3) and detected in at least two separate chip exposures with photometric uncertainties $<$0.2~mag. Otherwise, and if the object does not move fast (with proper motions over 100~mas~yr$^{-1}$), we adopt its warp photometry. In addition, the adopted warp photometry needs to have an uncertainty $<$0.2~mag and be calculated from at least two successful PSF fittings. Switching from chip to warp photometry helps improve our survey depth by $\approx 1$ mag, thereby lowering our mass sensitivity from $\approx 4$ M$_{\rm Jup}$ down to $\approx 3$ M$_{\rm Jup}$, according to the DUSTY evolutionary models of \cite{Chabrier+2000} and assuming a Taurus age of $1$~Myr.

We then cross-match the extracted PS1 objects with ancillary catalogs including AllWISE \citep{Cutri+2014}, 2MASS, and UGCS with matching radii of $5\arcsec$, $2\arcsec$, and $2\arcsec$, respectively. We use a larger matching radius between PS1 and {\it WISE} positions to compensate for {\it WISE}'s larger PSF size. Matching these databases eliminates transient objects (e.g., asteroids) in regions the surveys have only covered once.

\subsection{Photometric Criteria}
\label{subsec:phot_criteria}

Based on our PS1+AllWISE+2MASS+UGCS database, we select objects with good-quality detections in $y_{\rm P1}$, $W1$, $W2$, $J$, and $K$ bands. Good-quality $H$-band photometry are used as well. Each of the $J$, $H$, and $K$ bands is based on either 2MASS or MKO (UGCS detections) photometric systems. Objects selected in this way are limited to a volume within $\approx 1$~kpc. Good photometric quality for these bands is defined as follows (items in the parenthesis refer to flags within the corresponding catalog):

\begin{enumerate}[topsep=0pt,itemsep=-1ex,partopsep=1ex,parsep=1ex]
\item PS1 objects are detected in at least three separate frames in $y_{\rm P1}$ band ({\sf y:nphot} $\geqslant$ 3).
\item PS1 photometric signal-to-noise ratio (S/N) $>5$ in $y_{\rm P1}$ band (i.e., $\sigma_{y_{\rm P1}} < 0.2$~mag; {\sf y:err} $<$ 0.2~mag).
\item The $y_{\rm P1}$ detection is not saturated ($y_{\rm P1} \geqslant 12.5$~mag).
\item PS1 detections have clear PSF identification in $y_{\rm P1}$ band, not impacted by probable saturations or cosmic rays ({\sf y:flags} $=$ 16, 256, 512 or 1024).
\item AllWISE photometric S/N $>2$ in $W1$ and $W2$ bands ({\sf ph\_qual} $=$ {\sf A, B, }or {\sf C}).
\item The $W1$ and $W2$ detections are not saturated ($W1$ $\geqslant$ 8.1~mag and $W2$ $\geqslant$ 6.7~mag).
\item AllWISE detections have morphologies consistent with point sources ({\sf ext\_flg} = 0).
\item AllWISE detections are mostly likely not variables ({\sf var\_flg} $\leqslant$ 5) or data are insufficient to make determination of objects' possible variability ({\sf var\_flg} $=$ {\sf n}).
\item 2MASS photometric S/N$> 5$ in $J_{\rm 2MASS}$/$H_{\rm 2MASS}$/$K_{\rm 2MASS}$ bands ({\sf ph\_qual} $=$ {\sf A, B, }or {\sf  C}).
\item 2MASS detections are not saturated ($J_{\rm 2MASS} \geqslant 9.0$ mag, $H_{\rm 2MASS} \geqslant 8.5$ mag, and $K_{\rm 2MASS} \geqslant 8.0$ mag).
\item 2MASS detections have clear PSF identification, and therefore reliable derived photometry and astrometry ({\sf rd\_flg} $= 1, 2, $ or $3$).
\item 2MASS detections are unaffected by blending ({\sf bl\_flg} $= 1$) and any known artifacts ({\sf cc\_flg} $= 0$).
\item UGCS detections are unaffected by blending, nearby bright sources, and crosstalk artifact/contamination and diffraction spike contamination ({\sf Jflags, Hflags, Kflags} $= 0$).
\item UGCS detections are not saturated \citep[$J_{\rm MKO} \geqslant 11.0$ mag, $H_{\rm MKO} \geqslant 11.5$ mag, and $K_{\rm MKO} \geqslant 10.4$ mag;][]{Lodieu+2012}.
\item UGCS detections are stars instead of galaxies with a probability $\geqslant 90\%$ ({\sf Jcl, Hcl, Kcl} $= -2$ or $-1$).
\end{enumerate}

Then we extract brown dwarf candidates by applying the following photometric criteria, which are designed based on the locations of known $\geqslant$M6 Taurus members \citep[e.g.,][]{Best+2017, Luhman+2017} and field dwarfs (summarized by \citealt{Kraus+2007} and \citealt{Best+2018}) in color-color and color-magnitude diagrams.

\begin{figure*}[t]
\begin{center}
\includegraphics[height=3.in]{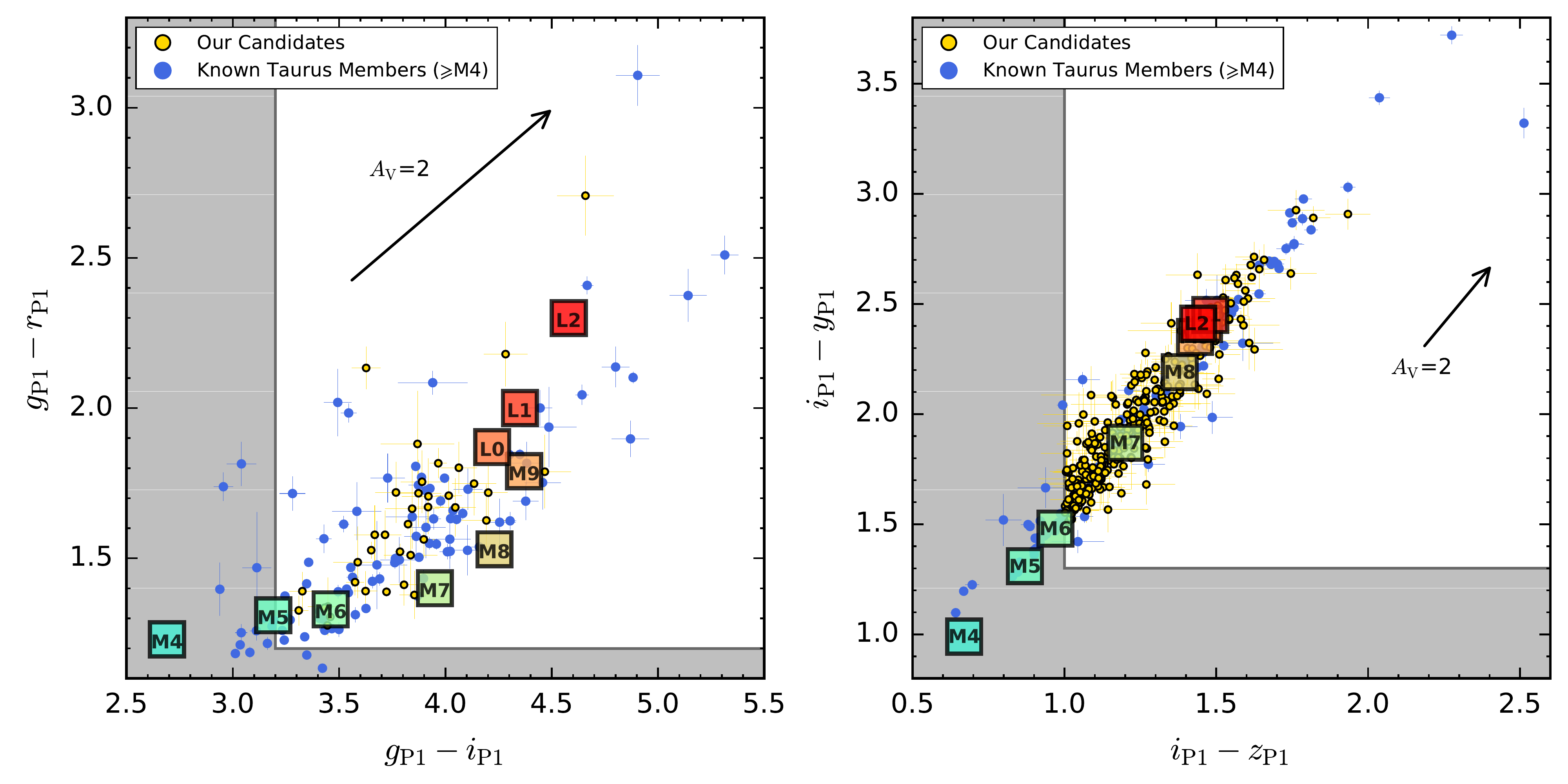}
\caption{PS1 color-color diagrams of our candidates ({\it gold}) and previously known $\geqslant$M4 members ({\it blue}) in Taurus. We only plot objects with good-quality photometry (defined in Section~\ref{subsec:phot_criteria}) in $g_{\rm P1}$/$r_{\rm P1}$/$i_{\rm P1}$ and $i_{\rm P1}$/$z_{\rm P1}$/$y_{\rm P1}$ in the left and the right panel, respectively. Colored squares show the median values of M4--L2 field dwarfs from \cite{Best+2018}. The extinction vector corresponds to $A_{\rm V} = 2$~mag, using the extinction law of \cite{Schlafly+2011}. The white region indicates our photometric criteria  (Section \ref{subsec:phot_criteria}).}
\label{fig:gr_gi_iy_iz}
\end{center}
\end{figure*}

\begin{figure*}[t]
\begin{center}
\includegraphics[height=3.in]{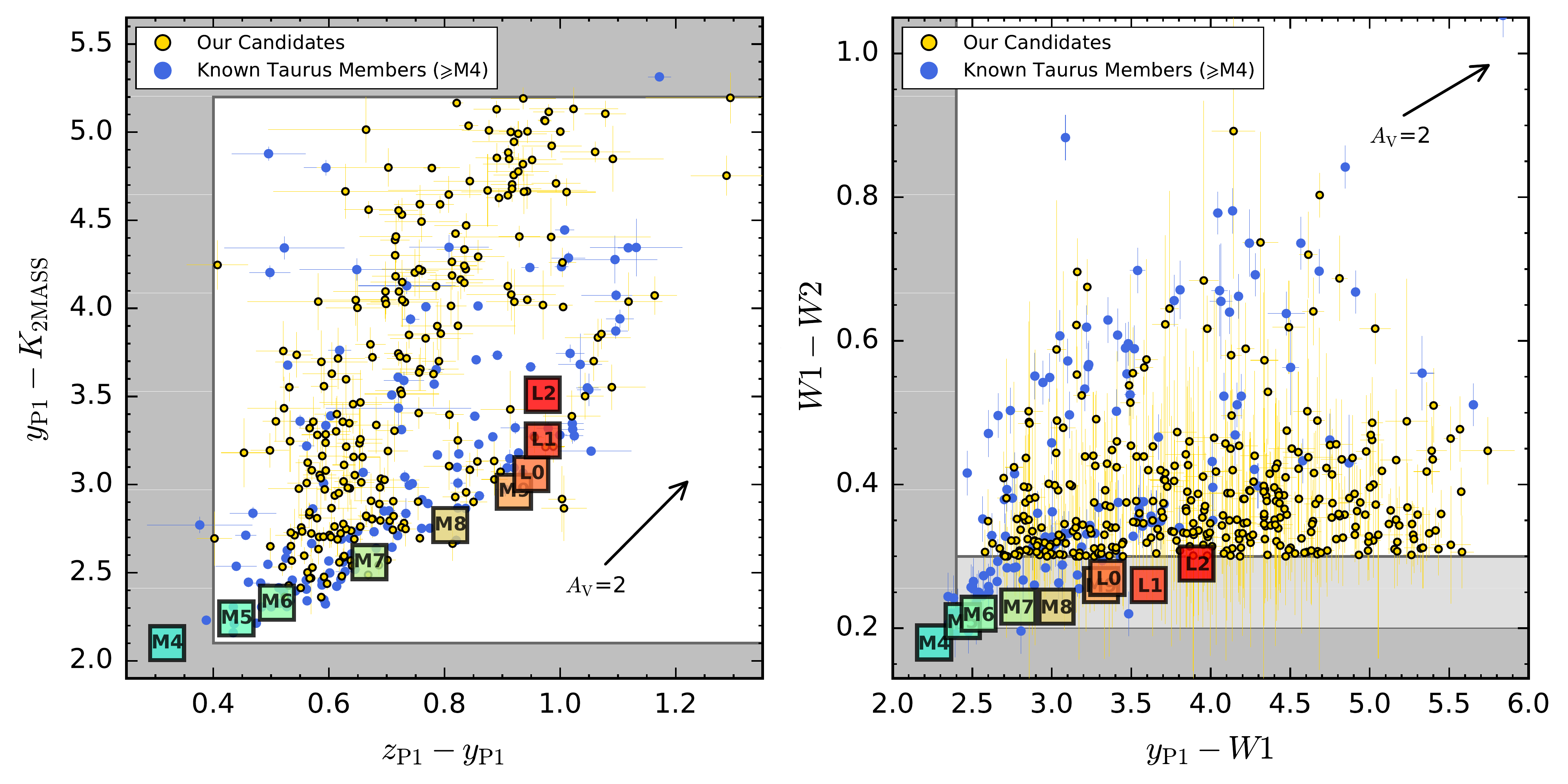}
\caption{PS1, AllWISE, and 2MASS color-color diagrams of our candidates ({\it gold}) and previously known $\geqslant$M4 members ({\it blue}) in Taurus, as described in the caption of Figure~\ref{fig:gr_gi_iy_iz}. We only plot objects with good-quality photometry in $z_{\rm P1}$/$y_{\rm P1}$/$K_{\rm 2MASS}$ and $y_{\rm P1}$/$W1$/$W2$ for the left and the right panel, respectively. While $\geqslant$M6 field dwarfs usually have $W1-W2$ colors redder than $\approx 0.2$~mag, we restrict our $W1 - W2$ color cut to 0.3~mag in our photometric criteria, as explained in Section~\ref{subsec:phot_criteria}.}
\label{fig:yK_zy_W1W2_yW1}
\end{center}
\end{figure*}

\begin{figure*}[t]
\begin{center}
\includegraphics[height=3.in]{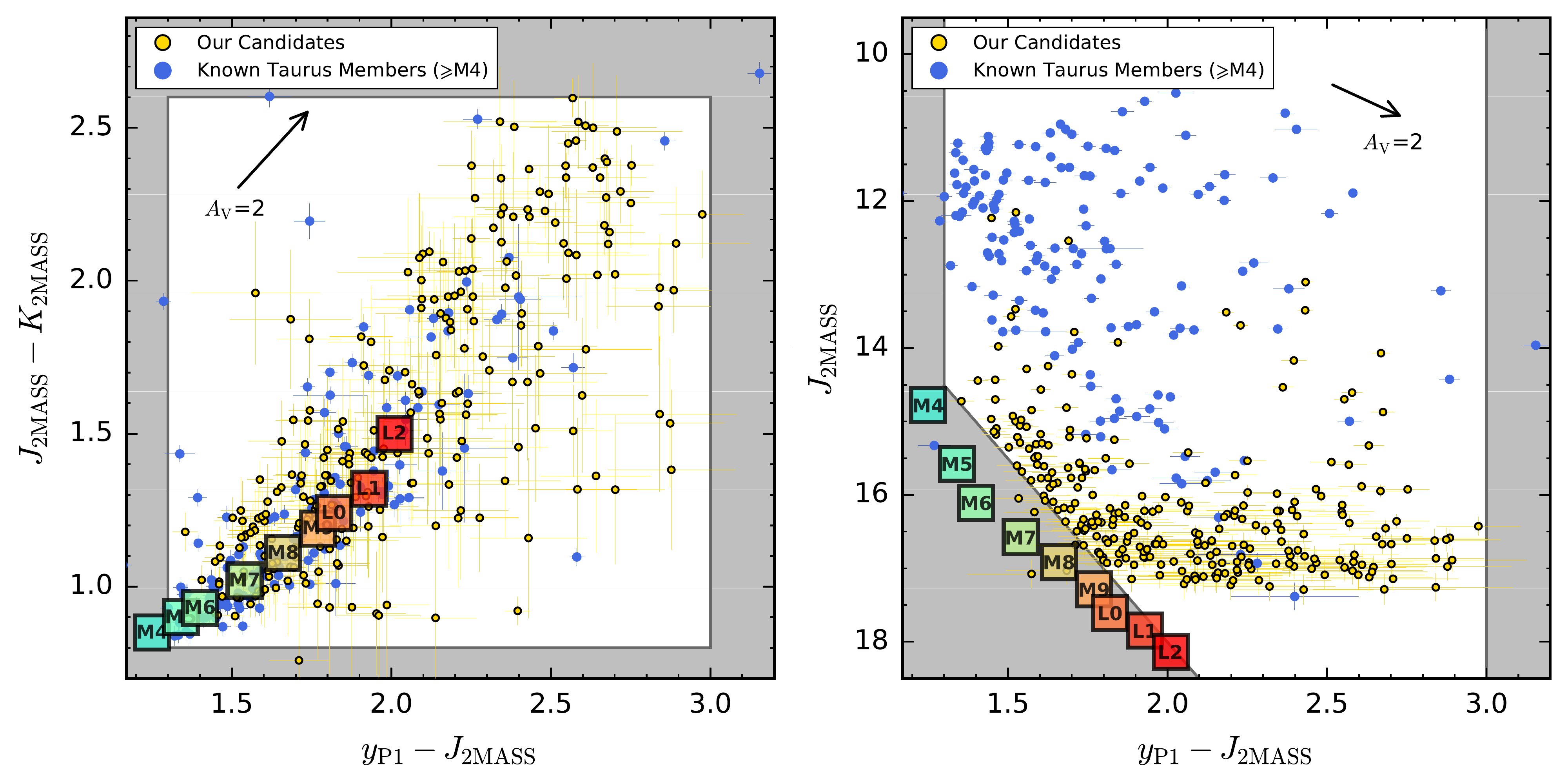}
\caption{PS1 and 2MASS color-color diagrams of our candidates ({\it gold}) and previously known $\geqslant$M4 members ({\it blue}) in Taurus, as described in the caption of Figure~\ref{fig:gr_gi_iy_iz}. We only plot objects with good-quality photometry in $y_{\rm P1}$/$J_{\rm 2MASS}$/$K_{\rm 2MASS}$ and $y_{\rm P1}$/$J_{\rm 2MASS}$ for the left and the right panel, respectively. $J_{\rm 2MASS}$ magnitudes of field dwarfs (colored squares) are from the 1Gyr isochrone of the BHAC15 models \citep{Baraffe+2015}.}
\label{fig:JK_yJ_J_yJ}
\end{center}
\end{figure*}

\begin{figure*}[t]
\begin{center}
\includegraphics[height=3.in]{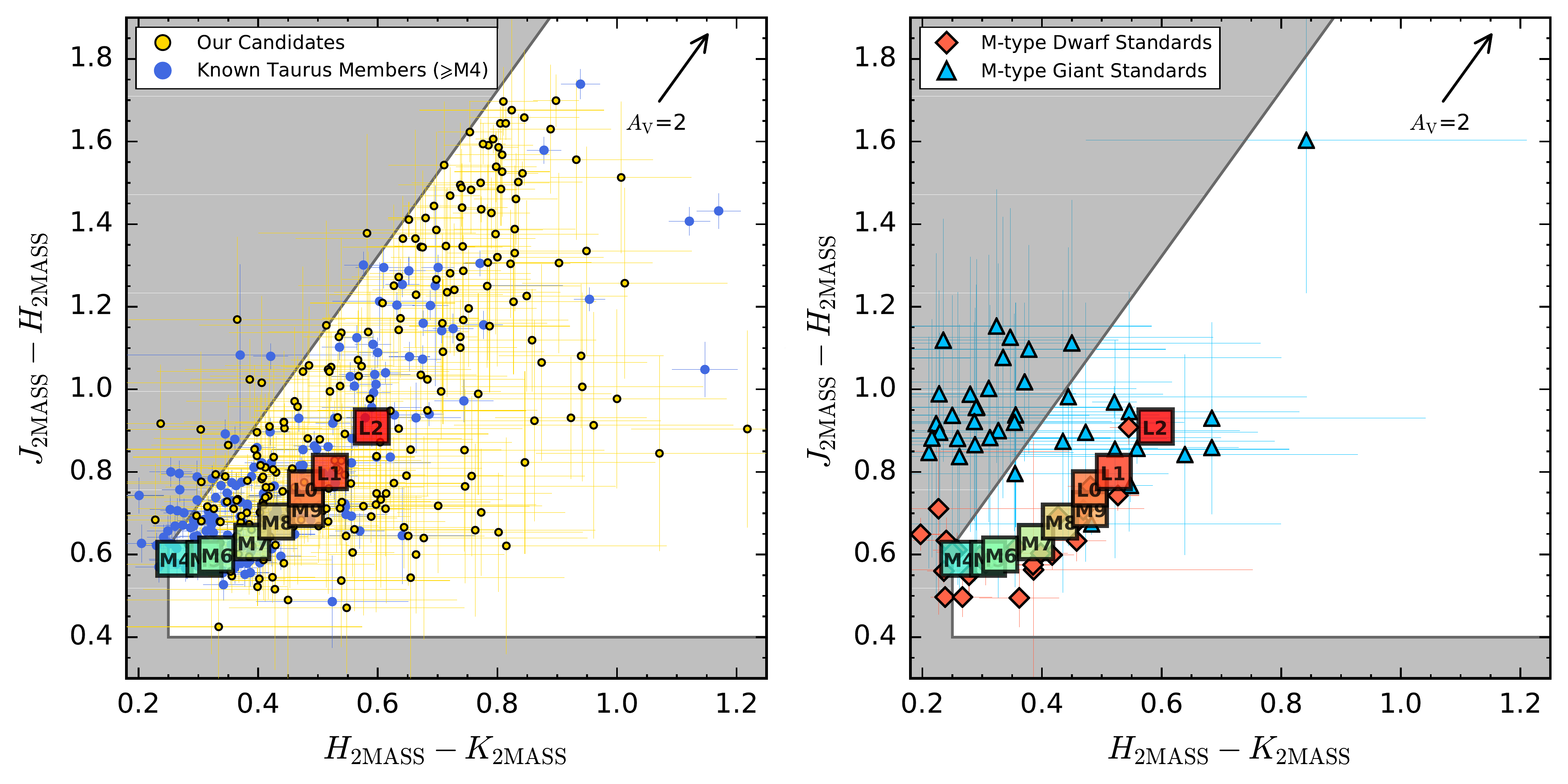}
\caption{Left: the 2MASS color-color diagram of our candidates ({\it gold}) and previously known $\geqslant$M4 members ({\it blue}) in Taurus, as described in the caption of Figure~\ref{fig:gr_gi_iy_iz}. We only plot objects with good-quality photometry in $J_{\rm 2MASS}$/$H_{\rm 2MASS}$/$K_{\rm 2MASS}$. Right: We plot the standards of M-type dwarfs ({\it orange}) and (super-)giants ({\it sky blue}) from the IRTF Spectral Library \citep{Cushing+2005, Rayner+2009} in the $JHK$ diagram, and overlay the sequence of field dwarfs provided by \cite{Best+2018}. Around 1/3 of the giant standards would also pass our JHK photometric criteria. }
\label{fig:JHK}
\end{center}
\end{figure*}

\begin{enumerate}[topsep=0pt,itemsep=-1ex,partopsep=1ex,parsep=1ex]
\item $g_{\rm P1} - r_{\rm P1} \geqslant 1.2$ mag (Figure~\ref{fig:gr_gi_iy_iz}). We only apply this criterion if the $g_{\rm P1}$ and $r_{\rm P1}$ detections of an object have the same quality standards required for the $y_{P1}$ band. The saturation limits of $g_{\rm P1}$ and $r_{\rm P1}$ photometry are both $14.5$~mag.  
\newline This color cut can also find strong H$\alpha$ emitters, as H$\alpha$ emission lines reside in the $r_{\rm P1}$ band and are signatures of disk-bearing young substellar objects with accretion activities \citep[e.g.,][]{Guieu+2006, Luhman+2010}.
\item $g_{\rm P1} - i_{\rm P1} \geqslant 3.2$ mag (Figure~\ref{fig:gr_gi_iy_iz}). Again, we only apply this criterion if the $g_{\rm P1}$ and $i_{\rm P1}$ detections of an object have the same quality standard required for the $y_{P1}$ band. The saturation limit of the $i_{\rm P1}$ band is $14.5$~mag. 
\item $i_{\rm P1} - z_{\rm P1} \geqslant 1.0$ mag (Figure~\ref{fig:gr_gi_iy_iz}). This criterion is only applied if the $i_{\rm P1}$ and $z_{\rm P1}$ detections of an object have the same quality standard required for the $y_{P1}$ band. The saturation limit of the $z_{\rm P1}$ band is $13.5$~mag.
\item $i_{\rm P1} - y_{\rm P1} \geqslant 1.3$ mag (Figure~\ref{fig:gr_gi_iy_iz}). We only apply this criterion if the $i_{\rm P1}$ detection of an object has the same quality standard required for the $y_{P1}$ band.
\item $z_{\rm P1} - y_{\rm P1} \geqslant 0.4$ mag (Figure~\ref{fig:yK_zy_W1W2_yW1}). We only apply this criterion if the $z_{\rm P1}$ detection of an object has the same quality standard required for the $y_{P1}$ band.
\item $y_{\rm P1} - W1 \geqslant 2.4$~mag (Figure~\ref{fig:yK_zy_W1W2_yW1}).
\item $W1-W2 \geqslant 0.3$ mag (Figure~\ref{fig:yK_zy_W1W2_yW1}). 
\newline While $\geqslant$M6 field dwarfs usually have $W1-W2$ colors redder than $\approx 0.2$~mag, here we restrict our $W1-W2$ color cut to 0.3~mag. The $W1-W2$ color is actually weakly dependent on spectral type from mid-M to mid-L objects as it changes by only $\approx 0.1$ mag from spectral type M6 ($W1-W2 \approx 0.22$~mag) to L4 \citep[$\approx 0.32$~mag;][]{Best+2018}. A cut of $0.2$~mag in $W1-W2$ would therefore bring more outliers of earlier-type (<M6) objects not of interest in this work. In addition, young objects have systematically redder $W1-W2$ colors compared to their field-age counterparts \citep{Best+2018}, as $\approx 90\%$ of $\geqslant$M6 known Taurus members are redder than 0.3~mag in $W1-W2$. Therefore we adopt $0.3$~mag as our $W1-W2$ color cut, acknowledging that $\approx 10\%$ of bona fide $\geqslant$M6 members in Taurus could be rejected by this criterion.
\item $0.8 \leqslant$ $J_{\rm 2MASS} -$ $K_{\rm 2MASS} \leqslant 2.6$~mag (Figure~\ref{fig:JK_yJ_J_yJ}).
\item $1.3 \leqslant$ $y_{\rm P1} -$ $J_{\rm 2MASS} \leqslant 3.0$~mag (Figure~\ref{fig:JK_yJ_J_yJ}).
\item $J_{\rm 2MASS} \leqslant 5 \times \left(y_{\rm P1} - J_{\rm 2MASS}\right) + 5 \times {\rm log}_{10} \left(d/10{\rm pc}\right)+ 2.2$~mag (Figure~\ref{fig:JK_yJ_J_yJ}), where $d=145$~pc is the adopted Taurus distance \citep{Zeeuw+1999}.
\newline We set the upper envelope of $J_{\rm 2MASS}$ magnitudes slightly brighter than field dwarfs but fainter than Taurus known objects from mid-M to early-L in spectral type, because young ultracool dwarfs over such a spectral type range are expected to be brighter than the field-age objects \citep{Liu+2016}.
\item $2.1 \leqslant$ $y_{\rm P1} -$ $K_{\rm 2MASS} \leqslant 5.2$ mag (Figure~\ref{fig:yK_zy_W1W2_yW1}).
\item (i) $J_{\rm 2MASS}-H_{\rm 2MASS} \leqslant 2.007 \times (H_{\rm 2MASS} - K_{\rm 2MASS}) + 0.118$~mag (Figure~\ref{fig:JHK}).
\newline (ii) $J_{\rm 2MASS}-H_{\rm 2MASS} \geqslant 0.4$ mag (Figure~\ref{fig:JHK}).
\newline (iii) $H_{\rm 2MASS}-K_{\rm 2MASS} \geqslant 0.25$ mag (Figure~\ref{fig:JHK}).
\newline The slope $2.007$ of the upper boundary of the $J_{\rm 2MASS}-H_{\rm 2MASS}$ color in the criterion (i) corresponds to the extinction vector in the $J_{\rm 2MASS}-H_{\rm 2MASS}$ vs. $H_{\rm 2MASS}-K_{\rm 2MASS}$ diagram based on the extinction law of \cite{Schlafly+2011}. These color-cuts could remove giant star contaminants and are designed by comparing the positions in the $JHK$ diagram between dwarf and (super)giant standards from the IRTF Spectral Library (\citealt{Cushing+2005,Rayner+2009}; see also \citealt{Kraus+2007, Lepine+2011}). Although around $1/3$ of M-type (super)giant standards could still pass these photometric cuts, they could be mostly removed by our further kinematic criteria (Section \ref{subsec:kine_criteria}). 
\item We apply criteria 8--11 if an object has good-quality photometry in $J_{\rm 2MASS}$ and $K_{\rm 2MASS}$ bands, and we additionally apply the criterion 12 when the $H_{\rm 2MASS}$ photometry has a good quality as well. For each of the $J$, $H$, and $K$ bands, if an object has both 2MASS and MKO photometries, we adopt the one with good-quality detection. If both photometric systems provide good detections, we prefer the MKO magnitudes, due to their smaller photometric uncertainty and fainter limiting magnitude. When the MKO photometry is used for any of $J$/$H$/$K$ bands, we adjust the boundary of the selection region in the 2MASS-based JHK diagram in criteria $8-12$ based on the transformation between the MKO and 2MASS photometric systems for $\geqslant$M6 dwarfs, as $J_{\rm MKO} - J_{\rm 2MASS} = - 0.05$~mag, $H_{\rm MKO} - H_{\rm 2MASS} = + 0.03$~mag, and $K_{\rm MKO} - K_{\rm 2MASS} = - 0.02$~mag. We obtain these conversions by comparing the differences of 2MASS and MKO magnitudes for the L and T dwarfs studied by \cite{Stephens+2004} and the M6$-$T9 dwarfs with measured parallaxes from \cite{Dupuy+2012}. The updated $JHK$ diagram could use a mixture of 2MASS and MKO photometries (e.g., $J_{\rm 2MASS} - H_{\rm MKO}$ versus $H_{\rm MKO} - K_{\rm MKO}$). In addition, for the criterion 12(i), we revise the slope of the upper boundary of the $J-H$ color to be the extinction vector in the updated $JHK$ diagram using the extinction law of \cite{Schlafly+2011}.
\end{enumerate}

We then test the kinematic properties of the selected candidates that pass all the above photometric criteria.

\subsection{Kinematic Criteria}
\label{subsec:kine_criteria}

Proper motions are enormously valuable to establishing membership in Taurus. Foreground field dwarfs and background reddened stars could pass our photometric criteria (Section~\ref{subsec:phot_criteria}). But they usually have inconsistent motions compared to Taurus and therefore could be removed from our list of candidates based on their kinematic information.

We use the PS1 proper motions described in \cite{Magnier+2016}. Based on PS1, 2MASS, and Gaia detections \citep{Gaia+2016, Lindegren+2016} spanning a 14--17~year baseline, \cite{Magnier+2016} computed the position, parallax, and proper motion of each PS1 object using iteratively reweighted least squares fitting with outlier clipping, and tied all astrometry to the Gaia DR1 reference frame. The median proper-motion uncertainty is $\approx 4$~mas~yr$^{-1}$ for known substellar ($\geqslant$M6) members in Taurus. Our search is the first to use proper motions of substellar candidates over such large area ($\approx 370$~deg$^{2}$) and long-time baseline with such high precision, enabling a more efficient candidate selection.

Following \cite{Best+2017}, the proper motion of a PS1 object is considered to have good quality if the object's $i_{\rm P1}$ and $y_{\rm P1}$ magnitudes are not saturated (Section~\ref{subsec:phot_criteria}) and if the reduced $\chi^{2}$ for its \cite{Magnier+2016} proper-motion fits satisfies $0.3 < \chi_{\nu}^{2} < 40$. We calculate the average motion of known Taurus members by including 181 objects with good quality proper-motion measurements and derive a weighted average value of $(\mu_{\alpha} {\rm cos} \delta, \mu_{\delta}) = (7.55 \pm 0.16, -17.44 \pm 0.16)$~mas~yr$^{-1}$ with a weighted rms of $4.90$~mas~yr$^{-1}$ and $6.37$~mas~yr$^{-1}$ in R.A. and Decl., respectively. We reject photometric candidates whose proper motions differ from the mean motion of Taurus by more than $2\sigma$ (Figure~\ref{fig:PM_analysis}). Around $92\%$ known Taurus members with good-quality proper motions could pass this criterion.

\subsection{Final Selection}
\label{subsec:add_constrain}

In addition to photometric and kinematic criteria, we visually check the PS1, 2MASS, and AllWISE images of each selected object in order to reject galaxies or other diffuse sources. We also utilize the SIMBAD webpage\footnote{\url{http://simbad.u-strasbg.fr/simbad/}.} \citep{Wenger+2000} to exclude previously known objects. We rediscover 83 previously known Taurus objects spanning M3--L2 in literature spectral types, including 54 out of 76 known substellar ($\geqslant$M6) members. We remove all known objects from our list of candidates as well.

We additionally include five objects as our candidates with discrepant ($>2\sigma$) proper motions from Taurus, which would be rejected by our current search criteria. They were selected as candidates during an earlier search attempt using preliminary proper motions from PS1, and our spectroscopic follow-up found they are M6--L0 low-gravity dwarfs. Given that their proper motions are not consistent with Taurus, they might be ejected brown dwarfs (Section~\ref{subsubsec:hiPM}), as predicted by dynamical models of brown dwarf formation \citep[e.g.,][]{Reipurth+2001}.

After applying photometric and kinematic criteria, as well as the above adjustments, we derive a list of 350 Taurus candidates.

\section{Near-infrared Spectroscopy}
\label{sec:NIRspec}

We used the NASA Infrared Telescope Facility (IRTF) to obtain near-infrared spectra for 83 candidates, among which 19 objects are located in the overlapping region between Taurus and Per~OB2. We use the facility spectrograph SpeX \citep{Rayner+2003} in the LowRes15 (prism) mode with the $0.8\arcsec$ slit (R$\approx 75$). A nearby A0V star with the airmass different from each target by $\lesssim$0.1 was observed contemporaneously for telluric correction (Appendix~\ref{sec:appendix_tell}). Table~\ref{tab:spex_log} lists the instrument configuration, integration times, and observation dates of our targets. We reduce the spectra in standard fashion using the version 4.1 of the Spextool software package \citep{Vacca+2003, Cushing+2004}.

We divide our entire candidate list into seven priority groups based on objects' magnitudes and proper motions. During our spectroscopic follow-up, we prioritize targets with brighter magnitudes and more Taurus-like proper motions. For the latter criterion, we choose targets with proper motions having S/N $>3$ and being consistent with the mean motion of Taurus (Section~\ref{subsec:kine_criteria}) within $1\sigma$. So far, our follow-up has been finished for $\approx 1/4$ of candidates, including $\approx 75\%$ candidates that have $J_{\rm 2MASS}\leqslant15.5$~mag. 

Around $80\%$ of the observed spectra have S/N$\gtrsim$30 per pixel in $J$ band, for which we can perform reliable spectral typing. Robust youth assessment based on gravity-sensitive spectral features is possible for spectra with S/N $\gtrsim 50$ ($\approx 60\%$ of our spectra satisfy this requirement). In addition, we observed 41 known Taurus members, with all the resulting spectra having $J$-band S/N$\geqslant$30 and $\approx 80\%$ with S/N$\geqslant$50. Combining our near-infrared spectra with those from previous studies \citep{Best+2017, Luhman+2017}, we have access to near-infrared spectra of all $\geqslant$M6 members in Taurus.

\section{A Unified Scheme of Reddening-free Spectral Classification, Extinction Determination, and Youth Assessment}
\label{sec:rfSPT_extinction}

Intrinsic magnitudes, colors, luminosities, and masses of our substellar candidates are essential to constructing empirical isochrones and IMFs. Precise determination of these characteristics depends on reliable spectral types and extinctions, which are hard to achieve due to degeneracy in photometry and spectroscopy. For (unreddened) field ultracool objects, spectral classification is typically done in two ways: (1) qualitative comparisons between observed spectra and established standards, which have no extinction (e.g., \citealt{Burgasser+2006}; \citealt{Kirkpatrick+2010}; \citealt{Allers+2013}, AL13 hereafter; \citealt{Cruz+2018}), and (2) quantitative measurements of near-infrared spectral features (e.g., H$_{2}$O indices for M and L dwarfs adopted by AL13, and H$_{2}$O and CH$_{4}$ indices for T dwarfs defined by \citealt{Burgasser+2006}). However, both of these methods cannot be directly applied to ultracool dwarfs in young and dusty star-forming regions, because extinction alters both overall continuum shape and specific spectral features. For instance, an M7 dwarf with extinction of $A_{\rm V}=7$ mag has a $J$-band continuum slope similar to an L0 dwarf. Without a precise spectral type, the extinction cannot be reliably measured (Section~\ref{subsec:extinction}). This also complicates gravity classification (Section~\ref{subsec:youth_logistics}). 

It is plausible to simultaneously derive both spectral types and extinctions by fitting the observed spectrum using libraries of standards based on visual comparisons or $\chi^{2}$-minimization. However, the heterogeneous colors of ultracool dwarfs at near-infrared wavelengths complicates selecting representative standards, as diverse physical properties of brown dwarfs (e.g., gravity, metallicity, and photospheric condensate variations) can cause a large spread in near-infrared colors at fixed optical spectral type (e.g., \citealt{Knapp+2004}; \citealt{Stephens+2009}; AL13). In addition, while standards have been proposed for old field dwarfs \citep[$>200$~Myr;][]{Kirkpatrick+2010}, young field dwarfs ($<30$~Myr; AL13), and young members of star-forming regions \citep[$\lesssim 10$~Myr;][]{Luhman+2017}, a comprehensive library of intermediate-age ($\approx 30-200$~Myr) standards is lacking, which inhibits robust classification based on spectral morphology.

In this section, we develop a new quantitative approach to classify brown dwarfs in dusty star-forming regions based on the AL13 classification system by determining reddening-free spectral types, extinctions, and gravity classifications.

\subsection{Revisiting the AL13 Spectral Classification}
\label{subsec:revisit}

AL13 employed low- and moderate-resolution near-infrared spectra of $73$ young ($\approx 10-300$~Myr) field ultracool dwarfs, $\approx 90\%$ of which were observed in prism and/or short-wavelength cross-dispersed (SXD) mode using IRTF. They measured four H$_{2}$O indices and then established a cubic polynomial relation between the optically determined spectral type and each H$_{2}$O index (their Figure~6 and Table~3). Their final near-infrared spectral type combines both qualitative and quantitative approaches and is the weighted average of the classifications determined using visual comparison and H$_{2}$O indices. The H$_{2}$O indices (H$_{2}$O, \citealt{Allers+2007}; H$_{2}$OD, \citealt{McLean+2003}; H$_{2}$O-1 and H$_{2}$O-2, \citealt{Slesnick+2004}) are defined as the flux ratios in two narrow bands:
\begin{equation}
\begin{split}
{\rm H}_{2}{\rm O} &\equiv F_{\lambda=1550-1560}  /  F_{\lambda=1492-1502}      \\
{\rm H}_{2}{\rm OD} &\equiv F_{\lambda=1951-1977}  /  F_{\lambda=2062-2088}    \\
{\rm H}_{2}{\rm O-1} &\equiv F_{\lambda=1335-1345}  /  F_{\lambda=1295-1305}    \\
{\rm H}_{2}{\rm O-2} &\equiv F_{\lambda=2035-2045}  /  F_{\lambda=2145-2155}
\end{split}
\label{eq:H2O_index}
\end{equation} 
where $F_{\lambda}$ is the average flux in a narrow band pass and the wavelengths are in units of nanometers. In our work, we redefine $F_{\lambda}$ in Equation~\ref{eq:H2O_index} as the integrated flux in narrow bands\footnote{The H$_{2}$O indices are traditionally defined as average (e.g., \citealt{Allers+2007}; AL13) or median \citep[e.g.,][]{McLean+2003} flux density in the narrow bands, which are equivalent or similar to our definitions, given that the numerators and denominators of H$_{2}$O indices in Equation~\ref{eq:H2O_index} share the same band width.} and convert their flux ratios into standard magnitude-based colors:
\begin{equation}
\begin{split}
W_{0} &\equiv -2.5\ {\rm log}_{10} \left( {\rm H}_{2}{\rm O}   \right)     \\
W_{D} &\equiv -2.5\ {\rm log}_{10} \left( {\rm H}_{2}{\rm OD} \right)   \\
W_{1} &\equiv -2.5\ {\rm log}_{10} \left( {\rm H}_{2}{\rm O-1} \right)   \\
W_{2} &\equiv -2.5\ {\rm log}_{10} \left( {\rm H}_{2}{\rm O-2} \right) 
\end{split}
\label{eq:H2O_color}
\end{equation}
We use $W_{z}$ to denote four H$_{2}$O index colors with $z$ being $0$~(H$_{2}$O), $D$~(H$_{2}$OD), $1$~(H$_{2}$O--1) and $2$~(H$_{2}$O--2) hereafter. In principle, $W_{z}$ could be contaminated by telluric absorption features due to the imperfect telluric correction. We provide a quantitative analysis of this issue in Appendix~\ref{sec:appendix_tell} and conclude that telluric contamination of H$_{2}$O indices is negligible for our work.

\begin{figure}[t]
\begin{center}
\includegraphics[height=4.5in]{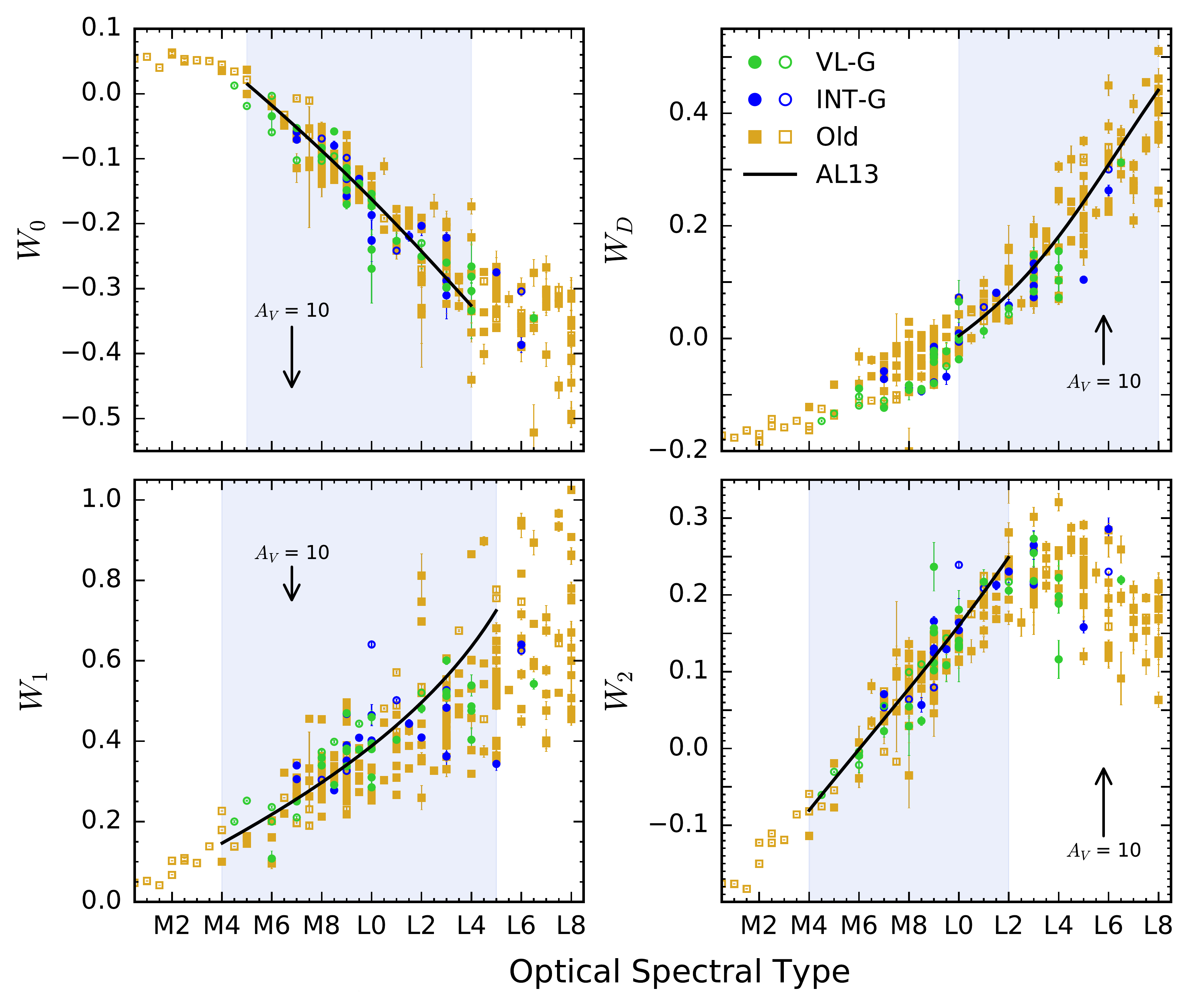}
\caption{ Relation between four H$_{2}$O indices ($W_{z}$; Equations \ref{eq:H2O_index} and \ref{eq:H2O_color}) and optical spectral types. Here we expand the original AL13 sample to include all M- and L-type dwarfs from the SpeX Prism Spectral Libraries \citep[e.g.,][]{Burgasser+2004, Chiu+2006, Kirkpatrick+2010} and the IRTF Spectral Library \citep{Cushing+2005, Rayner+2009}. Our sample contains 408 objects in total, and here we plot the 246 objects that have reported optical spectral types. Solid symbols are measurements from low-resolution spectra ($R \approx 100$), and open symbols are from moderate-resolution spectra ($R \approx 2000$). Errors in $W_{z}$ are calculated from the spectra in a Monte Carlo fashion. Uncertainties in spectral types are typically adopted as 1~subtype. {\it Green} and {\it blue} circles show young objects with low (\textsc{vl-g}) and intermediate (\textsc{int-g}) gravities. {\it Orange} squares show old objects with field gravities (\textsc{fld-g}) or no reported gravity. The spectral type calibration of the AL13 system for each H$_{2}$O index is overlaid as a solid line, and its applicable range is shown as {\it blue} shadow (see also Table~3 in AL13). Using the extinction law of \cite{Schlafly+2011}, we draw an extinction vector corresponding to $A_{\rm V}=10$ mag for each index-color $W_{z}$. }
\label{fig:AL13H2O}
\end{center}
\end{figure}

We reproduce the relations between $W_{z}$ and optical spectral types in Figure~\ref{fig:AL13H2O}, expanding the AL13 sample to include all M- and L-type ultracool dwarfs in the SpeX Prism Spectral Libraries\footnote{\url{http://pono.ucsd.edu/~adam/browndwarfs/spexprism}.} \citep[$R \approx 100$; e.g.,][]{Burgasser+2004, Chiu+2006, Kirkpatrick+2010} and the IRTF Spectral Library \citep[$R \approx 2000$;][]{Cushing+2005, Rayner+2009}. We exclude subdwarfs and companions to nearby stars. We remove from our sample 17 objects in IC~348 and Taurus studied by \cite{Muench+2007}, which could be reddened due to their membership in dusty star-forming regions. We additionally remove 3 reddened young field dwarfs studied by AL13: 2MASS~J04221413+1530525 (2M~0422+1530 hereafter), 2MASS~J04351455$-$1414468 (2M~0435$-$1414 hereafter), and 2MASS~J06195260$-$2903592 (2M~0619$-$2903 hereafter). Our sample consists of 408 objects in total, and in this section we focus on the 246 objects that have reported optical spectral types. In Figure~\ref{fig:AL13H2O}, we show these 246 objects, spanning M0--L8 and a mixture of surface gravities: $11\%$ objects have low gravity (\textsc{vl-g}), $9\%$ have intermediate gravity (\textsc{int-g}), and the remaining $80\%$ have field gravity (\textsc{fld-g}) or no reported gravity. Hereafter, we describe an object as ``young'' if its gravity classification is either \textsc{vl-g} or \textsc{int-g}, and as ``old'' if it has \textsc{fld-g} gravity or no previously reported gravity. All objects in our sample are located in the field and thus expected to have negligible extinction. If both low- and moderate-resolution spectra of the same objects are available, we use the low-resolution spectrum, leading to $\approx 82\%$ of our spectra being low-resolution. In addition, if there is more than one spectrum for the same object, we use the one with the highest S/N.

No clear distinction is seen in Figure~\ref{fig:AL13H2O} between objects with different resolution spectra and different surface gravities, again illustrating that the AL13 system is widely applicable for the near-infrared spectra of mid-M to L dwarfs. However, this classification method is not robust against reddening.  For instance, as shown in Figure~\ref{fig:AL13H2O}, a visual extinction of $A_{\rm V} \approx 10$~mag will result in the index-based AL13 spectral type being shifted later by $\approx 2$~subtypes. This change in spectral type would bring a young low-mass star (M5 spectral type with a mass of $\approx 80$~M$_{\rm Jup}$ and an age of $10$~Myr) into the substellar regime (M7 spectral type with a mass of $\approx 35$~M$_{\rm Jup}$), based on the DUSTY evolutionary models of \cite{Chabrier+2000} and the empirical effective temperature scales of \cite{Stephens+2009} and \cite{Herczeg+2014}. Additionally, a visual AL13 spectral type is difficult to obtain for highly reddened objects, since extinction alters the spectral morphology. Therefore, we are motivated to adapt the AL13 classification system for use in young star-forming regions such as Taurus ($A_{\rm V} \lesssim 30$~mag).

\subsection{Reddening-free Spectral Classification}
\label{subsec:rfspt}

The behavior of the H$_{2}$O spectral indices in the presence of reddening suggests a solution. Among four H$_{2}$O index colors, three of them, $W_{0}$, $W_{D}$, and $W_{2}$, are ``reddening-positive'' (i.e., mimicking later types with increasing reddening), while the other one, $W_{1}$, is ``reddening-negative'' (i.e., mimicking earlier types with increasing extinction). Though each index behaves differently as a function of reddening, the one reddening-negative index is overwhelmed by the other three reddening-positive indices when averaging to reach the final AL13 classification, which leads to a spectral type positively correlated with extinction.

Reddening effects can be cancelled out by combining \emph{one} reddening-positive color and \emph{one} reddening-negative color. We define three reddening-free indices by employing the same reddening-negative $W_{1}$:
\begin{equation}
\omega_{x} = \left( \frac{W_{x}}{A_{x}/A_{\rm V}} - \frac{W_{1}}{A_{1}/A_{\rm V}} \right) \left(  \frac{1}{|A_{x}/A_{\rm V}|} + \frac{1}{|A_{1}/A_{\rm V}|} \right)^{-1}
\label{eq:omegax}
\end{equation}
The subscript ``$x$'' here represents the three reddening-positive indices: H$_{2}$O ($x=0$), H$_{2}$OD ($x=D$), and H$_{2}$O$-2$ ($x=2$). The second term is invoked for normalization so that all three $\omega_{x}$ values roughly range from $0$ to $1$. Index colors ($W_{x}$ and $W_{1}$) are weighted by inverse extinction coefficients to cancel the extinction. Using the extinction law of \cite{Schlafly+2011}, extinction coefficients of the four H$_{2}$O-indices are $A_{0}/A_{\rm V}=-0.0105$, $A_{D}/A_{1}=+0.0099$, $A_{1}/A_{\rm V}=-0.0102$ and $A_{2}/A_{\rm V}=+0.0098$. Uncertainties for $\omega_{x}$ are propagated from the H$_{2}$O-index errors, which are calculated from the spectra in a Monte Carlo fashion. Our proposed $\omega_{x}$ is actually a general form of the reddening-free parameter $Q$ suggested by \citeauthor{Johnson+1953} (\citeyear{Johnson+1953}; see also \citealt{Hiltner+1956}; \citealt{Johnson+1958}), except that $Q$ is a combination of magnitudes in three bands ($U$, $B$, $V$) and our $\omega_{x}$ are composed of four near-infrared H$_{2}$O-bands. In the context of brown dwarf studies, reddening-free indices based on photometry have also been developed by \cite{Najita+2000b} and \cite{Allers+2010}.

\begin{figure}[t]
\begin{center}
\includegraphics[height=7.in]{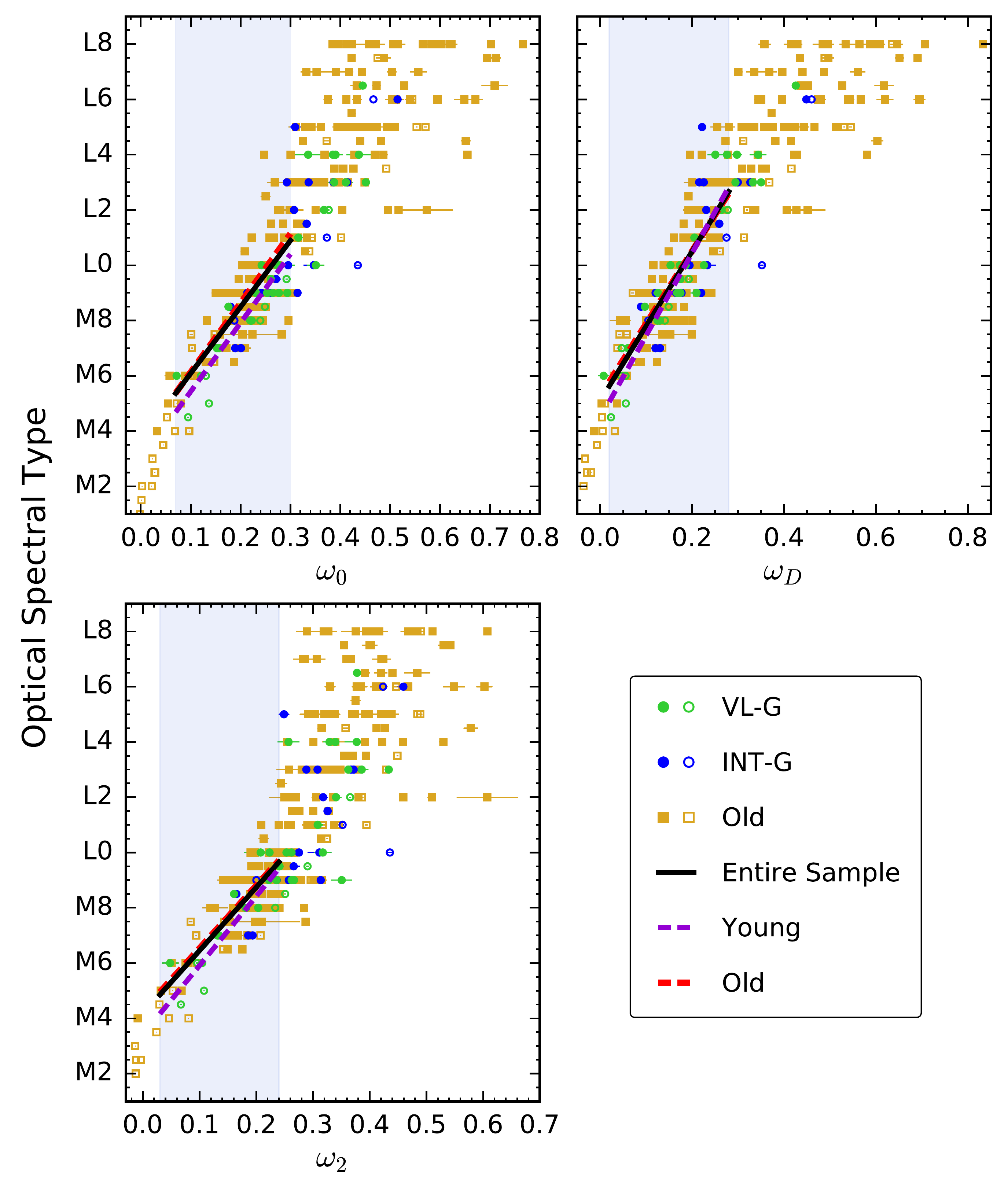}
\caption{Relation between optical spectral types and reddening-free indices $\omega_{x}$ (Equation \ref{eq:omegax}). Symbols are shown in the format of Figure \ref{fig:AL13H2O}. Spectral types pile up at $\lesssim$M4 and then monotonically increase with $\omega_{x}$, followed by a saturation for the latest type objects. Fitting results for our entire sample with reported optical spectral types ({\it black}) and two subgroups --- young (\textsc{vl-g} and \textsc{int-g}; {\it purple}) and old (\textsc{fld-g} or no reported gravity by any previous work; {\it red}) --- are overlaid. The polynomial fitting range for each $\omega_{x}$ is shown as a {\it blue} shadow. Polynomial parameters are tabulated in Table \ref{tab:omegaSpT}.  }
\label{fig:omega_SpT}
\end{center}
\end{figure}

Figure~\ref{fig:omega_SpT} examines the spectral type dependence of $\omega_{x}$ values. Optical spectral types pile up at early spectral types ($\lesssim$ M4) with similar $\omega_{x}$ values given that H$_{2}$O absorption features are weak for early-type objects. Then the optical types monotonically increase with $\omega_{x}$ followed by a saturation, indicating reddening-free spectral classification is possible as long as $\omega_{x}$ is not saturated. 

We fit polynomials to optical spectral types as a function of $\omega_{x}$, accounting for errors in both spectral types (adopted as 1 subtype) and $\omega_{x}$ by using Orthogonal Distance Regression (ODR), as implemented in the python module ``{\sf scipy.odr}.''\footnote{\url{https://docs.scipy.org/doc/scipy/reference/odr.html}.} This algorithm is more robust than normal least squares regression, which does not properly incorporate data uncertainties in the independent variables. We determine the fitting range in $\omega_{x}$ based on two factors: range width and rms about the fit. On the one hand, the fitting range of $\omega_{x}$ values should be as large as possible to be applicable for a wide range of spectral types. On the other hand, the fitting range should avoid values where $\omega_{x}$ is not well-correlated with spectral type, which would lead to a large rms of the data about the fit. Here we adopt a wide range as long as the resulting rms about the fit is $\lesssim 1$ subtype.

For each $\omega_{x}$, the order of its polynomial over the fitting range is decided by an F-test. We perform the polynomial fitting using three samples: our entire sample with reported optical types (246 objects), only the young objects (\textsc{int-g} or \textsc{vl-g}, 48 objects), and only the old objects (\textsc{fld-g} or no reported gravity, 198 objects). The spectral type uncertainty is computed by summing in quadrature the type uncertainty derived from the $\omega_{x}$ measurement uncertainties and the rms about the polynomial fit, which ascribe to a fundamental dispersion of the relation. We also tested the fitting by not incorporating optical spectral type uncertainties in the ODR algorithm (given that some objects do not have reported uncertainties in spectral type based on the SpeX Prism Spectral Libraries and the IRTF Spectral Library), which gave exactly the same results. Table~\ref{tab:omegaSpT} gives our resulting polynomial fits based on $\omega_{x}$, whose applicable range corresponds to $\approx$M5--L2 (Figure \ref{fig:omega_SpT}). This range is slightly narrower compared to the AL13 system, which covers M4--L7. For each $\omega_{x}$, the fitting results for all three samples are overall in agreement within the rms about the fits, and the typical difference between any two samples is $\lesssim 0.5$ subtype. Therefore, we recommend using the polynomial derived from the entire sample for spectral classification without distinguishing young and old targets.

We also tried a Monte Carlo method to incorporate uncertainties during the fitting by following \cite{Dupuy+2012}, instead of using the aforementioned ODR algorithm. We enlarged our data by drawing $10^{4}$ realizations for each data point, given its uncertainties, and then fitted this expanded sample of $N \times 10^{4}$ points using polynomials chosen by F-tests. Over the same fitting range in $\omega_{x}$, this approach differs from the ODR-based method by smaller than 0.5 subtype.  However, the ODR algorithm chooses a linear fit whose extrapolation follows the remaining data out of the fitting range for each $\omega_{x}$, while the Monte Carlo method chooses a $\geqslant 10^{\rm th}$--order of polynomial that quickly diverges at the edge of the applicable range. When we force the Monte Carlo polynomial to have the same order as the ODR one, their differences are typically smaller than 0.1 subtype. We adopt the ODR method to obtain the reddening-free spectral classification.

In principle, since our proposed $\omega_{x}$ is defined to cancel the extinction, the same purpose can be achieved by combining two reddening-positive indices (i.e., substituting $W_{1}$ and $A_{1}$ in Equation \ref{eq:omegax} with an reddening-positive index $W_{y}$ and $A_{y}$ with $y\neq x$). However, we found that the dependence between spectral types and the reddening-free indices defined in this way is too weak to establish a well-defined relation, and thus we do not include them in our method. 

We derive the final near-infrared spectral types and uncertainties from the weighted average of all $\omega_{x}$-based spectral types, as long as their $\omega_{x}$ are in the applicable fitting ranges (Table~\ref{tab:omegaSpT}). In addition, the irreducible error in our spectral types is described by
\begin{equation}
\sigma_{\rm floor} = {\rm min}\left({\rm rms}_{0}, {\rm rms}_{D}, {\rm rms}_{2}\right)
\end{equation}
where rms$_{x}$ is the rms about the polynomial fit of $\omega_{x}$ tabulated in Table~\ref{tab:omegaSpT}. If the spectral type uncertainty of an object computed from the weighted average is smaller than $\sigma_{\rm floor}$, then we will adopt $\sigma_{\rm floor}$ as the final uncertainty.

\subsection{Extinction Determination}
\label{subsec:extinction}

Measurements of extinction usually involve comparing observed colors \citep[e.g., $V-R_{C}$;][]{Gullbring+1998,Calvet+2004} or near-infrared spectral slopes \citep[e.g.,][]{Luhman+2017} that are representative of stellar photospheres with the intrinsic values at given spectral types defined by field-age and/or young dwarfs \citep[e.g.,][]{Strom+1994,Briceno+1998,Luhman+2000,White+2001,Pecaut+2013}. However, without the ability to determine a reddening-free spectral type, in principle these approaches could lead to an incorrect extinction. In addition, young late-M to early-L brown dwarfs are systematically brighter and/or redder than the field population \citep[e.g.,][]{Gizis+2012, Liu+2016, Best+2018}. Therefore, the common method for extinction determination may not be directly applicable for young brown dwarfs. As another approach, some authors fit the observed spectra with reddened spectral templates based on field dwarfs \citep[e.g.,][]{Rizzuto+2015}, but again this may not be ideal for fitting young lower-gravity targets.

Here we suggest two methods for extinction determination. One is based on color-color diagrams using H$_{2}$O indices, and the other is based on the intrinsic optical--near-infrared colors defined by our reddening-free spectral types.

\subsubsection{H$_{2}$O Color--Color Diagrams}
\label{subsubsec:WxCCD}

\begin{figure}[t]
\begin{center}
\includegraphics[height=5.in]{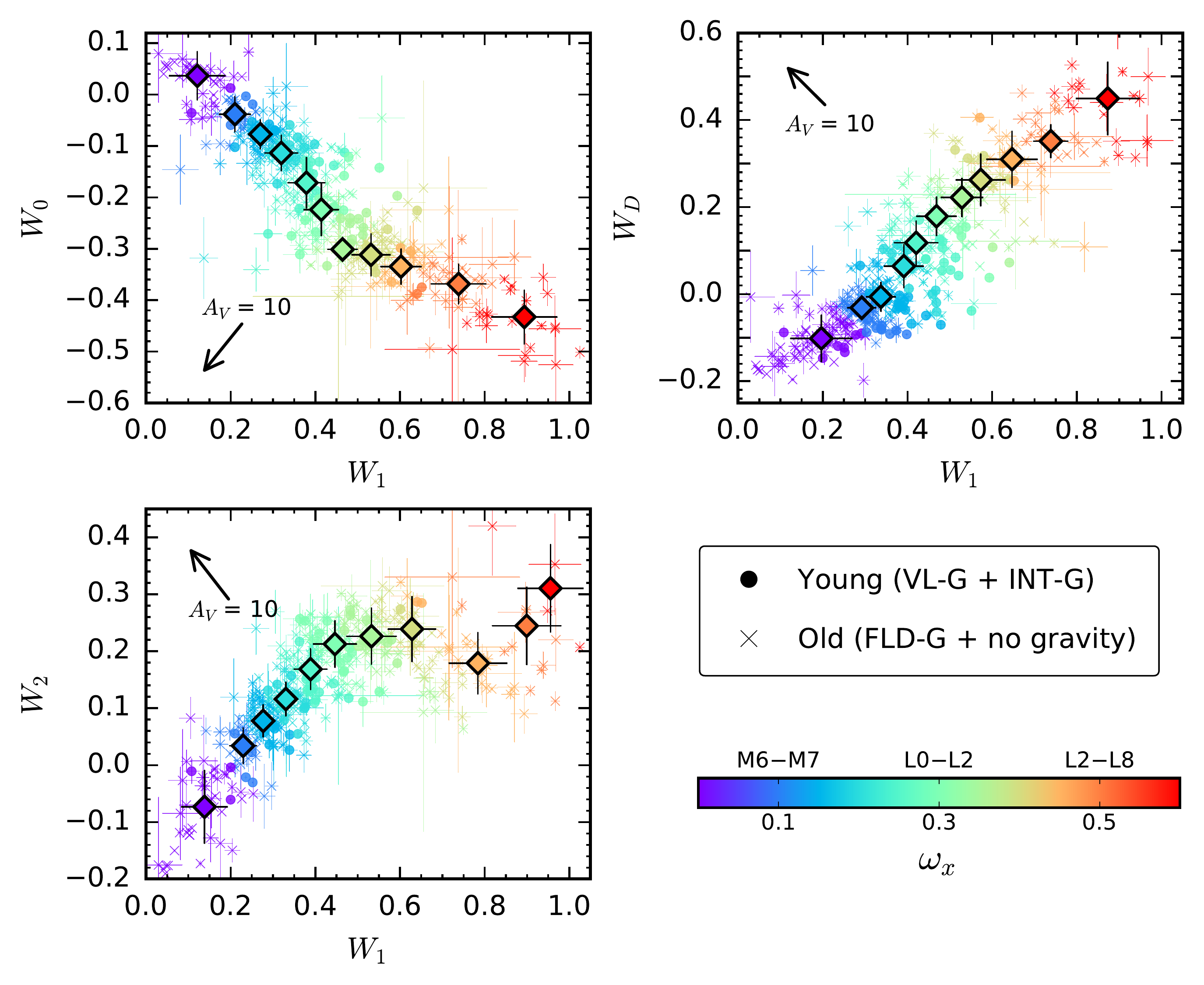}
\caption{H$_{2}$O color-color diagrams, using the entire sample (408 objects) described in Section~\ref{subsec:revisit}. Circles indicate young objects with low and intermediate gravity (\textsc{vl-g} and \textsc{int-g}), and crosses indicate old objects with field gravity (\textsc{fld-g}) or no reported gravity. Colors are encoded by $\omega_{x}$ values. We also show a rough conversion between $\omega_{x}$ and reddening-free spectral type (see also Figure~\ref{fig:omega_SpT}) in the color-bar. Colored diamonds with error bars show the intrinsic values and uncertainties of the $\omega_{x}$ sequences, and they are tabulated in Table \ref{tab:omegax_sequence}. Extinction vectors based on the extinction law of \cite{Schlafly+2011} are roughly perpendicular to the $\omega_{x}$ sequences, implying that the dereddening is possible. }
\label{fig:omegax_seq}
\end{center}
\end{figure}

Figure~\ref{fig:omegax_seq} presents the three H$_{2}$O color-color diagrams for our sample. Each of them is constructed with one reddening-positive color $W_{x}$ and one reddening-negative color $W_{1}$. The diagrams have well-defined intrinsic sequences of reddening-free index $\omega_{x}$, and the extinction vector is roughly perpendicular to the sequences, implying that these diagrams can be used to determine extinctions.

We first define a quantitative sequence, as shown in Table~\ref{tab:omegax_sequence}, for each color-color diagram as a function of $\omega_{x}$ only using old objects ($353$ objects) in our total sample ($408$ objects; Section~\ref{subsec:revisit}). We divide each sequence into bins based on $\omega_{x}$ values, with each bin spanning $0.05$ in $\omega_{x}$ except for two open bins at the tails, so that most bins contain $\gtrsim 20$ objects. A bin size of $0.05$ in $\omega_{x}$ corresponds to $\approx$1.0--1.5 subtypes (Table~\ref{tab:omegaSpT}). Each intrinsic sequence can be defined by three parameters, $\omega_{x}$, $W_{x}$, and $W_{1}$, with each parameter described by the median values in the corresponding bin. Uncertainties of the $W_{x}$ and $W_{1}$ values in each bin are computed from the standard deviations, whose typical value is $\approx 0.04$~mag in index-color and corresponds to a visual extinction of $A_{\rm V} \approx 4$~mag, which is the limiting uncertainty of this method. 

\begin{figure}[t]
\begin{center}
\includegraphics[height=8.in]{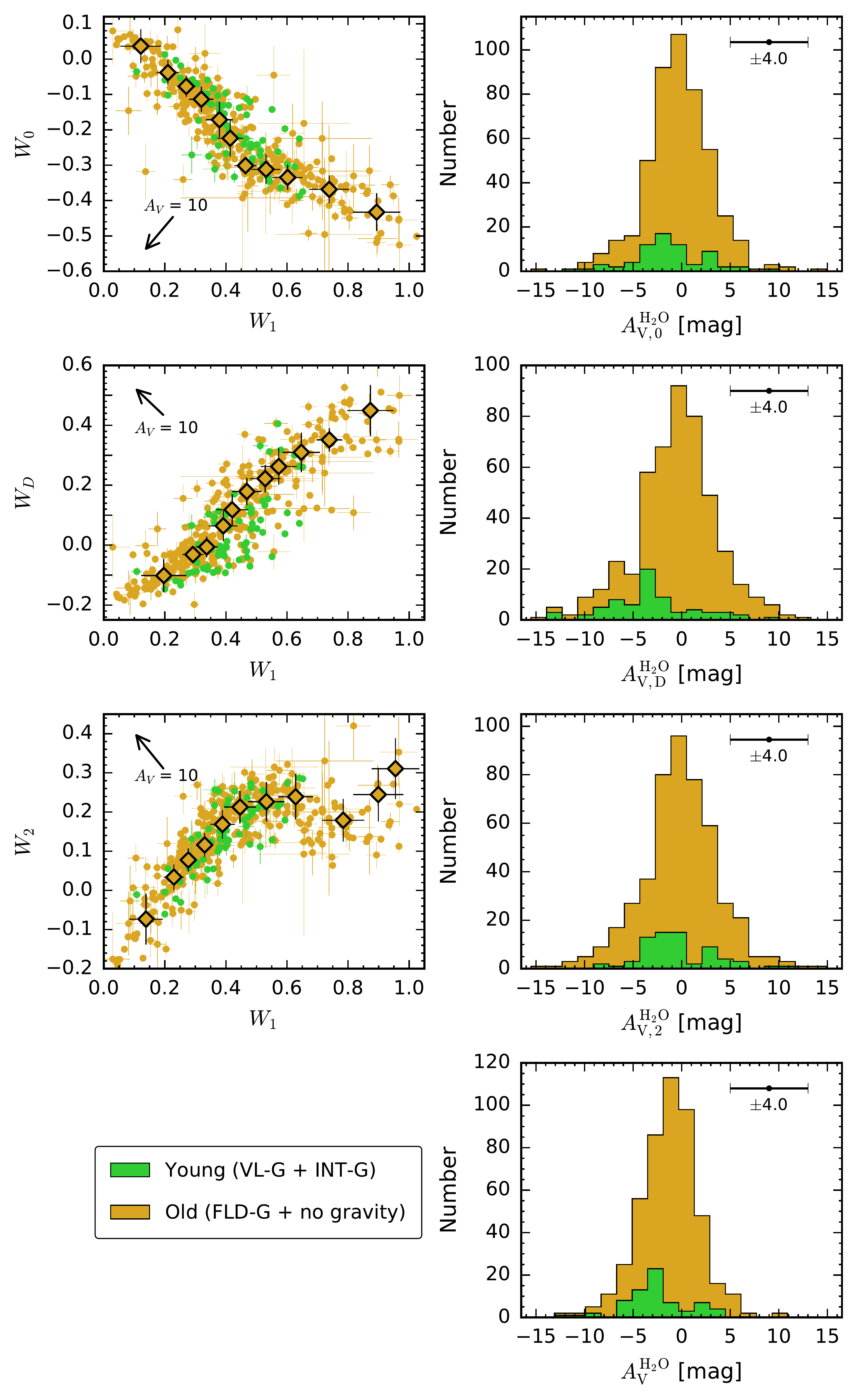}
\caption{Left: H$_{2}$O color-color diagrams, as shown in Figure~\ref{fig:omegax_seq}. We use {\it green} for young objects with low and intermediate gravity (\textsc{vl-g} and \textsc{int-g}) and use {\it orange} for old objects with field gravity (\textsc{fld-g}) or no reported gravity. Extinction vectors are based on the extinction law of \cite{Schlafly+2011}. The $\omega_{x}$ sequences in each diagram are shown as diamonds. Right: Histograms of extinctions ($A_{{\rm V,}x}$ and $A_{\rm V}^{\rm H_{2}O}$) for the entire sample (408 objects; Section~\ref{subsec:revisit}), as plotted in the left panels, derived from intrinsic $\omega_{x}$ sequences. The histogram of the final extinction $A_{\rm V}^{\rm H_{2}O}$ is shown at the bottom panel. A typical uncertainty of $4.0$~mag is shown at the top. Young objects have negative extinctions with a median of $\approx -1.97$~mag in their $A_{\rm V}^{\rm H_{2}O}$. We therefore add $2.0$~mag to $A_{\rm V}^{\rm H_{2}O}$ of young objects as a simple correction (Section~\ref{subsubsec:WxCCD}).}
\label{fig:omegax_seq_comp_young_old}
\end{center}
\end{figure}

As shown in Figure~\ref{fig:omegax_seq_comp_young_old}, young objects have intrinsically bluer H$_{2}$O-band spectral slopes than old objects, as they are mostly located blueward of the $\omega_{x}$ sequence relative to the extinction vector in each color-color diagram. However, young $\omega_{x}$ sequences cannot be reliably built, due to the relatively small number of young objects (55 in total) in our sample. Therefore, later we derive a simple correction factor for young objects.

To measure the extinction of an object, we first interpolate Table~\ref{tab:omegax_sequence} based on the object's measured $\omega_{x}$ to obtain the intrinsic H$_{2}$O indices and their uncertainties. Then we calculate the displacement $\overrightarrow{d_{x}}$ from the intrinsic $(W_{1,\rm int}, W_{x,\rm int})$ to the measured $(W_{1,\rm meas}, W_{x,\rm meas})$ values --- namely, $\overrightarrow{d_{x}} = (W_{1,\rm meas} - W_{1,\rm int}, W_{x,\rm meas} - W_{x,\rm int})$. We then project $\overrightarrow{d_{x}}$ into the direction of the corresponding extinction vector $\overrightarrow{a_{x}} = (A_{1}/A_{\rm V}, A_{x}/A_{\rm V})$, whose length corresponds to an extinction of $A_{\rm V} = 1$~mag using the extinction law of \cite{Schlafly+2011}, and thus compute the V-band extinction:
\begin{equation} \label{eq:Avx}
A_{V,x} = \frac{\overrightarrow{d_{x}} \cdot \overrightarrow{a_{x}}}{\left| \overrightarrow{a_{x}} \right|^{2}}
\end{equation}
The uncertainties of $\omega_{x}$, $W_{x}$ and $W_{1}$ values are incorporated in a Monte Carlo fashion into the extinction calculation. The final extinction ($A_{\rm V}^{\rm H_{2}O}$) and its uncertainty for an object are calculated from the weighted average of the reddenings $A_{V,x}$ computed from an object's three $\omega_{x}$ values. In addition, if the final extinction uncertainty is smaller than the irreducible error (i.e., $A_{\rm V} = 4$~mag), then we adopt the latter. We notice that the irreducible uncertainty is usually adopted for an object as long as its near-infrared spectrum has a S/N of $\geqslant 30$ per pixel in $J$ band. 

We compute the extinctions for all objects in our sample and compare the results between young and old subsets (Figure~\ref{fig:omegax_seq_comp_young_old}). Most young objects have negative $A_{V,x}$ with a median of $-1.97$~mag, again illustrating their slight intrinsic blueness relative to old objects. Therefore, we {\it add} $2.0$~mag to $A_{\rm V}^{\rm H_{2}O}$ values of young objects as a correction. If the youth of an object is unknown, then we assume a young age and then iterate, as described in Section~\ref{subsec:classification_recipe}.

As another possible approach, instead of dividing the sample into several $\omega_{x}$ bins, we also tried directly fitting $W_{x}$ as a polynomial function of $W_{1}$ in each H$_{2}$O color-color diagram to define the intrinsic sequence, using the ODR algorithm with F-tests. In each diagram, we compute the reddening of an object by shifting its measured $(W_{1,\rm meas}, W_{x,\rm meas})$ values back to the polynomial curve along the extinction vector, which involves solving a polynomial equation\footnote{\label{footnote:poly} For each H$_{2}$O color-color diagram, we assume the intrinsic polynomial sequence is $p_{x}(W_{1})$. The function $a(W_{1})$ expresses a straight line that passes through the measured $(W_{1,\rm meas}, W_{x,\rm meas})$ of an object and has a slope of $A_{x}/A_{1}$, corresponding to the extinction vector. Then the intrinsic $(W_{1,\rm int}, W_{x,\rm int})$ values for the object can be obtained by solving the polynomial equation:
\begin{equation}
p_{x}(W_{1}) - a(W_{1}) = 0
\label{eq:pW}
\end{equation}
By expressing the displacement from the intrinsic to the measured $(W_{1}, W_{x})$ values as $\overrightarrow{d_{x}^{\prime}} = (W_{1,\rm meas} - W_{1,\rm int}, W_{x,\rm meas} - W_{x,\rm int})$ and the extinction vector as $\overrightarrow{a_{x}} = (A_{1}/A_{\rm V}, A_{x}/A_{\rm V})$, we thus compute the V-band extinction as
\begin{equation}
A_{V,x}^{\prime} = \left| \overrightarrow{d_{x}^{\prime}} \right| \big/ \left| \overrightarrow{a_{x}} \right|
\end{equation}
For the purpose of the comparison with the $\omega_{x}$-based method, we use a first-order polynomial to fit all three $p_{x}(W_{1})$. However, an order of 4, 2, and 3, is found for $p_{0}(W1)$, $p_{D}(W_{1})$, and $p_{2}(W_{1})$, respectively, based on the F-tests. Solving the polynomial equation (Equation~\ref{eq:pW}) with the order over 2 would be very complicated in practice and thus we disfavor this approach.}.
We calculate the extinction uncertainty by incorporating the errors of both the H$_{2}$O-band index measurements and the polynomial coefficients in a Monte Carlo fashion. The final extinction and uncertainty are determined from the weighted average of values based on three $W_{x}$. The results from this method is consistent within uncertainties with the previous $\omega_{x}$-based approach. We therefore adopt the $\omega_{x}$-based approach to derive the reddening from H$_{2}$O color-color diagrams, because it only requires an interpolation and a dot product, rather than solving a polynomial equation.

\subsubsection{Intrinsic Optical -- Near-Infrared Colors}
\label{subsubsec:opt_NIR_color}

The extinction of an object can also be determined by comparing the observed optical--near-infrared colors with intrinsic values for unreddened objects with similar spectral types, assuming spectral types can be measured free of extinction effects. Here we use the red optical photometry from PS1 (i.e., $i_{\rm P1}$, $z_{\rm P1}$, and $y_{\rm P1}$). This is because (1) the optical data are more extinction-sensitive compared with the infrared data and are thus more robust indicators of reddening; (2) substellar SEDs peak at near-infrared wavelength, and thus bluer photometry ($g_{\rm P1}$ and $r_{\rm P1}$) is not always available, as objects are too faint; (3) the contamination by excess emission from magnetospheric accretion shocks is reduced at longer wavelengths \citep[e.g.,][]{Gullbring+1998, Najita+2000a}. We combine the PS1 red photometry with $J_{\rm 2MASS}$, as the latter minimizes the contamination by thermal emission from possible circumstellar disks.

We first build intrinsic optical--near-infrared color sequences for old and young dwarf populations, respectively, using $i_{\rm P1}-J_{\rm 2MASS}$,  $z_{\rm P1}-J_{\rm 2MASS}$ and $y_{\rm P1}-J_{\rm 2MASS}$, as functions of \emph{literature} spectral types. Intrinsic colors of old field M, L, and T dwarfs are provided by \cite{Best+2018}, who constructed a sample from DwarfArchives\footnote{\url{http://spider.ipac.caltech.edu/staff/davy/ARCHIVE/index.shtml}.}, \cite{West+2008}, and numerous literature sources over the span of 2012--2016.

\begin{figure}[t]
\begin{center}
\includegraphics[height=5.in]{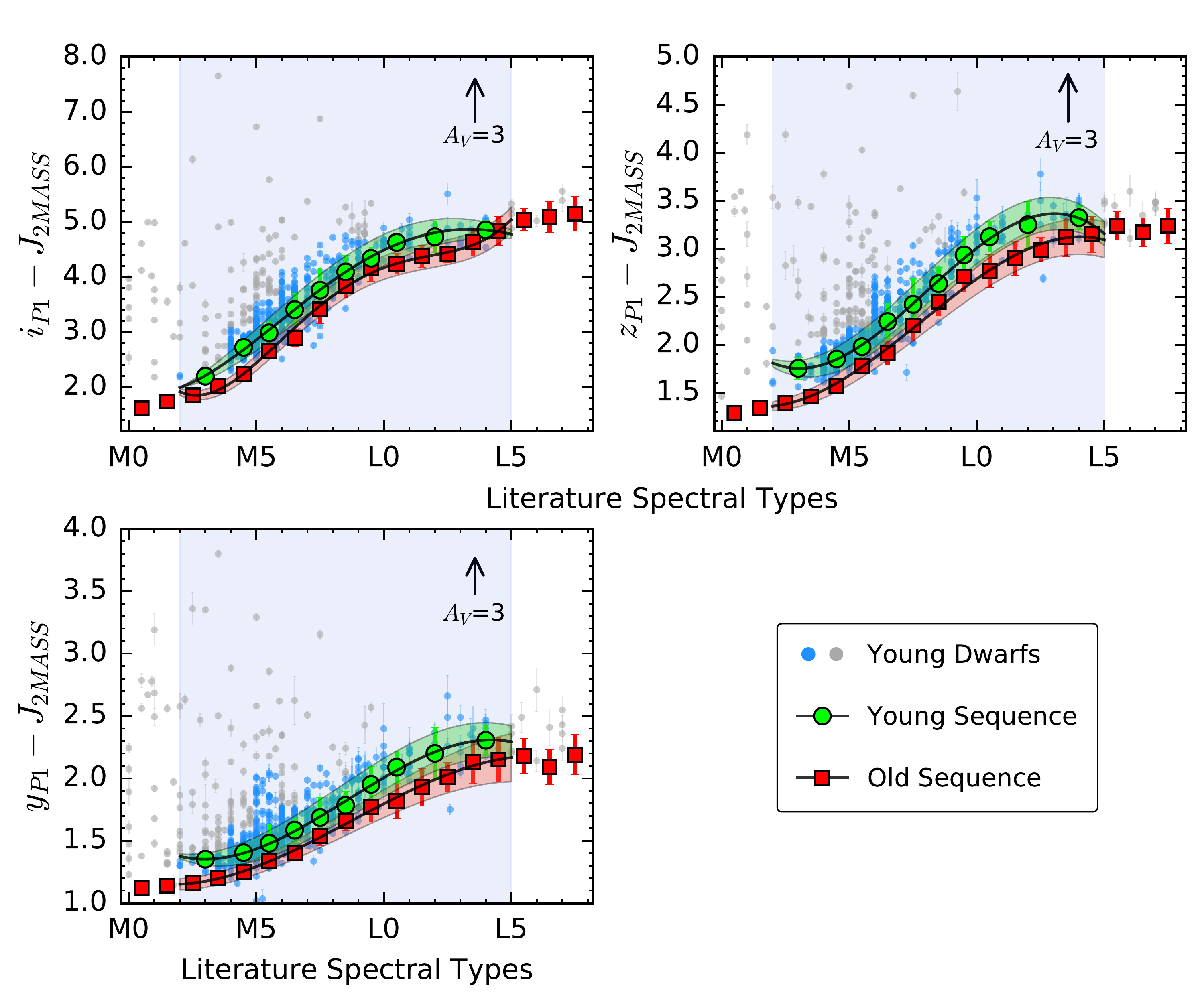}
\caption{ Optical--near-infrared colors vs. literature spectral type diagram for the young dwarf population described in Section \ref{subsubsec:opt_NIR_color}. {\it Green} circles show the intrinsic color sequence for young objects in each diagram with spectral types in [M2,L5) ({\it blue} shadow). Young objects that are included ({\it blue}) and excluded ({\it gray}) for defining the intrinsic color sequence are also shown. {\it Red} squares are for the field dwarfs established by \cite{Best+2018}. Polynomial fits of these two sequences are shown as solid lines, and their coefficients are tabulated in Table \ref{tab:instrinsic_color}. The typical difference between young and old color sequences is equivalent to a visual extinction of $A_{\rm V} \approx 1.3$ mag. The extinction vector corresponds to $A_{\rm V} = 3$ mag using the extinction law of \cite{Schlafly+2011}.}
\label{fig:color_seq}
\end{center}
\end{figure}

The young population is assembled from (1) known members of two star-forming regions, Taurus \citep[193 objects from][]{Best+2017, Luhman+2017} and Upper Scorpius \citep[629 objects from][]{Luhman+2012,Dawson+2014,Rizzuto+2015,Best+2017}; and (2) 95 field objects with reported youth but without any nearby stellar companions from the sample used by \cite{Best+2018}. All of the young objects with surface gravity of \textsc{vl-g} and \textsc{int-g} described in Section \ref{subsec:rfspt} are included here. We only select objects with M and L spectral types and with good-quality detections in $i_{\rm P1}/z_{\rm P1}/y_{\rm P1}$ and $J_{\rm 2MASS}$ bands (photometric qualities are defined in Section~\ref{subsec:phot_criteria}), leading to a sample of $917$ objects (Figure~\ref{fig:color_seq}). We divide the sample into different bins using their literature spectral types. There are relatively fewer young early-M and late-L objects in our sample, due to PS1 saturations and the rarity, respectively. Thus, we define color sequences spanning M2--L4 spectral types. The bin size is $1$ subtype for most bins but expanded to $2$ subtypes for objects in the [M2, M4) and [L3,L5), in order to include $\gtrsim 15$ objects per bin. 

\begin{figure}[t]
\begin{center}
\includegraphics[height=5.in]{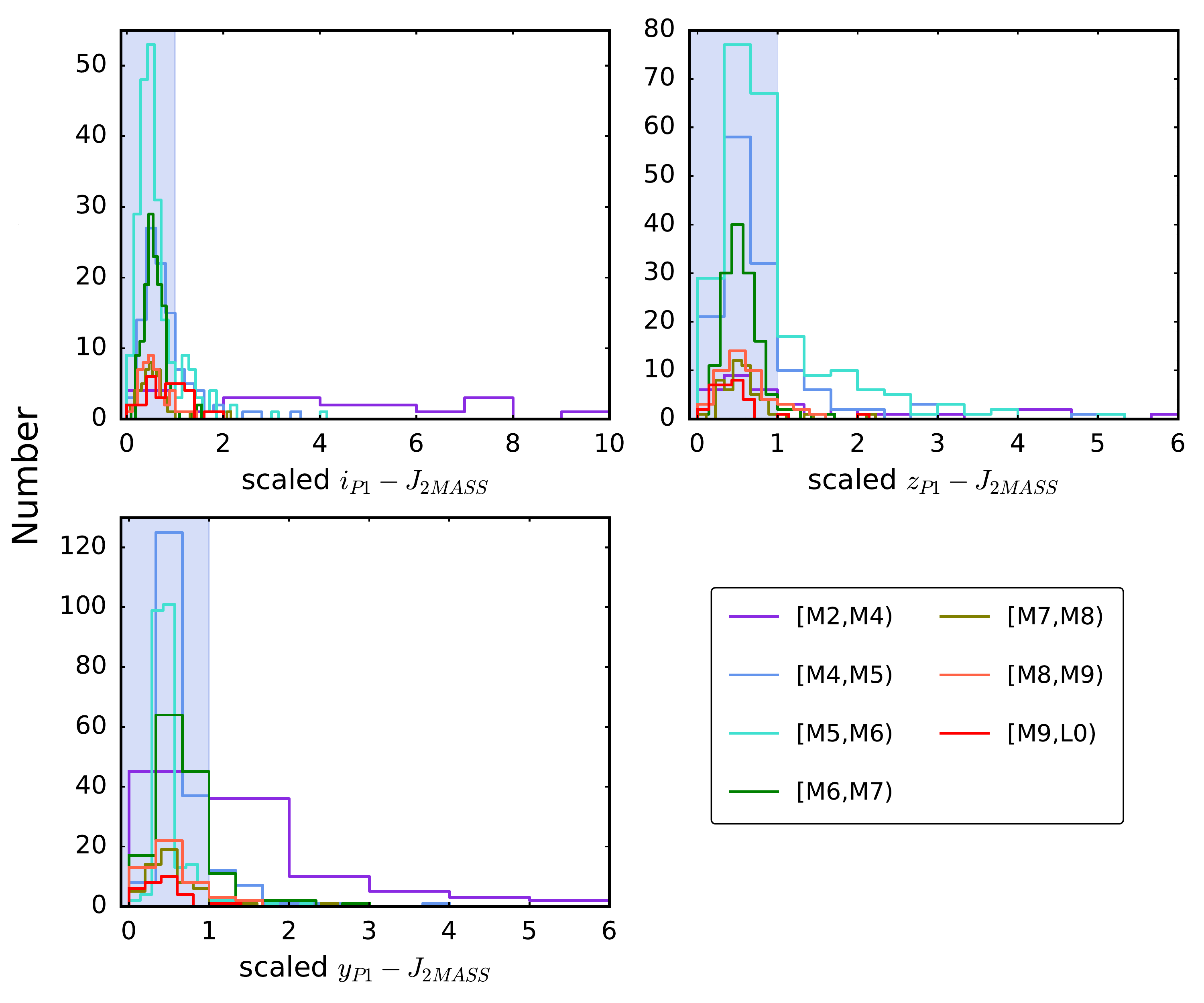}
\caption{ Stacked color distributions of young dwarf populations (described in Section \ref{subsubsec:opt_NIR_color}) in spectral type bins in [M2,L0). Colors are plotted as different spectral type bins. We consider the color distribution in each spectral type bin as a composite of a blue locus ({\it blue} shadow), located around the mode of the distribution, and red outliers, which result from the reddening in dusty star-forming regions. The blue locus in each bin is defined by a critical color $C_{\rm cr}$ (Section~\ref{subsubsec:opt_NIR_color}), and objects bluer than this color are used to construct the intrinsic color sequence for young objects (Figure \ref{fig:color_seq}). Here for each optical--near-infrared color, we scale the color distributions from different spectral type bins so that their minimal colors $C_{\rm min}$ are all zero, and their critical colors $C_{\rm cr}$ are all one.   }
\label{fig:hist_color}
\end{center}
\end{figure}

Since some of these young objects suffer from reddening, we need to pick up objects with no extinction to define the intrinsic young color sequences. For each spectral type bin in [M2,L0), we consider the color distribution as a composite of a blue locus, located around the mode of the distribution, and red outliers, which result from variable reddening in dusty star-forming regions (Figure~\ref{fig:hist_color}). Assuming the blue locus describes the intrinsic colors of young objects, we define the young sequences by choosing objects in each bin with colors bluer than a critical value ($C_{\rm cr}$), whose difference from the distribution mode ($C_{\rm mode}$) of that bin is the same as the difference between the mode and the minimal color ($C_{\rm min}$; i.e., $C_{\rm cr} - C_{\rm mode} = C_{\rm mode} - C_{\rm min}$). The intrinsic color and uncertainty are calculated as the median and the standard deviation of the blue locus. The calculated intrinsic color and the mode in each bin are consistent within uncertainties. For objects with spectral types of [L0, L5), we use the entire subsample in each bin to define corresponding intrinsic values, because there is no clear set of red outliers in the color distributions, and only $\lesssim 10\%$ of these objects are located in star-forming regions.

In Figure~\ref{fig:color_seq}, we plot the optical--near-infrared colors of the young and old populations as functions of spectral type. We fit polynomials to both young and old color sequences as a function of spectral type using the ODR algorithm, with the polynomial orders chosen by F-tests and incorporating the uncertainties in the intrinsic colors as described above and spectral types (adopted as the half width of the bin, 0.5 or 1 subtype). In addition, we fit the intrinsic color uncertainties as a function of spectral type, incorporating only the spectral type uncertainties during the fitting process (Table~\ref{tab:instrinsic_color}). The typical difference between young and old sequences for all three colors corresponds to an $A_{\rm V} \approx 1.4$~mag, which means that directly comparing the color of a young object to those of old field dwarfs, as is common in previous work, may have systematically overestimated the extinction. For each color sequence of both young and old population, the typical intrinsic color uncertainty is equivalent to an extinction of $A_{\rm V} \approx 0.85$~mag, and we adopt this as the irreducible error of this method. This error is $\gtrsim 3 \times$ larger than the rms about the polynomial fits of all three colors as a function of spectral type. Therefore, we ignore the fitting rms and only adopt the uncertainties computed from our polynomial fits for the extinction measurements.

The above intrinsic sequences are defined based on spectral types from literature. A conversion is still needed from our proposed reddening-free spectral classification (SpT$_{\omega}$; Section \ref{subsec:rfspt}) to the literature spectral types (SpT$_{\rm lit}$), so that one can derive the intrinsic colors. To determine such calibration, we employ (1) the total sample mentioned in Section \ref{subsec:revisit}, i.e., a combination of the AL13 sample, the SpeX Prism Library, and the IRTF Spectral Library; and (2) the objects with available near-infrared spectra of the young population used to define the intrinsic color sequences, i.e., the blue locus of [M2,L0) and all [L0,L5) dwarfs. We compute their reddening-free spectral types from their near-infrared spectra using the ``entire-sample'' polynomial tabulated in Table~\ref{tab:omegaSpT}. Then we only select the 324 objects with well-established SpT$_{\omega}$ ($\approx$M5--L2; i.e., spectral types with measured $\omega_{x}$ in applicable fitting ranges). In addition, if an object has both optical and near-infrared spectral types from literature, then we only adopt its optical type as SpT$_{\rm lit}$. By performing a ODR-based linear fitting, we obtain a conversion as
\begin{equation}
{\rm SpT}_{\rm lit} = 1.23 \times {\rm SpT}_{\omega} - 2.24, \quad {\rm rms}=1.04
\label{eq:SpT_lit_omega}
\end{equation}
where the numerical spectral type SpT is defined to be 0 for M0, 5 for M5, 10 for L0, and so on. This tight relation yields a systematic difference of $\lesssim 1$ subtype between SpT$_{\omega}$ and {\rm SpT}$_{\rm lit}$ in M5--L2. Since the sample we used here has no reddening, Equation~\ref{eq:SpT_lit_omega} confirms that our spectral classification is consistent with literature types in the zero-extinction case. When using Equation~\ref{eq:SpT_lit_omega} to convert the SpT$_{\omega}$ of an object into SpT$_{\rm lit}$, if the resulting uncertainty in SpT$_{\rm lit}$ is smaller than the rms about the polynomial fitting (i.e., 1.04 subtypes), then we adopt the fitting rms as the final uncertainty.

To determine the extinction of an object, we first measure its reddening-free spectral type and convert to literature type  (Equation~\ref{eq:SpT_lit_omega}). Then we determine the intrinsic colors of $i_{\rm P1}-J_{\rm 2MASS}$, $z_{\rm P1}-J_{\rm 2MASS}$, and $y_{\rm P1}-J_{\rm 2MASS}$ based on polynomials in Table~\ref{tab:instrinsic_color} corresponding to its youth. If its youth is unknown, then we assume a young age and iterate, as described in Section~\ref{subsec:classification_recipe}. An extinction is thus obtained from the difference between the intrinsic and the measured color. We incorporate the uncertainties of SpT$_{\rm lit}$, intrinsic colors, and observed colors using Monte Carlo method. The final extinction ($A_{\rm V}^{\rm OIR}$) of the object is calculated from the weighted average of all color-based extinctions. In addition, if the uncertainty from the weighted average is smaller than the irreducible error (i.e., $A_{\rm V}=0.85$ mag), then we adopt the latter. We notice that the irreducible error is usually adopted for an object as long as its near-infrared spectrum has an S/N of $\geqslant30$ per pixel in $J$ band and at least two out of $i_{\rm P1}$, $z_{\rm P1}$, and $y_{\rm P1}$ photometry have good qualities.

\subsubsection{Final Extinction}
\label{subsub:finalext}

For each object we compute two extinction measurements: (1) based on intrinsic $\omega_{x}$ sequences in H$_{2}$O color-color diagrams and (2) based on the intrinsic optical--near-infrared color as a function of spectral type. The first method does not require the spectral type of an object but produces a more uncertain extinction due to the large intrinsic scatter of the $\omega_{x}$ sequence, corresponding to $A_{\rm V} = 4$~mag (Section \ref{subsubsec:WxCCD}). The second method produces a more accurate extinction, due to a smaller intrinsic scatter of the color sequence of $A_{\rm V} = 0.85$~mag (Section \ref{subsubsec:opt_NIR_color}). However, it is only applicable for objects of $\approx$M5--L2 where our reddening-free spectral classification is well-defined (Section \ref{subsec:rfspt}). 

These two methods are actually suited for different observational datasets. The first method is applicable for an object if (1) only the near-infrared spectra are available, or (2) both near-infrared spectra and PS1+2MASS photometry are available but the reddening-free spectral type is ill-defined, since its $\omega_{x}$ values are all out of the valid fitting ranges (Table~\ref{tab:omegaSpT}). While the method based on intrinsic optical--near-infrared color sequences produces more precise extinctions, it is only usable if the reddening-free spectral type of an object is in $\approx$M5--L2. We recommend using the extinction measured from the intrinsic optical--near-infrared color sequences as long as it is available.

Negative values could be produced by both of our extinction determinations, since the zero-points of the reddening in our method are defined based on a statistical approach. We recommend keeping the negative reddening for the purpose of statistical comparisons (e.g., comparing $A_{\rm V}^{\rm H_{2}O}$ with $A_{\rm V}^{\rm OIR}$ and/or comparing our extinction measurements with literature values; see Section~\ref{subsubsec:extinction}), while we suggest replacing negative values with zeros when extinctions are used in astrophysical conditions (e.g., dereddening; see Section~\ref{subsec:color}).

\subsection{Gravity Classification}
\label{subsec:youth_logistics}

Following the AL13 classification system \citep[see also][]{Allers+2007}, we determine the youth of ultracool dwarfs based on five gravity-sensitive spectral indices measured from their dereddened near-infrared spectra: FeH$_{z}$ (0.99~$\mu$m), VO$_{z}$ (1.06~$\mu$m), FeH$_{J}$ (1.20~$\mu$m), K\textsc{i}$_{J}$ (1.24~$\mu$m), and H-cont ($H$-band continuum at 1.56~$\mu$m). These indices are defined as the flux ratios in and out of specific spectral features (Table~4 and Equation~1 in AL13). With lower gravity, young objects maintain a photosphere lying at lower pressure and therefore have weaker FeH and K\textsc{i} bands, stronger VO band, and distinctive triangular H-band continuum shapes. Based on the AL13 system, we assign each target with a gravity score of ``2'' for low gravity (\textsc{vl-g}, with ages $\lesssim 30$~Myr), ``1'' for intermediate gravity (\textsc{int-g}, with ages $\approx 30-200$~Myr), and ``0'' for field gravity (\textsc{fld-g}, with ages $\gtrsim 200$~Myr). Here we obtain the rough conversion from gravity classifications to ages based on Table~11 of AL13 and Figure~21 of \cite{Liu+2016}, both of which summarized the gravity classes of several young ultracool dwarfs with independent age measurements. We compute the uncertainty in gravity scores following \cite{Aller+2016} by propagating the errors of spectral indices, as well as extinctions, in a Monte Carlo fashion, and we allow negative extinctions for dereddening processes. Gravity scores are defined only for objects with spectral type $\geqslant$M6 (AL13). Again, \textsc{vl-g} and \textsc{int-g} objects are referred to as young objects, while \textsc{fld-g} are old objects (Section~\ref{subsec:revisit}).

\subsection{Implementation of Our Classification Scheme}
\label{subsec:classification_recipe}

\begin{figure}[t]
\begin{center}
\includegraphics[height=7.in]{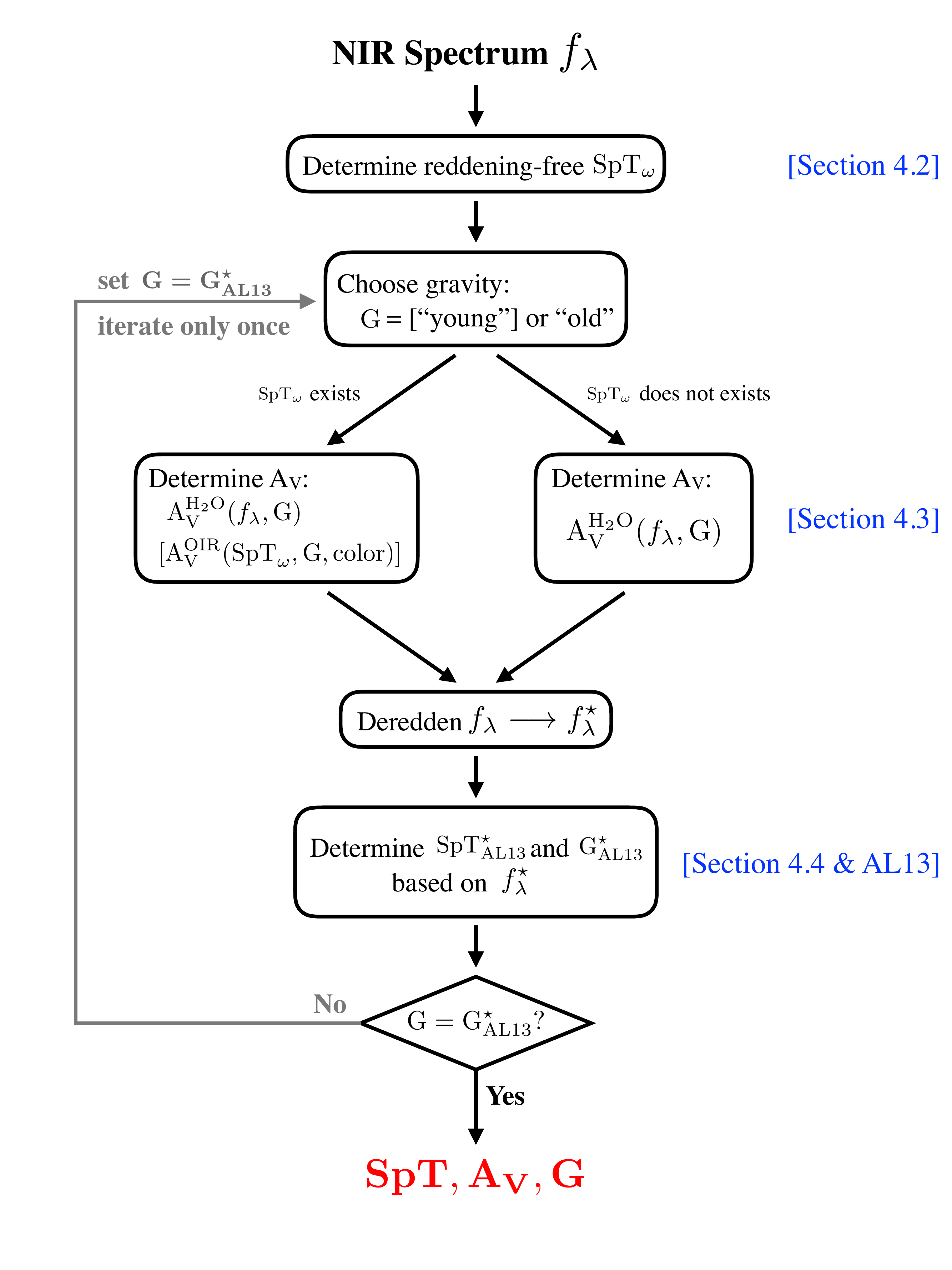}
\caption{Flowchart illustrating our new classification scheme, as described in Section~\ref{subsec:classification_recipe}. Variables are the same as in the text, with the addition of G for gravity classification and G$^{\star}_{\rm AL13}$ for gravity classification derived from the dereddened spectra using the AL13 system. We use ``[]'' to mark our recommended option for the initially assumed gravity and when choosing an extinction value for dereddening.}
\label{fig:flowchart}
\end{center}
\end{figure}

We summarize the implementation of our classification method as follows, with a flowchart given in Figure~\ref{fig:flowchart}.

\begin{enumerate}[topsep=0pt,itemsep=1ex,partopsep=1ex,parsep=0ex]
\item[1.] Compute the reddening-free spectral type SpT$_{\omega}$ by measuring our $\omega_{x}$ indices from the observed near-infrared spectra (Equation~\ref{eq:omegax}) and converting the $\omega_{x}$ values into SpT$_{\omega}$ based on Table~\ref{tab:omegaSpT}. The SpT$_{\omega}$ may not exist if all $\omega_{x}$ values are out of the valid fitting ranges.
\item[2.] Compute the extinction
\begin{enumerate}[topsep=0pt,itemsep=0ex,partopsep=1ex,parsep=0ex]
\item[(a)] $A_{\rm V}^{\rm H_{2}O}$ from the intrinsic H$_{2}$O color-color sequences as a function of $\omega_{x}$ (Table~\ref{tab:omegax_sequence}), and/or,
\item[(b)] $A_{\rm V}^{\rm OIR}$ from the intrinsic optical--near-infrared color sequences as a function of spectral type (Table~\ref{tab:instrinsic_color}). The $A_{\rm V}^{\rm OIR}$ extinction can only be computed when SpT$_{\omega}$ exists.
\end{enumerate}
\indent Note that youth information is needed for measuring extinction, as we need to correct $A_{\rm V}^{\rm H_{2}O}$ by $2$~mag for young objects (Section~\ref{subsubsec:WxCCD}) and we compare an object's observed optical--near-infrared colors to either young or old intrinsic color sequences to compute $A_{\rm V}^{\rm OIR}$ (Table~\ref{tab:instrinsic_color}). However, youth assessment (i.e., gravity classification) can only be obtained accurately after dereddening. Therefore, iteration might be needed for both of our extinction measurements. We suggest assuming a young age at first to derive the extinction, then determining the gravity classification and iterating if a contradiction occurs (i.e., if the resulting gravity classification is \textsc{fld-g}, rather than \textsc{vl-g} or \textsc{int-g} as initially assumed).
\item[3.] Deredden the spectrum (using $A_{\rm V}^{\rm OIR}$ when SpT$_{\omega}$ exists, otherwise using $A_{\rm V}^{\rm H_{2}O}$) and compute the AL13 spectral type SpT$^{\star}_{\rm AL13}$ and gravity classification from the dereddened spectra.
\item[4.] Adopt the spectral type from the reddening-free spectral type SpT$_{\omega}$ when it exists (SpT$_{\omega}$$\approx$M5--L2). Otherwise, adopt the spectral type from the AL13 spectral type SpT$^{\star}_{\rm AL13}$ measured from the dereddened spectra (SpT$^{\star}_{\rm AL13}$$\approx$M4--L7).
\item[5.] Adopt the extinction $A_{\rm V}^{\rm OIR}$ when SpT$_{\omega}$ exists. Otherwise, adopt $A_{\rm V}^{\rm H_{2}O}$ (Section~\ref{subsub:finalext}).
\item[6.] Adopt the gravity classification derived from the step~3, only if the youth assumption used in the step~2 is consistent with the final gravity classification before/after the iteration. Otherwise, adopt a null gravity classification.
\end{enumerate}

Note that while SpT$_{\omega}$ is applicable only for $\approx$M5--L2 dwarfs, our $A_{\rm V}^{\rm H_{2}O}$ method (step 2a) enables dereddening and thereby a reddening-free AL13 spectral type. Therefore, our spectral classification scheme works for mid-M to late-L ultracool dwarfs.

\section{Results}
\label{sec:application_discussion}

\subsection{Classifying Our Discoveries and Previously Known Objects}
\label{subsec:reclassification}

Using our new classification scheme, we obtain spectral types, extinctions, and gravity classifications for our 83 Taurus candidates with spectroscopic follow-up. We identify an object as a new member of Taurus, Pleiades, or Per~OB2, if its spectral type is $\geqslant$M6 and it has either very low (\textsc{vl-g}) or intermediate (\textsc{int-g}) surface gravity. For our [M4, M6) discoveries with no gravity classification, we tentatively include them as possible new Taurus members that worth passing further follow-up for membership assessment (see Section~\ref{subsec:FCE} for our estimate of field contamination). As a brief summary, among the 83 candidates, we have thus far discovered 58 new Taurus members, 1 new Pleiades member, 13 new Per~OB2 members, and 11 reddened early-type ($<$M4) objects without confirmed membership.  

Our 58 new Taurus members contain 14 [M4, M6) low-mass stars without gravity classification, and 36 brown dwarfs (M6--L1.6), including 25 objects with very low surface gravities (\textsc{vl-g}) and 11 objects with intermediate surface gravities (\textsc{int-g}). We thus for the first time discover \textsc{int-g} members of Taurus. 

The remaining eight ($= 58 - 14 - 36$) Taurus discoveries have too low S/N ($\lesssim 30$ per pixel in $J$ band) for robust spectral typing and gravity classification. We derive the eight objects' visual spectral types by qualitatively comparing their dereddened spectra (0.9--2.4$\mu$m) to the old and young spectral standards (\citealt{Kirkpatrick+2010}; AL13) and the members of young moving groups (AL13). Visual classifications of these eight low-S/N objects are all $\geqslant$M4 and consistent with our quantitative SpT$_{\omega}$ within 1~subtype. Therefore, we identify them as new members of Taurus, although reobservations are needed for more robust spectral classification and membership assessment. We thereby only include the remaining 50 ($= 58 - 8$) new Taurus members in our subsequent analysis.

Also, we identify one new Pleiades member, PSO~J058.8758+21.0194 (PSO~J058.8+21 hereafter), as its astrometry is more consistent with Pleiades rather than Taurus. This object has a \textsc{int-g} gravity classification, which is consistent with the Pleiades's age of $\approx$125~Myr \citep{Stauffer+1998b}. We discuss its membership assessment in Section~\ref{subsubsec:hiPM}.

In addition, among the 19 candidates located in the overlapping region between Taurus and Per~OB2, we identify 13 candidates as new Per~OB2 members (Section \ref{subsec:newPerOB2}; Figure~\ref{fig:ExtMap}), all of which span M6--M8 in spectral type and have \textsc{vl-g} gravity classification, consistent with Per~OB2's young age of $\lesssim 6-15$~Myr \citep{Zeeuw+1999, Bally+2008}. We assign the other 5 ($= 18 - 13$) objects to Taurus members (already included in our aforementioned 58 new Taurus members), based on their gravity classifications and HR diagram positions (Section \ref{subsec:newPerOB2}). We show near-infrared spectra of new members of Taurus, Pleiades, and Per~OB2 in Figure~\ref{fig:spec_Taurus} and Figure~\ref{fig:spec_Per}, and show sky positions of these new members as a function of gravity classification in Figure~\ref{fig:extmap_gravclass}.

Overall, our search to date has a success rate of at least $67\%$ ($= (36{\rm \ [Taurus]}+1 {\rm \ [Pleiades]}+13{\rm \ [Per~OB2]}) / (83{\rm \ [total]}-8{\rm \ [low S/N]})$) for finding substellar objects ($\geqslant$M6) in the Taurus area, and perhaps as high as $70\%$ ($=(36+1+13+8)/83$), depending on the aforementioned reobservations of the eight low-S/N objects. Our success rate is far better than previous searches in Taurus ($\lesssim 45\%$) over the same spectral type range (M6--L2), and therefore demonstrates the robustness of our selection method. 

We have also applied our new classification scheme to 212 known Taurus members with accessible near-infrared spectra. For each object, we adopt our classification results if the object meets the criteria that we used to identify new members from our candidates; otherwise, we keep its literature values. As a result, we have homogeneously reclassified 130 mid/late-M-type and L-type members, including all but one objects with literature spectral types $\geqslant$M6. The only exception is 2MASS~J04194657+2712552 (2M~0419+2712 hereafter). We derive its reddening-free spectral type of SpT$_{\omega} =$M$6.9\pm0.9$, consistent within uncertainties with the literature value of M7.5~$\pm 1.5$ \citep{Luhman+2009}. We estimate its V-band reddening based on the H$_{2}$O color-color sequences\footnote{2M~0419+2712 lacks good-quality J-band photometry from 2MASS and UGCS, therefore the extinction based on the intrinsic optical--near-infrared color sequences cannot be derived.} and derive $A_{\rm V} = 30\pm4$~mag, comparable with the previous measurement of $\approx 33$~mag by \cite{Luhman+2009}, based on the near-infrared spectral slope. However, since this object has too low S/N spectrum ($<$10 per pixel in $J$ band) for robust classifications based on our scheme, we adopt its literature values. 

Figure \ref{fig:spt_dist} shows the histogram of spectral types for all brown dwarfs ($\geqslant$M6) in Taurus, including our new discoveries. Based on our reclassification, the number of previously known brown dwarfs in Taurus has increased from 76 to 95, and the number of previously known L-type members (masses $\approx 5-10$~M$_{\rm Jup}$ assuming the Taurus age of $\approx1$~Myr, based on the DUSTY evolutionary models by \citealt{Chabrier+2000}; see also Figure~\ref{fig:HR}) has increased from three to eight. According to this updated census, our new discoveries have thus increased the substellar objects by $\approx 38\%$ and added three more L dwarfs in Taurus, constituting the largest single increase of young brown dwarfs found in Taurus to date.

AL13	 studied three young field dwarfs (i.e., 2M~0422+1530, 2M~0435$-$1414, and 2M~0619$-$2903) and suggested that these objects are reddened and 2M~0619$-$2903 is variable (discussed in Section~\ref{subsec:0619}). We reclassify them in this work and include them in subsequent analysis. Photometry, astrometry, and classification results of our new discoveries and reclassified known objects are tabulated in Table~\ref{tab:ps1_wise}$-$\ref{tab:pmtl}.

\subsection{Performance Investigation of Our Classification Scheme}
\label{subsec:performance}

We investigate the performance of our classification method and compare to the AL13 system and other literature values. We combine our 64 new discoveries with confirmed spectral classifications of $\geqslant$M4 (50 Taurus members with robust spectral classification, 1 Pleiades member, and 13 Per~OB2 members) and 133 reclassified known young objects (130 in Taurus and 3 reddened young dwarfs in the field).

\subsubsection{Spectral Types}
\label{sec:compspt}

Figure \ref{fig:spt_comp} compares our reddening-free spectral types (SpT$_{\omega}$) with the index-based AL13 spectral types derived from the observed (SpT$_{\rm AL13}$) and the dereddened (SpT$^{\star}_{\rm AL13}$) spectra, respectively. There is a linear correlation between SpT$_{\omega}$ and SpT$^{\star}_{\rm AL13}$, which we fit using the ODR algorithm mentioned in Section~\ref{subsec:rfspt}:
\begin{equation}
{\rm SpT_{\omega}} = 1.03 \times {\rm SpT^{\star}_{AL13}} + 0.21, \quad {\rm rms}=0.45
\label{eq:SpT_omega_AL13}
\end{equation}
This correlation is tight, as the rms about the fit ($0.45$~subtype) is smaller than the typical uncertainty in our SpT$_{\omega}$ ($\approx 1.0$~subtype). In addition, our SpT$_{\omega}$ is systematically later than SpT$^{\star}_{\rm AL13}$ by $\approx 0.3-0.6$ subtypes over the applicable range ($\approx$M5--L2). In contrast, the relation between SpT$_{\omega}$ and SpT$_{\rm AL13}$ is sensitive to reddening as expected. The low-extinction ($A_{\rm V} \lesssim 1$~mag) population follows the SpT$_{\omega}$--SpT$^{\star}_{\rm AL13}$ correlation, but objects with higher extinctions have much later SpT$_{\rm AL13}$ and deviate farther from the low-extinction locus, as expected. For instance, Figure~\ref{fig:spt_comp} shows that an extinction of $A_{\rm V} \approx 4$~mag can lead to a SpT$_{\rm AL13}$ around one subtype later than dereddened SpT$^{\star}_{\rm AL13}$.

In comparison, while our spectral classification is consistent with the index-based AL13 system in the low-extinction case or after dereddening, our SpT$_{\omega}$ is robust against the reddening and therefore provides a better spectral typing, especially for highly extincted objects in young, dusty star-forming regions.

\subsubsection{Extinctions}
\label{subsubsec:extinction}

In order to investigate the overall performance of our extinction measurements, we select four Taurus objects (two of our new discoveries and two previously known members) with high reddening ($A_{\rm V} \approx 3-10$~mag) and compare their observed and dereddened spectra (Figure~\ref{fig:spec_Taurus_dr}). They are M6--M9 objects with \textsc{vl-g} classifications. We compare their spectra to \textsc{vl-g} dwarf standards (AL13) with similar spectral types. For each object, while the observed spectrum has a very different overall shape from the reference spectrum, their spectral morphologies become much closer after dereddening. The comparison demonstrates the robustness of our classification method and again indicates the difficulties of qualitatively visual spectral typing, especially for highly reddened ultracool dwarfs.

We then compare the extinction values derived from our two methods. As shown in Figure~\ref{fig:Av_H2O_OIR}, the extinctions derived from the H$_{2}$O color-color diagrams ($A_{\rm V}^{\rm H_{2}O}$; Section~\ref{subsubsec:WxCCD}) and from the intrinsic optical--near-infrared colors ($A_{\rm V}^{\rm OIR}$; Section~\ref{subsubsec:opt_NIR_color}) are consistent within the uncertainties for most objects in our sample, except for seven outliers: 2MASS~J04135328+2811233 (2M~0413+2811 hereafter), 2MASS~J04185813+2812234 (2M~0418+2812 hereafter), 2MASS~J04295950+2433078 (2M~0429+2433 hereafter), 2MASS~J04355760+2253574 (2M~0435+2253 hereafter), 2MASS~J04382134+2609137 (2M~0438+2609 hereafter), 2MASS~J04381486+2611399 (2M~0438+2611 hereafter), and 2MASS~J04442713+2512164 (2M~0444+2512 hereafter). These outliers are all previously known Taurus members and they have higher $A_{\rm V}^{\rm H_{2}O}$ reddening than their $A_{\rm V}^{\rm OIR}$ values, by $\gtrsim 1\sigma-2 \sigma$. 

Among these outliers, three objects (2M~0418+2812, 2M~0438+2609, and 2M~0438+2611) have been suggested to possess circumstellar disks with $i \approx 60-80^{\circ}$ (where $90^{\circ}$ corresponds to edge-on), based on imaging and spectroscopic analysis \citep[e.g.,][]{White+2003, Luhman+2007, Andrews+2008, Herczeg+2008, Furlan+2011, Mayne+2012, Phan-Bao+2014}. The remaining four objects have disks with lower inclinations but intensive accretion activities \citep{Guieu+2006, Bouy+2008, Zasowski+2009, Mayne+2012, Ricci+2013, Ricci+2014, Li+2015}. On the one hand, disk emission results in a redder near-infrared spectrum and thereby a larger $A_{\rm V}^{\rm H_{2}O}$ reddening. On the other hand, high-inclination circumstellar disks could scatter and absorb light, so the observed optical--near-infrared colors are not indicative of stellar photospheres. Scattering by protoplanetary disks results in a bluer optical--near-infrared color, thereby a smaller $A_{\rm V}^{\rm OIR}$ reddening. The confluence of both facts leads to the significant difference between $A_{\rm V}^{\rm H_{2}O}$ and $A_{\rm V}^{\rm OIR}$.

In addition, 2M~0418+2812, 2M~0429+2433, 2M~0438+2609, and 2M~0444+2512 are variable in $i_{\rm P1}$, $z_{\rm P1}$, and $y_{\rm P1}$ bands (Figure~\ref{fig:phot_var}), probably due to their actively accreting disks, stellar spots, and/or variable extinction along the line of sight. Their optical light curves have peak-to-peak amplitudes of $\approx 0.5-1.5$~mag over the PS1 $3\pi$ Survey timeframe (2010 May$-$2014 December), equivalent to a change of $\approx 4-7$~mag in $A_{\rm V}^{\rm OIR}$, which could lead to a significant discrepancy between $A_{\rm V}^{\rm H_{2}O}$ (typical uncertainty $=4$~mag) and $A_{\rm V}^{\rm OIR}$ (typical uncertainty $=0.85$~mag). Indeed, variability would impact both $A_{\rm V}^{\rm H_{2}O}$ (derived from spectroscopy) and $A_{\rm V}^{\rm OIR}$ (derived from photometry). However, the extinction computed based on spectroscopy might be less impacted compared to those from (non-simultaneous) photometry, as suggested by \cite{Bozhinova+2016}. \cite{Bozhinova+2016} studied a small sample of seven young and highly variable M dwarfs, and noticed that their $I$-band magnitudes vary by $0.1-0.8$~mag over 5 years but spectra remain remarkably constant.

For all these seven outliers, we adopt their $A_{\rm V}^{\rm H_{2}O}$ values instead of $A_{\rm V}^{\rm OIR}$, as their optical--near-infrared colors are not photospheric and their near-infrared spectra are less vulnerable to variability. However, the $A_{\rm V}^{\rm H_{2}O}$ may not be accurate as well, if the near-infrared spectra are significantly contaminated by scattering and emission from circumstellar disks.

We also compare our extinction measurements of known objects with measurements by \cite{Luhman+2017} and AL13, who determined extinctions by comparing the observed colors \cite[e.g., $J-H$ or $J-K$; see also][]{Furlan+2011} and/or spectral slopes at $1\mu$m or longer wavelengths to the intrinsic values of standard objects. Our method produces extinctions systematically smaller than the literature, with a weighted mean difference of $A_{\rm V} = -0.89$~mag, though results from both sources are still consistent within uncertainties (Figure~\ref{fig:Av_ZJ_lit}). 

There are two outliers, 2M~0438+2611 and 2MASS~J04144158+2809583 (2M~0414+2809 hereafter), whose extinctions based on our method are too large or too small compared to the literature. 2M~0438+2611 (SpT$_{\omega} =$M8.5) has an extinction of $A_{\rm V}^{\rm H_{2}O} = 10.6\pm4.0$~mag based on our classification, larger than its literature value \citep[$A_{\rm V} = 0$~mag;][]{Luhman+2017} by $\approx 2.7\sigma$. Based on near-infrared spectroscopy and the disk SED models, \cite{Luhman+2007} suggested that its spectra cannot be reproduced by reddened substellar photospheres with the normal extinction law, and this object may possess an edge-on disk. Therefore, \cite{Luhman+2017} assigned a nominal zero extinction to 2M~0438+2611. In this work, we adopt our $A_{\rm V}^{\rm H_{2}O}$ value as a nominal extinction for this object to show its high reddening. 

2M~0414+2809 (SpT$_{\omega}=$L2.3) has an extinction of $A_{\rm V}^{\rm H_{2}O} = -3.9 \pm 4.0$~mag\footnote{2M~0414+2809 lacks good-quality $J$-band photometry from 2MASS and UGCS, therefore its extinction is derived based on the H$_{2}$O color-color sequence.} based on our method that is lower than the literature value \citep[$A_{\rm V} = 2.4\pm0.5$~mag;][]{Luhman+2017}. \cite{Luhman+2017} assigned a spectral type of M9.75 to this object and measured the extinction by comparing its spectral slope at $1\mu$m to that of young standards with similar spectral types. Our classification suggests a later spectral type of L2.3, the latest type discovered in Taurus so far, as its spectrum is closer to a L2 $\textsc{vl-g}$ standard (2MASS~J05361998$-$1920396; AL13) rather than a L0 \textsc{vl-g} standard (2MASS~J22134491$-$2136079; AL13). The different extinctions between our method and the literature value might result from the different adopted spectral types. L2 dwarfs have intrinsically redder $J$-band spectral slopes than L0 dwarfs, therefore using a reference spectrum of L0, instead of L2, could yield a higher reddening.

As an additional exploration, we compare our extinctions to the integrated reddening till $1$~kpc, based on the \cite{Green+2015} extinction map, which has a spatial resolution of $3\arcmin-14\arcmin$. As shown in Figure~\ref{fig:Av_ZJ_map}, most objects have smaller extinctions based on our method, consistent with being located in front of the dust along the line of sight. However, our measurements do produce higher extinctions than the map values for several objects, possibly due to the small-scale structure of reddening or background field contamination (see Section~\ref{subsec:FCE} for a detailed analysis of the field contamination).

\subsubsection{Gravity Classification}
\label{subsubsec:gravityscores}

We first compare our gravity classification of known objects with literature values. Using our classification method, we find that all previously known Taurus members with spectral type $\geqslant$M6 have \textsc{vl-g} gravity classes ($\lesssim 30$~Myr; AL13), consistent with the Taurus age of $\lesssim 5$~Myr suggested by previous studies \citep[e.g.,][]{Kraus+2009}. We also derive exactly the same gravity classifications for the three young field dwarfs, as reported in AL13 (see Section~\ref{subsec:0619} for more details about 2M~0619$-$2903). 

In fact, the above AL13 gravity classifications are determined by gravity-sensitive indices and SpT$_{\rm AL13}^{\star}$ spectral types measured from the dereddened spectra. However, we preferentially adopt our reddening-free spectral types (SpT$_{\omega}$) instead of SpT$_{\rm AL13}^{\star}$ for our discoveries and reclassified known objects. Therefore, it is necessary to examine the consistency of gravity classifications based on the two different spectral types, SpT$_{\omega}$ and SpT$_{\rm AL13}^{\star}$, though they are consistent within uncertainties (Section \ref{sec:compspt}). As \cite{Liu+2016} has pointed out, using a non-AL13-based spectral type to estimate the AL13 gravity classification could potentially cause discrepancies compared to using the AL13-based spectral type.

We therefore compute the gravity scores of our discoveries and reclassified known objects based on our SpT$_{\omega}$ and compare the results with the values derived from SpT$_{\rm AL13}^{\star}$. The gravity classifications based on two versions of spectral types are exactly the same for $73\%$ objects and are consistent for $8\%$ objects after considering uncertainties in their gravity scores. The remaining $19\%$ objects only have one gravity classification, as either their SpT$_{\omega}$ or SpT$_{\rm AL13}^{\star}$ is earlier than M6, where the gravity scores are not defined (AL13; Section \ref{subsec:youth_logistics}). Visually comparing the dereddened spectra of these objects with only one reported gravity class from two calculations yields consistent overall shapes. Therefore, in order to keep the self-consistency of the AL13 system, we use SpT$_{\rm AL13}^{\star}$ for gravity classifications.

\subsubsection{Initial Youth Assumption}
\label{subsubsec:order_iter}

We measure extinctions ($A_{\rm V}^{\rm H_{2}O}$ and $A_{\rm V}^{\rm OIR}$) for each of our discoveries and previously known objects by firstly assuming a young age. We then iterate if the gravity classification based on its dereddened spectrum contradicts with this initial assumption (Section~\ref{subsec:classification_recipe}). However, it is necessary to examine if our initial youth assumption would impact final classification results. 

For such test, we replace our initial assumption by an old age, determine extinctions and gravity classification for each object, and iterate if a contradiction occurs (i.e., if the resultant gravity class of an object is \textsc{vl-g} or \textsc{int-g}, rather than \textsc{fld-g}, as assumed). We compare the results derived in this way with those based on a young-age initial assumption. We obtain the same final extinctions for all $\geqslant$M6 dwarfs whose gravity classes are available, indicating that our initial youth assumption will not impact substellar ($\geqslant$M6) objects. 

However, results of [M4,M6) dwarfs depend on the initial youth assumption, as their gravity classes are not defined. For our work, we derive extinctions for [M4,M6) objects by assuming they are young. Compared to the old-age initial assumption, young [M4, M6) objects would have higher $A_{\rm V}^{\rm H_{2}O}$ by 2~mag  (Section~\ref{subsubsec:WxCCD}) and lower $A_{\rm V}^{\rm OIR}$ by $1.4$~mag (Section~\ref{subsubsec:opt_NIR_color}).

\subsection{Field Contaminants Among Our Taurus Candidates}
\label{subsec:FCE}

We estimate the number of interloping field dwarfs expected from our search based on the Besan\c{c}on Galactic model\footnote{\url{http://model2016.obs-besancon.fr}.} (BGM). We first adapt the BGM to the context of our Taurus survey and then compare the estimated field contamination to our entire candidate list ($350$ objects; Section~\ref{sec:selection}) and to our spectroscopic follow-up sample ($75 $ objects $= 83$ [total] $- 8$ [low S/N]; Section~\ref{subsec:reclassification}), respectively.

\subsubsection{Adapting the BGM}
\label{subsubsec:BGM}

BGM simulates four main stellar populations in our Galaxy: the thin disc, the thick disc, the bar, and the stellar halo \citep{Robin+2003, Robin+2012, Robin+2014}. Assuming a star-formation history, IMF, and stellar-density model, BGM computes ages and masses for synthetic stars and derives their proper motions, effective temperatures, and photometry \citep{Robin+2003, Robin+2017, Czekaj+2014, Lagarde+2017}, based on kinematic models \citep{Bienayme+2015, Robin+2017} and atmospheric models (e.g., BaSeL2.2, \citealt{Lejeune+1997, Lejeune+1998}; BaSeL3.1, \citealt{Westera+2002}; NextGen, \citealt{Hauschildt+1999}). However, the current version of BGM is not equipped with valid atmospheric models at temperatures of $\lesssim 3000$~K (spectral type $\geqslant$M5 based on the \citealt{Herczeg+2014} temperature scale) and thus is not prepared to produce effective temperatures and photometry for ultracool dwarfs (Annie Robin, private communication). For our work, we extend the BGM down to the substellar regime based on the BHAC15 models \citep{Baraffe+2015} and the DUSTY models \citep{Chabrier+2000}, and adapt it to the context of our brown dwarf survey in Taurus.

We firstly extract synthetic O0--M9 dwarfs from the BGM in a volume that spans our Taurus search area on the sky out to a distance of 1~kpc (Section~\ref{subsec:area}). We focus on the thin disc population ($\approx 1.6 \times 10^6$ objects in total), which has ages of $\leqslant 10$~Gyr \citep{Robin+2003} and represents $91\%$ dwarf stars in the considered volume. Since the thin-disc IMF in BGM is defined above $0.08$~M$_{\odot}$ \citep{Robin+2003}, which is more massive than most young ($\lesssim 200$~Myr) $\geqslant$M6 ultracool dwarfs (based on the BHAC15 models), we set a mass scatter of $0.1$~M$_{\odot}$ in BGM to simulate lower-mass objects. The resulting mass distribution of the synthetic low-mass dwarfs ($<0.08$~M$_{\odot}$) has a slope of $\Gamma = -1.4$ (with the IMF defined as $dN/d{\rm log}M \propto M^{-\Gamma}$), consistent with the observed \citep[e.g.,][]{Metchev+2008, Pinfield+2008} and analytic \citep[e.g.,][]{Chabrier+2003} substellar IMF in the field \citep[see also Figure~2 in][]{Bastian+2010}.

Then we interpolate the BHAC15 and DUSTY models to derive effective temperatures and $J/H/K$ absolute magnitudes in both 2MASS and MKO photometric systems, using the objects' ages and masses. We adopt the BHAC15 models for objects with $M > 0.06$~M$_{\odot}$ and the DUSTY models for $M \leqslant 0.06$~M$_{\odot}$. We only keep objects with ages and masses located in the convex envelope of the model grids, leading to $\approx$~$1.5 \times 10^6$ objects. 

In order to convert objects' effective temperatures into spectral types, we use the \cite{Herczeg+2014} temperature scale for ${\rm T}_{\rm eff} \geqslant 2980$~K ($\leqslant$M5), the  \cite{Stephens+2009} temperature scale for ${\rm T}_{\rm eff} \leqslant 2259$~K ($\geqslant$L0), and the average of both scales for effective temperatures in between (M6$-$M9). We assume an uncertainty of 1 spectral subtype for such conversion. The resulting BGM sample thereby has spectral types of F4$-$T7. Earlier- and later-type dwarfs are not included, because their ages and masses are outside of the BHAC15 and DUSTY model grids. 

For synthetic $\geqslant$M0 dwarfs, we convert our synthesized 2MASS $J/H/K$ photometry into PS1 and $W1$ and $W2$ absolute magnitudes, by using the median colors as a function of spectral type from \citet[][their Table~4]{Best+2018}. The uncertainty in the absolute magnitudes is composed of scatter in both colors and magnitudes, which are derived based on the rms values in colors and magnitudes at a given spectral type from \cite{Best+2018}\footnote{Since \cite{Best+2018} provides rms values in colors for $\geqslant$M0 dwarfs and in magnitudes for $\geqslant$M6 dwarfs, we assume the rms colors of $<$M0 objects are the same as that of M0 dwarfs and the rms magnitudes of $<$M6 objects are the same as that of M6 dwarfs. In addition, as \cite{Best+2018} provides rms magnitudes for 2MASS $J/H/K$ instead of MKO $J/H/K$, we assume the two photometric systems have the same rms magnitudes at a given spectral type \citep[also see][]{Liu+2016}.}.

In addition, the BGM adopts a diffuse extinction model with a typical $V$-band reddening of $0.7$~mag~kpc$^{-1}$ along the line of sight, which is not realistic for Taurus (e.g., Figure~\ref{fig:ExtMap})\footnote{BGM employs a three-dimensional extinction model by \cite{Marshall+2006} in a restricted region of $\lvert \ell \rvert <100^{\circ}$ and $\lvert b \rvert <10^{\circ}$ in Galactic coordinates, which does not cover the Taurus area.}. Instead we employ the three-dimensional Galactic reddening map by \cite{Green+2015}, derived from PS1 data to compute the extinctions and uncertainties, using the objects' coordinates and distances and the \cite{Schlafly+2011} extinction law. We then calculate the objects' apparent magnitudes using their absolute magnitudes, distances, and extinctions. Also, we use the objects' apparent $K_{\rm 2MASS}$ to estimate the uncertainties in their proper motions that would be expected from our survey data, as described in \citeauthor{Best+2017} (\citeyear{Best+2017}; their Section~5.1.4).

\subsubsection{Estimated Field Contaminantation}
\label{subsubsec:FC_Taurus}

We first estimate the number of field contaminants among our entire candidate list. Using our modified BGM (Section~\ref{subsubsec:BGM}), we run a Monte Carlo simulation to account for the uncertainties in the synthetic dwarfs' spectral types, photometry, and kinematics. For each Monte Carlo realization, we assign a good photometric quality for each object at a given band if its magnitude is brighter than the detection limit and is not saturated. In addition, objects that are not located in the $J/H/K$ coverage map of the UGCS DR9\footnote{\url{http://wsa.roe.ac.uk/coverage-maps.html}.} are assigned bad photometric quality in the corresponding band. We then apply our entire selection criteria (Section~\ref{sec:selection}) and count the synthetic dwarfs that would be selected as Taurus candidates. We quote the star counts and uncertainties based on the median and 16$-$84 percentile among the Monte Carlo trials. Also, we estimate field contamination in three bins of age, $\leqslant 30$~Myr, $(30, 200]$~Myr, and $>200$~Myr, in order to mimic the \textsc{vl-g}, \textsc{int-g}, and \textsc{fld-g} gravity classifications for $\geqslant$M6 dwarfs (Section~\ref{subsec:youth_logistics}). As a result, $156 \pm 13$ synthetic field dwarfs would pass our selection criteria and become contaminants among our 350 Taurus candidates (Table~\ref{tab:FCE}). In Figure~\ref{fig:allFCE_dist}, we compare the $J$-band magnitudes of the synthetic field contaminants with our entire Taurus candidates. Most field interlopers are faint, with $J \approx 15.5-18$~mag.

We estimate the number of field contaminants among our spectroscopic follow-up sample using a similar approach. As mentioned in Section~\ref{sec:NIRspec}, our follow-up targets were selected based on seven priority groups, defined by the objects' magnitudes and proper motions. We compute the spectroscopic follow-up fraction of each group based on our Taurus observations and apply this to the synthetic field dwarfs to mimic the follow-up. Specifically, after obtaining the BGM synthetic candidates in each Monte Carlo realization, we divide them into the seven priority groups and sum up the number of objects in each group multiplied by the follow-up fraction. As a result, we obtain a total of $32 \pm 4$ field contaminants among our 75 follow-up candidates (Table~\ref{tab:FCE}), leading to a model-predicted success rate of $52\%$ ($= 1 - [32+4]/75$) to $63\%$ ($= 1 - [32-4]/75$) for finding substellar objects in the Taurus area, close to the $67\%$ actually achieved (Section~\ref{subsec:reclassification}).

Also, we compare the estimated field contamination with our follow-up sample at different spectral types. As shown in Figure~\ref{fig:FCE}, the $<$M4 objects discovered by our survey are probably entirely early-type field interlopers. While most of our [M4, M6) discoveries could be field contaminants, we consider them, especially the [M5, M6) objects, as candidate Taurus members worth passing further follow-up. High precision parallaxes and proper motions are needed in order for a more robust membership assessment --- for example, from Gaia \citep{Gaia+2016} or infrared astrometry. Most notably, there is little field contamination predicted for [M6, L2), namely $8 \pm 2$ predicted contaminants compared to 50 actual discoveries, indicating that our substellar discoveries are probably bona fide members. Also, the lack of any $\geqslant$M6 \textsc{fld-g} discoveries in our spectroscopic follow-up is consistent with the zero \textsc{fld-g} ($>$200~Myr) interlopers predicted by BGM. Finally, our BGM-based modeling predicts no $\geqslant$L2 field dwarf would be in our follow-up sample.

\subsection{Magnitudes, Colors, and the HR Diagram: \\An Older Low-Mass Population in Taurus?}
\label{subsec:color}

In order to investigate the star formation history of Taurus, we compute the intrinsic magnitudes and colors of Taurus objects and compare with evolutionary models. Figures~\ref{fig:Dr_CMD} and \ref{fig:Dr_CCD} shows the observed and dereddened color-magnitude diagrams and color-color diagrams for all $\geqslant$M4 members of Taurus, combining our new members with reclassified known members. We assign zero extinction for objects computed to have negative extinctions by our dereddening process (Section~\ref{subsub:finalext}). For objects without detections in 2MASS or UKIDSS, we synthesize photometry from our spectra. 

The sequence of Taurus objects becomes much tighter after dereddening, again indicating the robustness of our extinction determinations. Also, while our newly identified members share similar intrinsic colors with previously known objects, they are typically fainter.

We compare our 50 new members with 415 previously known members in Taurus on the HR diagram (Figure~\ref{fig:HR}; Table~\ref{tab:pmtl}). There are in total 465 Taurus objects and we include 369 objects for the analysis in this section, as the other 96 previously known Taurus members either have no reported spectral types or have bad-quality $J$-band photometry, and thereby cannot be placed on the HR diagram.

In order to convert objects' near-infrared spectral types into effective temperatures, we combine the \cite{Stephens+2009} and the \cite{Herczeg+2014} temperature scales, as described in Section~\ref{subsubsec:BGM}. We assume an uncertainty of 100~K for such conversion. We caution that the temperature scale provided by \cite{Stephens+2009} was derived from revised versions of the \cite{Golimowski+2004} sample, assuming a field age of $\approx 3$~Gyr; therefore, the young L-type members in Taurus could be cooler than the \cite{Stephens+2009} relation by up to $300$~K \citep{Bowler+2013, Filippazzo+2015}. \cite{Luhman+2003} constructed a temperature scale for young M0--M9 objects, intermediate between the dwarf and giant sequences, and \cite{Luhman+2008} assigned $2200$~K for L0 dwarfs. The \cite{Luhman+2003, Luhman+2008} scale is mostly consistent with our adopted one within uncertainties. We do not adopt the \cite{Luhman+2003, Luhman+2008} temperature scale in this work, as it only goes down to L0.

To derive bolometric luminosities, we use objects' dereddened $J_{\rm 2MASS}$ magnitudes in order to minimize contributions from accretion (at shorter wavelengths) and circumstellar disk emission (at longer wavelengths). For $\leqslant$M7 objects, we use the $J$-band bolometric correction ($BC_{J}$) from \citeauthor{Herczeg+2015} (\citeyear{Herczeg+2015}; their Table~2), and assume an uncertainty of 0.05~mag. For $>$M7 dwarfs, we use the $BC_{J}$ from \citeauthor{Filippazzo+2015} (\citeyear{Filippazzo+2015}; their Table~10), as a polynomial function of spectral types, and incorporate their polynomial rms into uncertainties. \cite{Liu+2010} also provided polynomials to compute $BC_{J}$ of M6--T8.5 objects, which were again improved versions of \cite{Golimowski+2004}, designed based on the field-age sample. Since young early-L objects could have smaller $BC_{J}$ than field counterparts by up to 0.5~mag \citep{Filippazzo+2015}, we do not adopt the \cite{Liu+2010} $BC_{J}$ in this work.  

We assume a distance of $145 \pm 15$ pc \citep{Zeeuw+1999} for Taurus. Uncertainties in spectral types, $J$-band magnitudes, extinctions, bolometric corrections, and the distance are propagated into effective temperatures and luminosities in a Monte Carlo fashion. We overlay evolutionary models on the HR diagram by combining the BHAC15 models \citep{Baraffe+2015} for masses $M>0.06$~M$_{\odot}$ and the DUSTY models \citep{Chabrier+2000} for masses $M \leqslant 0.06$~M$_{\odot}$. These evolutionary models are subject to systematic uncertainties in the treatment of, for example, non-steady accretion \citep[e.g.,][]{Baraffe+2009, Baraffe+2012}, magnetic fields \citep[e.g.,][]{Feiden+2013, Feiden+2014, Feiden+2016}, and stellar rotation \citep[e.g.,][]{Somers+2015}. 

According to the HR diagram, our new members have masses from the stellar regime ($\approx 0.1$~M$_{\odot}$) down to the planetary-mass regime of $\approx 5$ M$_{\rm Jup}$, among the lowest-mass objects in Taurus found to date. Also, as shown in Figure~\ref{fig:comp_age_grav}, there is a general agreement between the gravity classifications from near-infrared spectroscopy and the ages inferred from HR diagram positions. \textsc{vl-g} objects are mostly younger than $30$~Myr, and \textsc{int-g} objects are on average older ($\gtrsim 30$~Myr). However, the correlation between gravity classes and ages is not exact, given that \textsc{vl-g} objects can be as old as $100$~Myr  (Figure~21 in \citealt{Liu+2016}), while \textsc{int-g} objects can be as young as $15$~Myr (Figure~14 in \citealt{Gagne+2015}).

Examining the HR diagram for the complete set of known objects and our new discoveries, we find that $93\%$ of known objects (297 objects) and $38\%$ of our newly identified Taurus members (19 objects) have model-based ages of $\leqslant10$~Myr (Figure~\ref{fig:HR}), consistent with the commonly adopted age of $\lesssim$5~Myr for the stellar members of Taurus \citep[e.g.,][]{Kraus+2009}. The remaining 53 objects have fainter luminosities and therefore are older, $>$10~Myr based on model isochrones. These fainter objects contain 16 previously known $<$M4 dwarfs, and contain 37 $\geqslant$M4 dwarfs where 31 objects are newly identified members by us. Here we propose several possible explanations for these fainter, apparently older objects in Taurus.

These 53 fainter objects could be field dwarfs that are not associated with Taurus. For a qualitative examination, we obtain all Taurus objects with good-quality proper motions (i.e., PS1 proper-motion fits with $0.3 < \chi_{\nu}^{2} < 40$; Section~\ref{subsec:kine_criteria}), divide them into different bins based on their model-derived ages, and compare their proper motions with the mean motion of Taurus (Section~\ref{subsec:kine_criteria}; Figure~\ref{fig:PM_analysis}). No clear distinction is seen among objects with different ages. The weighted average and weighted rms of proper motions of the younger Taurus objects (model-based ages $\leqslant$10~Myr) is $(\mu_{\alpha} {\rm cos} \delta, \mu_{\delta}) = (7.66 \pm 5.10, -17.82 \pm 5.87)$~mas~yr$^{-1}$, consistent with that of the older population ($>$10~Myr) of $(\mu_{\alpha} {\rm cos} \delta, \mu_{\delta}) = (5.41 \pm 5.34, -12.95 \pm 7.50)$~mas~yr$^{-1}$. Therefore, the fainter Taurus objects are unlikely to be mostly field contaminants, which would have systematically different kinematics compared to the younger Taurus population. 

For a quantitative examination, we estimate the number of field contaminants among the 37 Taurus objects that have $\geqslant$M4 spectral types and fainter bolometric luminosities (i.e., model-based ages $>$10~Myr on the HR diagram). In order to perform our BGM-based field contamination estimate (Section~\ref{subsec:FCE}), we focus on the 31 fainter, older objects discovered by us, including 4 out of 6 [M4, M5) objects, 9 out of 10 [M5, M6) objects, and 18 out of 21 $\geqslant$M6 objects. We first derive the expected HR diagram positions of the BGM synthetic candidates (selected from multiple Monte Carlo trials; Section~\ref{subsubsec:FC_Taurus}) in the same fashion as for real Taurus objects. We compute the objects' bolometric luminosities by using the Taurus distance (instead of the distances produced by BGM), the objects' $J_{\rm 2MASS}$ magnitudes dereddened via their $V$-band extinctions, and the same $J$-band bolometric corrections as for actual Taurus objects \citep{Filippazzo+2015, Herczeg+2015}. We compute the objects' effective temperatures using their spectral types and the combined temperature scales from \cite{Stephens+2009} and \cite{Herczeg+2014}. We then compare the synthetic dwarfs on the HR diagram with the 10~Myr isochrone combined from BHAC15 ($M > 0.06$~M$_{\odot}$) and DUSTY ($M \leqslant 0.06$~M$_{\odot}$). In the end, we find $5\pm1$ [M4, M5) field interlopers (as compared to our 4 real discoveries), $5\pm2$ [M5, M6) field interlopers (as compared to our 9 real discoveries), and $6\pm1$ $\geqslant$M6 field interlopers (as compared to 18 real discoveries) could be the fainter, older objects (model-based ages $>$10~Myr) in our spectroscopic follow-up sample, yielding a contamination fraction of $125\pm25\%$, $56\pm22\%$, and $33\pm5\%$ for our fainter [M4, M5), [M5, M6), and $\geqslant$M6 discoveries, respectively. Therefore, field contamination is unlikely to explain many of the fainter, older $\geqslant$M5 population in Taurus. 

As another alternative, the underluminous objects in Taurus might be those detected only in scattered light. A few known Taurus members are significantly fainter than the $30$~Myr isochrone. About $2/3$ of them (9 out of 14 objects) are suggested to have Class I envelopes or high-inclination circumstellar disks\footnote{These nine objects are 2MASS~J04153566+2847417 (2M~0415+2847 hereafter), 2MASS~J04202144+2813491, 2MASS~J04202583+2819237, 2MASS~J04220069+2657324, 2MASS~J04221568+2657060, 2MASS~J04290498+2649073, 2MASS~J04313747+1812244, 2MASS~J04331435+2614235, and 2MASS~J04333905+2227207.} \citep[e.g.,][]{Duchene+2010, Rebull+2010, Furlan+2011, Yen+2013, Li+2015}, with another one being diskless \citep[2MASS~J04373705+2331080, 2M~0437+2331 hereafter;][]{Luhman+2009} and no scenario proposed for the remaining 4 ($= 14 - 9 -1$) objects\footnote{These four objects are 2MASS~J04105425+2501266, 2MASS~J04345973+2807017, 2MASS~J04380191+2519266, and 2MASS~J04390525+2337450.} (Figure \ref{fig:HR}). Our classification method could be unreliable if the objects' detections are not photospheric, leading to unreliable measurements of spectral types and extinctions, and thus effective temperatures and bolometric luminosities. However, only 5 out of the 37 fainter Taurus objects show evidence of disk emission based on mid-infrared excess (Section~\ref{subsec:disks_Taurus}), suggesting that envelope/disk occultation cannot explain most of them. 

Overall, it is probable that many of these faint objects in Taurus are bona fide members with older ages. Therefore, our new members with model-derived ages of $>$10~Myr and spectral types of [M5, L0) (model-based masses of $\approx 0.1-0.02$~M$_\odot$; Figure~\ref{fig:HR}) might represent a newly identified low-mass population. A number of studies have suggested a spatially distributed older population of stars in Taurus. X-ray studies using ROSAT detected a very widely dispersed stellar population, $\approx 50$ pc ($\approx 20^{\circ}$) away from the central molecular clouds in Taurus \citep{Neuhaeuser+1995a, Neuhaeuser+1995b, Briceno+1997, Carkner+1997, Magazzu+1999}. One explanation is that these objects formed tens of million years before the well-studied young population. Also, based on lithium abundances, disk fractions, and kinematics of the known Taurus objects with masses $M\gtrsim 0.1$~M$_{\odot}$, recent studies have found evidence of an older stellar population within the main Taurus region with ages of $\approx$10--40~Myr \citep[e.g.,][]{Sestito+2008, Daemgen+2015, Kraus+2017}.

Though our survey is ongoing, most of our brighter candidates have been observed (including $\approx 75\%$ of objects that have $J_{\rm 2MASS} \leqslant 15.5$~mag; Section~\ref{sec:NIRspec}), so we conduct a preliminary investigation into the number ratio of stars with higher ($M\geqslant$0.1~M$_{\odot}$) and lower ($M<$0.1~M$_{\odot}$) model-based masses for the younger ($\leqslant$10~Myr) and older ($>$10~Myr) populations in Taurus. Among the younger population (316 objects), 191 objects have masses $\geqslant$0.1~M$_{\odot}$ and 125 objects have lower masses, leading to a high-to-low-mass ratio of $1.5\pm0.2$, with the uncertainty based on Poisson statistics. In contrast, the older population (53 objects) contains 20 objects with masses $\geqslant$0.1~M$_{\odot}$ and 33 objects with lower masses, leading to a ratio of $0.6 \pm 0.2$, which is $3\sigma$ lower than that of the younger population. 

We also compute these statistics by excluding the estimated field contaminants among our discoveries\footnote{While we should also exclude the field contaminants among the previously known Taurus members, such contamination is harder to predict based on our modified BGM, given that the previously discovered objects were selected by inhomogeneous criteria from various groups. However, the previously known members typically have more thorough followup thereby better membership assessment than our discoveries, which reduces the effects of field contamination.} and assuming that the 9 known Taurus objects with reported Class I envelopes or high-inclination circumstellar disks (see above) have ages $\leqslant 10$~Myr, instead of the $>$10~Myr as shown on the HR diagram. Our modified BGM predicts 19 field contaminants among our discoveries with $\geqslant$M4 spectral types (Figure~\ref{fig:FCE} and Table~\ref{tab:FCE}), including 3 younger objects (all with masses $<0.1$~M$_{\odot}$) and 16 older objects (5 with masses $\geqslant 0.1$~M$_{\odot}$ and 11 with masses $< 0.1$~M$_{\odot}$). Based on these adjustments for the younger population (322 objects $= 316 + 9 - 3$), we find 199 objects with masses $\geqslant$0.1~M$_{\odot}$ and 123 objects with lower masses, leading to a high-to-low-mass ratio of $1.6\pm0.2$. In comparison, the older population (28 objects $= 53 - 9 - 16$) possesses a much more bottom-heavy ratio of $0.3 \pm 0.1$, which is $6\sigma$ lower than that of the younger population, as 7 objects are more massive than 0.1~M$_{\odot}$ and 21 objects are less massive.

The discrepancy in IMFs between the younger and older populations could be caused if a fraction of older ($>$10~Myr) stellar members of Taurus were missed/discarded by previous searches. As another alternative, the younger and older Taurus population might have experienced different star formation processes, such as the in situ collapse \citep{Hennebelle+2009} and/or the dynamical ejection (\citealt{Kroupa+2003, Stamatellos+2009}; Section~\ref{subsubsec:sub_stellar_obj}). In order to better understand the star formation history of Taurus, spectroscopic follow-ups of our remaining brown dwarf candidates and Gaia astrometry \citep{Gaia+2016} of the higher-mass stellar members would be of importance.

Figure~\ref{fig:extmap_modelage} examines the spatial distribution of all Taurus objects as a function of model-derived ages. The younger population is mostly associated with high-extinction filaments and clumps. In contrast, the older objects are more dispersed, with their sky locations not closely following the younger members or extinction. These results are in accord with the results by \cite{Kraus+2017} for higher-mass members.

\subsection{Spatial and Kinematic Distribution}
\label{subsec:sp_kin_dist}

\subsubsection{Comparing Stellar and Substellar Spatial Distributions}
\label{subsubsec:sub_stellar_obj}

Comparing the spatial distributions of stellar and substellar objects in young star-forming regions opens an interesting window into the formation of substellar objects and free-floating planets. Based on modern theories, brown dwarfs can form via two ways: (1)~in situ collapse of single low-mass molecular cores \citep{Hennebelle+2009}, analogous to the formation of more massive T Tauri stars, and (2)~dynamical ejection from very young multiple systems \citep[e.g.,][]{Kroupa+2003} or from fragmenting circumstellar disks \citep{Stamatellos+2009}. If brown dwarfs are formed via dynamical ejection models with high velocity dispersion \citep[$\gtrsim 2$ km s$^{-1}$; e.g.,][]{Kroupa+2003}, then they could have a more dispersed distribution than stars in the same region. On the other hand, the distribution of these two populations could be similar if brown dwarfs are formed in situ, or from ejections with only modest velocities \citep{Bate+2003,Bate+2009}. Several studies have compared the distributions of the stellar and substellar populations in Taurus \citep[e.g.,][]{Briceno+2002, Guieu+2006}, but firm conclusion is lacking. To be specific, past work has measured the number ratios of substellar ($0.02\leqslant M/M_{\odot}\leqslant0.08$) to stellar ($0.08\leqslant M/M_{\odot}\leqslant10$) populations, and then examined this ratio as function of radii from the stellar aggregates in their search area. However, \cite{Luhman+2006} noted that this method depends on the adopted spectral classification system, as different classification methods could derive different spectral types for the same object and therefore change the number ratios between substellar and stellar objects. To avoid this problem, \cite{Luhman+2006} studied the stellar--substellar spatial distributions based on the angular distances to the nearest stellar neighbors. However, both approaches could be affected by the incompleteness of their adopted samples.

Figure \ref{fig:extmap_spt} shows the spatial distribution of all Taurus objects as a function of spectral type. Stars ($<$M6) are located in relatively compact regions associated with high-extinction filaments and clumps. While the distribution of substellar objects ($\geqslant$M6) is mostly consistent with the stellar population, some brown dwarfs are located in regions of lower stellar density. A quantitative analysis will be appropriate once our survey is finished.

\subsubsection{Young Brown Dwarfs with Non-Taurus Proper Motions: \\ A Probable New Pleiades Member and Candidate Ejected Brown Dwarfs}
\label{subsubsec:hiPM}

Figure~\ref{fig:PM_sigma} compares the PS1 proper motions of discoveries by our survey with the average motion of Taurus. Around $92\%$ of objects are consistent within $2\sigma$ with the mean Taurus motion. The outliers contain 5 of our discoveries and 14 known Taurus members (Table~\ref{tab:pm_sigma}), most of which have only mildly inconsistent proper motions ($2-4\sigma$) with Taurus and might still be members, while two objects, PSO~J075.9044+20.3854 (PSO~J075.9+20 hereafter; a new Taurus member) and 2MASS~J04313407+1808049 (a.k.a. L1551/IRS~5; 2M~0431+1808 hereafter; a known Taurus member), have very discrepant proper motions, $6\sigma$ and $13\sigma$ away from the mean Taurus motion, respectively.

The five proper-motion outliers (Figure~\ref{fig:PM_highPM}) from our discoveries were selected as candidates during our initial search, but their final proper motions do not fulfill our search criteria (Section~\ref{subsec:add_constrain}). They span M6--M9 in spectral type and are located in low-extinction regions of Taurus, consistent with our extinction measurements of $A_{\rm V} = -0.8$ to $0.4$~mag. Overall, four objects have \textsc{vl-g} gravity classification and a model-derived age of $\leqslant 30$~Myr based on their HR diagram positions. 

Our fifth proper-motion outlier, PSO~J058.8+21, has a spectral type of M9.25, intermediate surface gravity (\textsc{int-g}), and a model-derived age of $>$100~Myr (Section~\ref{fig:HR}). In fact, it is located on the sky between Taurus and the Pleiades. Pleiades has a distance of 136~pc \citep{Melis+2014} and an age of $\approx$125~Myr \citep{Stauffer+1998b}. The proper motion of PSO~J058.8+21, $(\mu_{\alpha} {\rm cos} \delta, \mu_{\delta}) = (12.04 \pm 7.31, -44.59 \pm 7.37)$~mas~yr$^{-1}$, is consistent with the mean Pleiades motion of $(\mu_{\alpha} {\rm cos} \delta, \mu_{\delta}) = (19.71, -49.82)$~mas~yr$^{-1}$ with a rms of 13~mas~yr$^{-1}$ \citep{Zapatero+2014}. Figure~\ref{fig:HR} shows PSO~J058.8+21 on the HR diagram assuming the Pleiades distance. This yields a model-derived age of $>$100~Myr, which is consistent with its \textsc{int-g} gravity classification and the Pleiades age. This agreement suggests that PSO~J058.8+21 is probably a new Pleiades member, missed by previous searches in the Pleiades \citep[][]{Steele+1995, Martin+1996, Martin+1998a, Martin+1998b, Martin+2000, Stauffer+1998a, Stauffer+1998b, Festin+1998, Pinfield+2003, Bihain+2006, Bihain+2010, Zapatero+1997, Zapatero+2014}.

Among the 14 previously known members with $>2\sigma$ discrepant proper motions, 2 objects have no reported spectral types, 7 objects are earlier than M6, and the remaining 5 objects have spectral types in M6--M9. The five $\geqslant$M6 known objects are all located in the high-extinction filaments in Taurus and have very low surface gravities (\textsc{vl-g}) based on our classification. In addition, 2 out of these 14 known objects were studied by \cite{Kraus+2017} and reassessed as confirmed members (Table~\ref{tab:pm_sigma}).

For the ten young $\geqslant$M6 dwarfs with discrepant proper motions (five of our discoveries, including PSO~J058.8+21, and five known Taurus substellar members; Figure~\ref{fig:ExtMap}, \ref{fig:extmap_gravclass}, \ref{fig:extmap_modelage}, \ref{fig:extmap_spt}, and \ref{fig:PM_highPM}), their non-Taurus velocities could result from dynamical ejection during their formation \citep[e.g.,][]{Reipurth+2001, Kroupa+2003, Stamatellos+2009}. High-precision measurements of their parallaxes and proper motions are warranted for their membership assessment --- for example, from Gaia \citep{Gaia+2016} or infrared astrometry. If membership in Taurus along with the discrepant proper motions are confirmed, then they would be the first strong candidates for ejected brown dwarfs and provide valuable benchmarks to test ejection models.

\subsection{New Per~OB2 Members}
\label{subsec:newPerOB2}

\subsubsection{Membership Assessment}
\label{subsubsec:membership}

We have found 19 young brown dwarfs located in the overlapping sky region between Taurus and another association, Per~OB2, which is northwest of Taurus and has a farther distance of 318~pc \citep{Zeeuw+1999, Bally+2008}. Among these 19 objects, PSO~J059.5+30 (SpT$_{\omega} =$M9.7; \textsc{vl-g}) is a candidate ejected brown dwarf with proper motions inconsistent with neither Taurus nor Per~OB2 (Section~\ref{subsubsec:hiPM}, Figure~\ref{fig:PM_highPM}, and Table~\ref{tab:pm_sigma}). We identify it as an ejected brown dwarf member in Taurus and investigate the membership of the remaining objects. 

The remaining 18 brown dwarfs span M6--M8 in spectral type, with low extinctions of $A_{\rm V}^{\rm OIR} = -0.8$ to $1.0$~mag based on our classification method. In addition, 13 of these objects have very low surface gravities ($\textsc{vl-g}$), and the other 5 objects have \textsc{int-g} gravity classes. All 18 objects are clustered in sky coordinates (Figure~\ref{fig:ExtMap} and \ref{fig:extmap_gravclass}).

Besides Per~OB2, there is another background star-forming region, IC~348 \citep{Herbst+2008}, which is located near these 18 discoveries, though not overlapping with our search area. IC~348 is embedded in the same Perseus Cloud as Per~OB2 and at the same distance of $315$~pc \citep[][]{Luhman+2003}, with a relatively compact size spanning $\alpha=3^{\rm h}40^{\rm m}$--$3^{\rm h}50^{\rm m}$ and $\delta=31^{\circ}30\arcmin$--$32^{\circ}30\arcmin$ (J2000.0). 

We investigate the membership of our 18 discoveries, which were not reported as known members of Taurus \citep{Best+2017, Luhman+2017}, Per~OB2 \citep[e.g.,][]{Azimlu+2015}, or IC~348 \citep[summarized by][]{Luhman+2016}. In Figure~\ref{fig:PM_Per}, we compare the proper motions of all 18 objects with the mean motion of Taurus (Section~\ref{subsec:kine_criteria}), Per~OB2, and IC~348. Per~OB2 has an average proper motion of $(\mu_{\alpha} {\rm cos} \delta, \mu_{\delta}) = (8.1, -8.4)$~mas~yr$^{-1}$ with a typical rms of $3$~mas~yr$^{-1}$, converted from the values given in the Galactic coordinates by \cite{Belikov+2002}. The typical proper motion of IC~348 is $(\mu_{\alpha} {\rm cos} \delta, \mu_{\delta}) = (1.9, -2.1)$~mas~yr$^{-1}$ with a rms of $3$~mas~yr$^{-1}$ \citep{Luhman+2016}. Proper motions of our 18 discoveries are consistent with any of the three regions, i.e., Taurus, Per~OB2, and IC~348. Therefore, no firm conclusion about their membership could be derived from kinematics.

We then compare the HR diagram positions of these 18 objects, assuming distances of Taurus and Per~OB2/IC~348 (Figure~\ref{fig:HR_Per}; Table~\ref{tab:pmtl}). We note that the 13 \textsc{vl-g} objects would have much younger model-derived ages ($\approx 1$~Myr, as compared to 10--100~Myr at the Taurus distance) when scaled at the Per~OB2/IC~348 distance, consistent with Per~OB2's age of $\lesssim 6-15$~Myr \citep{Zeeuw+1999, Bally+2008} and IC~348's age of $\approx$2--6~Myr \citep[e.g.,][]{Muench+2003, Bell+2013}. Therefore, these 13 objects could be new members of Per~OB2 or ejected young substellar members of IC~348 that are moving toward Per~OB2. As no firm evidence of ejected brown dwarfs have been discovered in star-forming regions, we tentatively suggest they are new Per~OB2 members. For the remaining 5 $\textsc{int-g}$ objects, membership in Per~OB2 would suggest ages of $\approx 1$~Myr, far too young compared to their intermediate surface gravities (\textsc{int-g}; $\approx$30--200~Myr). Therefore we favor membership in Taurus for these 5 objects.

However, it is still possible that all these 18 \textsc{vl-g} and \textsc{int-g} objects are located in Taurus with slightly older ages of 10--100~Myr, analogous to several other new \textsc{vl-g} Taurus members with model-derived ages of 10--30~Myr (Figure~\ref{fig:comp_age_grav}). Given their compactness on the sky, these objects may represent a young cluster or moving group. Direct distance measurements of these 18 objects are needed to refine their membership.

\subsubsection{An Extremely Wide Binary Brown Dwarf?}
\label{subsubsec:binary}

PSO~J060.9401+32.9790 and PSO~J060.9954+32.9996 are two new Per~OB2 members separated by $3.05$~arcmin, which is a projected separation of $58$~kAU assuming the Per~OB2 distance of $318$~pc \citep{Luhman+2003}. PSO~J060.9401+32.9790 has a spectral type of M6.6 with an extinction of $A_{\rm V}^{\rm OIR} = -0.82\pm0.85$~mag and PSO~J060.9954+32.9996 is a M8.2 dwarf with a comparably low extinction of $A_{\rm V}^{\rm OIR} = 0.38 \pm 0.85$~mag. Both of them have very low surface gravity (\textsc{vl-g}) and are lying near the $1$~Myr isochrone in the HR diagram (Figure~\ref{fig:HR_Per}). Their proper motions are different by only $0.8\sigma$ in R.A. and by $0.4\sigma$ in Dec. (Figure~\ref{fig:PM_Per}). In addition, they both show mid-infrared excesses and are circumstellar disk candidates (see Section~\ref{subsec:disks_PerOB2}).

We ran a simulation to estimate the probability that their proximity ($0.05^{\circ}$) results from pure chance. Since there are 13 Per~OB2 discoveries located in the $4^{\circ} \times 4^{\circ}$ overlapping region between Taurus and Per~OB2, we generated 13 test points with a uniform distribution in a $4^{\circ} \times 4^{\circ}$ box area. We produced $10^{7}$ ensembles and found $3.7\%$ of ensembles contain at least one ``binary'' with a separation no larger than $0.05^{\circ}$, indicating that our binary brown dwarf may be a true binary.

Although binaries with such large separation in the field are easily disrupted \citep{Tokovinin+2012}, a large number of candidate ultrawide ($\approx$10--100~kAU) binaries have been found \citep{Dhital+2010} awaiting explanations for their formation (e.g., dissolution of star clusters [\citealt{Kouwenhoven+2010}]; or enlarged semi-major axes due to N-body dynamical interactions [\citealt{Reipurth+2012}]). So far, only a handful of very wide separation ($\gtrsim 10$~kAU) binary/tertiary systems with M dwarf companions have been discovered, e.g., HR4796 \citep[$13.5$~kAU; A0$+$M2.5$+$M4.5;][]{Jura+1993, Kastner+2008}, TW~Hya \citep[$41$~kAU; K7$+$M8.5;][]{Scholz+2005, Teixeira+2008}, V4046~Sgr \citep[$12.35$~kAU; K5$+$K7$+$M1;][]{Torres+2006, Kastner+2011}, T~Cha \citep[$12-40$~kAU; K0$+$M3;][]{Kastner+2012}, NLTT~18587 \citep[$12.2$~kAU; M2$+$M7.5;][]{Deacon+2014}. However, no binary brown dwarf ($\geqslant$M6) system with a separation as large as our two Per~OB2 objects ($58$~kAU) has been previously reported.

Assuming PSO~J060.9401+32.9790 and PSO~J060.9954+32.9996 form a binary located at the distance of Per~OB2, we examine the stability of this system. Based on the HR diagram (Figure~\ref{fig:HR_Per}), each object has a mass of $\lesssim 0.04$~M$_{\odot}$, so their total mass is at most $\approx 0.08$~M$_{\odot}$. Based on \citeauthor{Close+2003} (\citeyear{Close+2003}; their Figure~15), the maximum separation of a gravitationally bound system with a total mass of $0.08$~M$_{\odot}$ is $\approx 10$~AU. The $58$~kAU separation in our candidate binary system is much wider than this maximum, indicating that the two objects are likely loosely bound or unbound. A more firm conclusion will be possible with astrometric follow-up of their distances, velocities, and ages.

\subsection{Circumstellar Disks around Our New Members}
\label{subsec:disks}

Circumstellar disks can be traced by emission at mid-infrared wavelengths that exceeds the amount expected from stellar photospheres. Following \cite{Esplin+2014}, we identify the mid-infrared excesses for our new members based on $K_{\rm 2MASS}$ and AllWISE photometry ($W2$, $W3$, $W4$). For objects with no 2MASS detection, we synthesize their $K_{\rm 2MASS}$ magnitudes using $K_{\rm MKO}$, as described in Section~\ref{subsec:color}. We measure the mid-infrared excess of an object by comparing its dereddened $K_{\rm 2MASS} - W2/3/4$ colors to the reddest colors expected for its photosphere given its spectral type, as derived from previously known Taurus members by \citeauthor{Esplin+2014} (\citeyear{Esplin+2014}; their Section~3). In addition, we visually check the PS1 images of our disk candidates and utilize the NED webpage\footnote{\url{https://ned.ipac.caltech.edu}.} in order to rule out the cases where the objects' mid-infrared excesses are contributed by nearby galaxies.

\subsubsection{Circumstellar Disk Candidates in Taurus}
\label{subsec:disks_Taurus}

Among our 50 new Taurus members, we have identified 5 objects that show mid-infrared excesses in at least one of the $W2/W3/W4$ bands with significance of $>2\sigma$ and therefore probably possess circumstellar disks (Figure~\ref{fig:IR_excess_Taurus} and Table~\ref{tab:IR_excess}). Our Taurus disk candidates span M6--M7 in spectral type and have extinctions of $A_{\rm V} = -0.6$ to $3.5$~mag. All objects have very low surface gravities (\textsc{vl-g}) with one exception, PSO~J059.3563+32.3043 (\textsc{int-g}; PSO~J059.3+32 hereafter). 

We compare the HR diagram positions (Figure~\ref{fig:HR_disk}) of our 5 disk candidates to the previously known disk population in Taurus. The latter contains circumstellar disks classified the same way as for our disk candidates, though with {\it Spitzer} photometry (IRAC [3.6$\mu$m]/[4.5$\mu$m]/[5.8$\mu$m]/[8.0$\mu$m] and MIPS [24$\mu$m]) considered as well (\citealt{Esplin+2014}; also see \citealt{Luhman+2012}). It is interesting that 3 of our disk candidates, PSO~J059.3+32, PSO~J065.6900+15.1818, and PSO~J069.3827+22.8857, and 2 known disk candidates, 2MASS~J04153566+2847417 (2M~0415+2847 hereafter) and 2MASS~J04345973+2807017 (2M~0434+2807 hereafter)\footnote{2M~0415+2847 \citep[SpT$_{\rm lit}=$M5.5;][]{Luhman+2017} is reclassified as SpT$_{\omega}=$M6.3 by our work (Table~\ref{tab:classification}). 2M~0434+2807 \citep[Spt$_{\rm lit}$=M5.75;][]{Luhman+2017} is not classified by our work due to the low-S/N of its near-infrared spectrum ($\lesssim 30$ per pixel in $J$ band). \label{footnote:2M0434+2807}}, are among the faintest objects in the substellar regime. Their reddening-corrected bolometric luminosities are fainter by a factor of $\approx 30$ than the other disk-bearing objects with similar spectral types. As a result, these five fainter disk candidates have ages of $\gtrsim 30$~Myr based on their HR diagram positions, much older than the typical disk lifetime of $\lesssim 10$~Myr \citep[e.g.,][]{Mamajek+2009, Williams+2011}. Here we consider two explanations for their faintness.

We could have underestimated the extinction. A higher extinction would lead to a brighter reddening-corrected bolometric luminosity, thereby a younger model-based age, and lead to bluer reddening-corrected $K_{\rm 2MASS} - W2/3/4$ colors, thereby a smaller or non-existent mid-infrared excess. These two effects could thus change our old disk candidates into young disk-bearing objects or young/old diskless objects. However, visually comparing the dereddened near-infrared spectra of these five disk-bearing brown dwarfs to the \textsc{vl-g} spectral standards from AL13 (M6: TWA~8B; M7: 2MASS~J03350208+2342356) does not show strong evidence of underestimated extinctions. In addition, the extinctions of our 3 fainter disk candidates and 2M~0415+2847 are measured based on the optical--near-infrared colors ($A_{\rm V}^{\rm OIR}$)\footnote{The other known object among the 5 faint disk candidates, 2M~0434+2807, have neither $A_{\rm V}^{\rm H_{2}O}$ nor $A_{\rm V}^{\rm OIR}$ extinction, because it is not classified by our work, see footnote~\ref{footnote:2M0434+2807}.}, which could be affected by non-photospheric emission due to disk occultation or accretion, while the extinctions derived from H$_{2}$O-band spectral indices ($A_{\rm V}^{\rm H_{2}O}$) are less vulnerable to disk-related processes (Section~\ref{subsubsec:extinction}). The $A_{\rm V}^{\rm H_{2}O}$ of these four fainter disk candidates are higher than their $A_{\rm V}^{\rm OIR}$ values by only $\lesssim 0.4\sigma$. If we switch to $A_{\rm V}^{\rm H_{2}O}$ and recompute the objects' reddening-free bolometric luminosities and mid-infrared excesses, they are still fainter/older disk candidates. Consequently, underestimated extinction is unlikely to explain the faintness of these objects.

Alternatively, these fainter disk candidates could be detected in scattered light. Among the two known disk candidates with fainter luminosities, 2M~0415+2847 was classified as a Class I object with bipolar outflows based on integrated intensity maps in (sub-)millimeter wavelengths \citep[e.g.,][]{Li+2015, Buckle+2015}. 2M~0434+2807 has no reported Class I envelope or circumstellar disk, while \cite{Luhman+2017} suggested the scattered-light detection to explain its faintness. However, the near-infrared spectra of all five faint disk candidates do not show signatures of disk scattering, as seen in the known Taurus edge-on disks, e.g., 2M~0438+2611, whose near-infrared spectra cannot be reproduced by reddened substellar photospheres with the normal extinction law \citep{Luhman+2007}. In addition, edge-on disks show significant excess emission in $K$ band \citep[e.g.,][]{Luhman+2004} and have anomalous colors in the optical--near-infrared color-color diagram (Figure~\ref{fig:yJK_disk}), with positions way off the sequence of other disk population. In comparison, the five faint disk candidates follow the sequence formed by the known disk population, with colors distinctively different from those of edge-on disks, suggesting that scattered-light detection does not explain the five fainter disks candidates.

Overall, it is probable that these 5 fainter disk-bearing objects in Taurus indeed have older ages. Therefore, we will need follow-up observations at (sub-)millimeter wavelengths to investigate the geometry, gas, and dust in these fainter disk candidates. Also, direct distance measurements of them are warranted to refine their membership.

\subsubsection{Circumstellar Disk Candidates in Per~OB2}
\label{subsec:disks_PerOB2}

Using the same dereddened $K_{\rm 2MASS} -W2/3/4$ colors, we find our new Pleiades member, PSO~J058.8+21, does not show any mid-infrared excess, but we identify 6 disk candidates among our 13 new Per~OB2 members, based on excess emission in $W2$, $W3$, and/or $W4$ with significance of $>2\sigma$ (Figure~\ref{fig:IR_excess_Other} and Table~\ref{tab:IR_excess}). These disk candidates span M6--M8 in spectral type and have extinctions of $A_{\rm V} = -0.8$ to $1.0$~mag, with all of them classified as $\textsc{vl-g}$ objects. 

It is interesting to note that 2 of our Per~OB2 disk candidates, PSO~J060.9401+32.9790 and PSO~J060.9954+32.9996, form a candidate very wide separation binary (Section~\ref{subsubsec:binary}). Although the circumstellar disk of each binary component could be truncated or disrupted due to dynamical interactions \citep{Artymowicz+1994}, we do not expect this phenomenon to occur in such a wide binary system, whose projected separation of $58$~kAU is far larger than the typical disk size of $\lesssim 200$~AU \citep[e.g.,][]{Hughes+2008b, Carpenter+2009, Andrews+2009, Andrews+2010}. Follow-up observations would be of interest to understanding the disk evolution and planet formation in wide binary systems.

\subsection{Notes on Selected Objects}
\label{subsec:notes}

\subsubsection{PSO~J065.8+19: A Bright L0 \textsc{vl-g} Dwarf}
\label{subsec:brightL0}

PSO~J065.8871+19.8386 (Figure \ref{fig:ExtMap}, \ref{fig:extmap_gravclass}, \ref{fig:extmap_modelage}, and \ref{fig:extmap_spt}), a L0 \textsc{vl-g} dwarf with an extinction of $A_{\rm V}^{\rm OIR} = -1.42\pm0.85$~mag, is identified as a new Taurus member. Its colors are consistent with the Taurus sequence after dereddening (Figure \ref{fig:Dr_CMD}, \ref{fig:Dr_CCD}). The proper motion of this object is consistent within $2\sigma$ of the mean motion of Taurus (Figure \ref{fig:PM_analysis}). Most notably, it is the brightest L dwarf that has been found in Taurus ($J_{\rm 2MASS} = 13.78\pm0.02$~mag), with its bolometric luminosity $\approx 5$ times higher than other L-type members, leading to a very young age ($<$1~Myr) based on its HR diagram position (Figure \ref{fig:HR}). In addition, based on our adapted Besan\c{c}on Galactic model (Section~\ref{subsec:FCE}), none of the L0 \textsc{vl-g} field contaminants could pass our selection criteria (Table~\ref{tab:FCE}), suggesting that PSO~J065.8+19 is a very young dwarf or binary system in Taurus as opposed of a foreground object. Astrometric follow-up is needed for a more robust membership assessment.

\subsubsection{2MASS~J0619$-$2903}
\label{subsec:0619}
2M~0619$-$2903 was discovered and optically classified as a M6 dwarf by \cite{Cruz+2003}. Two epochs of near-infrared spectra were then obtained by AL13 (epoch 2008~November) and \cite{Liu+2016} (epoch 2015~December), who derived a spectral type of M5 (with an extinction of $A_{\rm V} = 6.5$~mag) and M6, respectively. Applying our classification method to the two spectra of 2M~0619$-$2903, we derive spectral types of SpT$_{\omega} =$M5.3$\pm0.9$ for the first epoch and M5.8$\pm0.9$ for the second epoch, consistent with the results by AL13 and \cite{Liu+2016}, respectively. It has been suggested that this object has a young age of a few$\times$10 Myr based on the low-gravity signatures in optical and near-infrared spectra (\citealt{Cruz+2003}; AL13; \citealt{Liu+2016}) and its very bright absolute magnitude \citep{Liu+2016}. 

In addition, we note that this object's $A_{\rm V}^{\rm H_{2}O}$ extinction is larger than its $A_{\rm V}^{\rm OIR}$ value by $\approx 0.9\sigma$ for both epochs (Table~\ref{tab:classification}), probably because it possesses a circumstellar disk and is variable \citep[AL13;][]{Liu+2016}. We adopt the $A_{\rm V}^{\rm H_{2}O}$ values for both epochs since the extinctions based on H$_{2}$O-band spectral indices are less vulnerable to disk-related processes (Section~\ref{subsubsec:extinction}). We thereby obtain $A_{\rm V}^{\rm H_{2}O} = 5.5\pm4.0$~mag for the first epoch and $A_{\rm V}^{\rm H_{2}O} = 6.3\pm4.0$~mag for the second epoch, consistent with results by AL13 within uncertainties.

\section{Conclusions}
\label{sec:conclusion}
We have presented initial results from a multi-epoch survey of young brown dwarfs and free-floating planets in the Taurus star-forming region based on the PS1 3$\pi$ Survey. We have selected candidates based on photometry and proper motions and have obtained near-infrared spectra for 83 objects, including $\approx 75\%$ of our candidates that have $J_{\rm 2MASS} \leqslant 15.5$~mag.

Precise magnitudes, colors, luminosities and spectral types for both of our candidates and previously known Taurus members are of great importance for constructing empirical isochrones and the IMF. Such measurements are hampered in young, dusty star-forming regions, as extinction alters spectral morphologies and individual molecular features, thereby complicating spectral typing and gravity classification. To solve this problem, we have developed a new classification scheme based on the AL13 system to quantitatively determine reddening-free spectral types, extinctions, and gravity classification for M4--L7 ultracool dwarfs ($\approx 100-3$~M$_{\rm Jup}$ in Taurus) using low-resolution ($R \approx 100$) near-infrared spectra. Following AL13, our method uses H$_{2}$O-band spectral indices for classification. We find that imperfect telluric correction has a negligible effect on H$_{2}$O indices for typical observations (Appendix~\ref{sec:appendix_tell}).

Using our classification method, we identify new Taurus members from spectroscopic follow-up. We also homogeneously reclassify spectral types and extinctions of all previously known mid-M to early-L objects in Taurus. We have thus far found 14 [M4, M6) dwarfs, although most could be field interlopers based on our modified version of the Besan\c{c}on Galactic model. We have also found 36 new substellar ($\geqslant$M6) members in Taurus including 25 \textsc{vl-g} and 11 \textsc{int-g} objects, constituting the largest single increase of brown dwarfs found in Taurus to date. We have therefore for the first time discovered Taurus members with \textsc{int-g} gravities. We estimate little field contamination among these new $\geqslant$M6 discoveries. Overall, we have increased the substellar census in Taurus by $\approx 40\%$ and also added three more L-type members (masses $\approx 5-10$~M$_{\rm Jup}$). Our success rate of $\approx 70\%$ for finding substellar members in Taurus is better than previous searches and demonstrates the robustness of our selection method. 

Most notably, we have found an older low-mass population ($>$10~Myr; $\approx 0.1-0.02$~M$_{\odot}$) in Taurus from our newly identified members, supporting recent studies with similar conclusions for the stellar ($\gtrsim 0.1$~M$_{\odot}$) members, \citep[e.g.,][]{Sestito+2008, Daemgen+2015, Kraus+2017}. The mass function appears to differ between the younger and older Taurus populations, which could be caused if older stellar members of Taurus were missed by previous searches, or if the two populations have experienced different star formation processes. In addition, while the younger population is mostly associated with high-extinction filaments and clumps, the older objects are more dispersed, with their sky locations not closely following the known members or extinction. 

We have also discovered 1 new substellar member in the Pleiades, 13 new substellar members in Per~OB2, and 11 reddened $<$M4 stars which are probably background field dwarfs. Follow-up measurements are warranted for all of our discoveries to refine their membership, including high-precision astrometry \citep[e.g.,][]{Gaia+2016} and age measurements (e.g., lithium abundance measurements via optical spectroscopy). 

In addition, 5 of our discoveries (including 4 new $\geqslant$M6 Taurus members and 1 new Pleiades member) and 5 known $\geqslant$M6 Taurus members have proper motions inconsistent with Taurus by 2--13$\sigma$ and thus perhaps are ejected brown dwarfs. We have also found 1 unusually bright L0 \textsc{vl-g} dwarf, which likely represents a very young object in Taurus. Direct distance measurements are needed to assess the membership of these 11 ($=5+5+1$) objects and to establish a comprehensive picture of the Taurus association.

Among our 13 new members in Per~OB2, which overlaps with our search area, two objects have a separation of $58$~kAU with spectral types of M6.6 and M8.2, respectively. They share similar proper motions and model-derived ages based on the HR diagram and thus could be a very wide binary. 

We have identified 11 circumstellar disk candidates in Taurus (5 objects) and Per~OB2 (6 objects). Five disk candidates in Taurus (3 of our new disk candidates and 2 known disk candidates) have older model-based ages ($30-100$~Myr) compared to the typical disk lifetime ($\lesssim 10$~Myr). Two Per~OB2 disk candidates form the aforementioned candidate very wide binary. Follow-up observations at (sub)millimeter wavelengths are of interest to investigate the disk evolution in these systems.

So far, our near-infrared spectroscopic follow-up has been finished for most of the brighter candidates ($75\%$ of $J_{\rm 2MASS} \leqslant 15.5$~mag). Upon completion, our discoveries will help complete the Taurus IMF in the substellar and planetary-mass regime and deliver a more comprehensive picture of the Taurus star formation history.

%============================================================================

\acknowledgements
We thank the anonymous referee for constructive comments that improve the paper significantly. We also thank Katelyn Allers for insightful comments on this work and Annie Robin for discussions about the Besan\c{c}on Galactic models. We thank Kevin Luhman for providing his SpeX spectra of previously known Taurus members. The Pan-STARRS1 Surveys (PS1) have been made possible through contributions by the Institute for Astronomy, the University of Hawaii, the Pan-STARRS Project Office, the Max-Planck Society and its participating institutes, the Max Planck Institute for Astronomy, Heidelberg and the Max Planck Institute for Extraterrestrial Physics, Garching, The Johns Hopkins University, Durham University, the University of Edinburgh, the Queen's University Belfast, the Harvard-Smithsonian Center for Astrophysics, the Las Cumbres Observatory Global Telescope Network Incorporated, the National Central University of Taiwan, the Space Telescope Science Institute, and the National Aeronautics and Space Administration under Grant No. NNX08AR22G issued through the Planetary Science Division of the NASA Science Mission Directorate, the National Science Foundation Grant No. AST-1238877, the University of Maryland, Eotvos Lorand University (ELTE), and the Los Alamos National Laboratory. This publication makes use of data products from the Wide-field Infrared Survey Explorer, which is a joint project of the University of California, Los Angeles, and the Jet Propulsion Laboratory/California Institute of Technology, funded by the National Aeronautics and Space Administration. This publication makes use of data products from the Two Micron All Sky Survey, which is a joint project of the University of Massachusetts and the Infrared Processing and Analysis Center/California Institute of Technology, funded by the National Aeronautics and Space Administration and the National Science Foundation. UKIRT is owned by the University of Hawaii (UH) and operated by the UH Institute for Astronomy; operations are enabled through the cooperation of the East Asian Observatory. When (some of) the data reported here were acquired, UKIRT was operated by the Joint Astronomy Centre on behalf of the Science and Technology Facilities Council of the U.K. This research has made use of the SIMBAD database, the VizieR catalogue access tool, and ``Aladin sky atlas'' developed and operated at CDS, Strasbourg, France; the NASA/IPAC Extragalactic Database (NED), which is operated by the Jet Propulsion Laboratory, California Institute of Technology, under contract with the National Aeronautics and Space Administration; the NASA's Astrophysics Data System; the SpeX Prism Library, maintained by Adam Burgasser at \url{http://www.browndwarfs.org/spexprism}; and the catalog of photometry and proper motions of M, L, and T dwarfs from the PS1 $3\pi$ survey, maintained by William Best at \url{http://www.ifa.hawaii.edu/~wbest/Will_Best/PS1_MLT_Dwarfs.html}. This work was greatly facilitated by many features of the TOPCAT software written by Mark Taylor (\url{http://www.starlink.ac.uk/topcat/}). Finally, the authors wish to recognize and acknowledge the very significant cultural role and reverence that the summit of Maunakea has always had within the indigenous Hawaiian community.  We are most fortunate to have the opportunity to conduct observations from this mountain. 
\facilities{PS1, IRTF (SpeX), UKIRT (WFCAM)}
\software{IPython \citep{Perez+2007}, Numpy and Scipy \citep{Walt+2011}, Astropy \citep{Astropy+2013}, Matplotlib \citep{Hunter+2007}}

%----------------------------------------------------------------------------

%============================================================================

\newpage 

\appendix 

\section{Impact of Imperfect Telluric Correction on H$_{2}$O Indices}
\label{sec:appendix_tell}

Ground-based spectroscopic observations are affected by atmospheric (telluric) absorption especially in the infrared and (sub-)millimeter regime. Consequently, the observed near-infrared spectrum $O(\lambda)$ of an object is a combination of its intrinsic spectrum $I_{\rm intr}(\lambda)$ and the telluric absorption spectrum $T(\lambda)$:
\begin{equation}
O(\lambda) = I_{\rm intr}(\lambda) \cdot \left[ T(\lambda) \ast P(\lambda) \right] \cdot Q(\lambda)
\label{eq:Olambda}
\end{equation}
where ``$\ast$'' denotes a convolution, and $P(\lambda)$ and $Q(\lambda)$ are the instrumental profile and the instrumental throughput, respectively, following the notation of \citeauthor{Vacca+2003} (\citeyear{Vacca+2003}; their Section~2). However, measurements of the $T({\lambda})$ that corresponds to the science target are imperfect. The common method is to contemporaneously observe a nearby A0V standard star with an airmass difference of $\lesssim 0.1$, thereby obtain $O_{\rm std}(\lambda)$. Since the intrinsic spectrum of the A0V standard, $I_{\rm std}(\lambda)$, is known \citep[e.g., based on a model spectrum of Vega;][]{Vacca+2003}, the telluric absorption of the standard star $T_{\rm std}(\lambda)$ can be calculated based on Equation~\ref{eq:Olambda}. Thus we can derive the $I_{\rm intr}(\lambda)$ for the science target based on $O(\lambda)$, by assuming that the target and the A0V standard share the same telluric absorption spectrum, i.e., $T_{\rm obj}(\lambda) = T_{\rm std}(\lambda)$. The above procedure is usually termed ``telluric correction''.

However, the assumption of $T_{\rm obj}(\lambda) = T_{\rm std}(\lambda)$ is in fact not exactly true, given that telluric absorption varies with airmass (a function of sky position) and precipitable water vapor (pwv; a function of time). Such variations in $T(\lambda)$ cause imperfect telluric correction and leave some telluric features in the fully reduced spectrum, due to the different airmass and pwv between the science target and its A0V telluric standard. In this Appendix, we provide a quantitative analysis about the impact of such imperfect telluric correction on H$_{2}$O indices.

Considering that $T_{\rm obj}(\lambda) \neq T_{\rm std}(\lambda)$, the fully reduced spectrum $\widetilde I_{\rm intr} (\lambda)$ of the science target is  therefore not the intrinsic spectrum $I_{\rm intr} (\lambda)$ as expected but rather
\begin{equation}
\widetilde I_{\rm intr} (\lambda) = I_{\rm intr} (\lambda) \cdot \frac{T_{\rm obj}(\lambda;\rm am_{obj}, pwv_{obj}) \ast P(\lambda)}{T_{\rm std}(\lambda;\rm am_{std}, pwv_{std}) \ast P(\lambda)} \equiv I_{\rm intr} (\lambda) \cdot s(\lambda;\rm am_{obj}, pwv_{obj}; am_{std}, pwv_{std}) \label{eq:tilde_i}
\end{equation}
where $(\rm am_{obj}, pwv_{obj})$ and $(\rm am_{std}, pwv_{std})$ are the airmass and pwv corresponding to the science target and the standard star, respectively. The $s(\lambda) \equiv 1$ only if $\rm am_{obj} = am_{std}$ and $\rm pwv_{obj} = pwv_{std}$, which is infeasible in practice. The non-unity of $s(\lambda)$ causes the difference between the measured and intrinsic H$_{2}$O indices. Based on Equations~\ref{eq:H2O_index} and \ref{eq:H2O_color}, the measured H$_{2}$O indices $\widetilde W_{z}$ are
\begin{equation} 
\widetilde W_{z} = -2.5\ {\rm log}_{10}\left( \frac{\bigintsss_{\lambda\lambda_{z,\rm num}} \widetilde I_{\rm intr} (\lambda) d\lambda}{\bigintsss_{\lambda\lambda_{z,\rm den}} \widetilde I_{\rm intr} (\lambda) d\lambda} \right) = -2.5\ {\rm log}_{10}\left( \frac{\bigintsss_{\lambda\lambda_{z,\rm num}} I_{\rm intr} (\lambda) \cdot s(\lambda) d\lambda}{\bigintsss_{\lambda\lambda_{z,\rm den}} I_{\rm intr} (\lambda) \cdot s(\lambda) d\lambda} \right)
\end{equation} 
where $\lambda\lambda_{z,\rm num}$ and $\lambda\lambda_{z,\rm den}$ are the wavelength ranges of the numerators and denominators in definitions of $W_{z}$. For simplicity, we ignore the convolution between $T(\lambda)$ and the instrumental profile $P(\lambda)$ in Equation~\ref{eq:tilde_i} and thus directly compute $s(\lambda)$ as $T_{\rm obj}(\lambda) / T_{\rm std}(\lambda)$. Assuming $s(\lambda)$ is constant in narrow bands of $\lambda\lambda_{z,\rm num}$ and $\lambda\lambda_{z,\rm den}$ with a mean value of $s_{z, \rm num}$ and $s_{z, \rm den}$, respectively, we can then express $\widetilde W_{z}$ as:
\begin{equation} 
\widetilde W_{z} = -2.5\ {\rm log}_{10}\left( \frac{ \bigintsss_{\lambda\lambda_{z,\rm num}} I_{\rm intr} (\lambda) d\lambda}{ \bigintsss_{\lambda\lambda_{z,\rm den}} I_{\rm intr} (\lambda) d\lambda} \cdot \frac{s_{z, \rm num}}{s_{z, \rm den}} \right) = W_{z} - 2.5\ {\rm log}_{10}\left( \frac{s_{z, \rm num}}{s_{z, \rm den}} \right)
\end{equation} 
Therefore, the error in $W_{z}$ induced by imperfect telluric correction is
\begin{equation} 
\Delta W_{z} ({\rm am}_{\rm obj}, {\rm pwv}_{\rm obj}; {\rm am}_{\rm std}, {\rm pwv}_{\rm std}) = \left \lvert \widetilde W_{z} - W_{z} \right \rvert = \left \lvert -2.5\ {\rm log}_{10}\left( \frac{s_{z, \rm num}}{s_{z, \rm den}} \right) \right \rvert
\label{eq:DeltaWz}
\end{equation} 

In order to quantify $\Delta W_{z}$, we use Maunakea telluric absorption spectra with different airmass ($1.0$, $1.5$, $2.0$) and pwv ($1.0$, $1.6$, $3.0$, $5.0$~mm), which are generated by the ATRAN modeling software (Lord, S.D. 1992, NASA Technical Memor. 103957)\footnote{\url{http://www.gemini.edu/sciops/telescopes-and-sites/observing-condition-constraints/ir-transmission-spectra}.}. Then we assume the $T_{\rm std} (\lambda)$ in $s(\lambda)$ (Equation~\ref{eq:tilde_i}) corresponds to airmass$=$1.0 and pwv$=$1.0~mm, i.e., $T_{\rm std}(\lambda;\rm am_{std}=1.0, pwv_{std}=1.0mm)$, and use all 12 ($=$ 3 airmasses $\times$ 4 pwv's) ATRAN combinations to compute $\Delta W_{z} (\rm am_{obj}, pwv_{obj}; am_{std}=1.0, pwv_{std}=1.0mm)$ for $T_{\rm obj} (\lambda)$ with different airmass and pwv based on Equation~\ref{eq:DeltaWz}. Hereafter, we use $\Delta W_{z; \rm 1.0/1.0} (\rm am_{obj}, pwv_{obj})$ to denote $\Delta W_{z} (\rm am_{obj}, pwv_{obj}; am_{std}=1.0, pwv_{std}=1.0mm)$. By definition,
\begin{equation}
\Delta W_{z; \rm 1.0/1.0} (\rm am_{obj}=1.0, pwv_{obj}=1.0mm) \equiv 0
\end{equation}

We first investigate the impact of airmass differences between the science target and telluric standard. An airmass difference of $\lesssim 0.1$ between objects and standard stars is typical for telluric correction, with our observations having a median absolute airmass difference of $\approx 0.055$ (Table~\ref{tab:spex_log}). As presented in the left panels of Figure~\ref{fig:tell_H2O}, for a given pwv value, we perform a linear fit to $\Delta W_{z; \rm 1.0/1.0}$ as a function of airmass and obtain the slope $k_{z}^{\rm 0.05am}$ that corresponds to the change in $W_{z}$ when the airmass differs by $0.05$. Thus, $k_{z}^{\rm 0.05am}$ is the systematic error in $W_{z}$ purely induced by an airmass difference of 0.05 between the science target and the telluric standard. In comparison, based on our ultracool dwarf sample in Section~\ref{subsec:revisit} (a combination of the AL13 sample, the SpeX Prism Spectral Libraries, and the IRTF Spectral Library), we obtain typical measurement uncertainties ($\sigma_{z}^{\rm obs}$) in H$_{2}$O indices of $\sigma_{0}^{\rm obs} = 0.02$, $\sigma_{D}^{\rm obs} = 0.02$, $\sigma_{1}^{\rm obs} = 0.01$, and $\sigma_{2}^{\rm obs} = 0.02$. Comparing $k_{z}^{\rm 0.05am}$ and $\sigma_{z}^{\rm obs}$ demonstrates the relative importance of imperfect telluric correction on the measured $W_{z}$ due to airmass differences. As shown in Figure~\ref{fig:tell_H2O}, the telluric contamination is negligible for the four H$_{2}$O indices in this work, as their $k_{z}^{\rm 0.05am}$ are smaller than $\sigma_{z}^{\rm obs}$ by factors of $\approx 2-10$.

Next, we explore the impact of pwv differences between the science target and telluric standard. We estimate the pwv values on Maunakea using the $\tau_{\rm 225GHz}$ (opacity at 225~GHz) measured by the Caltech Submillimeter Observatory radiometer\footnote{\url{http://www.eao.hawaii.edu/weather/opacity/mk/archive/?C=M;O=D}.} and converting $\tau_{\rm 225GHz}$ into pwv based on \cite{Dempsey+2013}. According to the archived $\tau_{\rm 225GHz}$ values, pwv varies by $\approx 0.1$~mm per $30$~min during our observation dates (Table~\ref{tab:spex_log}). In the right panels of Figure~\ref{fig:tell_H2O}, for a given airmass, we perform a linear fit to $\Delta W_{z; \rm 1.0/1.0}$ as a function of pwv and obtain the slope $k_{z}^{\rm 0.1pwv}$ that indicates the change in $W_{z}$ when the pwv differs by $0.1$~mm. Since the typical integration time of our science targets is $\lesssim 30$~min (Table~\ref{tab:spex_log}), $k_{z}^{\rm 0.1pwv}$ is a representative systematic error in $W_{z}$ induced by the pwv variability. As shown in Figure~\ref{fig:tell_H2O}, $k_{z}^{\rm 0.1pwv}$ is smaller than $\sigma_{z}^{\rm obs}$ by factors of $\approx 2-500$ for the four H$_{2}$O indices. Therefore, variable pwv values do not cause significant errors in the measured $W_{z}$ presented in this work.

Finally, we provide an estimate of the composite systematic error in the H$_{2}$O indices $\Delta W_{z,\rm tell}$ caused by telluric variations in both airmass and pwv:
\begin{equation}
\Delta W_{z,\rm tell} = \sqrt{ \left[k_{z}^{\rm 0.05am}({\rm pwv}) \times \frac{\rm \Delta am}{0.05} \right]^{2} + \left[ k_{z}^{\rm 0.1pwv}({\rm am}) \times \frac{\rm \Delta pwv}{0.1\rm ~mm} \right]^{2} }
\label{eq:DelWz_tell}
\end{equation}
where $\Delta \rm am$ and $\Delta \rm pwv$ is the difference of airmass and pwv between the science target and telluric standard, respectively, and $k_{z}^{\rm 0.05am}({\rm pwv})$ and $k_{z}^{\rm 0.1pwv}({\rm am})$ can be obtained by interpolating Tables~\ref{tab:k_Wz_005am} and \ref{tab:k_Wz_01mm} based on the pwv and airmass corresponding to the observation, respectively. This additional error from Equation~\ref{eq:DelWz_tell} could be incorporated into the measured $W_{z}$ uncertainties when $\Delta W_{z,\rm tell}$ is comparable with or larger than $\sigma_{z}^{\rm obs}$.

\end{CJK*}

%============================================================================

\clearpage
\bibliographystyle{aasjournal}
\bibliography{Z18_PS1TauBD}

%%%%%%%%%%%%%%%%%%%%%%
%------------- Extinction Map --------------------
%%%%%%%%%%%%%%%%%%%%%%
%------------figure end-----------------

%%%%%%%%%%%%%%%%%%%%%%%%%%
%------------- g-r vs. g-i. &&  i-y vs. i-z--------------------
%%%%%%%%%%%%%%%%%%%%%%%%%%
%------------figure end-----------------

%%%%%%%%%%%%%%%%%%%%%%%%%%%%%%%
%------------- y-K vs. z-y  &&  W1-W2 vs. y-W1 --------------------
%%%%%%%%%%%%%%%%%%%%%%%%%%%%%%%
%------------figure end-----------------

%%%%%%%%%%%%%%%%%%%%%%%%%%%
%------------- J-K vs. y-J  &&  J vs. y-J --------------------
%%%%%%%%%%%%%%%%%%%%%%%%%%%
%------------figure end-----------------

%%%%%%%%%%%%%%%%%%%%
%--------------- J-H vs. H-K ----------------------
%%%%%%%%%%%%%%%%%%%%
%------------figure end-----------------

\clearpage
%%%%%%%%%%%%%%%%%%%%%%%%%
% ------------- AL13 H2O vs. OSpt --------------------
%%%%%%%%%%%%%%%%%%%%%%%%%
%------------figure end-----------------

%%%%%%%%%%%%%%%%%%%%%%%%%
% ------------- OSpt vs. ZJ H2O  --------------------
%%%%%%%%%%%%%%%%%%%%%%%%%
%------------figure end-----------------

\clearpage
%%%%%%%%%%%%%%%%%%%%%%%%%
% ------------- omegax sequence  --------------------
%%%%%%%%%%%%%%%%%%%%%%%%%
%------------figure end-----------------

%%%%%%%%%%%%%%%%%%%%%%%%%
% ------------- omegax sequence  --------------------
%%%%%%%%%%%%%%%%%%%%%%%%%
%------------figure end-----------------

%%%%%%%%%%%%%%%%%%%%%%%%%
% ------------- i/z/y-J vs. SpT  --------------------
%%%%%%%%%%%%%%%%%%%%%%%%%
%------------figure end-----------------

%%%%%%%%%%%%%%%%%%%%%%%%%%%%%%%%
% ------------- stack histograms i/z/y-J vs. SpT  --------------------
%%%%%%%%%%%%%%%%%%%%%%%%%%%%%%%%
%------------figure end-----------------

%%%%%%%%%%%%%%%%%%%%%%%%%%%%%%%%
% ------------- flowchart  --------------------
%%%%%%%%%%%%%%%%%%%%%%%%%%%%%%%%
%------------figure end-----------------

\clearpage
%%%%%%%%%%%%%%%%%%%%%%
% ------------- Taurus Spectra  --------------------
%%%%%%%%%%%%%%%%%%%%%%
%%% Taurus A
\begin{figure}[t]
\begin{center}
\includegraphics[height=7.in]{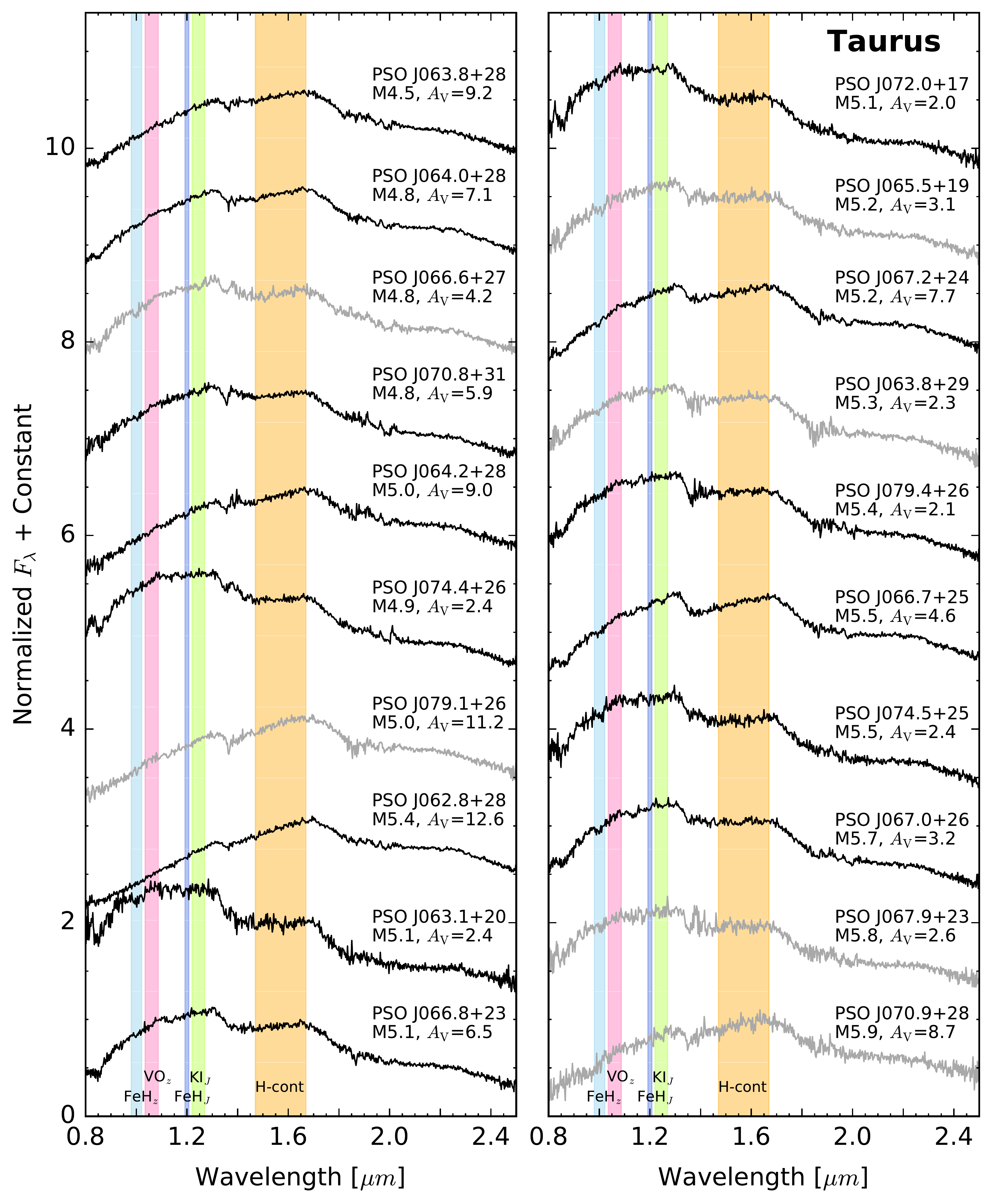}
\caption{Near-infrared spectra of our Taurus discoveries and one Pleiades discovery (PSO~J058.8+21), with their names, reddening-free spectral types, extinctions, and gravity classification (if available) noted. Gravity-sensitive features (i.e., FeH$_{z}$, VO$_{z}$, FeH$_{J}$, K\textsc{i}$_{J}$, and H-cont) used for youth assessment (Section~\ref{subsec:youth_logistics}) are shown as colored shadows. Eight objects with {\it gray} colors have low S/N ($\lesssim 30$ per pixel in $J$ band) and need better data for more robust spectral classification.}
\label{fig:spec_Taurus}
\end{center}
\end{figure}
%------------figure end-----------------
\renewcommand{\thefigure}{\arabic{figure}}
\addtocounter{figure}{-1}
%%% Taurus B
\begin{figure}[t]
\begin{center}
\includegraphics[height=7.in]{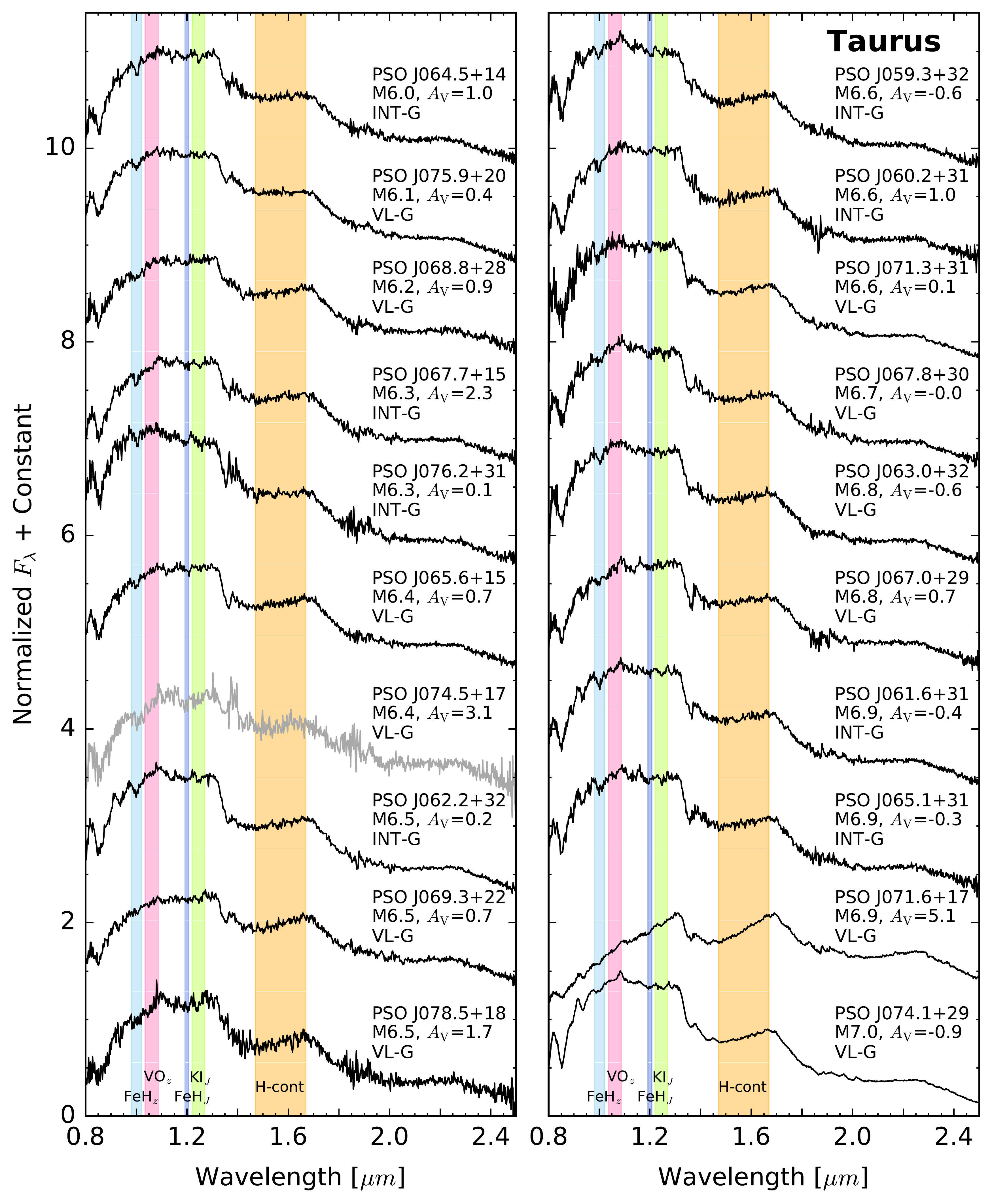}
\caption{Continued}
\end{center}
\end{figure}
%------------figure end-----------------
\addtocounter{figure}{-1}
%%% Taurus C
\begin{figure}[t]
\begin{center}
\includegraphics[height=7.in]{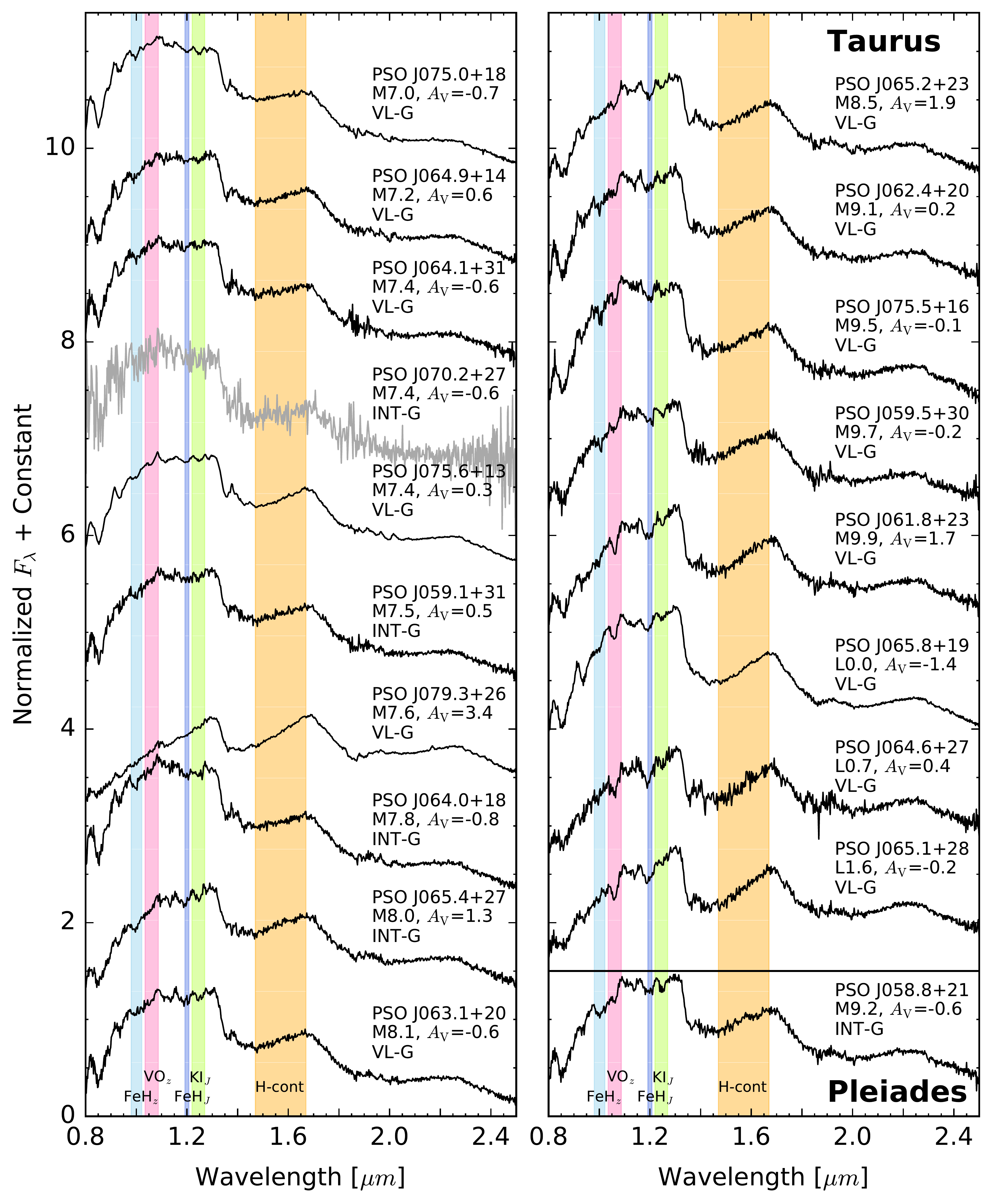}
\caption{Continued}
\end{center}
\end{figure}
%------------figure end-----------------

\clearpage
%%%%%%%%%%%%%%%%%%%%%%%%
% ------------- Per OB2 Spectra  --------------------
%%%%%%%%%%%%%%%%%%%%%%%%
\begin{figure}[t]
\begin{center}
\includegraphics[height=4.5in]{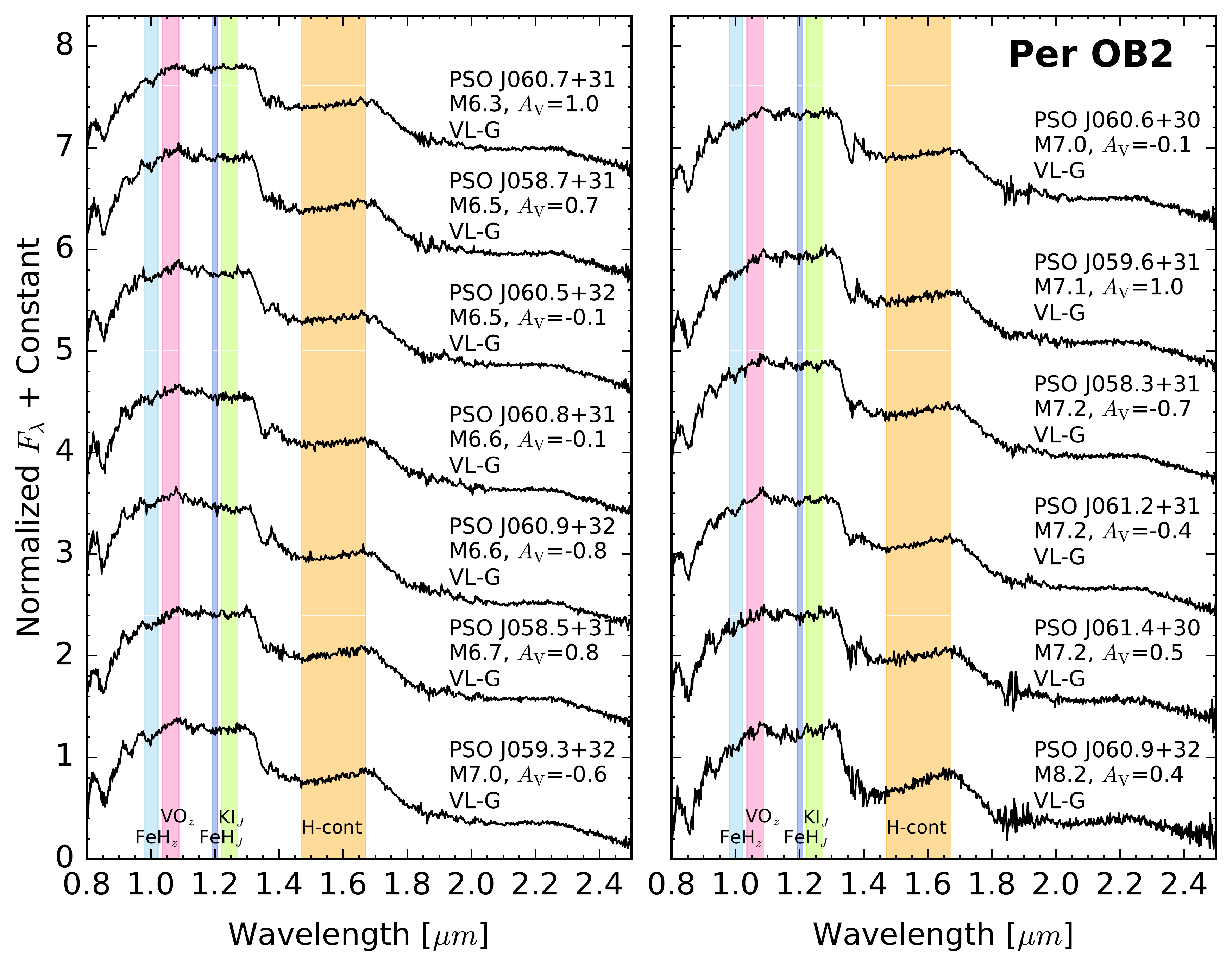}
\caption{Near-infrared spectra of newly confirmed brown dwarf members in Per~OB2, using the format of Figure~\ref{fig:spec_Taurus}.}
\label{fig:spec_Per}
\end{center}
\end{figure}
%------------figure end-----------------

\clearpage
%%%%%%%%%%%%%%%%%%%%%%%%%%%%%%
%------------------ spatial distribution (gravity classification) -------------------------
%%%%%%%%%%%%%%%%%%%%%%%%%%%%%%
\begin{figure}[t]
\begin{center}
\includegraphics[height=5.5in]{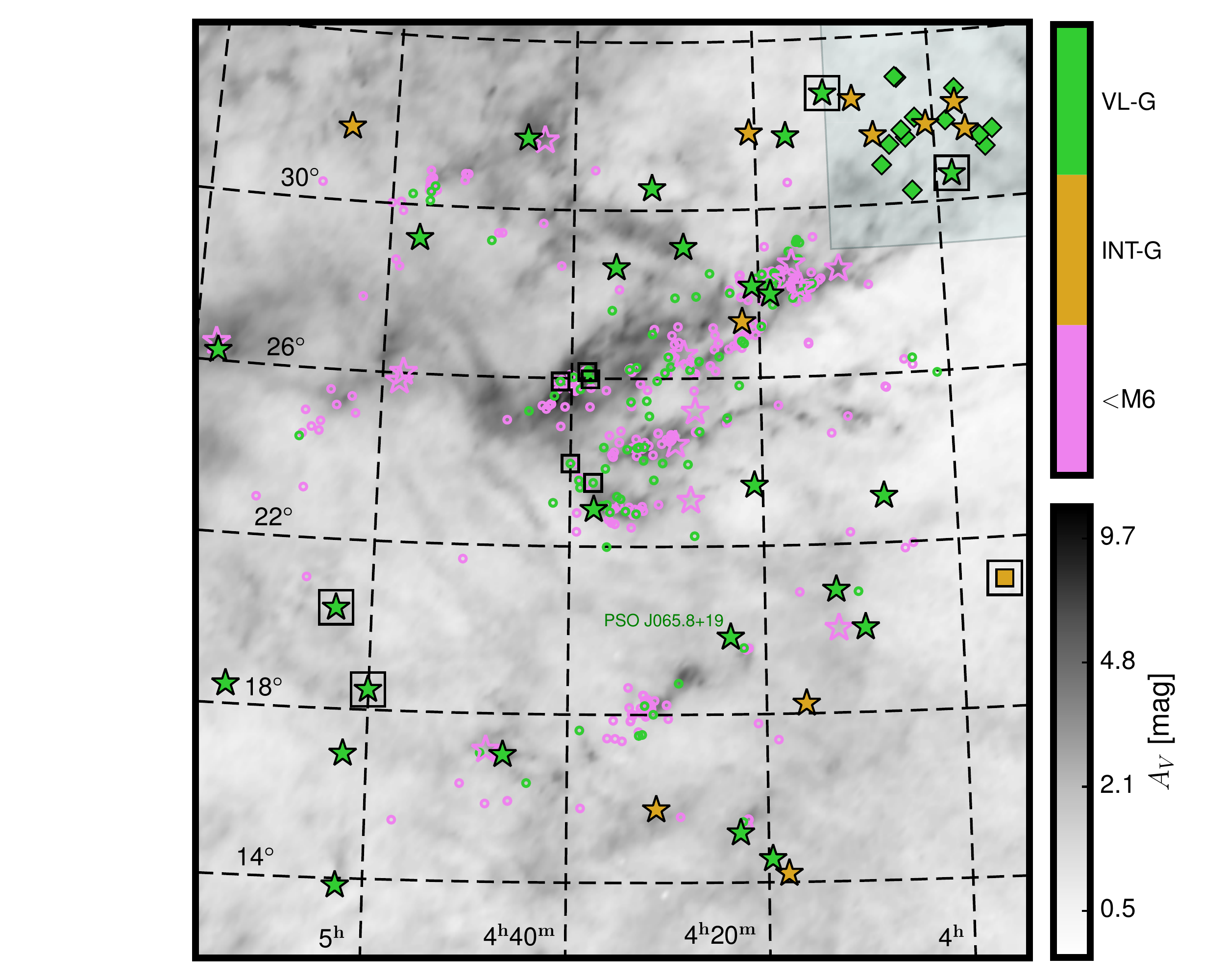}
\caption{Spatial distribution of our discoveries in Taurus (stars), Pleiades (square), and Per~OB2 (diamonds), and previously known objects (circles) in Taurus, as described in the caption of Figure~\ref{fig:ExtMap}. Here the plotting colors represent gravity classifications. {\it Green} colors are for \textsc{vl-g}, {\it gold} for \textsc{int-g}, and {\it pink} for no AL13 gravity classification, due to objects' too early spectral types ($<$M6) for the AL13 system (Section~\ref{subsec:youth_logistics}).}
\label{fig:extmap_gravclass}
\end{center}
\end{figure}
%------------figure end-----------------

\clearpage
%%%%%%%%%%%%%%%%%%%%%%%%%
%------------- SpT Distribution --------------------
%%%%%%%%%%%%%%%%%%%%%%%%%
\begin{figure}[t]
\begin{center}
\includegraphics[height=4.5in]{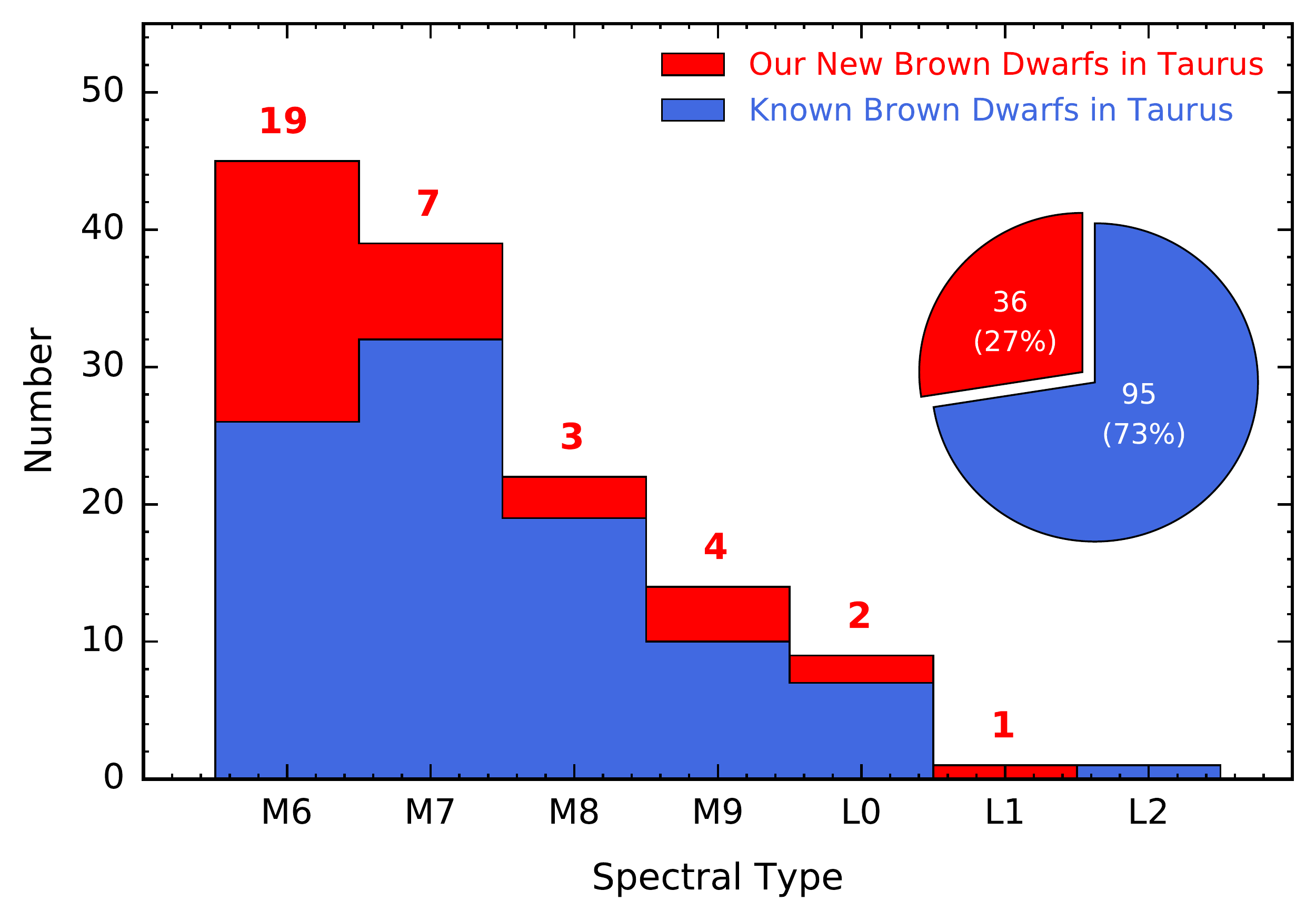}
\caption{The spectral type distributions of all brown dwarfs ($\geqslant$M6) in Taurus. The pie chart shows the relative numbers and fractions of objects in different categories. {\it Blue} regions show the 95 previously known Taurus brown dwarfs, using spectral types from our reclassificaiton. {\it Red} regions show our 36 newly identified brown dwarfs with robust spectral classification. The red number above each spectral type bin indicates the number of our newly confirmed members. Our discoveries so far have increased the current substellar census ($\geqslant$M6) in Taurus by $\approx 40\%$ and added three more L dwarfs ($\approx 3-10$~M$_{\rm Jup}$), constituting the largest single increase of brown dwarfs found in Taurus to date.   }
\label{fig:spt_dist}
\end{center}
\end{figure}
%------------figure end-----------------

\clearpage
%%%%%%%%%%%%%%%%%%%%%%%%%
%------------- SpT Comparison --------------------
%%%%%%%%%%%%%%%%%%%%%%%%%
\begin{figure}[t]
\begin{center}
\includegraphics[height=5.5in]{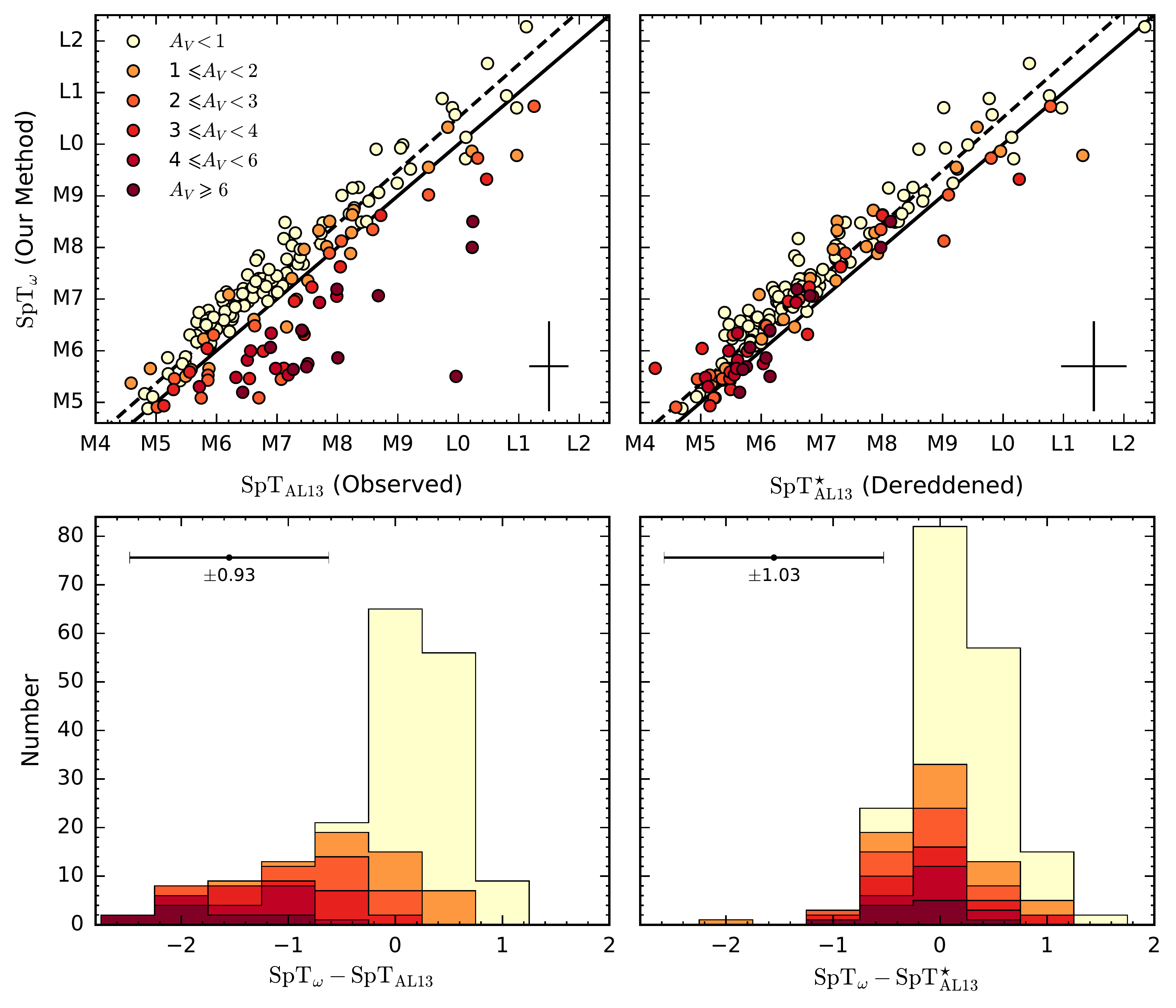}
\caption{Comparison between our reddening-free spectral types (SpT$_{\omega}$; Section \ref{subsec:rfspt}) and the index-based AL13 spectral types derived from observed (SpT$_{\rm AL13}$; {\it left}) and dereddened (SpT$^{\star}_{\rm AL13}$; {\it right}) spectra. The plotted sample contains our 64 new discoveries in Taurus (50 objects), Pleiades (1 object), and Per~OB2 (13 objects), and 133 reclassified previously known objects, including 130 Taurus M- and L-type members and three reddened young field dwarfs studied by AL13. While 2M~0619$-$2903 (AL13) has two epochs of near-infrared spectra (2008 November and 2015 December; see Section~\ref{subsec:0619}), here we only include the second epoch (2015), which has a higher $J$-band S/N. We divide the sample into different colors based on extinctions computed by our new classification scheme. We use redder colors for objects with higher reddening. Top: the black solid line is the one-to-one relation, and the black dashed line is the fitted SpT$_{\omega}$--SpT$^{\star}_{\rm AL13}$ correlation (Equation \ref{eq:SpT_omega_AL13}). Typical uncertainties of the measurements are shown at the lower right corner. Bottom: Histograms of ${\rm SpT}_{\omega} - {\rm SpT}_{\rm AL13}$ and ${\rm SpT}_{\omega} - {\rm SpT}^{\star}_{\rm AL13}$. Typical uncertainties of the spectral type differences are shown at the top. }
\label{fig:spt_comp}
\end{center}
\end{figure}
%------------figure end-----------------

\clearpage
%%%%%%%%%%%%%%%%%%%%%%%%%%%%%%%%%%
% ------------- Taurus Spectra - reddening comparison  --------------------
%%%%%%%%%%%%%%%%%%%%%%%%%%%%%%%%%%
\begin{figure}[t]
\begin{center}
\includegraphics[height=9.in]{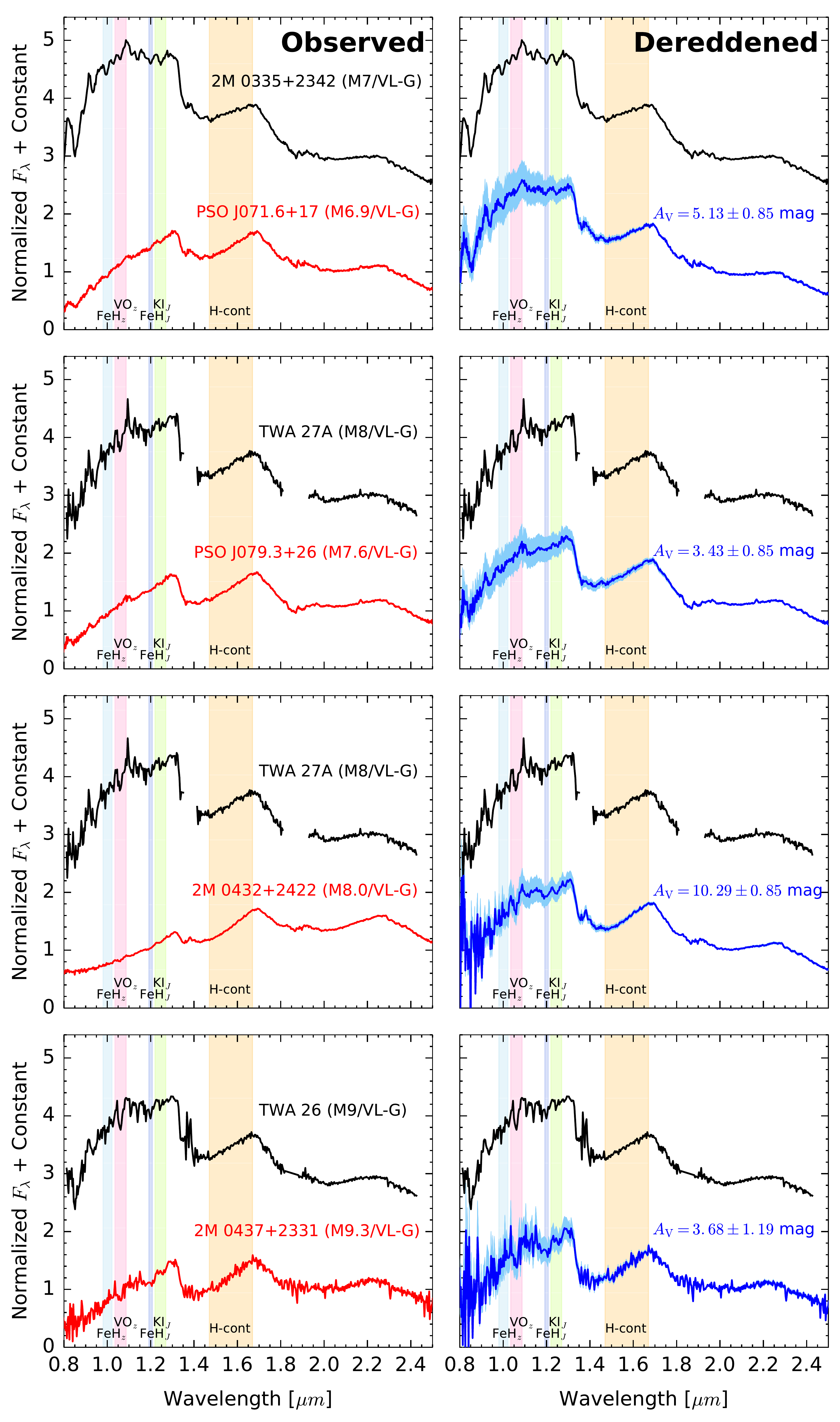}
\caption{Observed ({\it left, red}) and dereddened ({\it right, blue}) near-infrared spectra of four Taurus objects, compared to the \textsc{vl-g} dwarf standards ({\it black}; AL13) with similar spectral types, including 2MASS~J03350208+2342356 (M7, \textsc{vl-g}), TWA 27A (M8, \textsc{vl-g}), and TWA 26 (M9, \textsc{vl-g}). The {\it top} two objects are our new discoveries and the {\it bottom} two are previously known Taurus members. The spectra dereddened using $A_{\rm V}$ values within uncertainties are shown as {\it light blue}. }
\label{fig:spec_Taurus_dr}
\end{center}
\end{figure}
%------------figure end-----------------

\clearpage
%%%%%%%%%%%%%%%%%%%%%%%%%
%------------- Av H2O vs. OIR --------------------
%%%%%%%%%%%%%%%%%%%%%%%%%
\begin{figure}[t]
\begin{center}
\includegraphics[height=3.5in]{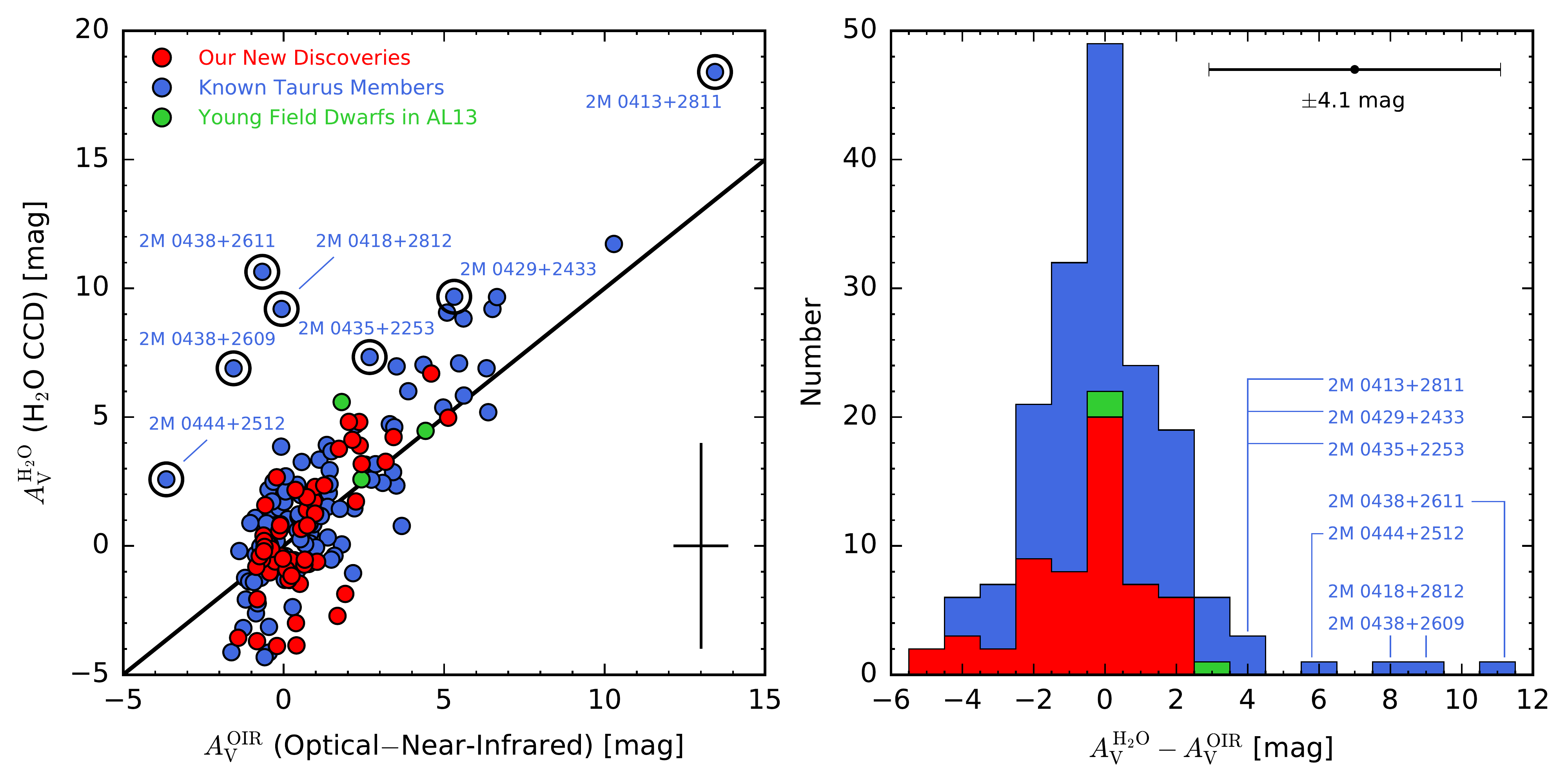}
\caption{Comparison between extinctions derived from H$_{2}$O color-color diagrams ($A_{\rm V}^{\rm H_{2}O}$; Section \ref{subsubsec:WxCCD}) and from intrinsic optical--near-infrared colors ($A_{\rm V}^{\rm OIR}$; Section \ref{subsubsec:opt_NIR_color}), for our 64 new discoveries ({\it red}) in Taurus (50 objects), Pleiades (1 object), and Per~OB2 (13 objects), and for 133 reclassified previously known objects, including 130 Taurus M- and L-type members ({\it blue}) and 3 reddened young field dwarfs studied by AL13 ({\it green}). As described in the caption of Figure~\ref{fig:spt_comp}, the results of 2M~0619$-$2903 (AL13) shown in the figure are based on the second epoch (2015) of its near-infrared spectrum (Section~\ref{subsec:0619}). The solid line is the one-to-one relation, and the typical measurement uncertainties are shown in the lower right corner. We use black circles to mark the objects with notable discrepancies ($>$1--2$\sigma$) between $A_{\rm V}^{\rm H_{2}O}$ and $A_{\rm V}^{\rm OIR}$. They have circumstellar disks with high inclinations and accretion activity, and/or are photometrically variable, so that their optical--near-infrared colors do not provide robust reddening measurements (Section \ref{subsubsec:extinction}). Right: Histogram of differences between $A_{\rm V}^{\rm H_{2}O}$ and $A_{\rm V}^{\rm OIR}$ with the typical uncertainty shown at the top.  Our two methods of extinction determination produce consistent results within uncertainties.}
\label{fig:Av_H2O_OIR}
\end{center}
\end{figure}
%------------figure end-----------------

\clearpage
%%%%%%%%%%%%%%%%%%%%%%%%%
%------------- photomtric variability --------------------
%%%%%%%%%%%%%%%%%%%%%%%%%
\begin{figure}[t]
\begin{center}
\includegraphics[height=3.in]{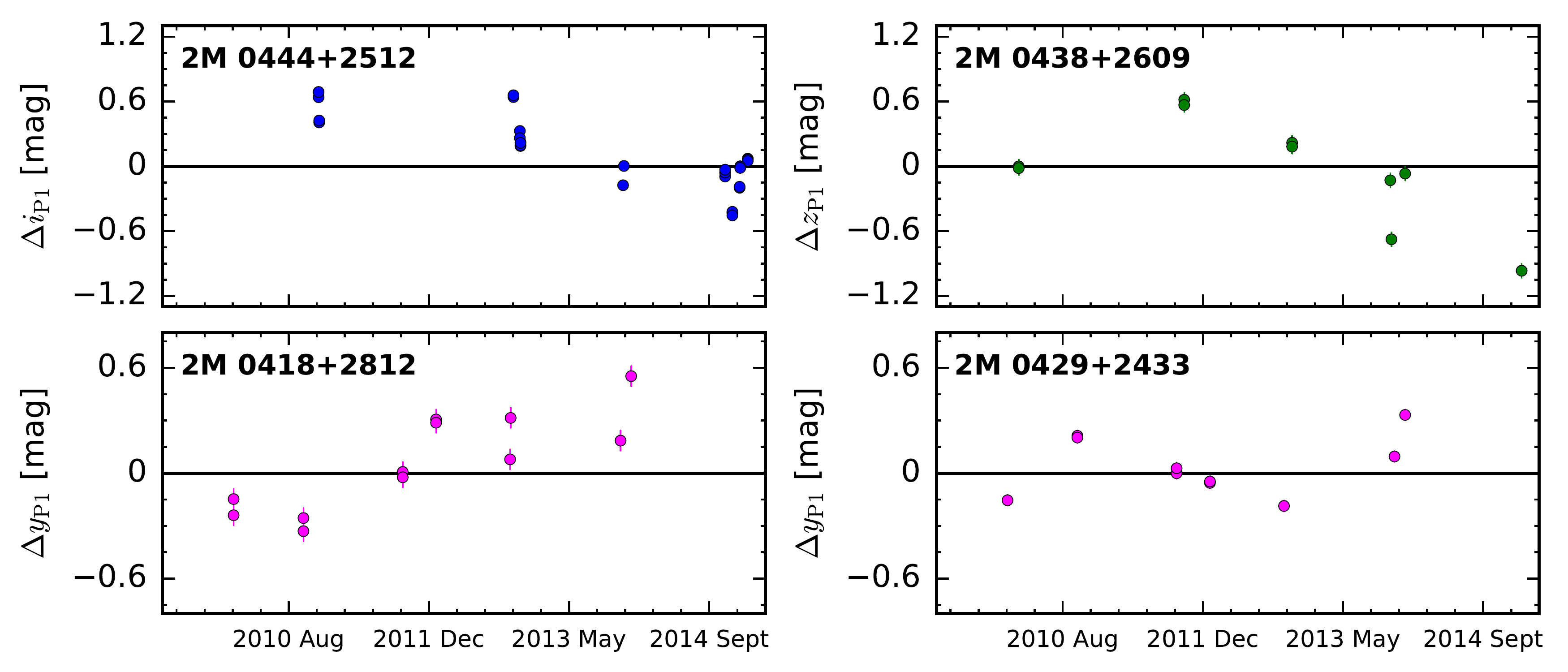}
\caption{Light curves of 2M~0418+2812, 2M~0429+2433, 2M~0438+2609, and 2M~0444+2512, in $i_{\rm P1}$ ({\it top left}), $z_{\rm P1}$ ({\it top right}), and $y_{\rm P1}$ ({\it bottom}) bands. The horizontal lines at $0$ correspond to objects' averaged magnitudes over all epochs. Uncertainties in magnitude are shown if they exceed the size of the symbols. These objects are variable with peak-to-peak amplitudes of $\approx 0.5-1.5$~mag over the PS1 3$\pi$ Survey timeframe (2010 May$-$2014 December). }
\label{fig:phot_var}
\end{center}
\end{figure}
%------------figure end-----------------

%%%%%%%%%%%%%%%%%%%%%%%%%
%------------- Av Our vs. Literature --------------------
%%%%%%%%%%%%%%%%%%%%%%%%%
\begin{figure}[t]
\begin{center}
\includegraphics[height=3.5in]{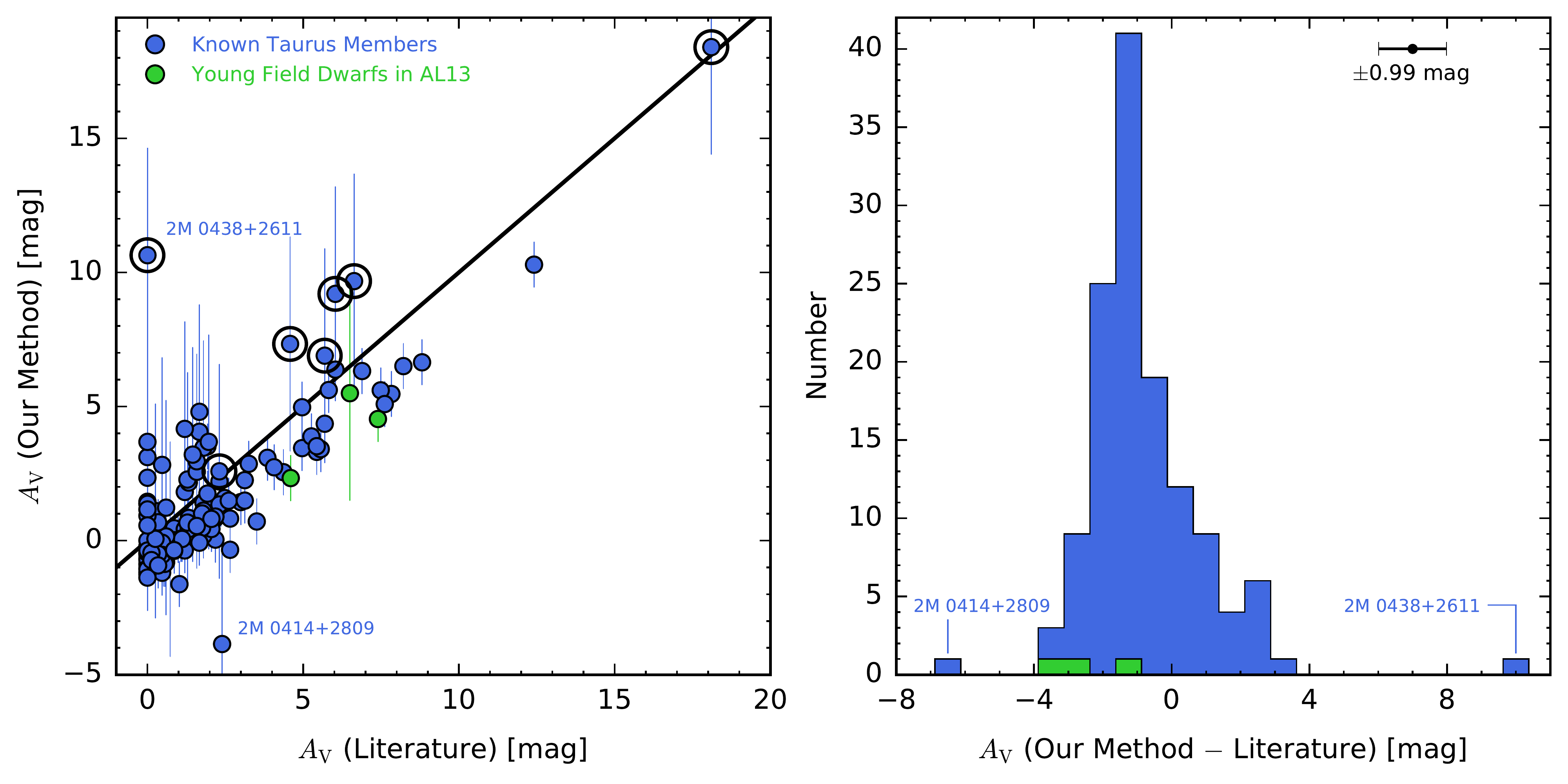}
\caption{Comparison between our final extinctions and the values from literature using 133 reclassified known objects, including 130 Taurus M- and L-type members ({\it blue}) and 3 reddened young field dwarfs studied by AL13 ({\it green}). As described in the caption of Figure~\ref{fig:spt_comp}, the results of 2M~0619$-$2903 (AL13) shown in the figure are based on the second epoch (2015) of its near-infrared spectrum (Section~\ref{subsec:0619}). Left: The solid line corresponds to the one-to-one relation. We use black circles to mark the objects with notable discrepancies ($>$1--2$\sigma$) between $A_{\rm V}^{\rm H_{2}O}$ and $A_{\rm V}^{\rm OIR}$, as described in the caption of Figure~\ref{fig:Av_H2O_OIR}. Right: Histogram of the differences between the extinctions derived based on our method and in the literature. Outliers are noted and discussed in Section~\ref{subsubsec:extinction}. Our $V$-band extinctions are systematically smaller than the literature with a weighted mean difference of $0.89$~mag, though comparable with the typical uncertainty of $\approx 0.99$~mag noted at the upper right corner. We assume a $A_{\rm V}$ uncertainty of 0.5~mag \citep[corresponding to an error of $\approx$0.1--0.3~mag error in $A_{\rm J}$;][]{Luhman+2017} for the literature values.}
\label{fig:Av_ZJ_lit}
\end{center}
\end{figure}
%------------figure end-----------------

%%%%%%%%%%%%%%%%%%%%%%%%%
%------------- Av Our vs. ExtMap --------------------
%%%%%%%%%%%%%%%%%%%%%%%%%
\begin{figure}[t]
\begin{center}
\includegraphics[height=4.in]{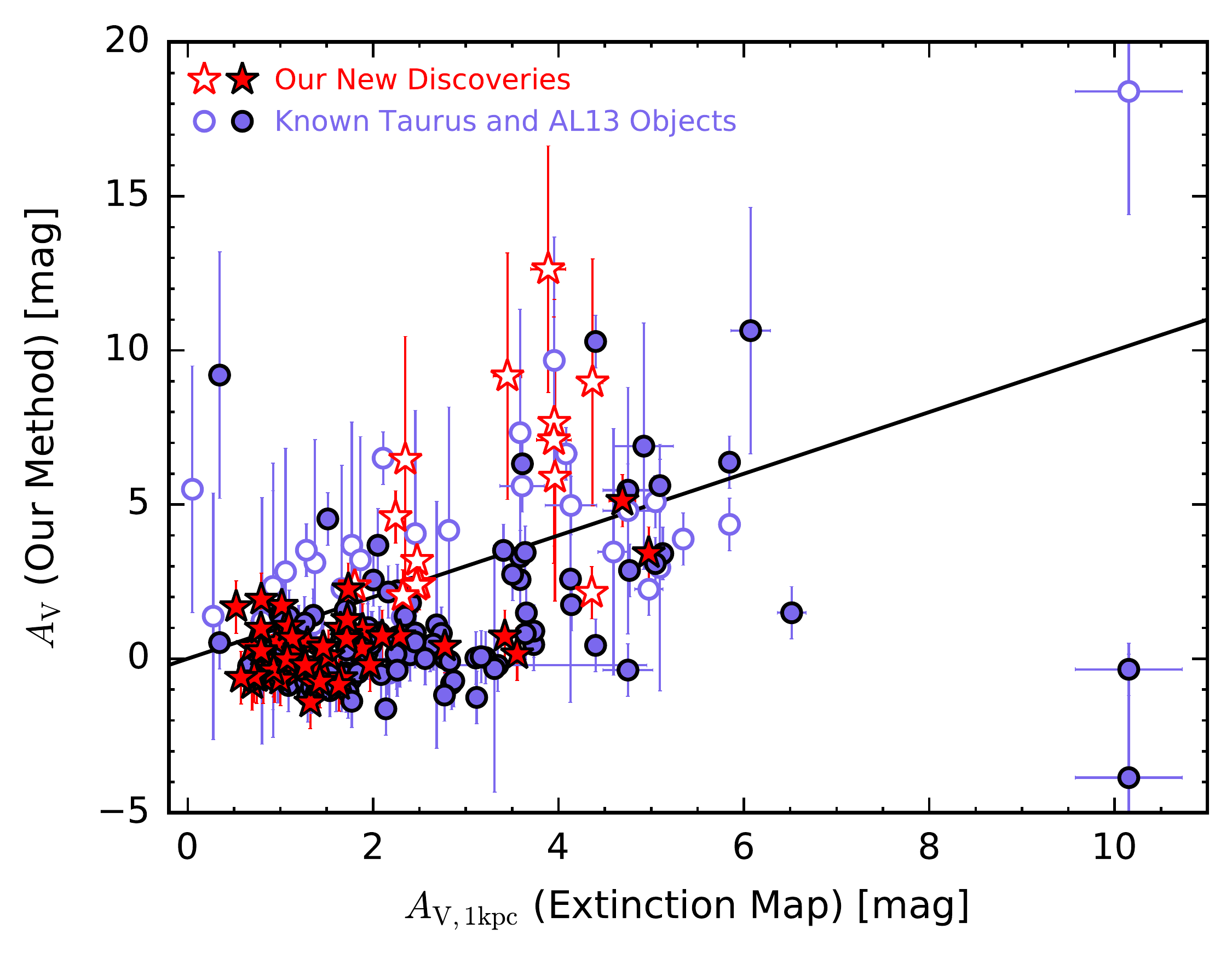}
\caption{Comparison between our extinctions and the integrated reddening till $1$~kpc based on the \cite{Green+2015} extinction map, for our 64 new discoveries ({\it red stars}) in Taurus (50 objects), Pleiades (1 object), and Per~OB2 (13 objects), and for 133 reclassified known objects ({\it slate blue circles}), including 130 Taurus M- and L-type members and 3 reddened young field dwarfs studied by AL13. As described in the caption of Figure~\ref{fig:spt_comp}, the results of 2M~0619$-$2903 (AL13) shown in the figure are based on the second epoch (2015) of its near-infrared spectrum (Section~\ref{subsec:0619}). We use filled symbols for $\geqslant$M6 objects. We use open symbols for [M4, M6) objects that do not have gravity classifications based on near-infrared spectra, although previously known [M4, M6) members of Taurus mostly have youth as indicated via optical spectroscopy. Uncertainties are shown if they exceed the size of the symbols. The solid line corresponds to the one-to-one relation. }
\label{fig:Av_ZJ_map}
\end{center}
\end{figure}
%------------figure end-----------------

\clearpage
%%%%%%%%%%%%%%%%%%%%%%%%%
%------------- Jmag distribution of Field Contaminants --------------------
%%%%%%%%%%%%%%%%%%%%%%%%%
\begin{figure}[t]
\begin{center}
\includegraphics[height=3.3in]{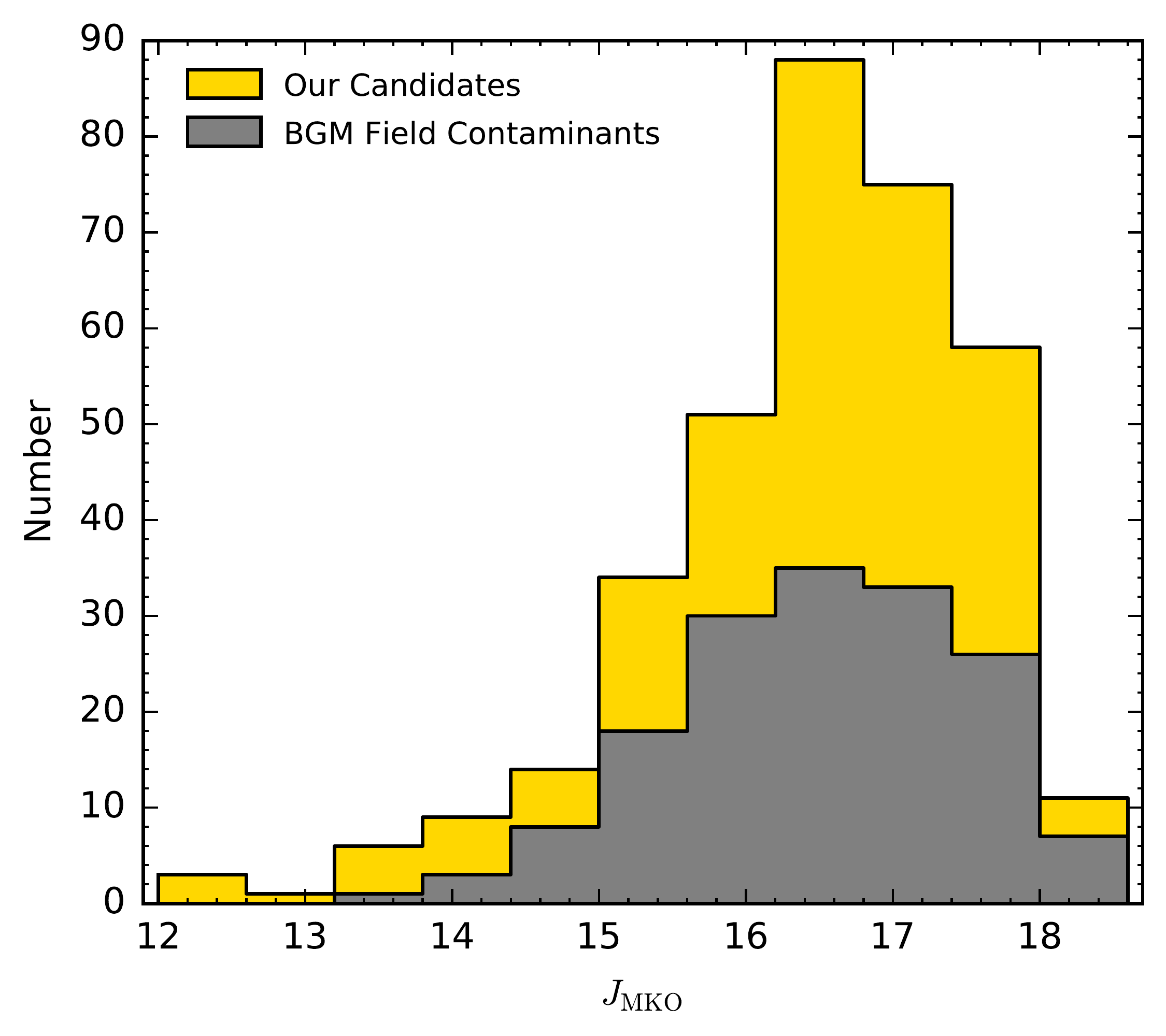}
\caption{Distributions of $J_{\rm MKO}$-band magnitudes of our entire Taurus candidates ({\it gold}) and the expected field contaminants ({\it gray}) predicted by our BGM-based modeling (Section~\ref{subsec:FCE}). We plot only objects with good photometric quality, as defined in Section~\ref{subsec:phot_criteria} for our candidates and defined in Section~\ref{subsec:FCE} for BGM objects. For objects without good-quality $J_{\rm MKO}$ photometry, we synthesize $J_{\rm MKO}$ based on $J_{\rm 2MASS}$ using near-infrared spectra obtained in this work or using the scaling $J_{\rm MKO} - J_{\rm 2MASS} = - 0.05$~mag (Section~\ref{subsec:phot_criteria}). Most field interlopers are faint, with $J$-band magnitudes of $\approx 15.5-18$~mag.}
\label{fig:allFCE_dist}
\end{center}
\end{figure}
%------------figure end-----------------

\clearpage
%%%%%%%%%%%%%%%%%%%%%%%%%
%------------- Field Contamination Estimate --------------------
%%%%%%%%%%%%%%%%%%%%%%%%%
\begin{figure}[t]
\begin{center}
\includegraphics[height=4.3in]{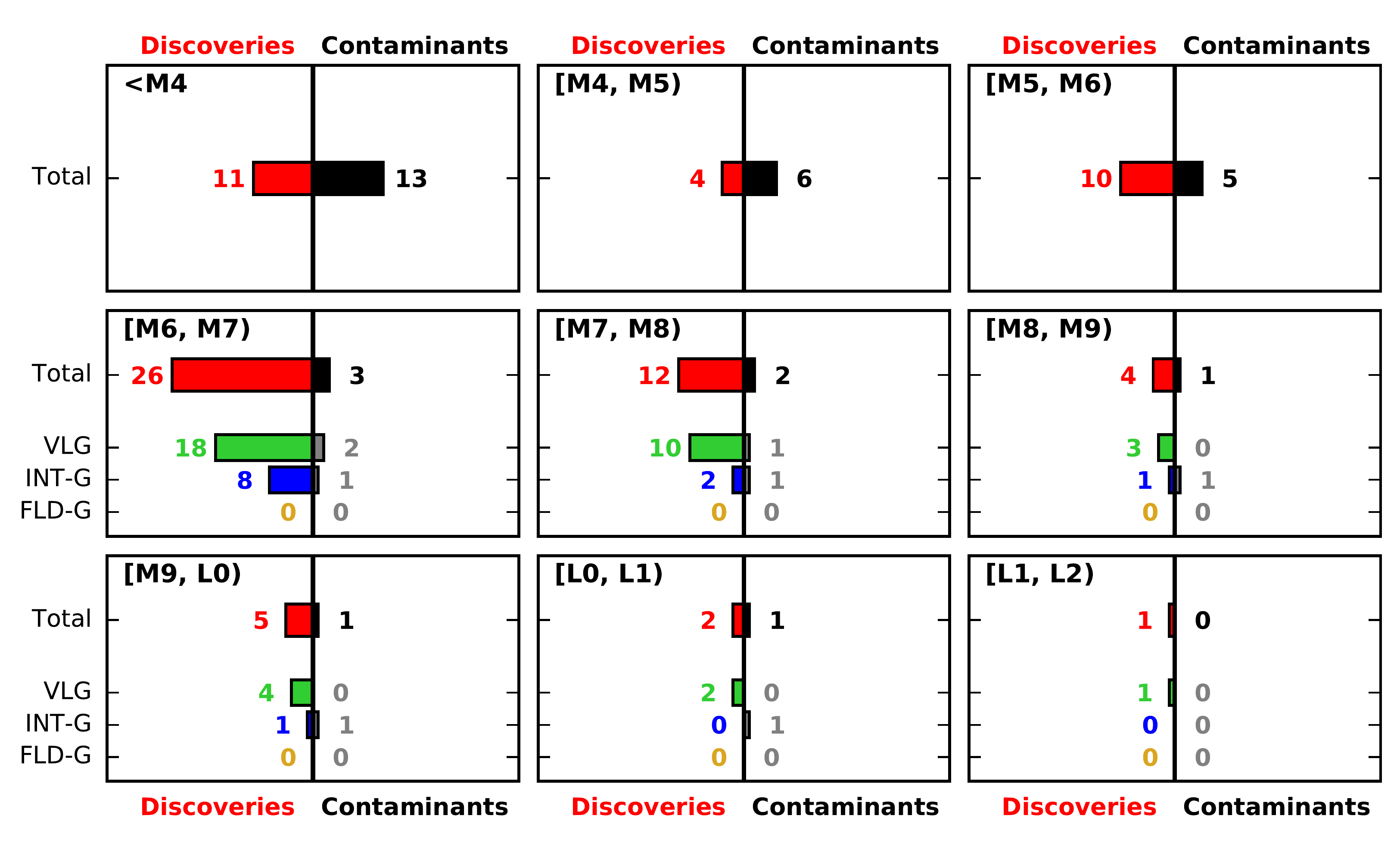}
\caption{Comparison between our spectroscopic follow-up sample ({\it left} in each panel) and the estimated field contamination ({\it right} in each panel; Section~\ref{subsec:FCE} and Table~\ref{tab:FCE}). Our follow-up sample contains 75 objects, including 11 early-type ($<$M4) stars, 50 new Taurus members with robust spectral classification, 1 new Pleiades member, and 13 new Per~OB2 members. In each spectral type range, we compare the total number of discoveries ({\it red} histogram) with the total number of field contaminants ({\it black} histogram). For [M6, L2) dwarfs, we also compare the numbers in each gravity class between our discoveries (\textsc{vl-g}, {\it green}; \textsc{int-g}, {\it blue}; \textsc{fld-g}, {\it orange}) and estimated field contaminants ({\it gray}). Overall, most of our $<$M4 and [M4, M6) discoveries could be field interlopers, but our substellar discoveries are probably bona fide members of star-forming regions. For [L0, L1), the only one predicted field contaminant has $\textsc{int-g}$ gravity class, so our discovered 2 [L0, L1) $\textsc{vl-g}$ dwarfs are unlikely field interlopers. }
\label{fig:FCE}
\end{center}
\end{figure}
%------------figure end-----------------

\clearpage
%%%%%%%%%%%%%%%%%
%------------- CMD --------------------
%%%%%%%%%%%%%%%%%
\begin{figure}[t]
\begin{center}
\includegraphics[height=7in]{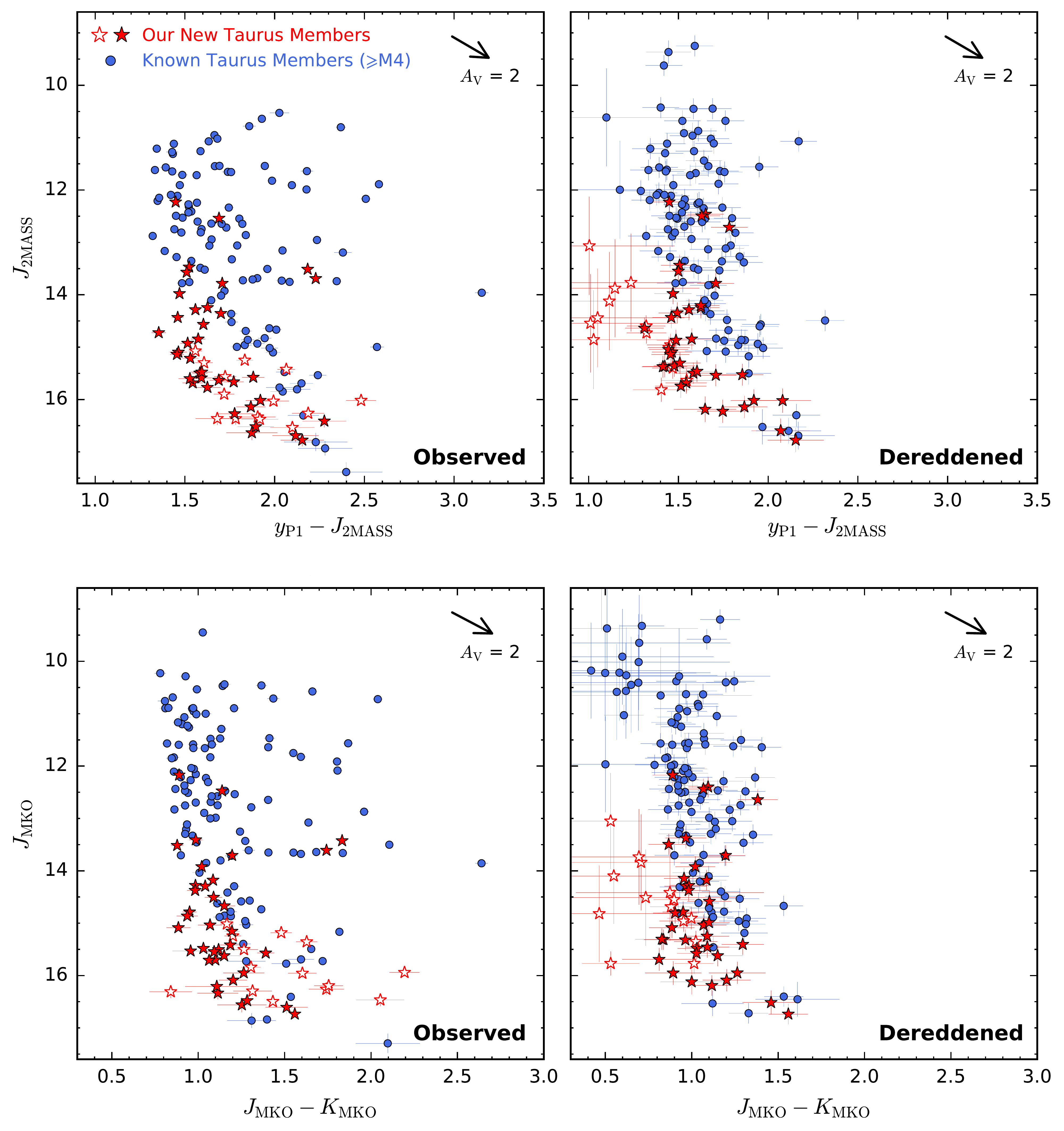}
\caption{Taurus color-magnitude diagrams of $J_{\rm 2MASS}$~vs.~$y_{\rm P1} - J_{\rm 2MASS}$ ({\it top}) and $J_{\rm MKO}$~vs.~$J_{\rm MKO} - K_{\rm MKO}$ ({\it bottom}) for reclassified known $\geqslant$M4 objects ({\it blue} circles) and our new members with robust spectral classification ({\it red} stars) with their observed ({\it left}) and dereddened ({\it right}) photometry. We use open stars to show our [M4,M6) discoveries, as they lack the gravity classifications needed for a firm membership assessment. For objects without detections in 2MASS ($J_{\rm 2MASS}$) or UKIDSS ($J_{\rm MKO}$ and $K_{\rm MKO}$), we synthesize photometry from our spectra. Only objects with good-quality photometry, as defined in Section \ref{subsec:phot_criteria}, are plotted. Photometric uncertainties are shown if they exceed the size of the symbols. The extinction vector corresponds to $A_{\rm V} = 2$~mag using the extinction law of \cite{Schlafly+2011}. The sequence becomes tighter after dereddening, indicating the robustness of our extinction determinations. Some known Taurus objects in the bottom right panel with $J_{\rm MKO} \approx 10$~mag have relatively large extinction errors ($\approx 4$~mag in $A_{\rm V}$), because their spectral types $<$M5 and thereby their extinction values are from $A_{\rm V}^{\rm H_{2}O}$ with worse precisions (as $A_{\rm V}^{\rm OIR}$, with smaller uncertainties, are not accessible; Section~\ref{subsub:finalext}). These early-type objects have bad-quality detections in $y_{\rm P1}$ band, so they do not show up in the top panels. }
\label{fig:Dr_CMD}
\end{center}
\end{figure}
%------------figure end-----------------

\clearpage
%%%%%%%%%%%%%%%%%
%------------- CCD --------------------
%%%%%%%%%%%%%%%%%
\begin{figure}[t]
\begin{center}
\includegraphics[height=7in]{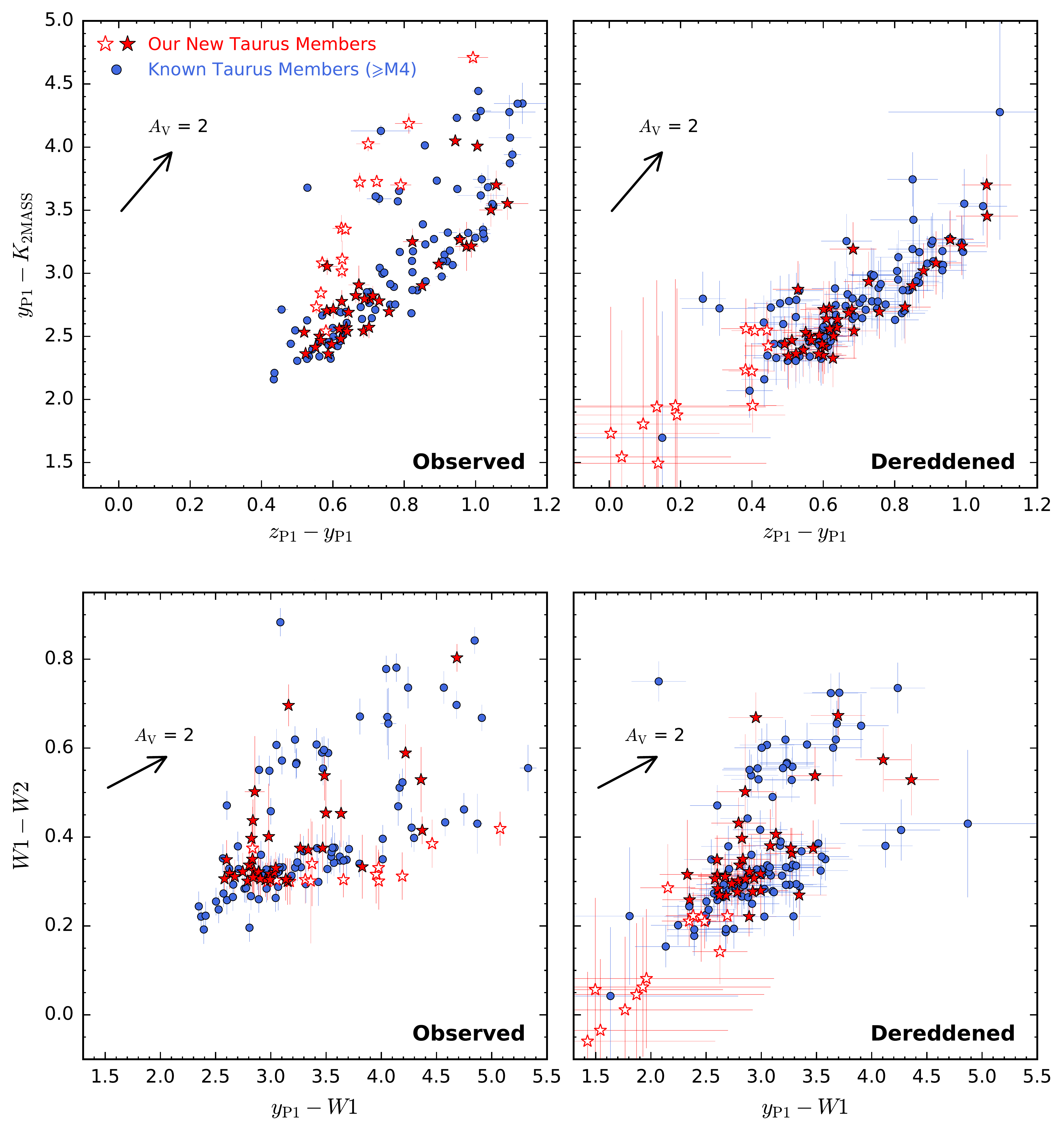}
\caption{Taurus color-color diagrams of $y_{\rm P1} - K_{\rm 2MASS}$~vs.~$z_{\rm P1} - y_{\rm P1}$ and $W1 - W2$~vs.~$y_{\rm P1} - W1$ using the same format as Figure~\ref{fig:Dr_CMD}. The color sequences of the objects are much tighter after dereddening.}
\label{fig:Dr_CCD}
\end{center}
\end{figure}
%------------figure end-----------------

\clearpage
%%%%%%%%%%%%%%%%%%%%%%
%--------------- HR - Diagram ---------------------
%%%%%%%%%%%%%%%%%%%%%%
\begin{figure}[t]
\begin{center}
\includegraphics[height=7.5in]{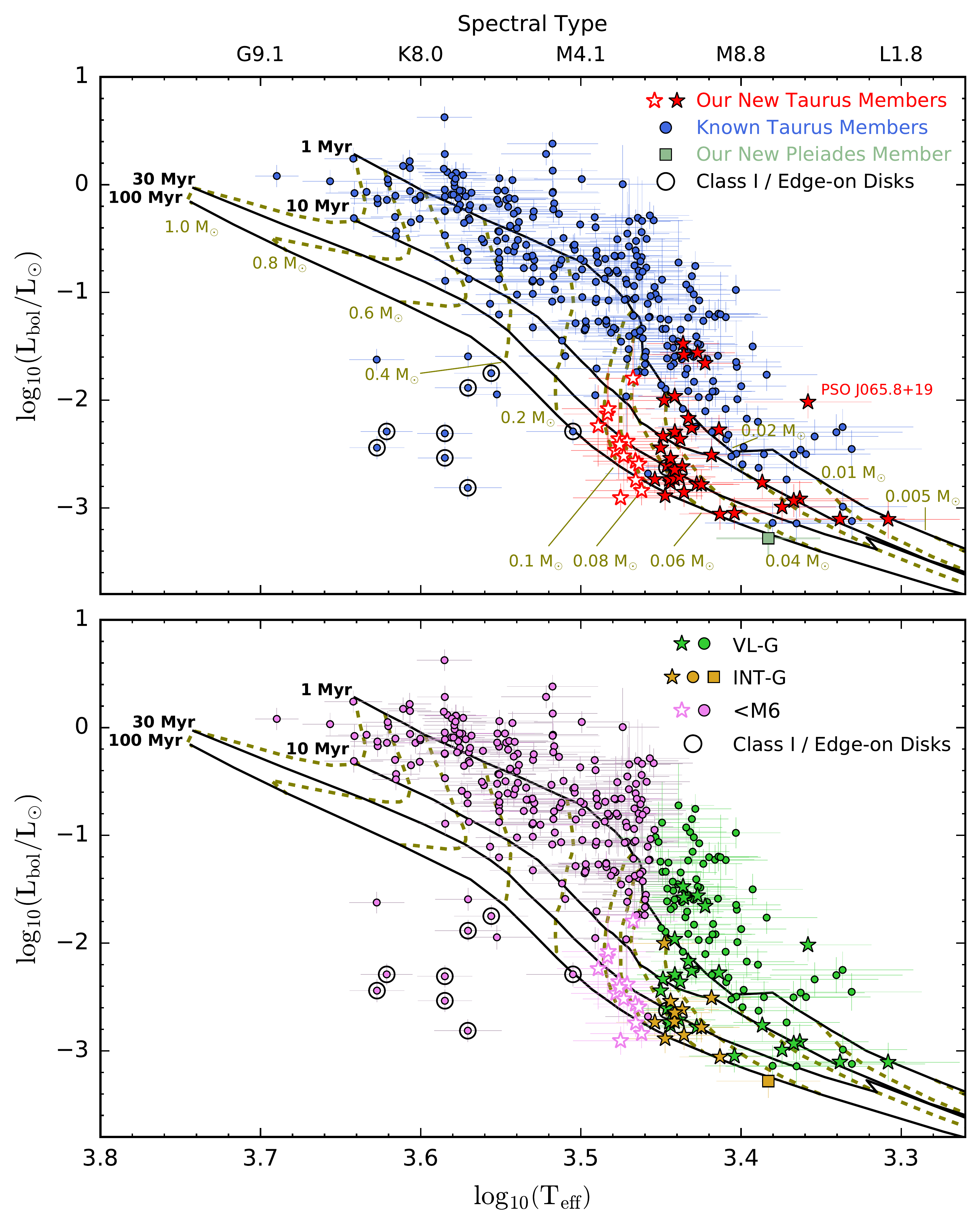}
\caption{Top: HR diagram for our discoveries ({\it red} stars) and known members ({\it blue} circles) in Taurus, with spectral type plotted as the top axis based on the \cite{Stephens+2009} and the \cite{Herczeg+2014} temperature scales (Section~\ref{subsubsec:BGM} and \ref{subsec:color}). The overlaid evolutionary tracks and isochrones are based on the BHAC15 models of \cite{Baraffe+2015} for $M>0.06$~M$_{\odot}$ and the DUSTY models of \cite{Chabrier+2000} for $M\leqslant 0.06$~M$_{\odot}$. We use open stars to show our [M4,M6) discoveries, as they lack the gravity classifications needed for a firm membership assessment. Only objects with good-quality $J$-band magnitudes, thereby reliable bolometric luminosities, are plotted. Among known Taurus objects with model-derived ages of $>30$~Myr based on their HR diagram positions, we use black open circles to mark the ones with reported Class I envelopes or high-inclination circumstellar disks (Section~\ref{subsec:color}). We also note the position of our unusually bright L0 \textsc{vl-g} dwarf, PSO~J065.8+19 (Section \ref{subsec:brightL0}). In addition, we overlay our newly identified Pleiades member ({\it dark green} squares), scaled to the Pleiades distance ($d=136$~pc). Bottom: The same objects and evolutionary models are plotted as the top panel but now with their gravity classifications color-encoded. {\it Green} colors are for the \textsc{vl-g} classification, {\it gold} for \textsc{int-g}, and {\it pink} for no AL13 gravity classification due to objects' too early spectral types ($<$M6) for the AL13 system (Section \ref{subsec:youth_logistics}).}
\label{fig:HR}
\end{center}
\end{figure}
%------------figure end-----------------

\clearpage
%%%%%%%%%%%%%%%%%%%%%%%%%%%%
%--------------- gravity vs. age histogram ---------------------
%%%%%%%%%%%%%%%%%%%%%%%%%%%%
\begin{figure}[t]
\begin{center}
\includegraphics[height=4in]{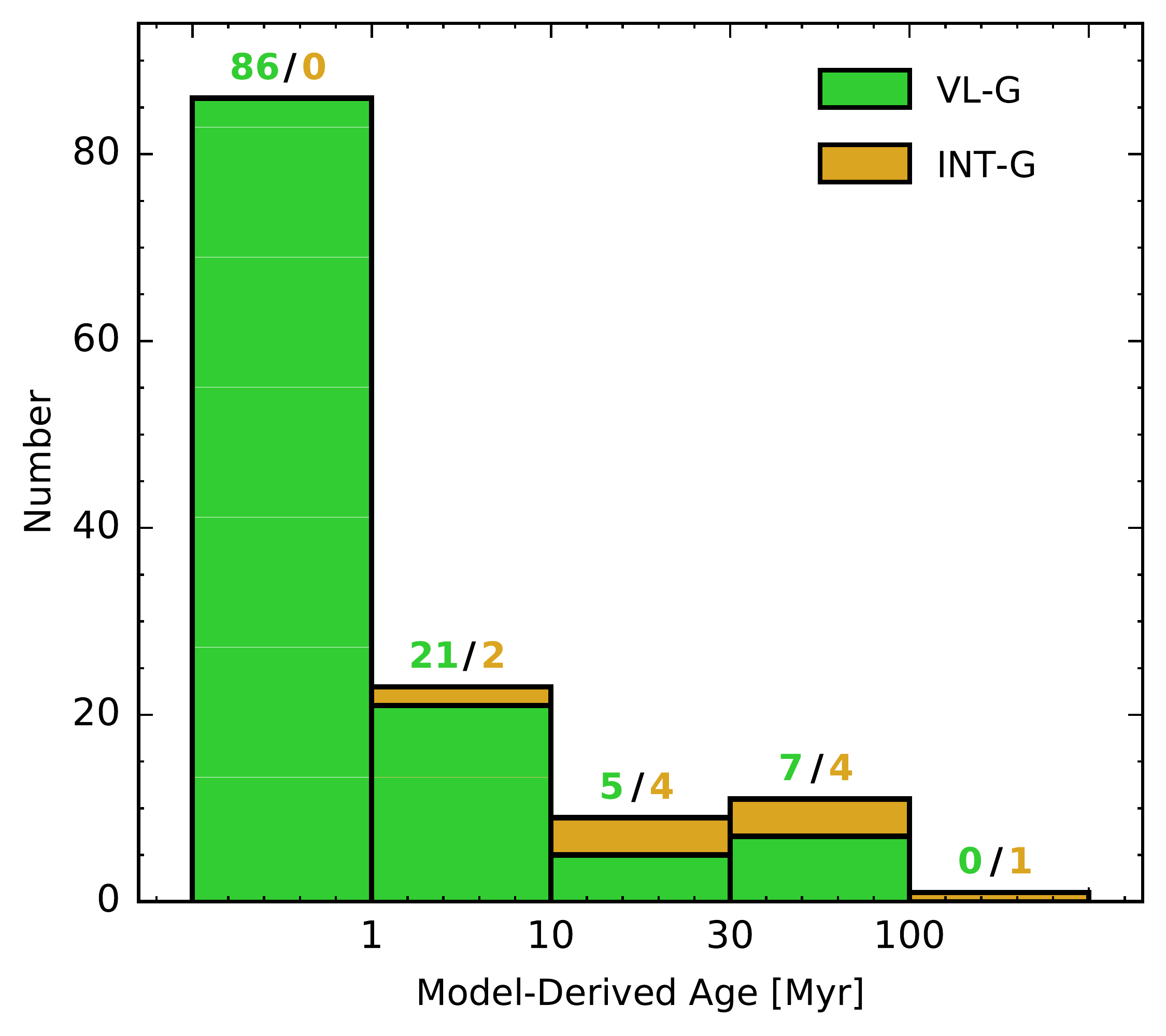}
\caption{Histogram of model-derived ages for all Taurus members with \textsc{vl-g} ({\it green}) and \textsc{int-g} ({\it orange}) gravity classifications. Numbers above each age bin indicate the number of objects with different surface gravities. The first bin from the left is for $\leqslant$1~Myr and the last bin is for $>$100~Myr. }
\label{fig:comp_age_grav}
\end{center}
\end{figure}
%------------figure end-----------------

\clearpage
%%%%%%%%%%%%%%%%%%%%%%
%--------------- PM - Diagram ---------------------
%%%%%%%%%%%%%%%%%%%%%%
\begin{figure}[t]
\begin{center}
\includegraphics[height=3.5in]{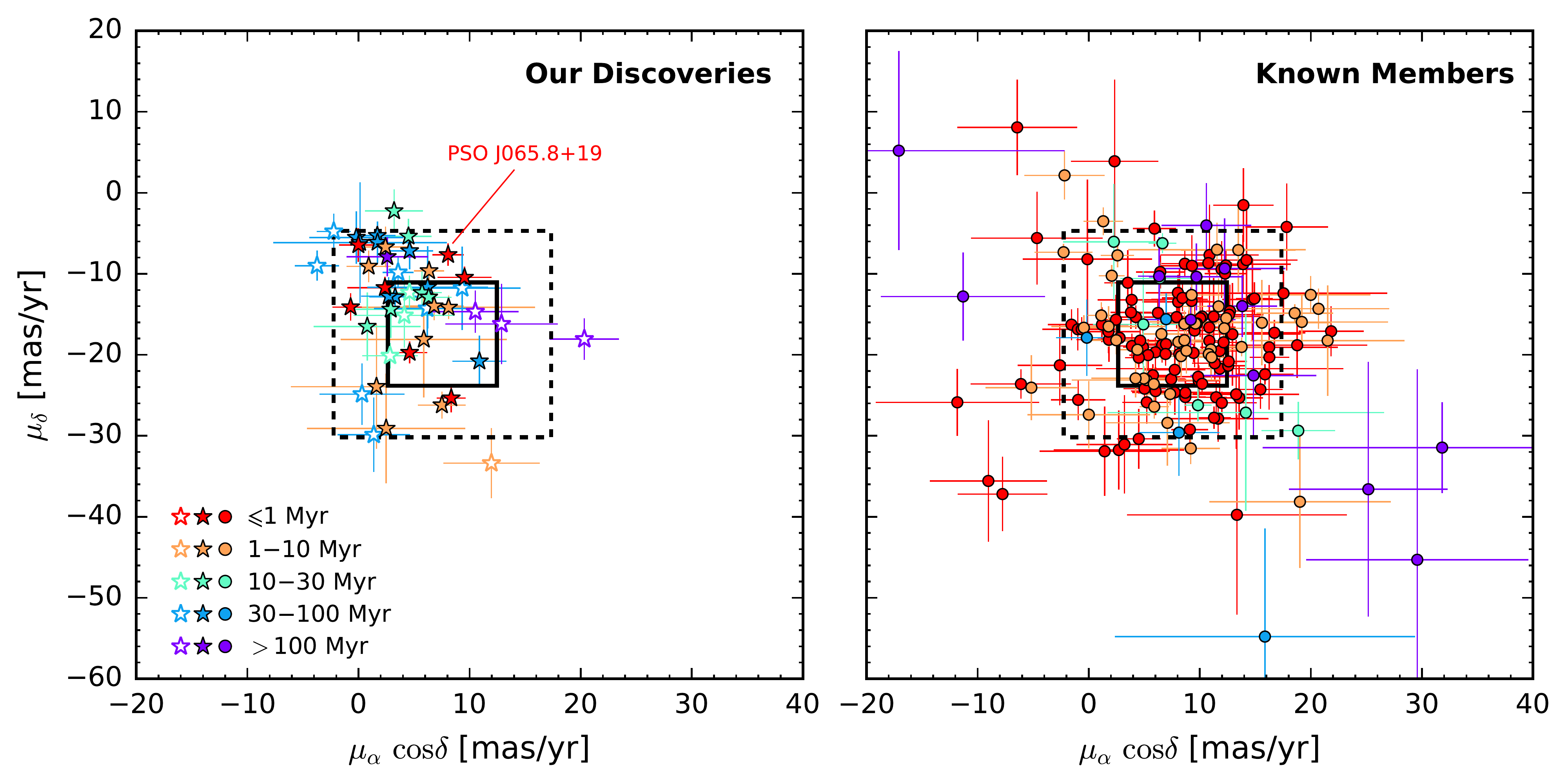}
\caption{Proper motion distribution of our discoveries ({\it left}) and previously known objects ({\it right}) in Taurus. Only objects with good-quality proper motion measurements (defined in Section \ref{subsec:kine_criteria}) are plotted, which includes our 50 new members and 181 known members in Taurus. We use open stars to show our [M4,M6) discoveries, as they lack the gravity classifications needed for a firm membership assessment. The colors of the plotting symbols represent the model-derived ages based on their positions on the HR diagram (Figure~\ref{fig:HR}). We also mark the position of our unusually bright L0 $\textsc{vl-g}$ dwarf discovery, PSO~J065.8+19 (Section \ref{subsec:brightL0}). Its proper motion is consistent with the mean Taurus motion within $2\sigma$. }
\label{fig:PM_analysis}
\end{center}
\end{figure}
%------------figure end-----------------

\clearpage
%%%%%%%%%%%%%%%%%%%%%%%%%%%%%%
%------------------ spatial distribution (modelage) -------------------------
%%%%%%%%%%%%%%%%%%%%%%%%%%%%%%
\begin{figure}[t]
\begin{center}
\includegraphics[height=5.5in]{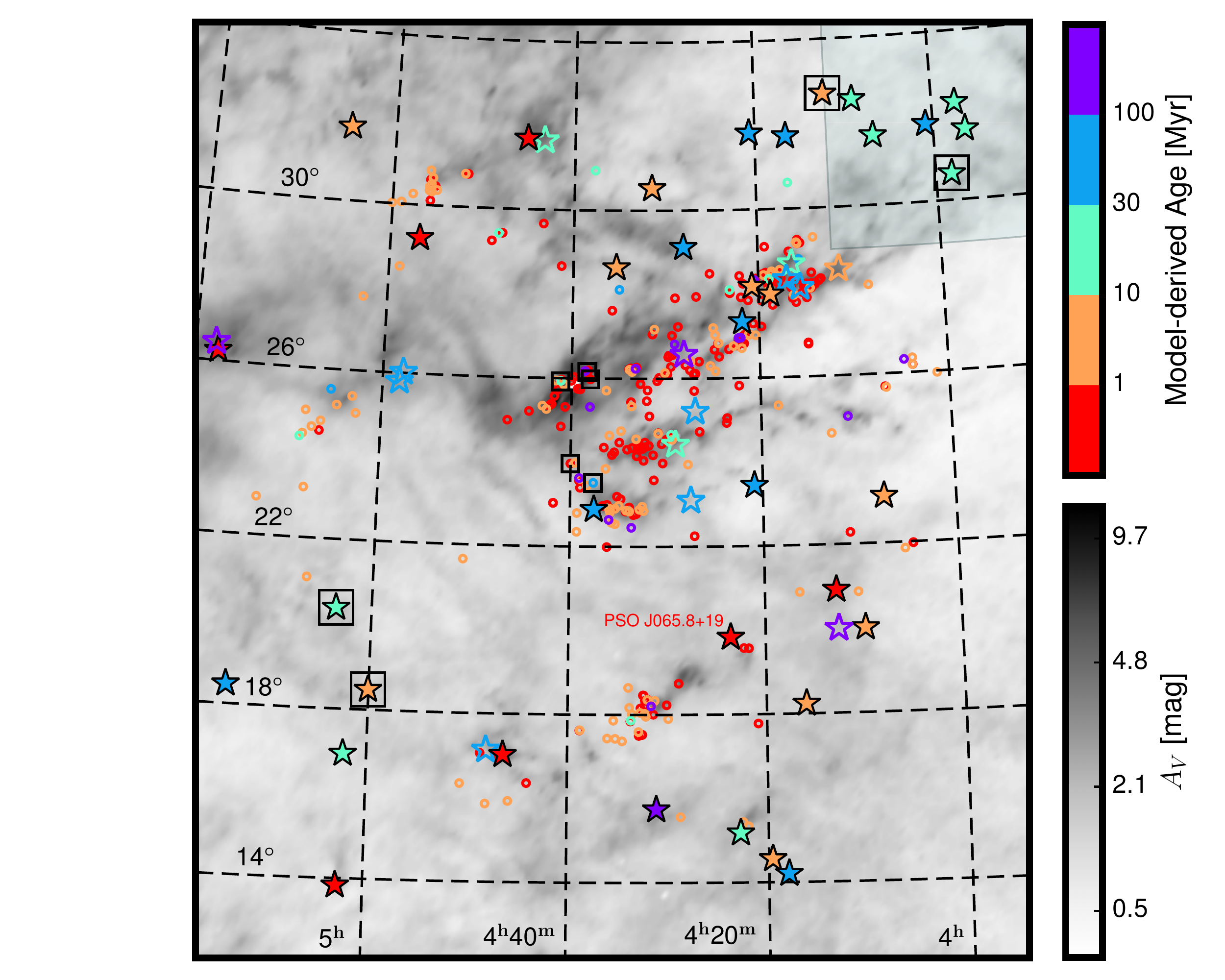}
\caption{Spatial distribution of our new members (stars) and previously known members (circles) in Taurus. Here the plotting colors indicate ages of objects derived from their HR diagram positions. We use open stars to show our [M4,M6) discoveries, as they lack the gravity classifications needed for a firm membership assessment. The ten objects (5 of our new discoveries, including the one probable new Pleiades member, and 5 known $\geqslant$M6 members) with proper motion different from the mean Taurus motion by $>2\sigma$ (Section~\ref{subsubsec:hiPM}) are shown by open squares. We label our unusually bright L0 $\textsc{vl-g}$ dwarf discovery, PSO~J065.8+19 (Section \ref{subsec:brightL0}). While younger ($\leqslant$10~Myr) objects in Taurus are associated with high reddening, older ($>$10~Myr) objects are more dispersed, with their sky locations not closely following the younger members or extinction. }
\label{fig:extmap_modelage}
\end{center}
\end{figure}
%------------figure end-----------------

\clearpage
%%%%%%%%%%%%%%%%%%%%%%%%%%%%%%
%------------------ spatial distribution (spt) -------------------------
%%%%%%%%%%%%%%%%%%%%%%%%%%%%%%
\begin{figure}[t]
\begin{center}
\includegraphics[height=5.5in]{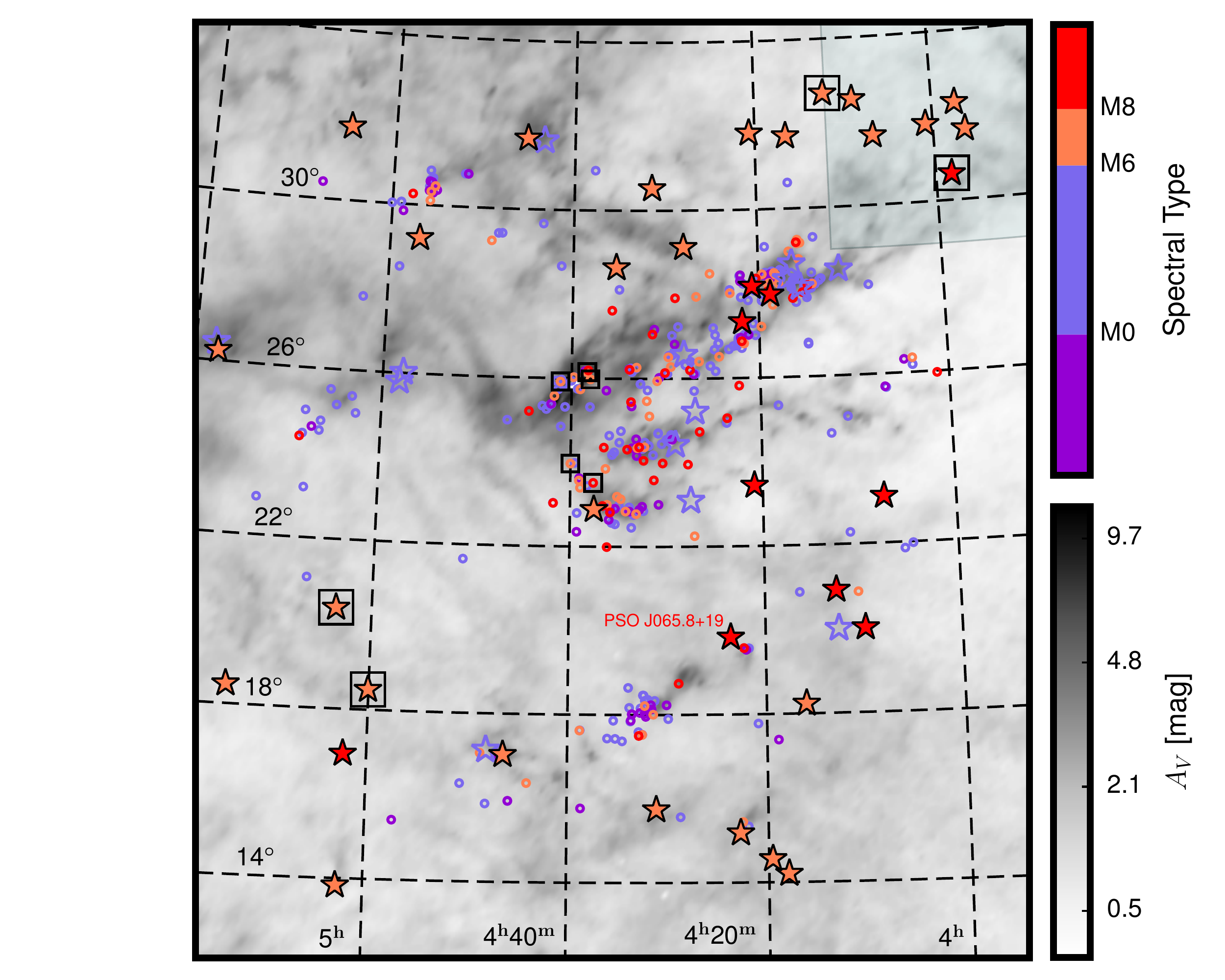}
\caption{Spatial distribution of our discoveries (stars) and previously known objects (circles) in Taurus, as described in the caption of Figure~\ref{fig:extmap_modelage}. Here the plotting colors represent spectral type. The locations of the stars ($<$M6) are mostly associated with regions of high extinction. While most substellar ($\geqslant$M6) objects are consistent with the stellar population, some brown dwarfs are located in lower stellar-density regions.}
\label{fig:extmap_spt}
\end{center}
\end{figure}
%------------figure end-----------------

\clearpage
%%%%%%%%%%%%%%%%%%%%%%
%--------------- PM - sigma histogram ---------------------
%%%%%%%%%%%%%%%%%%%%%%
\begin{figure}[t]
\begin{center}
\includegraphics[height=4in]{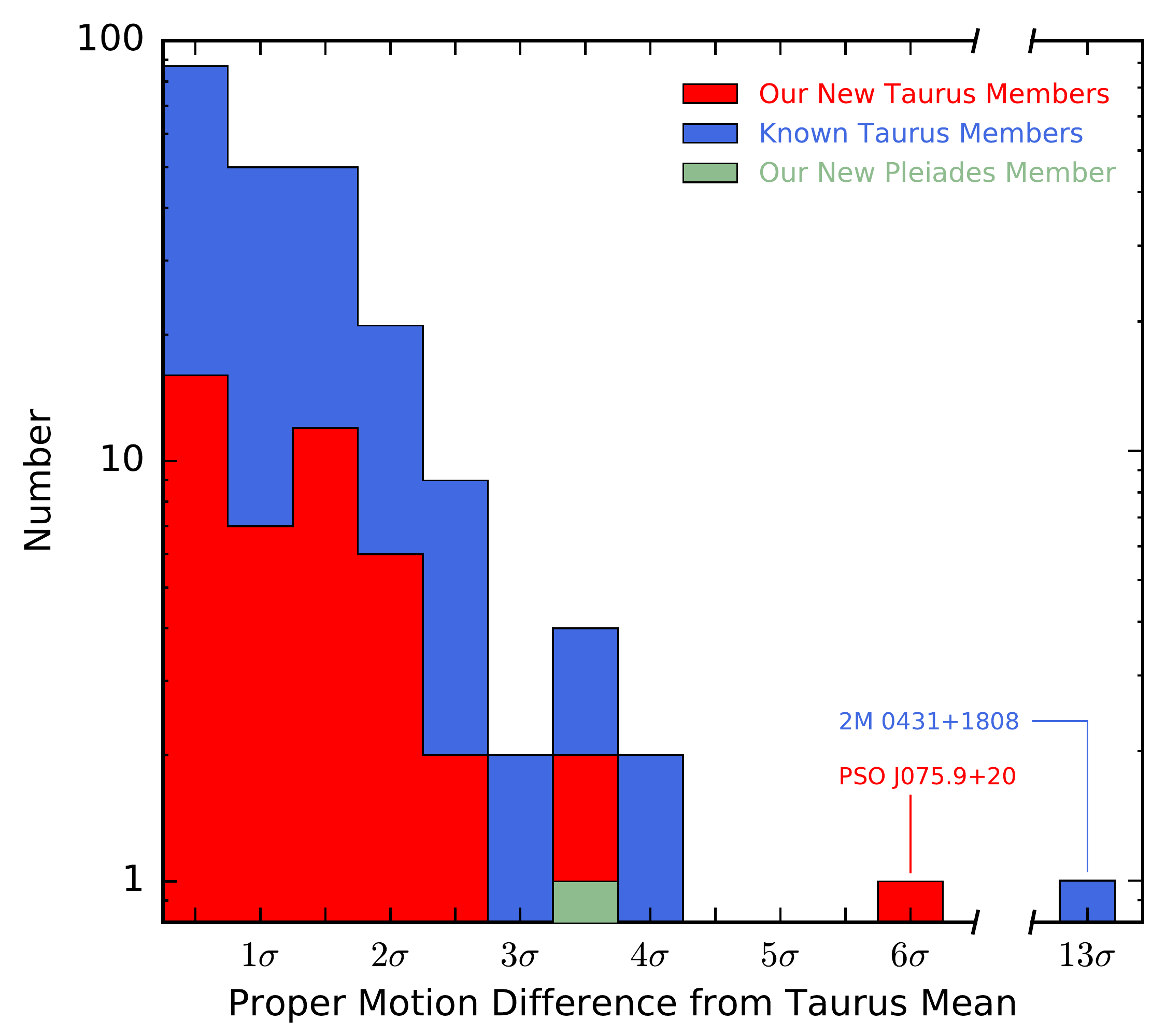}
\caption{Comparisons of the PS1 proper motions of new ({\it red}) and known ({\it blue}) Taurus members and one probable new Pleiades member (PSO~J058.8+21; {\it dark green}), to the mean Taurus motion (Section~\ref{subsec:kine_criteria}). We break the x-axis to provide a compact plotting configuration. Most ($92\%$) Taurus objects have proper motions consistent with Taurus within $2\sigma$. Two outliers with significantly discrepant proper motions are also noted. }
\label{fig:PM_sigma}
\end{center}
\end{figure}
%------------figure end-----------------

\clearpage
%%%%%%%%%%%%%%%%%%%%%%
%--------------- PM - highPM ---------------------
%%%%%%%%%%%%%%%%%%%%%%
\begin{figure}[t]
\begin{center}
\includegraphics[height=4in]{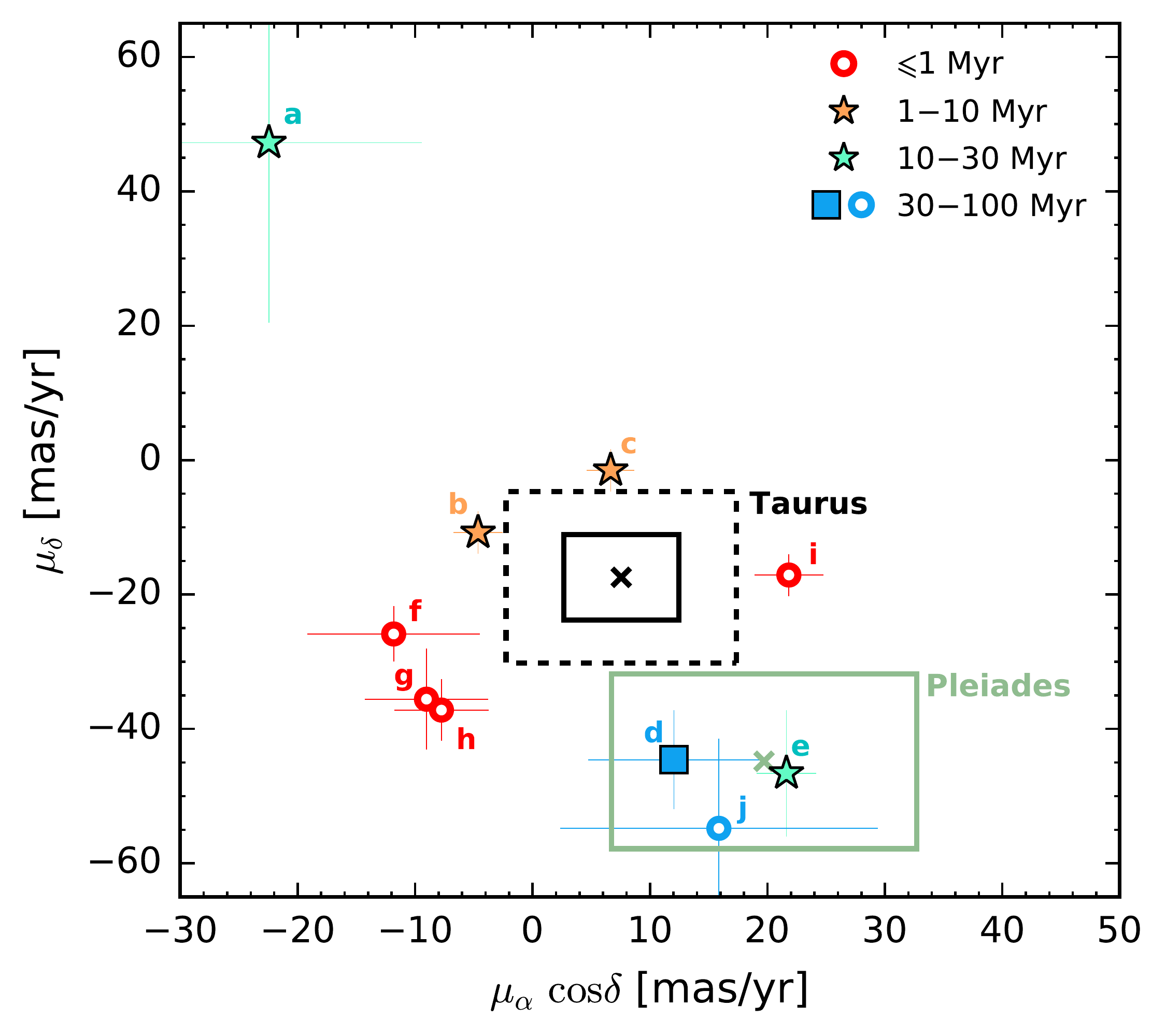}
\caption{Proper motions of our 5 discoveries (four $\textsc{vl-g}$ Taurus members as stars, and one \textsc{int-g} Pleiades member as a square) and 5 known Taurus $\geqslant$M6 members (circles) with non-Taurus proper motions, as tabulated in Table~\ref{tab:pm_sigma}. The colors of the plotting symbols represent the model-derived ages based on their positions on the HR diagram. The average proper motion of Taurus and its $1\sigma$ and $2\sigma$ confidence limits are shown as black cross and black squares with solid and dashed boundaries, respectively. As a comparison, we overlay the typical proper motion of the Pleiades and its $1\sigma$ confidence level \citep{Zapatero+2014} as an dark green cross and solid square. The objects' labels are described in the caption of Figure~\ref{fig:ExtMap}. }
\label{fig:PM_highPM}
\end{center}
\end{figure}
%------------figure end-----------------

\clearpage
%%%%%%%%%%%%%%%%%%%%%%
%--------------- PM - PerOB2 ---------------------
%%%%%%%%%%%%%%%%%%%%%%
\begin{figure}[t]
\begin{center}
\includegraphics[height=4in]{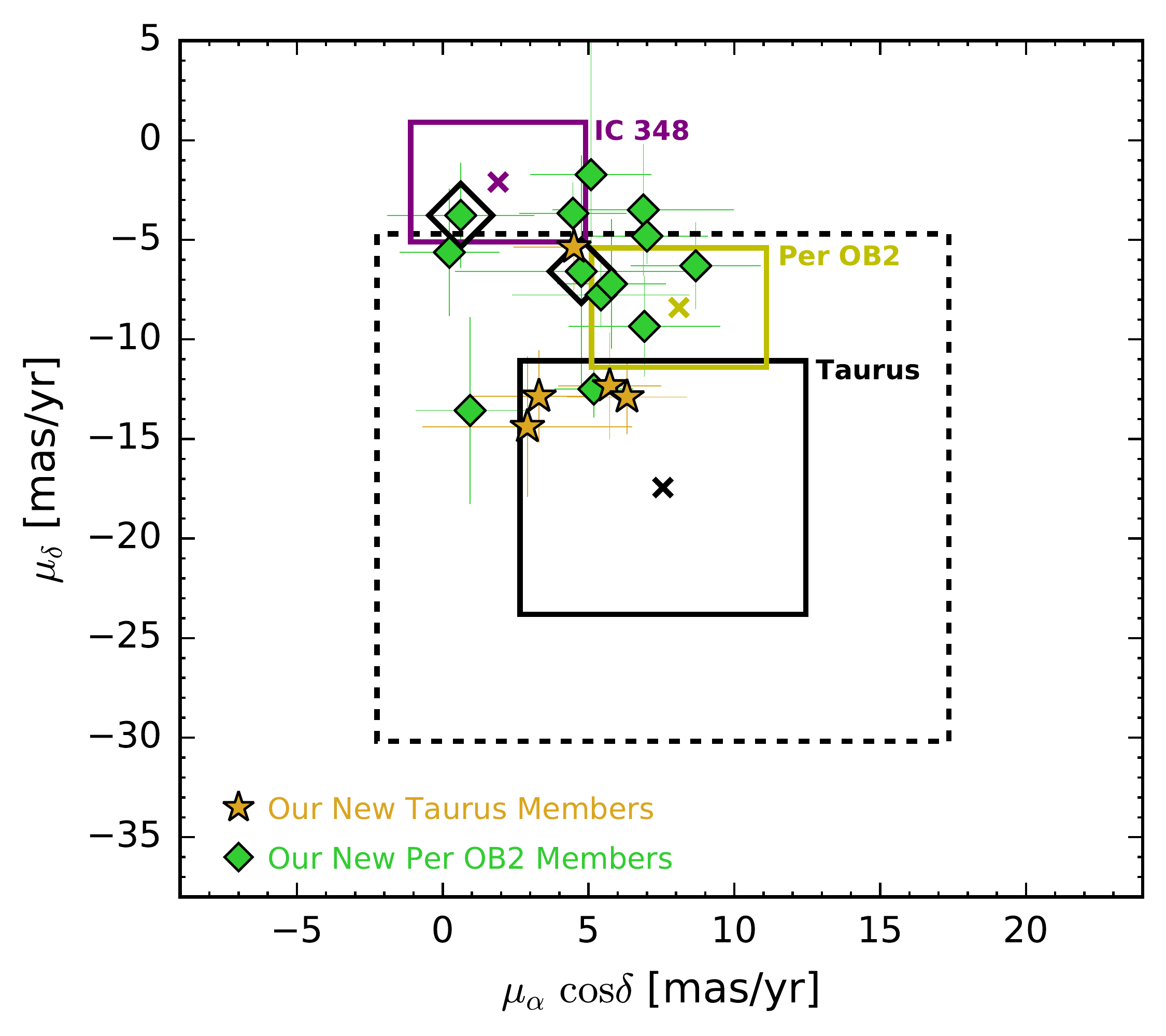}
\caption{Proper motion distribution of our 18 discoveries in the overlapping region between Taurus and Per~OB2. Five are new Taurus members with intermediate gravity classification (\textsc{int-g}; {\it gold} stars) and the other 13 objects are new Per~OB2 members with very low surface gravity ($\textsc{vl-g}$; {\it green} diamonds). We use crosses to indicate the average proper motions of Taurus ({\it black}), Per~OB2 ({\it olive}), and IC~348 ({\it purple}), with solid boundaries corresponding to their $1\sigma$ confidence levels. We additionally show the $2\sigma$ confidence level of Taurus as a dashed box. The open solid diamonds mark our two Per~OB2 discoveries that form a candidate very wide separation ($58$~kAU) binary (Section~\ref{subsubsec:binary}).}
\label{fig:PM_Per}
\end{center}
\end{figure}
%------------figure end-----------------

\clearpage
%%%%%%%%%%%%%%%%%%%%%%
%--------------- HR - Diagram - PerOB2 ---------------------
%%%%%%%%%%%%%%%%%%%%%%
\begin{figure}[t]
\begin{center}
\includegraphics[height=3.5in]{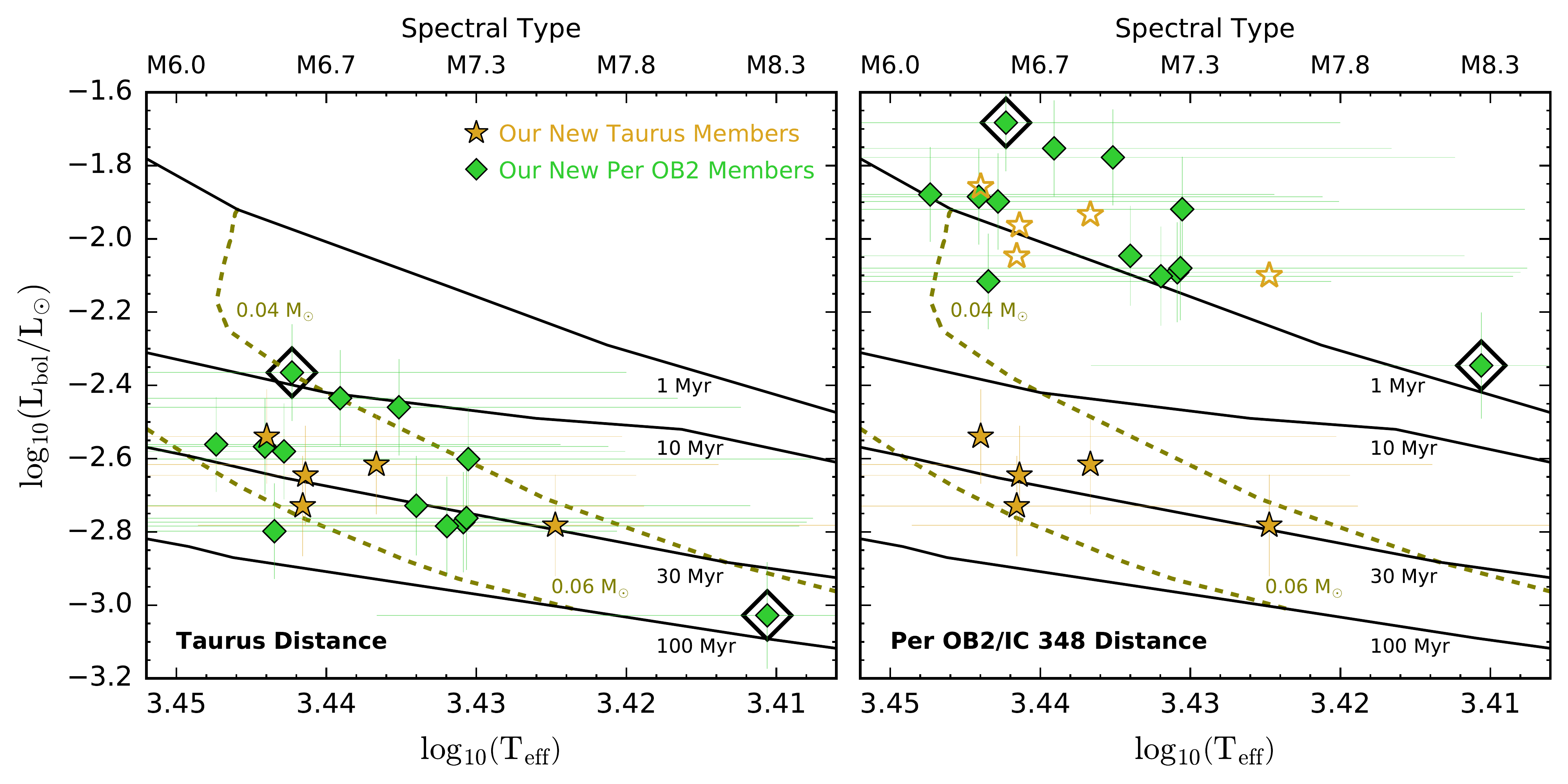}
\caption{HR diagram for 18 discoveries in the overlapping region between Taurus and Per~OB2, with spectral type plotted as the top axis based on the combined \cite{Herczeg+2014} and \cite{Stephens+2009} temperature scales (Section~\ref{subsubsec:BGM} and \ref{subsec:color}). Five are new Taurus members with intermediate gravity classification (\textsc{int-g}; {\it gold} stars) and the other 13 objects are new Per~OB2 members with very low surface gravity ($\textsc{vl-g}$; {\it green} diamonds). The overlaid evolutionary tracks and isochrones are based on the BHAC15 models of \cite{Baraffe+2015} for $M>0.06$~M$_{\odot}$ and the DUSTY models of \cite{Chabrier+2000} for $M\leqslant 0.06$~M$_{\odot}$. We compare the HR diagram positions of these objects when placed at the Taurus distance ($145$~pc; {\it left}) and at the Per~OB2/IC~348 distance ($318$~pc; {\it right}). In the right panel, we use open {\it gold} stars to show the positions of the 5 \textsc{int-g} objects when scaled to the Per~OB2/IC~348 distance, and filled {\it gold} stars when scaled to Taurus for comparison. The 5~\textsc{int-g} objects are probably members of Taurus, since their model-derived ages ($\approx 1$~Myr) would be too young compared to their intermediate gravity classification if they are located in Per~OB2. The open diamonds mark our two Per~OB2 discoveries that form a candidate very wide separation ($58$~kAU) binary (Section~\ref{subsubsec:binary}).}
\label{fig:HR_Per}
\end{center}
\end{figure}
%------------figure end-----------------

\clearpage
%%%%%%%%%%%%%%%%%%%%%%
%--------------- K-W colors: Taurus ---------------------
%%%%%%%%%%%%%%%%%%%%%%
\begin{figure}[t]
\begin{center}
\includegraphics[height=5.5in]{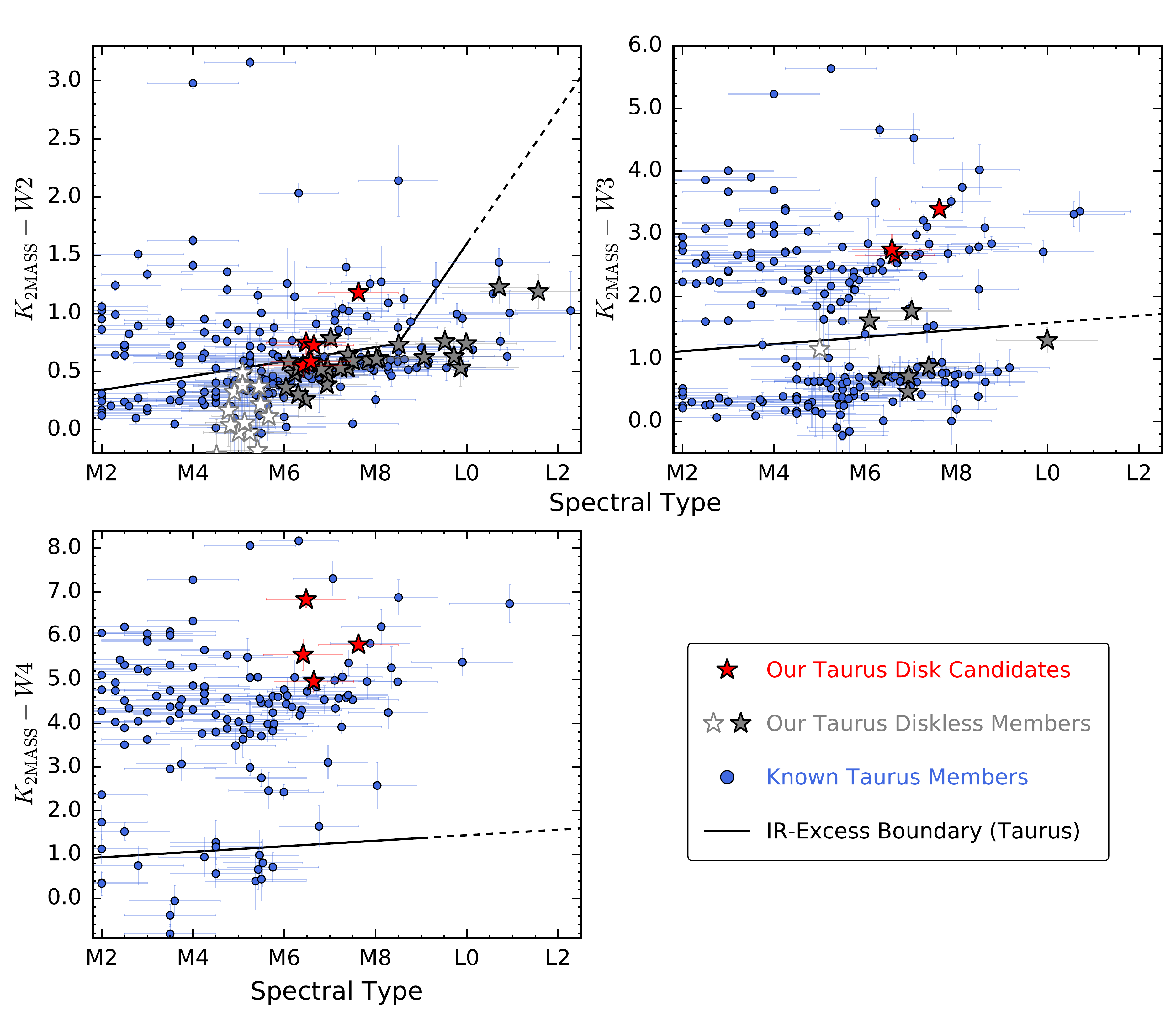}
\caption{Dereddened $K_{\rm 2MASS} - W2/3/4$ colors of our 5 disk candidates ({\it red} stars), our diskless members ({\it grey} stars), and previously known members ({\it blue} circles) in Taurus. We use open stars to show our [M4,M6) discoveries, as they lack gravity classifications for a firm membership assessment. For objects with no detection in 2MASS, we synthesize their $K_{\rm 2MASS}$ magnitudes using $K_{\rm MKO}$, as described in Section~\ref{subsec:color}. Only objects with good-quality photometry, as defined in Section~\ref{subsec:phot_criteria}, are plotted, with good quality for $W3$ and $W4$ defined the same way as for $W1$ and $W2$. Objects' uncertainties are shown if they exceed the size of the symbol. We overlay the reddest colors ({\it black} solid line) expected for stellar photospheres as a function of spectral type, based on previously known Taurus members by \cite{Esplin+2014}. We use {\it black} dashed lines for linear extrapolations of the boundary. The dereddened $K_{\rm 2MASS} - W2/3/4$ colors and the significance of objects' mid-infrared excesses are tabulated in Table~\ref{tab:IR_excess}. }
\label{fig:IR_excess_Taurus}
\end{center}
\end{figure}
%------------figure end-----------------

\clearpage
%%%%%%%%%%%%%%%%%%%%%%
%--------------- HR Diagram - Taurus disks ---------------------
%%%%%%%%%%%%%%%%%%%%%%
\begin{figure}[t]
\begin{center}
\includegraphics[height=4in]{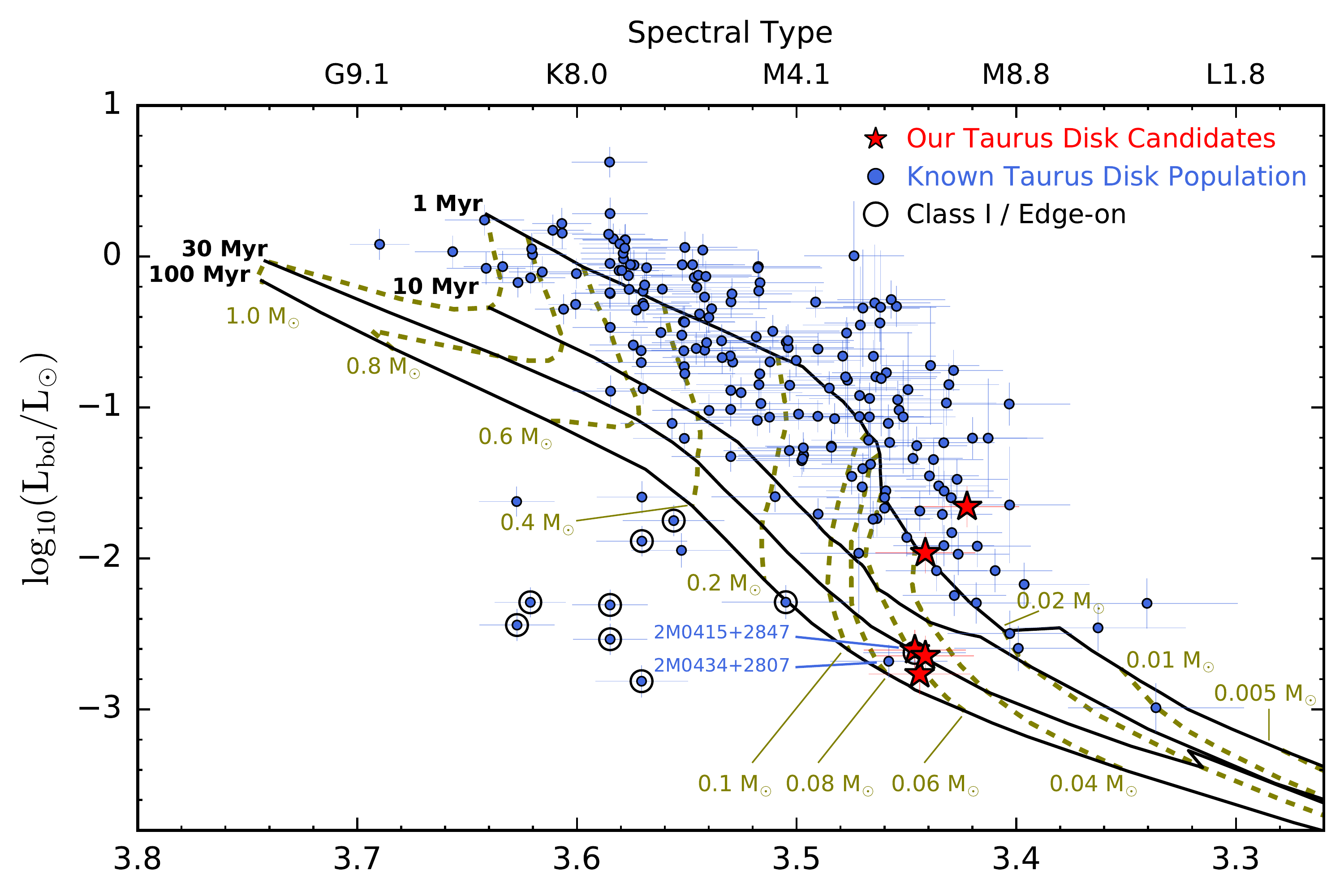}
\caption{HR diagram for our 5 disk candidates ({\it red} stars) and previously known disk population ({\it blue} circles) in Taurus, using the same format as Figure~\ref{fig:HR}. Among known Taurus objects with model-derived ages of $>30$~Myr based on their HR diagram positions, we use black open circles to mark the ones with reported Class I envelopes or high-inclination circumstellar disks (Section~\ref{subsec:color}). Five disk candidates are among the faintest objects, including 3 of our disk candidates and 2 previously known disk candidates, 2M~0415+2847 (Class~I; hidden behind our disk candidates in the figure) and 2M~0434+2807 (no reported Class I envelope or circumstellar disk). Their reddening-corrected bolometric luminosities are fainter by a factor of $\approx 30$ than the rest disk population in the same spectral type range. }
\label{fig:HR_disk}
\end{center}
\end{figure}
%------------figure end-----------------

\clearpage
%%%%%%%%%%%%%%%%%%%%%%
%--------------- yJK diagram ---------------------
%%%%%%%%%%%%%%%%%%%%%%
\begin{figure}[t]
\begin{center}
\includegraphics[height=3.5in]{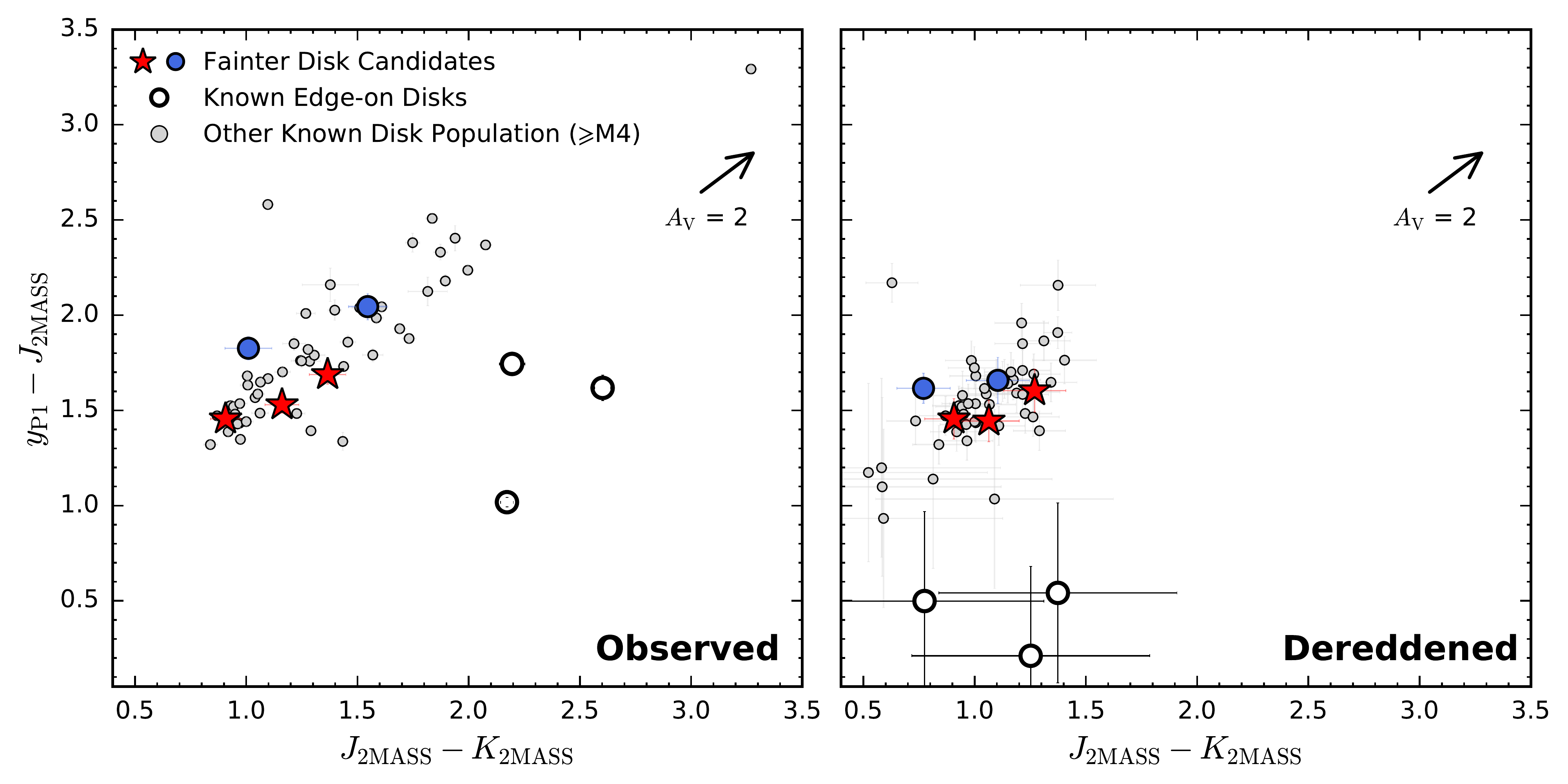}
\caption{Color-color diagram of $y_{\rm P1} - J_{\rm 2MASS}$~vs~$J_{\rm 2MASS} - K_{\rm 2MASS}$ for $\geqslant$M4 disk candidates in Taurus, using their observed ({\it left}) and dereddened ({\it right}) photometry. The spectral types, extinctions, and gravity classifications of these objects are (re)classified by our work. We highlight the three of our ({\it red}) and two of previously known ({\it blue}) disk candidates that have faint reddening-corrected bolometric luminosities compared to the other disk candidates with similar spectral types. We use {\it black} open circles to show previously known edge-on disks in Taurus, 2M~0418+2812, 2M~0438+2609, and 2M~0438+2611 (discussed in Section~\ref{subsubsec:extinction}), and use {\it light grey} circles for the remaining known disk candidates. Extinctions of these known edge-on disks are actually nominal values measured based on the H$_{2}$O color-color diagrams (Table~\ref{tab:classification}) given that they are detected in scattered light and thereby their reddening cannot be accurately estimated (Section~\ref{subsubsec:extinction}). Only objects with good-quality photometry, as defined in Section~\ref{subsec:phot_criteria}, are plotted. Photometric uncertainties are shown if they exceed the size of the symbols. The extinction vector corresponds to $A_{\rm V} = 2$~mag using the extinction law of \cite{Schlafly+2011}. While edge-on disks show significant excess emission in $K$ band, with positions way off the sequence formed by other disk population, the five fainter disk candidates follow the sequence before/after dereddening and have colors distinctively different from those of edge-on disks.}
\label{fig:yJK_disk}
\end{center} 
\end{figure} 
%------------figure end-----------------

\clearpage
%%%%%%%%%%%%%%%%%%%%%%
%--------------- K-W colors: Other---------------------
%%%%%%%%%%%%%%%%%%%%%%
\begin{figure}[t]
\begin{center}
\includegraphics[height=5.5in]{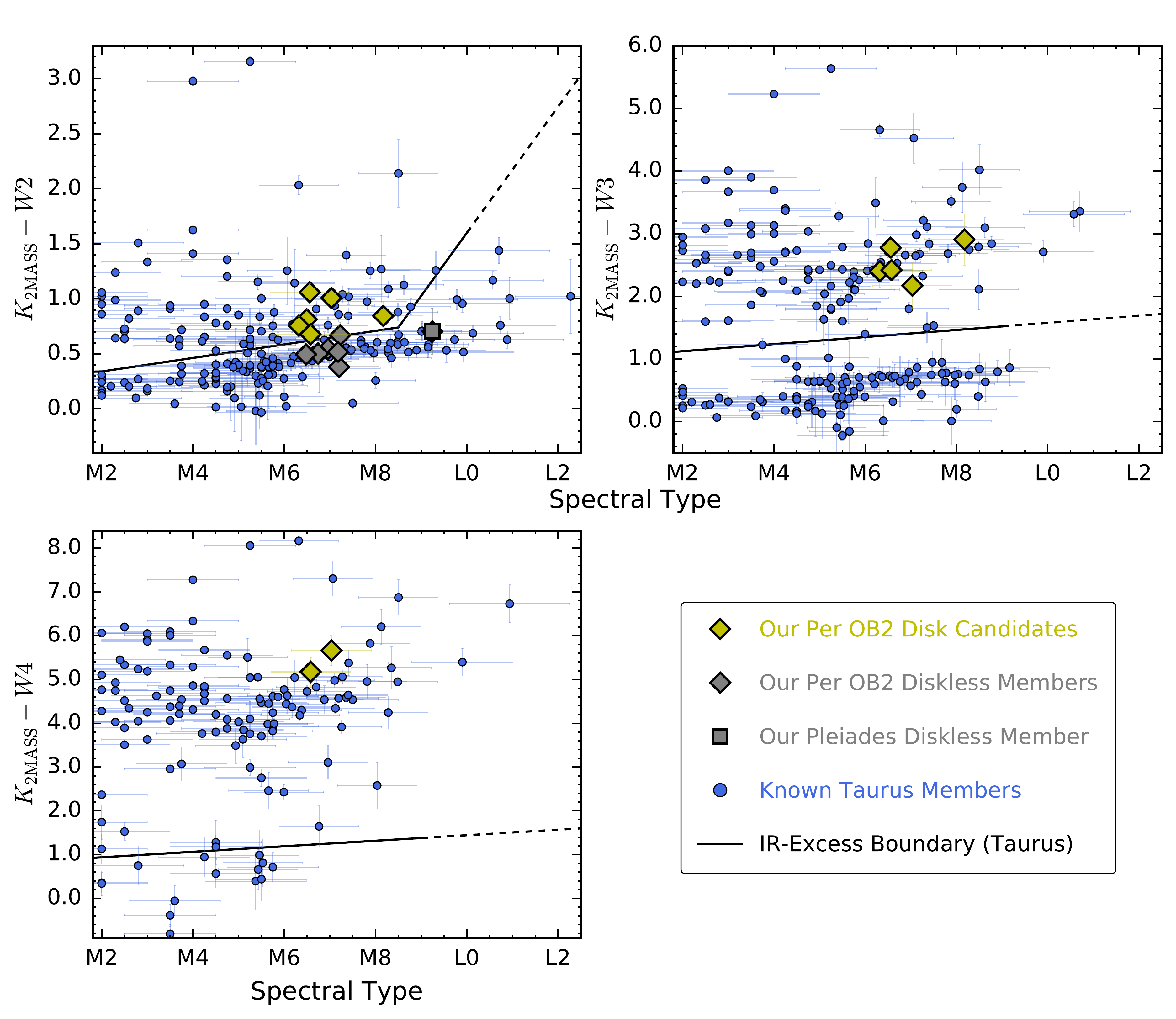}
\caption{Dereddened $K_{\rm 2MASS} - W2/3/4$ colors of our 6 disk candidates in Per~OB2 ({\it olive} diamonds), our diskless members in Per~OB2 ({\it grey} diamonds) and Pleiades ({\it grey} square), and previously known Taurus members ({\it blue} circles), using the same format as Figure~\ref{fig:IR_excess_Taurus}. The dereddened $K_{\rm 2MASS} - W2/3/4$ colors and the significance of objects' mid-infrared excesses are tabulated in Table~\ref{tab:IR_excess}. }
\label{fig:IR_excess_Other}
\end{center}
\end{figure}
%------------figure end-----------------

\clearpage
%%%%%%%%%%%%%%%%%%%%%%
%--------------- telluric H2O ---------------------
%%%%%%%%%%%%%%%%%%%%%%
\begin{figure}[t]
\begin{center}
\includegraphics[height=5.5in]{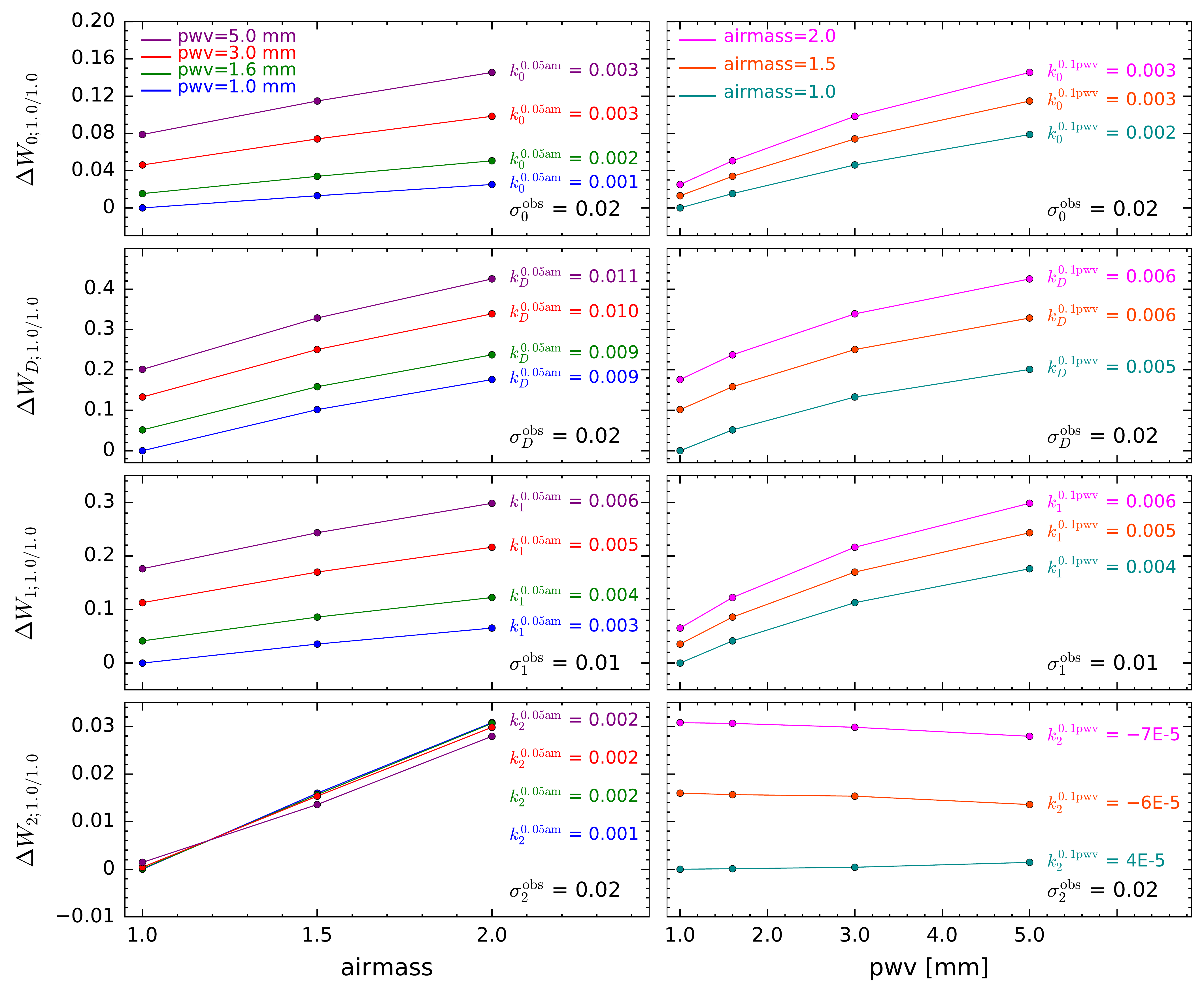}
\caption{The impact of imperfect telluric correction on H$_{2}$O indices $W_{z}$. We use $\Delta W_{z}$ to denote the change in $W_{z}$ due to different telluric absorption spectra between the science target and standard star, and the data points in the figure are the $\Delta W_{z}$ values when the airmass and pwv of the standard star is fixed at $1.0$ and $1.0$~mm, respectively ($\Delta W_{z; \rm1.0/1.0}$; Equation~\ref{eq:DeltaWz}). We list our typical measurement uncertainties $\sigma_{z}^{\rm obs}$ of $W_{z}$ in the lower right of each subplot for comparison. Left: We show $\Delta W_{z; \rm1.0/1.0}$ as a function of airmass for different pwv. For each pwv value, we perform a linear fit to to obtain the slope $k_{z}^{\rm 0.05am}$, which corresponds to the change in $W_{z}$ when the airmass differs by $0.05$. Since $k_{z}^{\rm 0.05am}$ is smaller than $\sigma_{z}^{\rm obs}$ by factors of $\approx 2-10$ for the four H$_{2}$O indices, the telluric contamination due to airmass differences is negligible in this work. Right: We show $\Delta W_{z; \rm1.0/1.0}$ as a function of pwv for different airmasses, and for a given airmass, we perform a linear fit to obtain the slope $k_{z}^{\rm 0.1pwv}$, which corresponds to the change in $W_{z}$ when the pwv differs by $0.1$~mm. Variable pwv values do not cause significant contamination in the measured $W_{z}$ presented in this work, as $k_{z}^{\rm 0.1pwv}$ is smaller than $\sigma_{z}^{\rm obs}$ by factors of $\approx 2-500$ for the four H$_{2}$O indices.}
\label{fig:tell_H2O}
\end{center}
\end{figure}
%------------figure end-----------------

%%%%%%%%%%%%%%%%%%
%%%%%%   TABLES   %%%%%%
%%%%%%%%%%%%%%%%%%
\tabletypesize{\scriptsize}

%----------------------------- 
%------- SpeX Log -------
%-----------------------------

\clearpage
\startlongtable
\begin{deluxetable*}{llcccclr}
\tablewidth{0pc}
\tablecaption{IRTF/SpeX Observations\label{tab:spex_log}}
\tablehead{
\multicolumn{1}{l}{Object}    &\multicolumn{1}{l}{Date}      &\colhead{Slit}                &\colhead{R}                   &\colhead{t$_{\rm int}$}       &\colhead{S/N}                 &\multicolumn{1}{l}{A0V Standard}&\multicolumn{1}{l}{$\Delta$airmass} \\
\colhead{}                    &\multicolumn{1}{l}{(UT)}      &\colhead{($\arcsec$)}         &\colhead{($\lambda/\Delta \lambda$)}&\colhead{(s)}                 &\colhead{}                    &\colhead{}                    &\colhead{}                     }
\startdata
\hline
\multicolumn{8}{c}{New Taurus Members} \\
\hline
PSO J059.1103+31.6643                              & 2016-10-13           & 0.8 & 75 & 533 & 45 & HD 23258                                           & $-0.080$    \\
PSO J059.3563+32.3043                              & 2016-10-13           & 0.8 & 75 & 177 & 46 & HD 23258                                           & $-0.080$    \\
PSO J059.5714+30.6327                              & 2016-11-24           & 0.8 & 75 & 1434 & 34 & HD 25175                                           & $0.042$    \\
PSO J060.2075+31.8384                              & 2016-11-24           & 0.8 & 75 & 956 & 42 & HD 25175                                           & $0.042$    \\
PSO J061.6961+31.6537                              & 2016-10-12           & 0.8 & 75 & 237 & 55 & HD 25175                                           & $-0.077$    \\
PSO J061.8714+23.1104                              & 2016-10-13           & 0.8 & 75 & 1434 & 33 & HD 21038                                           & $-0.090$    \\
PSO J062.2433+32.5380                              & 2016-10-12           & 0.8 & 75 & 237 & 64 & HD 25175                                           & $-0.077$    \\
PSO J062.4648+20.0118                              & 2016-10-12           & 0.8 & 75 & 237 & 39 & HD 25175                                           & $-0.014$    \\
PSO J062.8220+28.5315                              & 2016-10-12           & 0.8 & 75 & 597 & 38 & HD 21038                                           & $-0.059$    \\
PSO J063.0534+32.7055                              & 2016-11-17           & 0.8 & 75 & 237 & 46 & HD 21038                                           & $-0.079$    \\
\enddata
\tablecomments{S/N is the median $J$-band signal-to-noise ratio per pixel. The $\Delta$airmass is the average airmass difference between the A0V standard star and the science target during exposures (i.e., standard $-$ target). The eight new Taurus members that need reobservations (Section~\ref{subsec:reclassification}) have ``$\dagger$'' behind their PS1 names.  \\(This table is available in its entirety in machine-readable form.)}
\end{deluxetable*}
\clearpage

\clearpage
\begin{deluxetable*}{ccccccccccccc}
\tablewidth{0pc}
\tablecaption{L\lowercase{inear} F\lowercase{its} \lowercase{to} R\lowercase{eddening-free} I\lowercase{ndices} \lowercase{versus} O\lowercase{ptical} S\lowercase{pectral} T\lowercase{ypes} \label{tab:omegaSpT}}
\tablehead{
\colhead{}                    & \colhead{}                    & \multicolumn{3}{c}{Entire Sample}                 & \colhead{}                    & \multicolumn{3}{c}{Young Sample}                  & \colhead{}                    & \multicolumn{3}{c}{Old Sample}                    \\
\cline{3-5}                   \cline{7-9}                   \cline{11-13}                 
\colhead{$\omega_{x}$}        & \colhead{range}               & \colhead{c$_{0}$}             & \colhead{c$_{1}$}             & \colhead{rms$_{x}$}           & \colhead{}                    & \colhead{c$_{0}$}             & \colhead{c$_{1}$}             & \colhead{rms$_{x}$}           & \colhead{}                    & \colhead{c$_{0}$}             & \colhead{c$_{1}$}             & \colhead{rms$_{x}$}           }
\startdata
\vspace{0.07cm}  $\omega_{0}$                  & $[0.07, 0.30]$ & $3.68642$ & $24.02588$ & 1.11&                                & $2.94584$ & $24.85340$ & 0.83&                                & $3.62483$ & $25.05236$ & 1.13\\
\vspace{0.07cm}  $\omega_{D}$                  & $[0.02, 0.28]$ & $5.08633$ & $27.05023$ & 1.22&                                & $4.45581$ & $29.92918$ & 1.16&                                & $5.25325$ & $26.26717$ & 1.23\\
   $\omega_{2}$                  & $[0.03, 0.24]$ & $4.16702$ & $22.85994$ & 0.87&                                & $3.41198$ & $25.03094$ & 0.79&                                & $4.26868$ & $22.68608$ & 0.87\\
\enddata
\tablecomments{Our spectral classification is based on the reddening-free index $\omega_{x}$ (Equation \ref{eq:omegax}) for our entire sample and its young (\textsc{vl-g} or \textsc{int-g}) and old (\textsc{fld-g} or no previously reported gravity) subgroups (see also Figure~\ref{fig:omega_SpT}). The applicable fitting range in $\omega_{x}$ and the rms$_{x}$ about each polynomial fit are also shown. The spectral type (SpT) of an object is calculated as SpT$=c_{0} + c_{1} \omega_{x}$ and the numerical SpT is defined to be 0 for M0, 5 for M5, 10 for L0, and so on. For each $\omega_{x}$, the typical difference between the polynomials of any two samples is $\lesssim 0.5$ subtype. Therefore, we recommend using the polynomial defined by the entire sample for spectral classification without distinguishing young and old targets. }
\end{deluxetable*}
\clearpage

\clearpage
\begin{rotate}
\begin{deluxetable*}{cccrrcccrrcccrrc}
\tablewidth{0pc}
\tablecaption{ Sequence of H$_{2}$O Color-Color Diagrams \label{tab:omegax_sequence} }
\tablehead{
\colhead{}                    & \colhead{}                    & \multicolumn{4}{c}{H$_{2}$O}                      & \colhead{}& \multicolumn{4}{c}{H$_{2}$OD}                     & \colhead{}                    & \multicolumn{4}{c}{H$_{2}$O$-2$}                    \\
\cline{3-6}                   \cline{8-11}                  \cline{13-16}                 
\colhead{$\omega_{x}$ range}  & \colhead{}                    & \colhead{$\omega_{0}$}        & \colhead{$W_{0}$}             & \colhead{$W_{1}$}             & \colhead{$N_{1}$}             & \colhead{}                    & \colhead{$\omega_{D}$}        & \colhead{$W_{D}$}             & \colhead{$W_{1}$}             & \colhead{$N_{D}$}             & \colhead{}                    & \colhead{$\omega_{2}$}        & \colhead{$W_{2}$}             & \colhead{$W_{1}$}             & \colhead{$N_{2}$}             }
\startdata
\vspace{0.08 cm}$< 0.1$ &   & 0.054 & $0.037\pm0.048$ & $0.121\pm0.067$ & 36 &   & 0.042 & $-0.102\pm0.055$ & $0.197\pm0.073$ & 77 &   & 0.032 & $-0.073\pm0.065$ & $0.138\pm0.055$ & 43\\
\vspace{0.08 cm}$[0.10, 0.15)$ &   & 0.130 & $-0.038\pm0.035$ & $0.210\pm0.037$ & 23 &   & 0.126 & $-0.031\pm0.027$ & $0.292\pm0.034$ & 67 &   & 0.127 & $0.034\pm0.032$ & $0.229\pm0.033$ & 25\\
\vspace{0.08 cm}$[0.15, 0.20)$ &   & 0.177 & $-0.077\pm0.029$ & $0.270\pm0.030$ & 54 &   & 0.171 & $-0.006\pm0.034$ & $0.337\pm0.035$ & 44 &   & 0.177 & $0.078\pm0.029$ & $0.277\pm0.027$ & 57\\
\vspace{0.08 cm}$[0.20, 0.25)$ &   & 0.222 & $-0.114\pm0.035$ & $0.319\pm0.040$ & 60 &   & 0.229 & $0.064\pm0.050$ & $0.391\pm0.047$ & 36 &   & 0.221 & $0.116\pm0.030$ & $0.330\pm0.027$ & 50\\
\vspace{0.08 cm}$[0.25, 0.30)$ &   & 0.275 & $-0.172\pm0.050$ & $0.379\pm0.044$ & 34 &   & 0.271 & $0.118\pm0.051$ & $0.421\pm0.052$ & 21 &   & 0.276 & $0.169\pm0.036$ & $0.389\pm0.040$ & 43\\
\vspace{0.08 cm}$[0.30, 0.35)$ &   & 0.318 & $-0.224\pm0.051$ & $0.414\pm0.042$ & 38 &   & 0.325 & $0.179\pm0.045$ & $0.469\pm0.048$ & 20 &   & 0.319 & $0.213\pm0.042$ & $0.446\pm0.052$ & 41\\
\vspace{0.08 cm}$[0.35, 0.40)$ &   & 0.386 & $-0.301\pm0.021$ & $0.464\pm0.036$ & 14 &   & 0.369 & $0.222\pm0.045$ & $0.529\pm0.050$ & 18 &   & 0.380 & $0.226\pm0.050$ & $0.533\pm0.060$ & 39\\
\vspace{0.08 cm}$[0.40, 0.45)$ &   & 0.422 & $-0.312\pm0.041$ & $0.531\pm0.046$ & 36 &   & 0.422 & $0.262\pm0.061$ & $0.573\pm0.059$ & 26 &   & 0.424 & $0.239\pm0.058$ & $0.628\pm0.057$ & 28\\
\vspace{0.08 cm}$[0.45, 0.50)$ &   & 0.473 & $-0.335\pm0.035$ & $0.602\pm0.049$ & 21 &   & 0.480 & $0.310\pm0.066$ & $0.647\pm0.061$ & 17 &   & 0.470 & $0.179\pm0.055$ & $0.784\pm0.069$ & 22\\
\vspace{0.08 cm}$[0.50, 0.60)$ &   & 0.553 & $-0.368\pm0.039$ & $0.738\pm0.065$ & 27 &   & 0.545 & $0.351\pm0.039$ & $0.739\pm0.041$ & 15 &   & 0.538 & $0.245\pm0.069$ & $0.899\pm0.082$ & 8\\
 $\geqslant 0.6$ &   & 0.648 & $-0.433\pm0.053$ & $0.894\pm0.078$ & 17 &   & 0.643 & $0.450\pm0.085$ & $0.873\pm0.075$ & 19 &   & 0.607 & $0.311\pm0.077$ & $0.955\pm0.078$ & 4\\
\enddata
\tablecomments{ The intrinsic sequence of the H$_{2}$O color-color diagrams ($W_{x}$ versus $W_{1}$, Figure \ref{fig:omegax_seq}). The $\omega_{x}$ range describes the range of $\omega_{x}$ bins for each H$_{2}$O index. The $\omega_{x}$, $W_{x}$ and $W_{1}$ columns tabulate the median of $\omega_{x}$ values, and the median and standard deviations of $W_{x}$ and $W_{1}$ in each $\omega_{x}$ bin. $N_{x}$ is the number of objects in each bin.}
\end{deluxetable*}
\end{rotate}
\clearpage

\clearpage
\begin{deluxetable*}{lccrrrrrrrrr}
\tablewidth{0pc}
\tablecaption{ Intrinsic Optical--Near-Infrared Color Sequence as a Function of Literature Spectral Type \label{tab:instrinsic_color} }
\tablehead{
\colhead{}                    & \colhead{}                    & \colhead{}                    & \multicolumn{5}{c}{Intrinsic Colors vs. SpT$_{\rm lit}$}& \colhead{}                    & \multicolumn{3}{c}{Uncertainties vs. SpT$_{\rm lit}$}\\
\cline{4-8}                   \cline{10-12}                 
\colhead{Color}               & \colhead{Population}          & \colhead{}                    & \colhead{$c_{0}$}             & \colhead{$c_{1}$}             & \colhead{$c_{2}$}             & \colhead{$c_{3}$}             & \colhead{$c_{4}$}             & \colhead{}                    & \colhead{$c_{0}$}             & \colhead{$c_{1}$}             & \colhead{$c_{2}$}             }
\startdata
\vspace{-0.25cm} & Young&  & $0.00015$& $-0.00735$& $0.09778$& $-0.14118$& $1.94003$&  & $-0.00589$& $0.10589$& $-0.18551$\\  \vspace{-0.25cm}
$i_{\rm P1} - J_{\rm 2MASS}$  & & & & & & & & & & & \\
\vspace{0.25cm} & Old&  & $0.00081$& $-0.02900$& $0.34459$& $-1.27276$& $3.30290$&  & $-0.00131$& $0.03459$& $-0.00067$\\
\vspace{-0.25cm} & Young&  & $-0.00321$& $0.07746$& $-0.38162$& $2.28808$& \multicolumn{1}{c}{...}&  & $-0.00271$& $0.05228$& $-0.05275$\\  \vspace{-0.25cm}
$z_{\rm P1} - J_{\rm 2MASS}$  & & & & & & & & & & & \\
\vspace{0.25cm} & Old&  & $-0.00182$& $0.04279$& $-0.12156$& $1.44695$& \multicolumn{1}{c}{...}&  & $-0.00067$& $0.02229$& $0.00410$\\
\vspace{-0.25cm} & Young&  & $-0.00134$& $0.03462$& $-0.17012$& $1.58831$& \multicolumn{1}{c}{...}&  & $-0.00170$& $0.03724$& $-0.04888$\\  \vspace{-0.25cm}
$y_{\rm P1} - J_{\rm 2MASS}$  & & & & & & & & & & & \\
  & Old&  & $-0.00072$& $0.01895$& $-0.05733$& $1.19601$& \multicolumn{1}{c}{...}&  & $0.00061$& $0.00093$& $0.04065$\\
\enddata
\tablecomments{ Intrinsic color sequences of $i_{\rm P1}-J_{\rm 2MASS}$, $z_{\rm P1}-J_{\rm 2MASS}$, and $y_{\rm P1}-J_{\rm 2MASS}$, as functions of literature spectral types (SpT$_{\rm lit}$) for both young and old populations (see also Figure~\ref{fig:color_seq}). The intrinsic colors and uncertainties are described by polynomials $\sum\limits_{i} c_{i} \times {\rm SpT}_{\rm lit}^{i}$, where SpT$=0$ for M0, 5 for M5, 10 for L0, and so on.  }
\end{deluxetable*}
\clearpage

%--------------------------------- 
%------- PS1 + WISE -------
%---------------------------------

\clearpage
\begin{longrotatetable}
\begin{deluxetable*}{lcccccccccccccc}
\tablewidth{0pc}
\tablecaption{PS1 and AllWISE Photometry of Our Discoveries and Previously Known Objects\label{tab:ps1_wise}}
\tablehead{
\multicolumn{1}{l}{Object}    &\colhead{$g_{\rm P1}$}        &\colhead{$g_{\rm P1, qual}$}  &\colhead{$r_{\rm P1}$}        &\colhead{$r_{\rm P1, qual}$}  &\colhead{$i_{\rm P1}$}        &\colhead{$i_{\rm P1, qual}$}  &\colhead{$z_{\rm P1}$}        &\colhead{$z_{\rm P1, qual}$}  &\colhead{$y_{\rm P1}$}        &\colhead{$y_{\rm P1, qual}$}  &\colhead{$W1$}                &\colhead{$W1_{\rm qual}$}     &\colhead{$W2$}                &\colhead{$W2_{\rm qual}$}      \\
\colhead{}                    &\colhead{(AB mag)}            &\colhead{}                    &\colhead{(AB mag)}            &\colhead{}                    &\colhead{(AB mag)}            &\colhead{}                    &\colhead{(AB mag)}            &\colhead{}                    &\colhead{(AB mag)}            &\colhead{}                    &\colhead{(mag)}               &\colhead{}                    &\colhead{(mag)}               &\colhead{}                     }
\startdata
\hline
\multicolumn{15}{c}{New Taurus Members} \\
\hline
PSO J059.1103+31.6643                               &  $22.97\pm0.13$  &  0  &  $21.86\pm0.05$  &  1  &  $19.41\pm0.01$  &  1  &  $18.15\pm0.01$  &  1  &  $17.43\pm0.01$  &  1  &  $14.28\pm0.03$  &  1  &  $13.98\pm0.04$  &  1    \\
PSO J059.3563+32.3043                               &  $21.81\pm0.10$  &  1  &  $20.23\pm0.01$  &  1  &  $18.14\pm0.01$  &  1  &  $17.12\pm0.01$  &  1  &  $16.59\pm0.01$  &  1  &  $13.99\pm0.03$  &  1  &  $13.65\pm0.04$  &  1    \\
PSO J059.5714+30.6327                               &  $22.49\pm0.60$  &  0  &  $20.52\pm0.60$  &  0  &  $20.43\pm0.03$  &  1  &  $18.93\pm0.01$  &  1  &  $17.94\pm0.02$  &  1  &  $14.47\pm0.03$  &  1  &  $14.09\pm0.04$  &  1    \\
PSO J060.2075+31.8384                               &  $20.58\pm0.60$  &  0  &  $21.62\pm0.05$  &  1  &  $19.08\pm0.01$  &  1  &  $17.88\pm0.01$  &  1  &  $17.18\pm0.01$  &  1  &  $14.29\pm0.03$  &  1  &  $13.97\pm0.05$  &  1    \\
PSO J061.6961+31.6537                               &  $22.44\pm0.10$  &  1  &  $20.69\pm0.02$  &  1  &  $18.30\pm0.01$  &  1  &  $17.18\pm0.01$  &  1  &  $16.56\pm0.01$  &  1  &  $13.77\pm0.03$  &  1  &  $13.47\pm0.03$  &  1    \\
PSO J061.8714+23.1104                               &  $22.66\pm0.60$  &  0  &  $20.76\pm0.60$  &  0  &  $21.39\pm0.07$  &  1  &  $19.73\pm0.02$  &  1  &  $18.69\pm0.02$  &  1  &  $14.86\pm0.04$  &  1  &  $14.53\pm0.06$  &  1    \\
PSO J062.2433+32.5380                               &  $21.78\pm0.06$  &  1  &  $20.39\pm0.04$  &  1  &  $18.16\pm0.01$  &  1  &  $17.06\pm0.01$  &  1  &  $16.44\pm0.00$  &  1  &  $13.70\pm0.03$  &  1  &  $13.37\pm0.03$  &  1    \\
PSO J062.4648+20.0118                               &  $21.75\pm0.60$  &  0  &  $21.87\pm0.19$  &  0  &  $19.87\pm0.02$  &  1  &  $18.35\pm0.01$  &  1  &  $17.46\pm0.01$  &  1  &  $14.12\pm0.03$  &  1  &  $13.75\pm0.04$  &  1    \\
PSO J062.8220+28.5315                               &  $27.97\pm0.60$  &  0  &  $24.28\pm0.60$  &  0  &  $20.97\pm0.02$  &  1  &  $19.49\pm0.03$  &  1  &  $18.50\pm0.03$  &  1  &  $13.42\pm0.03$  &  1  &  $13.00\pm0.03$  &  1    \\
PSO J063.0534+32.7055                               &  $21.11\pm0.04$  &  1  &  $19.69\pm0.02$  &  1  &  $17.53\pm0.00$  &  1  &  $16.46\pm0.00$  &  1  &  $15.89\pm0.00$  &  1  &  $13.22\pm0.03$  &  1  &  $12.91\pm0.03$  &  1    \\
\hline
\enddata
\tablecomments{We use ``1'' to indicate a good quality in each photometric band, as defined in Section~\ref{subsec:phot_criteria}, and use ``0'' otherwise. The eight new Taurus members that need reobservations (Section~\ref{subsec:reclassification}) have ``$\dagger$'' behind their PS1 names. \\(This table is available in its entirety in machine-readable form.) }
\end{deluxetable*}
\clearpage
\end{longrotatetable}

%------------------------- 
%------- UGCS -------
%-------------------------

\clearpage
\startlongtable
\begin{deluxetable*}{llcccccc}
\tablewidth{0pc}
\tablecaption{UGCS Photometry of Our Discoveries and Previously Known Objects\label{tab:ugcs}}
\tablehead{
\multicolumn{1}{l}{Object}    &\multicolumn{1}{l}{UGCS Name} &\colhead{$J_{\rm MKO}$}       &\colhead{$J_{\rm MKO, qual}$} &\colhead{$H_{\rm MKO}$}       &\colhead{$H_{\rm MKO, qual}$} &\colhead{$K_{\rm MKO}$}       &\colhead{$K_{\rm MKO, qual}$}  \\
\colhead{}                    &\colhead{}                    &\colhead{(mag)}               &\colhead{}                    &\colhead{(mag)}               &\colhead{}                    &\colhead{(mag)}               &\colhead{}                     }
\startdata
\hline
\multicolumn{8}{c}{New Taurus Members} \\
\hline
PSO J059.1103+31.6643                               &  J035626.49+313951.8                                 &  $15.701\pm0.009$  &  1  &  $15.084\pm0.007$  &  1  &  $14.602\pm0.008$  &  1    \\
PSO J059.3563+32.3043                               &  ...                                                 &  ...  &  ...  &  ...  &  ...  &  ...  &  ...    \\
PSO J059.5714+30.6327                               &  ...                                                 &  ...  &  ...  &  ...  &  ...  &  ...  &  ...    \\
PSO J060.2075+31.8384                               &  ...                                                 &  ...  &  ...  &  ...  &  ...  &  ...  &  ...    \\
PSO J061.6961+31.6537                               &  ...                                                 &  ...  &  ...  &  ...  &  ...  &  ...  &  ...    \\
PSO J061.8714+23.1104                               &  J040729.14+230637.9                                 &  $16.548\pm0.014$  &  0  &  ...  &  ...  &  $15.222\pm0.020$  &  1    \\
PSO J062.2433+32.5380                               &  ...                                                 &  ...  &  ...  &  ...  &  ...  &  ...  &  ...    \\
PSO J062.4648+20.0118                               &  J040951.55+200042.8                                 &  ...  &  ...  &  ...  &  ...  &  $14.387\pm0.008$  &  1    \\
PSO J062.8220+28.5315                               &  ...                                                 &  ...  &  ...  &  ...  &  ...  &  ...  &  ...    \\
PSO J063.0534+32.7055                               &  ...                                                 &  ...  &  ...  &  ...  &  ...  &  ...  &  ...    \\
\hline
\enddata
\tablecomments{We use ``1'' to indicate a good quality in each photometric band, as defined in Section~\ref{subsec:phot_criteria}, and use ``0'' otherwise. The eight new Taurus members that need reobservations (Section~\ref{subsec:reclassification}) have ``$\dagger$'' behind their PS1 names. Reddened field dwarfs studied by AL13 (i.e., 2M~0422+1530, 2M~0435$-$1414, 2M~0619$-$2903), do not have UGCS detections and thus are not included in the table. \\(This table is available in its entirety in machine-readable form.)}
\end{deluxetable*}
\clearpage

% ------------------
% --- 2MASS ---
% ------------------

\clearpage
\begin{longrotatetable}
\begin{deluxetable*}{llcccccc}
\tablewidth{0pc}
\tablecaption{2MASS Photometry of Our Discoveries and Previously Known Objects\label{tab:2mass}}
\tablehead{
\multicolumn{1}{l}{Object}    &\multicolumn{1}{l}{2MASS/Literature Name}&\colhead{$J_{\rm 2MASS}$}     &\colhead{$J_{\rm 2MASS, qual}$}&\colhead{$H_{\rm 2MASS}$}     &\colhead{$H_{\rm 2MASS, qual}$}&\colhead{$K_{\rm 2MASS}$}     &\colhead{$K_{\rm 2MASS, qual}$} \\
\colhead{}                    &\colhead{}                    &\colhead{(mag)}               &\colhead{}                    &\colhead{(mag)}               &\colhead{}                    &\colhead{(mag)}               &\colhead{}                     }
\startdata
\hline
\multicolumn{8}{c}{New Taurus Members} \\
\hline
PSO J059.1103+31.6643                               &  2MASS J03562649+3139518                             &  $15.663\pm0.055$  &  1  &  $14.985\pm0.064$  &  1  &  $14.614\pm0.079$  &  1    \\
PSO J059.3563+32.3043                               &  2MASS J03572552+3218157                             &  $15.139\pm0.038$  &  1  &  $14.515\pm0.046$  &  1  &  $14.232\pm0.056$  &  1    \\
PSO J059.5714+30.6327                               &  2MASS J03581711+3037586                             &  $16.017\pm0.080$  &  1  &  $15.372\pm0.090$  &  1  &  $14.721\pm0.090$  &  1    \\
PSO J060.2075+31.8384                               &  2MASS J04004981+3150186                             &  $15.588\pm0.057$  &  1  &  $14.993\pm0.072$  &  1  &  $14.608\pm0.086$  &  1    \\
PSO J061.6961+31.6537                               &  2MASS J04064707+3139137                             &  $15.095\pm0.048$  &  1  &  $14.364\pm0.049$  &  1  &  $14.000\pm0.051$  &  1    \\
PSO J061.8714+23.1104                               &  2MASS J04072912+2306381                             &  $16.412\pm0.120$  &  1  &  $15.701\pm0.149$  &  1  &  $15.187\pm0.130$  &  1    \\
PSO J062.2433+32.5380                               &  2MASS J04085839+3232170                             &  $14.926\pm0.040$  &  1  &  $14.298\pm0.047$  &  1  &  $13.961\pm0.047$  &  1    \\
PSO J062.4648+20.0118                               &  2MASS J04095154+2000428                             &  $15.575\pm0.061$  &  1  &  $14.930\pm0.075$  &  1  &  $14.383\pm0.046$  &  1    \\
PSO J062.8220+28.5315                               &  2MASS J04111728+2831540                             &  $16.017\pm0.080$  &  1  &  $14.529\pm0.050$  &  1  &  $13.789\pm0.040$  &  1    \\
PSO J063.0534+32.7055                               &  2MASS J04121282+3242199                             &  $14.433\pm0.032$  &  1  &  $13.789\pm0.034$  &  1  &  $13.425\pm0.034$  &  1    \\
\hline
\enddata
\tablecomments{We use ``1'' to indicate a good quality in each photometric band, as defined in Section~\ref{subsec:phot_criteria}, and use ``0'' otherwise. The eight new Taurus members that need reobservations (Section~\ref{subsec:reclassification}) have ``$\dagger$'' behind their PS1 names. We show 2MASS designations for our discoveries and literature names for previously known objects. \\(This table is available in its entirety in machine-readable form.)}
\end{deluxetable*}
\clearpage
\end{longrotatetable}

% --------------------------------
% ------ Classification ------
% --------------------------------

\clearpage
\begin{longrotatetable}
\begin{deluxetable*}{lcccccccccccc}
\tablewidth{0pc}
\tablecaption{Classification Results of Our Discoveries and Previously Known Objects\label{tab:classification}}
\tablehead{
\colhead{}                    &\colhead{}                    &\multicolumn{4}{c}{Spectral Type}&\colhead{}                    &\multicolumn{4}{c}{Extinction (mag)}&\colhead{}                    &\colhead{}                    \\
\cline{3-6} \cline{8-11}
\multicolumn{1}{l}{Object}    &\colhead{}                    &\colhead{SpT$_{\omega}$}      &\colhead{SpT$_{\rm AL13}$}    &\colhead{SpT$_{\rm AL13}^{\star}$}&\colhead{SpT$_{\rm lit}$}     &\colhead{}                    &\colhead{$A_{\rm V}^{\rm H_{2}O}$}&\colhead{$A_{\rm V}^{\rm OIR}$}&\colhead{$A_{\rm V}^{\rm lit}$}&\colhead{Adopted}             &\colhead{}                    &\colhead{Gravity}             }
\startdata
\hline
\multicolumn{13}{c}{New Taurus Members} \\
\hline
PSO J059.1103+31.6643                               &  & M7.5$\pm 0.9$  &  M7.2$\pm 0.5$  &  M7.0$\pm 0.8$  &  ...  & &  $0.7\pm4.0$ & $0.54\pm0.85$ & ... & $0.54\pm0.85$  & &  \textsc{int-g}     \\
PSO J059.3563+32.3043                               &  & M6.6$\pm 0.9$  &  M5.8$\pm 0.5$  &  M6.0$\pm 0.7$  &  ...  & &  $-0.0\pm4.0$ & $-0.59\pm0.85$ & ... & $-0.59\pm0.85$  & &  \textsc{int-g}     \\
PSO J059.5714+30.6327                               &  & M9.7$\pm 1.1$  &  L0.1$\pm 0.6$  &  L0.2$\pm 0.7$  &  ...  & &  $2.7\pm4.0$ & $-0.21\pm0.85$ & ... & $-0.21\pm0.85$  & &  \textsc{vl-g}     \\
PSO J060.2075+31.8384                               &  & M6.6$\pm 0.9$  &  M6.3$\pm 0.6$  &  M6.3$\pm 1.0$  &  ...  & &  $1.3\pm4.0$ & $0.98\pm0.85$ & ... & $0.98\pm0.85$  & &  \textsc{int-g}     \\
PSO J061.6961+31.6537                               &  & M6.9$\pm 0.9$  &  M6.2$\pm 0.4$  &  M6.3$\pm 0.7$  &  ...  & &  $-0.1\pm4.0$ & $-0.38\pm0.85$ & ... & $-0.38\pm0.85$  & &  \textsc{int-g}     \\
PSO J061.8714+23.1104                               &  & M9.9$\pm 1.3$  &  L0.2$\pm 0.6$  &  L0.0$\pm 1.0$  &  ...  & &  $-2.7\pm4.0$ & $1.68\pm0.85$ & ... & $1.68\pm0.85$  & &  \textsc{vl-g}     \\
PSO J062.2433+32.5380                               &  & M6.5$\pm 0.9$  &  M5.8$\pm 0.6$  &  M5.5$\pm 0.6$  &  ...  & &  $-1.2\pm4.0$ & $0.25\pm0.85$ & ... & $0.25\pm0.85$  & &  \textsc{int-g}     \\
PSO J062.4648+20.0118                               &  & M9.1$\pm 0.9$  &  M8.7$\pm 0.5$  &  M8.7$\pm 0.8$  &  ...  & &  $-0.5\pm4.0$ & $0.21\pm0.85$ & ... & $0.21\pm0.85$  & &  \textsc{vl-g}     \\
PSO J062.8220+28.5315                               &  & M5.4$\pm 1.0$  &  M6.5$\pm 0.4$  &  M5.4$\pm 1.2$  &  ...  & &  $12.6\pm4.0$ & ... & ... & $12.6\pm4.0$  & &  ...     \\
PSO J063.0534+32.7055                               &  & M6.8$\pm 0.9$  &  M6.2$\pm 0.5$  &  M6.3$\pm 0.7$  &  ...  & &  $-0.2\pm4.0$ & $-0.61\pm0.85$ & ... & $-0.61\pm0.85$  & &  \textsc{vl-g}     \\
\hline
\enddata
\tablecomments{We present our reddening-free spectral type (SpT$_{\omega}$; adopted), AL13 spectral type based on the observed spectra (SpT$_{\rm AL13}$) and dereddened spectra (SpT$_{\rm AL13}^{\star}$), extinction values ($A_{\rm V}^{\rm H_{2}O}$ and $A_{\rm V}^{\rm OIR}$), and gravity classifications. We also show the literature spectral types (SpT$_{\rm lit}$) and extinctions ($A_{\rm V}^{\rm lit}$) for previously known objects in Taurus and the field from \cite{Luhman+2017} and AL13, respectively. Two classification results are shown for 2M~0619$-$2903, based on its near-infrared spectrum obtained in 2008 November and 2015 December. The eight new Taurus members that need reobservations (Section~\ref{subsec:reclassification}) have ``$\dagger$'' behind their PS1 names. \\(This table is available in its entirety in machine-readable form.)}
\end{deluxetable*}
\clearpage
\end{longrotatetable}

% -------------------------
% ------ PM, T, L ------
% -------------------------

\clearpage
\startlongtable
\begin{deluxetable*}{lcccccccc}
\tablewidth{0pc}
\tablecaption{Proper Motions, Effective Temperatures, and Bolometric Luminosities of Our Discoveries and Previously Known Objects\label{tab:pmtl}}
\tablehead{
\colhead{}                    &\colhead{}                    &\multicolumn{3}{c}{Proper Motion}&\colhead{}                    &\colhead{}                    &\colhead{}                    &\colhead{}                    \\
\cline{3-5}
\multicolumn{1}{l}{Object}    &\colhead{}                    &\colhead{$\mu_{\alpha}$cos$\delta$}&\colhead{$\mu_{\delta}$}      &\colhead{$\chi_{\nu}^{2}$}    &\colhead{}                    &\colhead{T$_{\rm eff}$}       &\colhead{log$_{10}({\rm L}_{\rm bol}/{\rm L}_{\odot})$}&\colhead{Model-Derived Age}   \\
\colhead{}                    &\colhead{}                    &\colhead{(mas~yr$^{-1}$)}     &\colhead{(mas~yr$^{-1}$)}     &\colhead{}                    &\colhead{}                    &\colhead{(K)}                 &\colhead{(dex)}               &\colhead{(Myr)}               }
\startdata
\hline
\multicolumn{9}{c}{New Taurus Members} \\
\hline
PSO J059.1103+31.6643                               &  & $5.72\pm1.76$  &  $-12.34\pm2.67$  &  $0.7$  & &  $2659\pm146$ & $-2.78\pm0.14$ & $[10,30)$    \\
PSO J059.3563+32.3043                               &  & $4.50\pm2.07$  &  $-5.37\pm2.18$  &  $1.1$  & &  $2763\pm140$ & $-2.65\pm0.14$ & $[10,30)$    \\
PSO J059.5714+30.6327                               &  & $21.61\pm2.55$  &  $-46.59\pm9.42$  &  $5.1$  & &  $2329\pm218$ & $-2.93\pm0.15$ & $[10,30)$    \\
PSO J060.2075+31.8384                               &  & $3.30\pm2.33$  &  $-12.86\pm2.33$  &  $0.7$  & &  $2764\pm145$ & $-2.73\pm0.14$ & $[30,100)$    \\
PSO J061.6961+31.6537                               &  & $2.90\pm3.60$  &  $-14.38\pm3.53$  &  $6.1$  & &  $2733\pm144$ & $-2.62\pm0.14$ & $[10,30)$    \\
PSO J061.8714+23.1104                               &  & $5.88\pm7.48$  &  $-18.10\pm7.16$  &  $0.9$  & &  $2308\pm229$ & $-2.92\pm0.17$ & $[1,10)$    \\
PSO J062.2433+32.5380                               &  & $6.32\pm2.06$  &  $-12.90\pm1.85$  &  $0.9$  & &  $2780\pm152$ & $-2.54\pm0.13$ & $[10,30)$    \\
PSO J062.4648+20.0118                               &  & $6.81\pm2.66$  &  $-13.99\pm1.77$  &  $0.7$  & &  $2437\pm176$ & $-2.76\pm0.15$ & $[1,10)$    \\
PSO J062.8220+28.5315                               &  & $11.96\pm4.34$  &  $-33.38\pm4.33$  &  $0.6$  & &  $2934\pm187$ & $-1.80\pm0.38$ & $[1,10)$    \\
PSO J063.0534+32.7055                               &  & $6.65\pm2.03$  &  $-1.52\pm3.17$  &  $23.5$  & &  $2742\pm143$ & $-2.36\pm0.13$ & $[1,10)$    \\
\hline
\enddata
\tablenotetext{a}{The bolometric luminosities of these objects could be unreliable, since they have bad-quality $J$-band photometry.\\(This table is available in its entirety in machine-readable form.)}
\end{deluxetable*}
\clearpage

% ---------------------------------------------------
% ----- Field Contamination Estimate -----
% ---------------------------------------------------

\clearpage
\begin{longrotatetable}
\begin{deluxetable*}{lllllllllllllllllll}
\tablewidth{0pc}
\tablecaption{Estimated Field Contamination in Taurus\label{tab:FCE}}
\tablehead{
\colhead{}                    &\colhead{}                    &\multicolumn{11}{c}{BGM-Predicted Field Contamination}&\colhead{}                    &\multicolumn{5}{c}{Observed Spectroscopic Follow-up}  \\
\cline{3-13}
\colhead{}                    &\colhead{}                    &\multicolumn{5}{c}{Entire Candidate List}&\colhead{}                    &\multicolumn{5}{c}{Spectroscopic Follow-up Sample}&\colhead{}                    &\multicolumn{5}{c}{}            \\
\cline{3-7}   \cline{9-13}   \cline{15-19}
\multicolumn{1}{l}{Spectral Type}&\colhead{}                    &\multicolumn{1}{l}{$\leqslant$30 Myr}&\multicolumn{1}{l}{$30-200$ Myr}&\multicolumn{1}{l}{$>$200 Myr}&\colhead{}                    &\multicolumn{1}{l}{Total}     &\colhead{}                    &\multicolumn{1}{l}{$\leqslant$30 Myr}&\multicolumn{1}{l}{$30-200$ Myr}&\multicolumn{1}{l}{$>$200 Myr}&\colhead{}                    &\multicolumn{1}{l}{Total}     &\colhead{}                    &\multicolumn{1}{l}{\textsc{vl-g}}&\multicolumn{1}{l}{\textsc{int-g}}&\multicolumn{1}{l}{\textsc{fld-g}}&\colhead{}                    &\multicolumn{1}{l}{Total}      }
\startdata
$<$M4                    & &  $12 \pm 3$  &  $14 \pm 4$  &  $49 \pm 7$  & &   $75 \pm 9$    & &  $3 \pm 1$  &  $2 \pm 1$  &  $8 \pm 2$  & &   $13 \pm 2$    & & ... & ... & ... & &  11 ($118\%\pm45\%$)   \\ 
 {\rm [M4, M5)}           & &  $12 \pm 3$  &  $5 \pm 2$  &  $9 \pm 3$  & &   $26 \pm 5$    & &  $4 \pm 1$  &  $1 \pm 1$  &  $2 \pm 1$  & &   $6 \pm 2$    & & ... & ... & ... & &   4 ($150\%\pm50\%$)   \\ 
 {\rm [M5, M6)}           & &  $11 \pm 4$  &  $3 \pm 2$  &  $3 \pm 2$  & &   $17 \pm 4$    & &  $4 \pm 1$  &  $1 \pm 1$  &  $1 \pm 1$  & &   $5 \pm 2$    & & ... & ... & ... & &  10 ($50\%\pm20\%$)   \\ 
 {\rm [M6, M7)}           & &  $7 \pm 2$  &  $2 \pm 2$  &  $1 \pm 2$  & &   $10 \pm 3$    & &  $2 \pm 1$  &  $1 \pm 1$  &  $0 \pm 0$  & &   $3 \pm 1$    & & 18 &  8 &  0 & &  26 ($11\%\pm 3\%$)   \\ 
 {\rm [M7, M8)}           & &  $4 \pm 2$  &  $2 \pm 2$  &  $2 \pm 2$  & &   $8 \pm 3$    & &  $1 \pm 1$  &  $1 \pm 1$  &  $0 \pm 0$  & &   $2 \pm 1$    & & 10 &  2 &  0 & &  12 ($16\%\pm 8\%$)   \\ 
 {\rm [M8, M9)}           & &  $2 \pm 1$  &  $3 \pm 2$  &  $2 \pm 2$  & &   $7 \pm 3$    & &  $0 \pm 0$  &  $1 \pm 1$  &  $0 \pm 0$  & &   $1 \pm 1$    & &  3 &  1 &  0 & &   4 ($25\%\pm25\%$)   \\ 
 {\rm [M9, L0)}           & &  $2 \pm 2$  &  $3 \pm 2$  &  $2 \pm 2$  & &   $7 \pm 3$    & &  $0 \pm 0$  &  $1 \pm 1$  &  $0 \pm 0$  & &   $1 \pm 1$    & &  4 &  1 &  0 & &   5 ($20\%\pm20\%$)   \\ 
 {\rm [L0, L1)}           & &  $1 \pm 1$  &  $2 \pm 2$  &  $1 \pm 1$  & &   $4 \pm 2$    & &  $0 \pm 0$  &  $1 \pm 0$  &  $0 \pm 0$  & &   $1 \pm 1$    & &  2 &  0 &  0 & &   2 ($50\%\pm50\%$)   \\ 
 {\rm [L1, L2)}           & &  $0 \pm 0$  &  $1 \pm 1$  &  $0 \pm 0$  & &   $1 \pm 1$    & &  $0 \pm 0$  &  $0 \pm 0$  &  $0 \pm 0$  & &   $0 \pm 0$    & &  1 &  0 &  0 & &   1 ($ 0\%\pm 0\%$)   \\ 
 $\geqslant$L2            & &  $0 \pm 0$  &  $1 \pm 1$  &  $0 \pm 0$  & &   $1 \pm 1$    & &  $0 \pm 0$  &  $0 \pm 0$  &  $0 \pm 0$  & &   $0 \pm 0$    & &  0 &  0 &  0 & &   0   \\ 
 \vspace{-0.3cm} \\ 
 Total     & &  &  &  & &  $156 \pm 13$   & & & & & &  $32 \pm 4$   & & & & & & 75   \\ 
\enddata
\tablecomments{Estimated numbers of field contaminants among our entire selected Taurus candidates and among our spectroscopic follow-up sample (Section~\ref{subsec:FCE}), as compared to the number of real discoveries from our survey. In each spectral type bin, we show the total number of field contaminants, as well as those with three age ranges of $\leqslant$30~Myr, (30, 200]~Myr, and $>$200~Myr, which are analogous to the near-infrared gravity classifications of \textsc{vl-g}, \textsc{int-g}, and \textsc{fld-g} for $\geqslant$M6 dwarfs. Gravity classifications are not defined for M4 and M5 dwarfs. The star counts and uncertainties are quoted based on the median and 16-84 percentile of the Monte Carlo trials, and are rounded off to integers. Our adapted BGM predicts $156\pm13$ field interlopers among our 350 selected Taurus candidates and predicts $32\pm4$  field interlopers among our spectroscopic follow-up sample of 75 objects. A contamination fraction for our discoveries (i.e., total number of expected contaminants divided by the total number of real discoveries) is also shown for each spectral type bin in the last column. }
\end{deluxetable*}
\end{longrotatetable}
\clearpage

% -------------------------
% ----- PM sigma -----
% -------------------------

\clearpage
\begin{longrotatetable}
\begin{deluxetable*}{llcccccccc}
\tablewidth{0pc}
\tablecaption{Young Objects with Non-Taurus Proper Motions\label{tab:pm_sigma}}
\tablehead{
\colhead{}                    &\colhead{}                    &\colhead{}                    &\multicolumn{3}{c}{Proper Motion}&\colhead{}                    &\colhead{}                    &\colhead{}                    &\colhead{}                     \\
\cline{4-6}
\multicolumn{1}{l}{Object}    &\multicolumn{1}{l}{2MASS/Literature Name}&\colhead{}                    &\colhead{$\mu_{\alpha}$cos$\delta$}&\colhead{$\mu_{\delta}$}      &\colhead{$\chi_{\nu}^{2}$}    &\colhead{}                    &\colhead{PM Discrepancy}      &\colhead{Spectral Type\tablenotemark{a}}&\colhead{SpT References\tablenotemark{b}} \\
\colhead{}                    &\colhead{}                    &\colhead{}                    &\colhead{(mas yr$^{-1}$)}     &\colhead{(mas yr$^{-1}$)}     &\colhead{}                    &\colhead{}                    &\colhead{}                    &\colhead{}                    &\colhead{}                     }
\startdata
\hline
\multicolumn{10}{c}{New Taurus Members} \\
\hline
PSO J059.5714+30.6327                              & 2MASS J03581711+3037586                            &  &  $21.61\pm2.55$       &  $-46.59\pm9.42$      &  $5.1$                 &  & 3.5$\sigma$           &  M9.7                   & 6                       \\
PSO J063.0534+32.7055                              & 2MASS J04121282+3242199                            &  &  $6.65\pm2.03$        &  $-1.52\pm3.17$       &  $23.5$                &  & 2.5$\sigma$           &  M6.8                   & 6                       \\
PSO J075.0118+18.4800                              & 2MASS J05000284+1828484                            &  &  $-4.64\pm2.10$       &  $-10.17\pm3.14$      &  $11.4$                &  & 2.5$\sigma$           &  M7.0                   & 6                       \\
PSO J075.9044+20.3854                              & 2MASS J05033708+2023076                            &  &  $-22.45\pm13.04$     &  $47.23\pm26.80$      &  $6.9$                 &  & 6.0$\sigma$           &  M6.1                   & 6                       \\
\hline
\multicolumn{10}{c}{New Pleiades Member} \\
\hline
PSO J058.8758+21.0194                              & 2MASS J03553018+2101102                            &  &  $12.04\pm7.31$       &  $-44.59\pm7.37$      &  $1.1$                 &  & 3.5$\sigma$           &  M9.2                   & 6                       \\
\hline
\multicolumn{10}{c}{Known Taurus Members} \\
\hline
2MASS J04194148+2716070                            & IRAS 04166+2708                                    &  &  $28.99\pm3.27$       &  $-19.35\pm2.81$      &  $5.9$                 &  & 4.0$\sigma$           &  $<$M0                  & 2                       \\
2MASS J04202583+2819237                            & IRAS 04173+2812                                    &  &  $25.17\pm7.15$       &  $-36.60\pm15.72$     &  $35.0$                &  & 2.5$\sigma$           &  M4                     & 3                       \\
2MASS J04215851+1520145                            &                                                    &  &  $-2.17\pm3.61$       &  $2.15\pm2.98$        &  $5.6$                 &  & 3.0$\sigma$           &  M5.4                   & 6                       \\
2MASS J04220069+2657324                            & Haro 6-5B                                          &  &  $-11.31\pm7.39$      &  $-12.80\pm5.44$      &  $7.0$                 &  & 2.5$\sigma$           &  K5                     & 1                       \\
2MASS J04275730+2619183                            & IRAS 04248+2612                                    &  &  $-6.44\pm5.39$       &  $8.08\pm5.92$        &  $20.6$                &  & 3.5$\sigma$           &  M5.7                   & 6                       \\
2MASS J04313407+1808049                            & L1551/IRS5                                         &  &  $-79.05\pm25.65$     &  $-28.49\pm12.64$     &  $39.2$                &  & 13.0$\sigma$          &  K0                     & 5                       \\
2MASS J04373705+2331080$^{\star}$                  &                                                    &  &  $15.86\pm13.52$      &  $-54.79\pm13.35$     &  $0.5$                 &  & 4.0$\sigma$           &  M9.3                   & 6                       \\
2MASS J04375670+2546229                            & ITG 1                                              &  &  $-4.61\pm1.59$       &  $7.56\pm3.40$        &  $4.9$                 &  & 3.5$\sigma$           &  ...                    & 2                       \\
2MASS J04380083+2558572$^{\star}$                  & ITG 2                                              &  &  $-9.03\pm5.27$       &  $-35.58\pm7.51$      &  $15.0$                &  & 2.5$\sigma$           &  M7.0                   & 6                       \\
2MASS J04380191+2519266                            &                                                    &  &  $29.58\pm10.00$      &  $-45.31\pm23.51$     &  $34.6$                &  & 2.5$\sigma$           &  M1                     & 5                       \\
2MASS J04382134+2609137                            &                                                    &  &  $21.82\pm2.93$       &  $-17.10\pm3.12$      &  $5.7$                 &  & 2.5$\sigma$           &  M6.1                   & 6                       \\
2MASS J04400067+2358211                            & GM Tau                                             &  &  $-7.76\pm4.03$       &  $-37.19\pm4.59$      &  $8.7$                 &  & 2.5$\sigma$           &  M7.1                   & 6                       \\
2MASS J04411078+2555116                            & ITG 34                                             &  &  $-11.83\pm7.35$      &  $-25.87\pm4.13$      &  $19.9$                &  & 2.5$\sigma$           &  M6.3                   & 6                       \\
2MASS J04591661+2840468                            &                                                    &  &  $3.06\pm3.26$        &  $3.06\pm2.31$        &  $1.1$                 &  & 3.0$\sigma$           &  ...                    & 4                       \\
\hline
\enddata
\tablecomments{PM discrepancy shows the difference in proper motions of these objects compared to the mean Taurus motion (Section~\ref{subsec:kine_criteria}; Figure~\ref{fig:PM_sigma} and \ref{fig:PM_highPM}). 2MASS~J04373705+2331080 and 2MASS~J04380083+2558572, with ``$\star$'' behind their PS1 names, are the only two objects in this list that were studied by \cite{Kraus+2017}. They were both assessed as confirmed Taurus members by \cite{Kraus+2017}.}
\tablenotetext{a}{We use our reddening-free spectral type (SpT$_{\omega}$), identified by their decimal spectral types, for our discoveries and reclassified known members, and adopt the literature integer spectral type for the remaining ones.}
\tablenotetext{b}{(1)~\cite{White+2004}; (2)~\cite{Luhman+2006}; (3)~\cite{Rebull+2010}; (4)~\cite{Esplin+2014}; (5)~\cite{Luhman+2017}; (6)~this work.}
\end{deluxetable*}
\end{longrotatetable}
\clearpage

\clearpage
\begin{deluxetable*}{lcccccccccc}
\tablewidth{0pc}
\tablecaption{New Members with Mid-Infrared Excesses\label{tab:IR_excess}}
\tablehead{
\multicolumn{1}{l}{Object}     & \colhead{SpT$_{\omega}$}       & \colhead{$A_{\rm V}$}          & \colhead{Gravity}              & \colhead{Model-derived Age}    & \colhead{$K_{\rm 2MASS} - W2$} & \colhead{$K_{\rm 2MASS} - W3$} & \colhead{$K_{\rm 2MASS} - W4$} & \colhead{$W2$ Excess}          & \colhead{$W3$ Excess}          & \colhead{$W4$ Excess}         \\
\multicolumn{1}{l}{}           & \colhead{}                     & \colhead{(mag)}                & \colhead{}                     & \colhead{(Myr)}                & \colhead{(mag)}                & \colhead{(mag)}                & \colhead{(mag)}                & \colhead{}                     & \colhead{}                     & \colhead{}                     }
\startdata
\hline
\multicolumn{11}{c}{Taurus} \\
\hline
PSO J059.3563+32.3043                               &  M6.6$\pm 0.9$  &  $-0.59\pm0.85$  &  \textsc{int-g}  &  $[10,30)$  &  $0.59\pm0.09$  &  $2.74\pm0.24$  &  $5.40\pm0.50$  &  N  &  $5.6$$\sigma$  &  ?\\
PSO J065.6900+15.1818                               &  M6.4$\pm 0.9$  &  $0.73\pm0.85$  &  \textsc{vl-g}  &  $[10,30)$  &  $0.55\pm0.10$  &  $<2.38$  &  $5.57\pm0.36$  &  N  &  ?  &  $12.2$$\sigma$\\
PSO J069.3827+22.8857                               &  M6.5$\pm 0.9$  &  $0.72\pm0.85$  &  \textsc{vl-g}  &  $[30,100)$  &  $0.75\pm0.10$  &  $<1.82$  &  $6.83\pm0.19$  &  $1.3$$\sigma$  &  ?  &  $29.6$$\sigma$\\
PSO J071.3189+31.6888                               &  M6.6$\pm 0.9$  &  $0.15\pm0.85$  &  \textsc{vl-g}  &  $<$1  &  $0.72\pm0.08$  &  $2.66\pm0.11$  &  $4.96\pm0.26$  &  $1.3$$\sigma$  &  $11.7$$\sigma$  &  $14.3$$\sigma$\\
PSO J079.3986+26.2455                               &  M7.6$\pm 0.9$  &  $3.43\pm0.85$  &  \textsc{vl-g}  &  $<$1  &  $1.18\pm0.07$  &  $3.39\pm0.09$  &  $5.80\pm0.10$  &  $7.0$$\sigma$  &  $21.7$$\sigma$  &  $47.3$$\sigma$\\
\hline
\multicolumn{11}{c}{Per~OB2} \\
\hline
PSO J060.5031+32.0075                               &  M6.5$\pm 0.9$  &  $-0.10\pm0.85$  &  \textsc{vl-g}  &  $<$1  &  $0.81\pm0.09$  &  $1.68\pm0.42$  &  $<5.45$  &  $2.3$$\sigma$  &  ?  &  ?\\
PSO J060.6881+30.2903                               &  M7.0$\pm 0.9$  &  $-0.13\pm0.85$  &  \textsc{vl-g}  &  $<$1  &  $1.01\pm0.09$  &  $2.17\pm0.37$  &  $5.66\pm0.34$  &  $4.0$$\sigma$  &  $2.1$$\sigma$  &  $12.8$$\sigma$\\
PSO J060.7891+31.5527                               &  M6.3$\pm 0.9$  &  $0.98\pm0.85$  &  \textsc{vl-g}  &  $<$1  &  $0.76\pm0.08$  &  $2.40\pm0.26$  &  $5.12\pm0.43$  &  $1.8$$\sigma$  &  $4.0$$\sigma$  &  ?\\
PSO J060.8968+31.7282                               &  M6.6$\pm 0.9$  &  $-0.07\pm0.85$  &  \textsc{vl-g}  &  $<$1  &  $1.06\pm0.08$  &  $2.78\pm0.17$  &  $<4.99$  &  $5.5$$\sigma$  &  $8.1$$\sigma$  &  ?\\
PSO J060.9401+32.9790$^{\tablenotemark{a}}$         &  M6.6$\pm 0.9$  &  $-0.82\pm0.85$  &  \textsc{vl-g}  &  $<$1  &  $0.68\pm0.08$  &  $2.42\pm0.17$  &  $5.17\pm0.33$  &  $0.7$$\sigma$  &  $5.9$$\sigma$  &  $11.9$$\sigma$\\
PSO J060.9954+32.9996$^{\tablenotemark{a}}$         &  M8.2$\pm 0.9$  &  $0.38\pm0.85$  &  \textsc{vl-g}  &  $<$1  &  $0.84\pm0.13$  &  $2.91\pm0.41$  &  $<6.42$  &  $0.9$$\sigma$  &  $3.5$$\sigma$  &  ?\\
\enddata
\tablecomments{Our new disk candidates in Taurus and Per~OB2 that show mid-infrared excesses with significance of $>2\sigma$ (also see Figure~\ref{fig:IR_excess_Taurus} and \ref{fig:IR_excess_Other}). We present the objects' reddening-free spectral types, extinctions, gravity classifications, and model-based ages. We show the significance of objects' mid-infrared excesses in columns of $W2/3/4$ Excess. We use ``N'' for colors that do not show mid-infrared excesses. We use ``?'' if the corresponding AllWISE photometry, i.e., $W1/W2/W3$, does not have good quality (quality of the $W3$ photometry is defined the same way as for $W1$ and $W2$, described in Section~\ref{subsec:phot_criteria}) or is not reliable based on our visual inspection of the $WISE$ images, as the object is not distinct from the noise. }
\tablenotetext{a}{The two Per~OB2 disk candidates that form a candidate very wide binary (Section~\ref{subsubsec:binary}). }
\end{deluxetable*}
\clearpage

\clearpage
\begin{deluxetable*}{crrrr}
\tablewidth{0pc}
\tablecaption{ $k_{z}^{\rm 0.05am}$ as a function of pwv \label{tab:k_Wz_005am}}
\tablehead{
\colhead{pwv}                 & \colhead{$k_{0}^{\rm 0.05am}$}& \colhead{$k_{D}^{\rm 0.05am}$}& \colhead{$k_{1}^{\rm 0.05am}$}& \colhead{$k_{2}^{\rm 0.05am}$}\\
\colhead{(mm)}                & \colhead{($\times 10^{-3}$)}  & \colhead{($\times 10^{-3}$)}  & \colhead{($\times 10^{-3}$)}  & \colhead{($\times 10^{-3}$)}  }
\startdata
1.0       &$1.3$&$8.8$&$3.3$&$1.54$\\
1.6       &$1.8$&$9.3$&$4.0$&$1.53$\\
3.0       &$2.6$&$10.3$&$5.2$&$1.47$\\
5.0       &$3.3$&$11.2$&$6.1$&$1.32$\\
\enddata
\tablecomments{$k_{z}^{\rm 0.05am}$ is the change in $W_{z}$ when the airmass differs by 0.05 between the science target and telluric standard. The $k_{z}^{\rm 0.05am}$ that corresponds to a pwv value between $1.0$~mm and $5.0$~mm can be derived by interpolation. We provide more decimals of $k_{2}^{\rm 0.05am}$ to avoid the same values under pwv$= 1.0$, $1.6$, $3.0$~mm.}
\end{deluxetable*}
\clearpage

\clearpage
\begin{deluxetable*}{crrrr}
\tablewidth{0pc}
\tablecaption{ $k_{z}^{\rm 0.1pwv}$ as a function of airmass \label{tab:k_Wz_01mm}}
\tablehead{
\colhead{airmass}             & \colhead{$k_{0}^{\rm 0.1pwv}$}& \colhead{$k_{D}^{\rm 0.1pwv}$}& \colhead{$k_{1}^{\rm 0.1pwv}$}& \colhead{$k_{2}^{\rm 0.1pwv}$}\\
\colhead{}                    & \colhead{($\times 10^{-4}$)}  & \colhead{($\times 10^{-4}$)}  & \colhead{($\times 10^{-4}$)}  & \colhead{($\times 10^{-4}$)}  }
\startdata
1.0       &$19.6$&$49.1$&$43.3$&$0.4$\\
1.5       &$25.2$&$55.5$&$51.0$&$-0.6$\\
2.0       &$29.7$&$61.0$&$57.1$&$-0.7$\\
\enddata
\tablecomments{$k_{z}^{\rm 0.1pwv}$ is the change in $W_{z}$ when pwv differs by $0.1$~mm between the science target and telluric standard. The $k_{z}^{\rm 0.1pwv}$ that corresponds to an airmass value between $1.0$ and $2.0$ can be derived by interpolation.}
\end{deluxetable*}
\clearpage

\vfill
\eject
\end{document}